\documentclass[12pt,letterpaper]{article}

\usepackage[margin=1in]{geometry}
\usepackage{amsthm, amsfonts}

\usepackage{newtxtext}
\usepackage{newtxmath}

\usepackage{setspace}
\usepackage{float}
\usepackage{graphicx}
\usepackage{bbm}
\usepackage{url}
\usepackage{caption}
\usepackage{subcaption}
\usepackage{natbib}
\usepackage{verbatim}
\usepackage{mathrsfs}

\input{mymacros.sty}

\begin{document}

\begin{center}

{\Large\bfseries
FireGen: Quantifying Wildfire Risk by Simulation
}

\vspace{0.25cm}

{\bfseries
Allyson Hineman\footnote{Department of Applied Mathematics,
University of Colorado Boulder.
Author e-mail: \texttt{allyson.hineman@colorado.edu}},
William Kleiber\footnote{Department of Applied Mathematics,
University of Colorado Boulder.
Author e-mail: \texttt{william.kleiber@colorado.edu}},
Stephan Sain\footnote{Jupiter Intelligence.
Author e-mail: \texttt{steve.sain@jupiterintel.com}},
and Alexis Hoffman\footnote{Jupiter Intelligence.
Author e-mail: \texttt{alexis.hoffman@jupiterintel.com}}
}

\vspace{0.25cm}

{\bfseries July 20, 2026}

\end{center}


\begin{abstract}
Wildfire regimes in Northern California exhibit strong spatio-temporal variability, heavy-tailed fire size distributions, and sensitivity to climatic conditions. 
We develop a hierarchical statistical framework to model wildfire occurrence, geometry, and burned area in Northern California from 1984-2023. 
Fire centroids are modeled as a spatio-temporal point process with covariate-driven intensity, extended by a multiplicative gamma-based stochastic shock to better capture the distribution of monthly fire counts. 
Conditional on fire location, burned area polygons are represented using a parametric ellipse model that separates scale, shape, and orientation, with ellipse parameters estimated from standardized fire perimeters. 
Total burned area is modeled using a heteroskedastic lognormal regression incorporating climatic and spatial covariates, including vapor pressure deficit anomalies. 
The framework enables Monte Carlo simulation of wildfire processes and provides a probabilistic tool for assessing wildfire risk and cumulative burned area under observed climate variability. 

\bigskip
\noindent
{\sc Keywords: Wildfire modeling; spatio-temporal point processes; hierarchical models; ellipse approximation; lognormal regression; Monte Carlo simulation} 
\end{abstract}

\section{Introduction}  \label{sec:intro}

Wildfires are a major ecological and societal hazard throughout many regions of the world, with particularly severe impacts in the western United States. 
In California, wildfire activity has increased substantially in recent decades, contributing to widespread ecological damage, infrastructure loss, smoke-related air quality issues, and threats to human health and safety \citep{westerling2006warming, abatzoglou2016impact, williams2019observed}.
The Intergovernmental Panel on Climate Change (IPCC) and the U.S. National Climate Assessment identify wildfire as a key climate-sensitive hazard in western North America, with increasing fire weather and expanding exposure jointly amplifying risk \citep{ipcc2021ar6, nca52023}.

A substantial body of work attributes increasing wildfire activity to warming temperatures, increased atmospheric moisture demand, prolonged drought, declining snowpack, and shifts in vegetation and fuel aridity \citep{abatzoglou2016impact, holden2018decreasing, williams2019observed}. These changes have lengthened fire seasons and increased fire-conducive conditions, while continued expansion of the wildland--urban interface has increased human exposure to wildfire hazards \citep{dong2022fireseason, radeloff2018rapid}. As a result, both wildfire hazard and societal exposure are increasing across much of the western United States. 

Wildfire behavior emerges from complex interactions among climate, fuels, vegetation, topography, ignition processes, and human activity across multiple spatial and temporal scales \citep{moritz2005wildfires, littell2009climate}. Consistent with this complexity, fire size distributions are highly skewed and heavy-tailed, with a small number of large fires accounting for a disproportionate share of total burned area \citep{strauss1989one, malamud1998forest}. 
This inherent variability makes probabilistic methods well suited for wildfire risk assessment \citep{reed2002from}, highlighting that aggregate wildfire impacts are dominated by extreme events.

\subsection{Mechanistic and Statistical Wildfire Modeling}

Mechanistic approaches to wildfire modeling explicitly simulate the processes governing wildfire spread. 
These models typically integrate fuels, weather, and topographic information to represent fire behavior \citep{finney2006flammap}. 
Mechanistic modeling encompasses physical, empirical, and stochastic representations of fire dynamics \citep{sullivan2009wildland1, sullivan2009wildland2, sullivan2009wildland3}.
Recent developments have incorporated probabilistic representations of uncertainty through Bayesian and stochastic fire spread models \citep{yoo2024levelset, grieshop2024cellular}.
Related frameworks such as FSim address large-scale risk assessment by simulating thousands of hypothetical fire seasons, but similarly require extensive environmental data and computational resources \citep{finney2011fsim}.

Statistical models have played a central role in large-scale wildfire analysis by relating wildfire occurrence and burned area to climatic and environmental drivers.
Such approaches have demonstrated strong links between climate variability, anthropogenic warming, and wildfire activity across diverse ecosystems \citep{preisler2004statistical, littell2009climate, turco2018exacerbated}.
These studies show that statistical models can effectively capture broad climate-fire relationships without explicitly resolving fire spread dynamics. 

More recently, machine learning and deep learning methods have been applied to wildfire prediction and climate-fire interaction modeling, including work by \citet{richards2023insights} on extreme Mediterranean wildfires. 
However, many such approaches prioritize predictive accuracy over generative consistency and are not designed to produce coherent stochastic realizations of wildfire events or full probabilistic representations of uncertainty.

A complementary statistical framework for wildfire occurrence is based on spatial and spatio-temporal point processes. 
Inhomogeneous Poisson point process models allow wildfire intensity to vary continuously over space and time as a function of environmental and climatic drivers \citep{diggle2013statistical, baddeley2015spatial}, and such models have been widely applied to wildfire ignition and occurrence data \citep{hering2009modeling, serra2014spatio}. 
Notably, wildfire occurrence often exhibits clustering and variability beyond standard Poisson assumptions, so stochastic extensions such as Cox processes, log-Gaussian Cox processes, and hierarchical Bayesian models have been developed to represent latent environmental variability and multiscale dependence \citep{cox1955some, moller1998log, diggle2013lgcp, wikle2003hierarchical, banerjee2014hierarchical}. 

Separate statistical models have also been developed for wildfire burned area distributions. Due to the inherent variability, uncertainty, and complex behavior of wildfire growth, burned area is commonly modeled using statistical distributions with heavy tails to capture the extreme skewness of observed fire sizes \citep{ricotta1999flaming, reed2002power, schoenberg2003distribution, cumming2001parametric, joseph2019spatiotemporal}. 
Further work shows that incorporating climate variability and drought indices improves explanation of regional fire size dynamics \citep{barbero2014climate, wang2020projected}. In particular, vapor pressure deficit has emerged as a key driver linking atmospheric dryness to both ignition probability and fire growth potential \citep{abatzoglou2016impact, williams2019observed}. Despite these advances, wildfire occurrence and burned area are often modeled separately rather than within a unified generative framework.

In addition, relatively few statistical wildfire models explicitly generate spatial burned area polygons suitable for simulation-based risk analysis. 
Many instead reduce fire outcomes to scalar summaries such as counts or total burned area, limiting their usefulness for exposure and impact modeling. 
Simulation frameworks that generate ensembles of full wildfire realizations provide a natural alternative, preserving spatial dependence while enabling Monte Carlo estimation of downstream risk \citep{williams2026wulffss}.

\subsection{Overview of This Study}

This study focuses on {\em generative} models of wildfire characteristics, including locations, burned area and burn pattern geometry.  
This should be contrasted against {\em predictive} approaches that focus on wildfire spread that can be used in real-time forecasting. 
Our goal is to capture distributional behavior of fire that can be used in offline risk assessment efforts. 
Wildfire occurrence is modeled as a spatio-temporal point process whose intensity depends on environmental covariates and seasonal forcing. 
To account for extreme wildfire months beyond what is captured by deterministic climate covariates alone, we introduce a multiplicative stochastic shock process that varies over time. 
Conditional on wildfire occurrence, burned area polygons are represented through a simplified geometric approximation in which fire perimeters are modeled as randomly-oriented ellipses with stochastic area and shape parameters. 
This representation separates wildfire geometry into interpretable components, including orientation, elongation, and scale, while remaining computationally tractable for large-scale simulation. 

The framework is illustrated using wildfire perimeter data together with climate covariates. It supports direct simulation of complete wildfire realizations under observed or future climate conditions, enabling probabilistic characterization of wildfire risk and its associated uncertainty over time.
Our approach complements recent stochastic simulators such as WULFFSS \citep{williams2026wulffss}, which simulate wildfire activity at a monthly grid-cell level, by modeling individual wildfire events and jointly generating their occurrence, burned area, and spatial geometry within a unified hierarchical statistical model.
By generating ensembles of plausible wildfire realizations, our model captures both typical wildfire behavior and rare high-impact events while preserving event-level spatial structure relevant for risk assessment and impact modeling.

The remainder of this paper is organized as follows.  
Section \ref{sec:data} introduces the study region and wildfire and other datasets used in the analysis. 
Section \ref{sec:methods} presents the hierarchical statistical framework for wildfire occurrence, burned area, and geometric fire representation together with estimation procedures. 
Section \ref{sec:results} evaluates model performance and illustrates simulation behavior across Northern California. 
Finally, Section \ref{sec:discussion} concludes with discussion of the framework, limitations, and directions for future work.

\section{Data}  \label{sec:data}

This section describes the study region, wildfire dataset, and covariate data used in our analysis.

\subsection{Study Region}  
\label{subsec:study_region}
Our analysis focuses on a portion of Northern California bounded by longitudes $-124.5^o$ to $-120^o$ and latitudes $38.3^o$ to $42^o$. This bounding box encompasses a diversity of landscapes, including coastal ranges, interior valleys, and forested mountain regions, capturing the range of conditions under which wildfires occur in the region. 

\begin{figure}[htbp]
    \centering
    \includegraphics[width=0.8\textwidth]{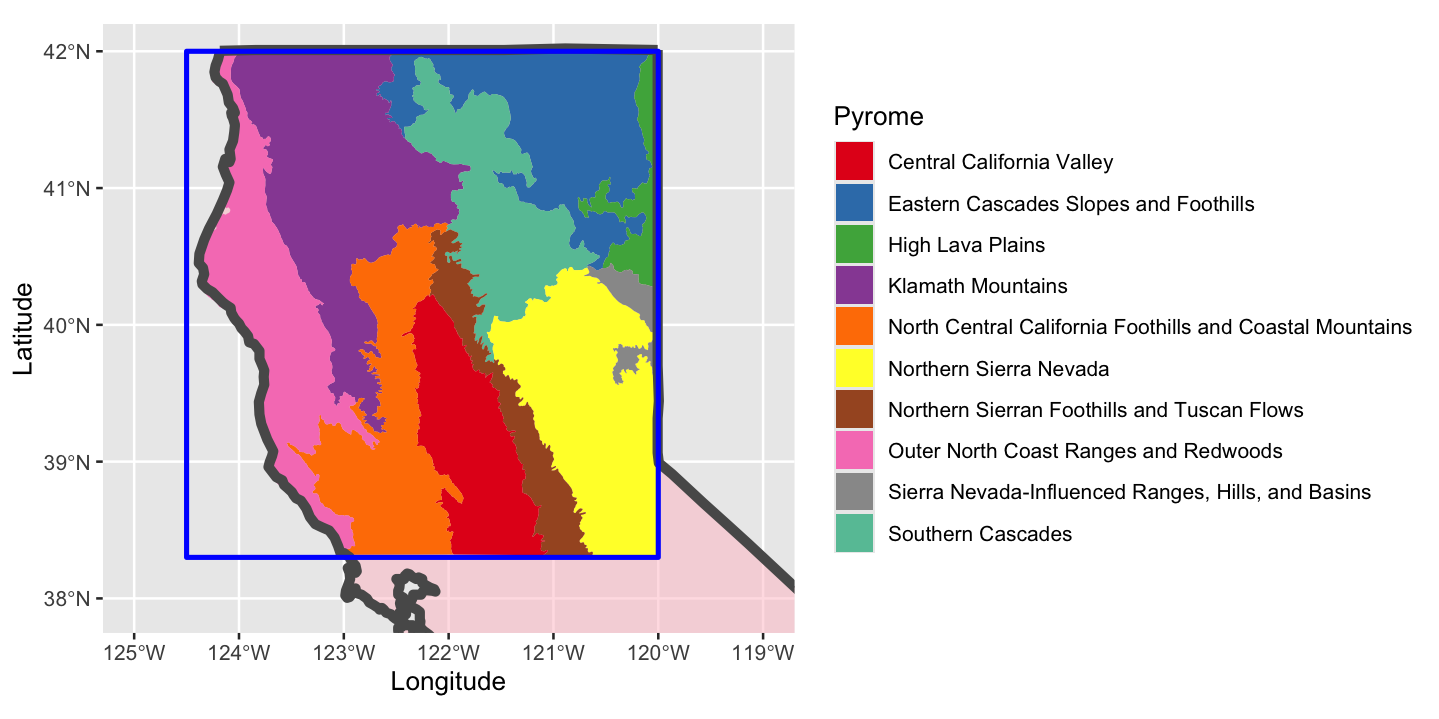}
    \caption{Pyrome boundaries within the Northern California study region.}
    \label{fig:norcal_pyromes}
\end{figure}

To contextualize fire occurrence within ecological and fire management zones, we intersected the bounding box with pyrome boundaries, as shown in Figure \ref{fig:norcal_pyromes}. 
Pyromes are spatial regions defined by similar historical wildfire regimes, based on observed patterns of fire frequency, size, and behavior \citep{short2020pyromes}. 
Only the portions of each pyrome that fall within the study region are retained. 
Fires are assigned to a pyrome if the centroid of the burned area polygon falls within the corresponding pyrome region intersecting the study bounding box. 

\subsection{MTBS}  

Wildfire occurrence and burned area data are obtained from the Monitoring Trends in Burn Severity (MTBS) dataset, which provides spatially explicit fire perimeter polygons and ignition dates for large fires across the United States \citep{mtbs_burned_areas_2025}. 
In the western U.S., MTBS defines a fire as an event burning at least 1,000 acres (approximately 405 ha), while in the eastern U.S., the threshold is 500 acres (approximately 202 ha). 

For this study, we restrict fires to the Northern California subregion defined in Section \ref{subsec:study_region} during 1984--2023.
Only fires classified by MTBS as \textit{wildfire} or \textit{unknown} are retained; \textit{prescribed burns} and \textit{wildland fire use} incidents are excluded to focus on unplanned wildfire events. 
Wildfire characteristics include ignition date, burned area and location defined by the polygon's centroid.
 
\subsection{ERA5}  

Climate covariates are obtained from the ECMWF Reanalysis 5th Generation ERA5 product, a global atmospheric dataset produced by the European Centre for Medium-Range Weather Forecasts \citep{ERA5}.  
Monthly fields of mean temperature, maximum temperature, wind speed, total precipitation, mean and minimum relative humidity, standardized precipitation evapotranspiration index (SPEI), and vapor pressure deficit (VPD) are extracted at $0.25^\circ \times 0.25^\circ$ spatial resolution and aggregated to each pyrome by averaging grid-cell values within the pyrome boundary. 

For each climate variable, monthly climatologies are calculated by averaging values from 1983--2024 within each calendar month. 
Climate anomalies are then defined as departures from these monthly climatologies. In particular, VPD anomalies represent deviations from the long-term monthly mean, providing a measure of anomalous atmospheric moisture demand. To account for delayed wildfire responses, lagged climate metrics are also computed as averages over the preceding one-, two-, and three-month periods.

To assess sensitivity to different representations of atmospheric moisture demand, additional candidate predictors include multiple summaries of monthly VPD, including the mean, minimum, maximum, median, and 25th and 75th percentile values.

\subsection{Elevation}  

Elevation is included as a spatial covariate to capture topographic influences on wildfire occurrence and behavior. Higher or lower elevations can affect vegetation types, fuel availability, and microclimate conditions, all of which may influence fire risk and spread. 
Gridded elevation data are obtained from the MERIT digital elevation model and resampled to 100m spatial resolution \citep{yamazaki2017merit}. 


\section{Methods}   \label{sec:methods}

In this section, we detail a statistical model for wildfire burned areas, a proposed estimation procedure, and how the model can be used to simulate burned area polygons under different climate conditions. 
The model includes distinct components for fire occurrence (location) and fire size (burned area).  
We refer to `burned area' as the spatial extent of a wildfire event, represented as a polygon in geospatial data. 

\subsection{Model}  

We propose a hierarchical approach to modeling wildfire burned area polygons, consistent with general hierarchical spatial modeling frameworks \citep{wikle2003hierarchical, banerjee2014hierarchical}.  
At the first level, we model the spatio-temporal distribution of fire event locations represented by the centroids of burned area polygons. 
For each event, we assume the resulting burned area polygon is approximately elliptical, a simplifying assumption sometimes used in wildfire spread modeling \citep{anderson1983predicting}. 
Conditional on time and location, the second level specifies a simple geometric model for the shape of the burned area polygon consistent with this elliptical assumption, and the final layer simulates the total area burned.

\subsubsection{Fire Location Model}  

The first level of the hierarchical model concentrates on the location of fires. 
We represent wildfire occurrences as a spatio-temporal point pattern observed over a spatial domain $\mathcal{S} \subset \real^2$ and a temporal interval $\mathcal{T} \subset \real$, where $\real$ denotes the set of real numbers and $\real_+ = (0,\infty)$ the positive real line. Point process models are widely used for wildfire occurrence data and related environmental applications \citep{diggle2013statistical, baddeley2015spatial}.
Points are interpreted as the centroid (i.e., the geometric center) of a burned area polygon. 
Let $X = \{(\bs_i,t_i)\}_{i=1}^N$ denote the points within $\mathcal{S} \times \mathcal{T}$ where $\bs_i \in \mathcal{S}$ is the spatial location of the $i$-th wildfire centroid, $t_i \in \mathcal{T}$ denotes the time of occurrence, and $N$ is the total number of wildfires within $\mathcal{S} \times \mathcal{T}$.

We start with a baseline model for for the space-time point process $X$, assuming it is an inhomogeneous Poisson point process, characterized by an intensity function $\lambda(\bs,t):\mathcal{S}\times\mathcal{T}\rightarrow\real_+$ \citep{diggle2013statistical}. The intensity function defines the expected rate of wildfire occurrence across space and time, with higher values indicating regions and periods with greater expected fire activity.

The baseline intensity is specified using a log-linear form 
\begin{equation*}
    \lambda(\bs,t) = \exp \left( 
    \beta_0 
    + \sum_{j=1}^J \beta_{Z,j} \, Z_j(\bs) 
    + \sum_{k=1}^K \beta_{Y,k} \, Y_k(t) 
    \right) 
\end{equation*}
where $\{Z_j(\bs)\}_{j=1}^J$ represent spatial covariates defined on $\mathcal{S}$, $\{Y_k(t)\}_{k=1}^K$ represent temporal covariates defined on $\mathcal{T}$, and $\beta_0, \{\beta_{Z,j}\}_{j=1}^J,$ and $\{\beta_{Y,k}\}_{k=1}^K$ are parameters to be estimated. 
The log-linear specification is standard in spatio-temporal point process modeling \citep{diggle2013statistical} and includes a separable intensity function with distinct spatial and temporal components. 
The spatial covariates control relative variation across space while temporal covariates scale the spatial intensity surface over time. 
Although nonseparable intensity functions can be imagined, we find a separable model works well for our example below.

While the baseline model allows for spatial and temporal variation in the mean event rate, it maintains the Poisson property that variance equals the mean conditional on the intensity. 
Introducing stochastic behavior in the intensity surface is a common approach, which are broadly referred to as Cox models \citep{cox1955some}. 
We consider a simple extension of the baseline model specific to our application by introducing a multiplicative stochastic shock term to the intensity surface that varies in time. 
Specifically, the full model has intensity function $U_t \, \lambda(\bs,t)$, where 
\begin{equation}
    U_t = G_t^\gamma, 
    \qquad 
    G_t \sim \mathrm{Gamma}(\alpha, \alpha),
    \label{eq:shock}
\end{equation}
with rate $\alpha > 0$  controlling the variability in fire activity over time and $\gamma \ge 0$ controlling how strongly the shock influences the intensity. 
We model the variables $\{U_t | t \in \mathcal{T}\}$ as independent over the index $t$. 
Conditional on $U_t$, the process remains Poisson, but 
this 
relaxes the strict mean--variance relationship implied by the baseline model (see Supplement Section S1). Such excess temporal variability and departures from simple Poisson assumptions are commonly observed in wildfire occurrence data \citep{littell2009climate, turco2018exacerbated}. The shock term can therefore accommodate temporal fluctuations in wildfire activity that are not explained by the observed covariates alone. 

The scientific motivation for introducing the time-varying (but not spatially-varying) stochastic shock term is that the climate variables in our application are monthly averages, whereas wildfire occurrence is often influenced by higher-frequency local weather variation and other unobserved factors. Consequently, $U_t$ may be interpreted as a latent temporal adjustment capturing residual variability in wildfire activity beyond that explained by the observed temporal covariates $\{Y_k(t)\}$. 

\subsubsection{Shape and Area Model}  

Conditional on the location of fire occurrence, we turn to modeling burned area and polygon shape. 
Burned area polygons exhibit complicated edge patterns that depend on local land characteristics and firefighter response behavior. 
Mechanistic fire spread models such as FARSITE are widely used to simulate these dynamics in physically based settings \citep{finney1998farsite, sullivan2009wildland1, sullivan2009wildland2, sullivan2009wildland3}

In this work we consider a simplified representation of burned area polygons that aims to capture the area and general orientation of burns. 
Specifically, we approximate the complex polygonal geometries using parametric ellipses capturing key geometric features including orientation, eccentricity, and scale, following classical wind-driven fire shape approximations \citep{anderson1983predicting}. 
To this end, we develop separate models for the area and shape of the burned area. 

An ellipse can be parameterized geometrically by its semi-major axis $a>0$, semi-minor axis $b>0$ with $a>b$, centroid $(x_0,y_0)$, and rotational angle $\phi\in[0,\pi)$, where $\phi$ denotes the angle between the semi-major axis and the positive $x$ axis. 
In this parameterization, the area of an ellipse is $A = \pi a b$, so scaling the axes directly affects area, while axis ratio and orientation (rotation, center, and elongation) determine the shape. Equivalently, area can be cleanly separated from the shape and elongation of the ellipse. 
To separate shape from scale, we fit ellipses to unit-area standardized polygons, yielding {\em standardized} axes $a',b'$ satisfying $\pi a' b' = 1$. Under this constraint, $b'$ is uniquely determined by $b' = 1/(\pi a')$, so the shape (up to rotation) can be summarized by the single axis parameter $a'$, with orientation captured separately by $\phi$. 

We therefore model the triplet $(\phi, a', A)$, representing orientation, shape, and scale, respectively. The full ellipse is obtained by rescaling the unit-area ellipse to area $A$, which implies semi-axes $a = a' \sqrt{A}$ and $b = b' \sqrt{A}$. Further, we assume each triplet $(\phi, a', A)$ is independent across fires. 

Under the constraint $\pi a'b' = 1$, we develop a stochastic model for the (standardized) semi-major axis $a'$ and rotational angle $\phi$. 
We propose using a shifted gamma distribution for $a'$, with space-time constant parameters. 
Given the unit area constraint and the ordering $a' \geq b'$, $a'$ is bounded below by $L = 1/\sqrt{\pi}$, so we model the shifted variable $a'- L$ with a gamma distribution. 
This formulation respects the geometric lower bound, ensures positive axis lengths, and allows for right-skewed variability in ellipse elongation. 
The rotational angle $\phi$ summarizes the predominant orientation of each fire. 
We propose the simple model $\phi \sim U(0,\pi)$, which works well for our data domain, although more flexible Beta models can be envisioned.

The total burned area $A$ is modeled as a spatio-temporal random variable indexed by spatial unit (pyrome) $p$ and time $t$. To account for the strong right-skew typically observed in burned area, we assume that the shifted response follows a lognormal model:
\[
    \log(A(p,t) - c) \sim \mathcal{N}(\mu(p,t), \sigma^2(p))
\]
where $c$ is a positive constant shift applied prior to the transformation, $\mu(p,t)$ is the conditional mean structure, and $\sigma^2(p)$ is a group-specific residual variance parameter. 
This formulation aligns with distributional approaches emphasizing heavy-tailed wildfire size behavior \citep{reed2002from, reed2002power, joseph2019spatiotemporal}. 
The mean structure is specified as:
\[
    \mu(p,t) = \eta_0 + \sum_{k=1}^K \eta_k Y_k(t) + \omega(p),
\]
where $Y_k(t)$ denotes a set of temporally varying covariates, $\omega(p)$ represents spatially varying group effects, and $\boldsymbol{\eta} = (\eta_0, \eta_1, \dots, \eta_K)^\top$ is the vector of regression coefficients. 

This formulation separates temporal and spatial sources of variability in the burned area process. Temporal variation is represented through time-dependent covariates in the mean structure, allowing the expected burned area to evolve systematically in response to seasonal and climatic forcing. Spatial heterogeneity is incorporated through group-specific mean effects and through heteroskedastic residual variances $\sigma^2(p)$, permitting different regions to exhibit distinct levels of unexplained variability. 
The resulting model provides a stochastic representation of burned area that captures systematic spatio-temporal structure while preserving residual stochastic variability.

\subsection{Estimation}  

The parameters to estimate are those associated with the point process model and the shape model. 
In particular, assume data are available with fire centroid locations $X = \{(\bs_i,t_i)\}_{i=1}^n$, and associated burned area polygons. 

\subsubsection{Point Process Model Estimation} \label{sec:point_process_estimation}

We fit the model separately for each pyrome \citep{banerjee2014hierarchical, cressie1993statistics}. 

Recall that there are two competing point process models -- the baseline model as an inhomogeneous Poisson process, and the full model that includes a stochastic shock amplification to the baseline's intensity function. 
Both baseline and full models begin with estimating parameters controlling the space--time intensity function $\lambda(\bs,t)$ by maximum likelihood. 
Let $\mathcal{W} \subseteq \mathcal{S} \times \mathcal{T}$ denote the spatio-temporal observation window, then the corresponding log-likelihood function for $\lambda(s,t)$ takes the standard form \citep{diggle2013statistical, baddeley2015spatial}: 
\[
    \ell(\boldsymbol{\beta}) = \sum_{i=1}^n \log \lambda(\bs_i,t_i) - \int_{\mathcal{W}} \lambda(\bs,t) \, {\rm d}\bs\, {\rm d}t
\]
where $\boldsymbol{\beta}$ includes all regression parameters of the log intensity surface, $\lambda$. All continuous covariates are standardized prior to estimation by subtracting their sample mean and dividing by their sample standard deviation. This scaling improves numerical stability and allows coefficient magnitudes to be compared on a common scale.

The integral term in the likelihood does not have a closed-form expression; we approximate it numerically using a fine-pixel quadrature scheme, in which $\mathcal{W}$ is discretized into small space-time pixels of equal area $q$ and the intensity is evaluated at pixel centroids \citep{moller1998log, baddeley2015spatial}. 
In particular, 
\[
    \ell(\boldsymbol{\beta}) \approx \sum_{i=1}^M [ n_i \log \lambda(\bs_i,t_i) - q \, \lambda(\bs_i,t_i) ]
\]
where $n_i$ is the number of data points falling in the $i$-th pixel (either 0 or 1) and $M$ is the number of pixels. 
The resulting approximate log-likelihood can be maximized using a L-BFGS quasi-Newton optimization algorithm \citep{liu1989limited, nocedal2006numerical}.

The full model includes a temporal shock process, whose parameters must also be estimated. 
The genesis of including such a shock process is to capture fire distribution features for which the standard Poisson process framework is inadequate (e.g., excess variability and frequent large counts), consistent with Cox process formulations \citep{cox1955some, moller1998log, diggle2013lgcp}. 
In particular, the baseline Poisson model imposes a rigid coupling between the mean and variance of event counts, so that increased variability cannot be accommodated without distorting the mean intensity.

Introducing a stochastic multiplicative shock relaxes this constraint by allowing the intensity to fluctuate over time. 
This decouples the mean and variance by attributing part of the dispersion to latent changes in intensity rather than the deterministic covariate-driven structure. 
However, these distributional qualities are difficult to investigate analytically due to the latent intensity, motivating a simulation-based framework conditional on $\hat{\boldsymbol{\beta}}$.

For a particular $(\alpha,\gamma)$ pair in the stochastic shock model \eqref{eq:shock}, we propose simulating many space-time trajectories of fire counts over $\mathcal{S} \times \mathcal{T}$. 
Then, calculate a weighted composite goodness-of-fit score defined as 
\[
  GOF_{\text{total}}(\alpha,\gamma) = w_1 \Tilde{D}_Q + w_2 \tilde{D}_C + w_3 \Tilde{D}_0 + w_4 \Tilde{D}_1 + w_5 \Tilde{D}_{max}
\]
where $\Tilde{D}$ denotes a rescaled version of each discrepancy statistic to place all components on a comparable scale. 
In particular, $D_Q$ is rescaled by its mean over the parameter grid, while $D_C, D_0, D_1,$ and $D_{\max}$ are rescaled using the observed total count, observed proportions, and observed maximum count, respectively. 
The estimates of $(\alpha,\gamma)$ are those that minimize $GOF_{\text{total}}(\alpha,\gamma)$.  
Due to sampling variability, we recommend calculating the composite goodness-of-fit over a grid of plausible parameter pairs, $\{(\alpha_i,\gamma_i)\}_{i=1}^N$ from which a minimizing value can be approximated by fitting a thin plate spline to  $\{GOF_{\text{total}}(\alpha_i,\gamma_i)\}_{i=1}^N$. 
This provides a computationally efficient approach to optimization by avoiding an exhaustive search over a dense parameter grid.  

The component error statistics include a weighted quantile distance, $D_Q$, which compares the empirical quantiles of the simulated and observed counts, following extreme-value calibration approaches \citep{reed2002from, reed2002power, joseph2019spatiotemporal}. 
For quantiles $p_k \in \{0.25, 0.5, 0.75, 0.90, 0.95, 0.99\}$, the discrepancy is 
\[
  D_Q = \sum_{k=1}^6 q_k (Q_{obs}(p_k) - Q_{sim}(p_k))^2, 
\]
where the weights $q_k = \{1,1,2,3,4,7\}$ emphasize the upper quantiles, giving additional importance to extreme events. 
To ensure the overall magnitude of fire activity is captured, the total count error is defined as 
\[
  D_C = \bigg| \sum_t N_{sim, t} - \sum_t N_{obs, t} \bigg|, 
\]
where $N_{sim,t}$ and $N_{obs,t}$ denote the simulated and observed counts over time. 
To capture the frequency of low activity time periods, we define the proportions of time with exactly $k$ fires in the observed and simulated data as 
\[
    \text{PropObs}(k) = \frac{1}{T}\sum_{t=1}^T \indicator\{ N_{\text{obs},t} = k \}, \qquad 
    \text{PropSim}(k) = \frac{1}{T}\sum_{t=1}^T \indicator\{ N_{\text{sim},t} = k \}
\]
where $N_{\text{obs},t}$ and $N_{\text{sim},t}$ denote the observed and simulated number of wildfires in time period $t$, respectively, and $\indicator\{\cdot\}$ is the indicator function. The absolute discrepancies in these proportions are 
\[
  D_0 = |\text{PropSim}(0) - \text{PropObs}(0)|, \qquad D_1 = |\text{PropSim}(1) - \text{PropObs}(1)|.
\]
Finally, to account for months with unusually high wildfire activity, we also include the maximum count error  
\[
  D_{max} = \bigg| \max_t N_{sim,t} - \max_t N_{obs, t} \bigg|.
\]

Each discrepancy metric is included in the goodness-of-fit score, motivated by diagnostic comparisons between simulated and observed fire-count distributions. Initial calibration based solely on weighted quantile distance produces simulations that inadequately reproduce other important distributional characteristics, including total counts, sparsity, and tail behavior. Additional discrepancy metrics are incorporated sequentially until the simulated distributions provide a better representation of the observed distribution across extreme-event frequency, central tendency, sparsity, and tail behavior \citep{malamud1998forest, reed2002power, schoenberg2003distribution}. 

We treat the weights $(w_1, \dots, w_5)$ in $GOF_{\text{total}}(\alpha,\gamma)$ as spatially-dependent hyperparameters. 
To propose different weight vectors per spatial region, we use a space-filling design, producing a diverse set of candidate weight vectors. 
For each candidate vector, we evaluate model performance using 4-fold cross-validation across decades: for each fold, we simulate counts over a grid of plausible $(\alpha, \gamma)$ values and compute the normalized discrepancy components. 
These components are combined according to the candidate weights to yield a fold-specific fit score. 
Aggregating the mean total fire count errors across folds and across multiple data subsets, we have a grid of plausible parameters to which we fit a thin plate spline and select the weight vector that minimizes the average discrepancy in total fire counts. 
These methods are consistent with spatio-temporal validation practices \citep{wikle2003hierarchical}. 

\subsubsection{Burned Area Extent Model Estimation} \label{subsec:burn.estimation}

Estimation of the burned area extent parameters, those governing the ellipse and the fire area model, proceeds in three main steps: polygon standardization, empirical ellipse fitting, followed by estimation of model parameters including those of the random ellipse and total burned area. 
These components are then combined to simulate elliptical burned area polygons that preserve the statistical structure of fire shape and size. 

\paragraph{Polygon Standardization.}

To separate shape from size, each fire polygon is first standardized to unit area and centered at the origin. 
For a polygon with area $A$ and coordinates $\mathbf{C}$ which is a set of spatial coordinate pairs defining the polygon boundary, the standardized polygon is 
\[
  \Tilde{\mathbf{C}} = \frac{\mathbf{C} - \mathbf{c}}{\sqrt{A}}.
\]
where $\mathbf{c} = (c_x,c_y)$ is the centroid of the polygon. For a simple polygon with vertices $(x_i,y_i)$, $i = 1,\dots,n$, the centroid is 
\[
    c_x = \frac{1}{6A} \sum_{i=1}^n (x_i + x_{i+1})(x_i y_{i+1} - x_{i+1} y_i), \qquad 
    c_y = \frac{1}{6A} \sum_{i=1}^n (y_i + y_{i+1})(x_i y_{i+1} - x_{i+1} y_i)
\]
with $(x_{n+1}, y_{n+1}) = (x_1,y_1)$. Some burned area polygons consist of multiple disjoint parts. For such multipolygons (sets of one or more polygons represented as as single feature), the centroid is computed as the area-weighted average of the centroids of the component polygons.  

Scaling and translation are applied to every point, ensuring that the centroid is at the origin and the polygon has area 1, following classical spatial geometry methods \citep{cressie1993statistics}. 

For the purpose of ellipse fitting, we treat each fire polygon as a single object, applying the standardization and fitting procedure to all parts collectively, and ultimately only fitting one ellipse per fire regardless of the number of multipolygon components. This ensures that even complex geometries are summarized by a single parametric shape while preserving the overall spatial extent and orientation of the fire.  

\paragraph{Empirical Estimation of Ellipses.}

Each standardized polygon is approximated by an ellipse using a constrained nonlinear least-squares optimization procedure implemented via an augmented Lagrangian method. This is consistent with nonlinear constrained optimization methods \citep{nocedal2006numerical}. 

Although each polygon is first centered at the origin during standardization, the optimization still includes the centroid location $(x_0,y_0)$ as free parameters to improve the fit to irregular boundary geometries and to reduce residual mismatch after scaling. The semi-axes are both estimated subject to a unit-area constraint, $a' b' - 1/\pi = 0$, which is enforced during optimization using an equality-constrained augmented Lagrangian formulation.

The optimization is performed over the parameter vector $\btheta = (x_0, y_0, a', b', \phi)$, representing the standardized ellipse centroid, semi-axis lengths, and rotational angle, respectively. This ensures that all fitted ellipses preserve unit area while still allowing variation in shape and orientation. 

Let $\mathbf{C} = \{(x_i,y_i)\}_{i=1}^{N_C}$ denote the coordinates of the standardized polygon boundary points. 
Polygon points are first translated and rotated into the ellipse-aligned coordinate system: 
\[
  \begin{pmatrix} x_{\text{rot},i} \\ y_{\text{rot,i}} \end{pmatrix}
  =
  \begin{pmatrix}
  \cos \phi & \sin \phi \\
      -\sin \phi & \cos \phi
      \end{pmatrix}
      \begin{pmatrix}
      x_i - x \\ y_i - y
      \end{pmatrix}
\]
for $i=1,\ldots,N_C$.  
Residuals are defined as 
\[
  r_i = \left(\frac{x_{\text{rot}, i}}{a}\right)^2 + \left(\frac{y_{\text{rot}, i}}{b}\right)^2 - 1.
\]
so that a perfect elliptical polygon yields zero residuals. The objective function is to minimize the sum of squared residuals, 
\[
  f(\btheta) = \sum_{i=1}^{N_C} r_i^2 .
\]
Optimization is performed using a gradient-based augmented Lagrangian method in \texttt{R} via the \texttt{nloptr} package \citep{nocedal2006numerical}, with analytic gradients supplied to improve convergence (see supplement Section S2). 

In cases where the initial ellipse fitting procedure fails to converge, we apply a single retry step to improve numerical stability. Failures typically occur for geometrically complex or high-vertex-count polygons that produce unstable optimization gradients. The polygon is first simplified using \texttt{st\_simplify} in \texttt{R}'s \texttt{sf} package, after which the constrained nonlinear least squares fit is rerun. This additional pass resolves previously failed fits without further iteration.

\paragraph{Estimation of Model Parameters.}

We treat the empirical estimates of 
$\{\hat{a}^\prime_i\}$
from all wildfires from the prior step as independent samples from a shifted gamma distribution, whose parameters are estimated by maximum likelihood.

The burned-area model parameters, including regression coefficients $\boldsymbol{\eta}$, group-specific mean effects $\omega(p)$, and heteroskedastic residual variance parameters $\sigma^2(p)$, are estimated by maximum likelihood methods within the generalized least squares framework, treating observed fire areas as conditionally independent given the covariates and variance structure.  

Because the burned area model is specified using a lognormal distribution, some simulated realizations produce unrealistically large areas in the upper tail. 
To prevent simulated fires from exceeding the spatial extent of the modeling domain, burned area is capped above by the total area of the pyrome in which the fire occurred. 
In addition, simulated burned area polygons are cropped to intersect pyrome boundaries, ensuring that simulated perimeters remained spatially consistent with the environmental regions used to construct the model. 
We view these constraints as ecologically reasonable and necessary for maintaining realistic large-fire behavior in simulation.

\section{Results}   \label{sec:results}

In this section we detail the specific modeling choices for the northern California data and illustrate performance on each component of the model hierarchy. 

\subsection{Point Process Model Feature Selection}

Modeling pyromes as independent allows baseline wildfire intensity and covariate effects to vary across fire regimes, while avoiding the assumption of a single homogeneous process over the entire study region. 

Table \ref{tab:pyrome_summary} below summarizes wildfire occurence across the ten pyromes during the 1984--2023 study period, and lists pyrome abbreviations used throughout the paper. 
The summaries show substantial regional variation in average fire activity and extreme monthly counts, with highly active pyromes such as KM contrasting with lower-activity regions such as SNIRHB and CCV. These differences motivate the inclusion of spatial structure in the intensity model.

The intensity model includes two distinct classes of covariates. 
Spatial covariates defined over the domain, $\mathcal{S}$, capture geographic variation in wildfire occurrence and include latitude, longitude, their interaction and quadratic terms, and elevation. 
These covariates vary across space but are constant in time. We also consider incorporating land use and land cover classifications as spatial covariates. However, these variables are ultimately excluded from the final model because they do not substantially improve predictive performance and introduce several interpretability challenges. In particular, wildfire activity occasionally occurs in seasonally wet areas, leading to misleading associations between water-related classes and fire occurrence. Additionally, while forested regions account for a large proportion of wildfire activity in Northern California overall, a simple indicator for forest cover is not consistently informative across pyromes, especially in less forested regions where other environmental and spatial factors play a larger role. 

Temporal covariates, defined over the temporal domain $\mathcal{T}$, include large-scale climate conditions that vary from month to month and drive the intensity surface up and down. 
In addition, monthly harmonic terms specified as cyclical sine and cosine functions of month encode the smooth annual fire seasonality and capture recurring within-year temporal patterns in wildfire activity. 
Monthly climate variables are spatially averaged over the pyrome domain to produce a single value per month, reflecting region-wide atmospheric and moisture conditions (for example). 

\begin{figure}[htbp]
    \centering    \includegraphics[width=0.8\textwidth]{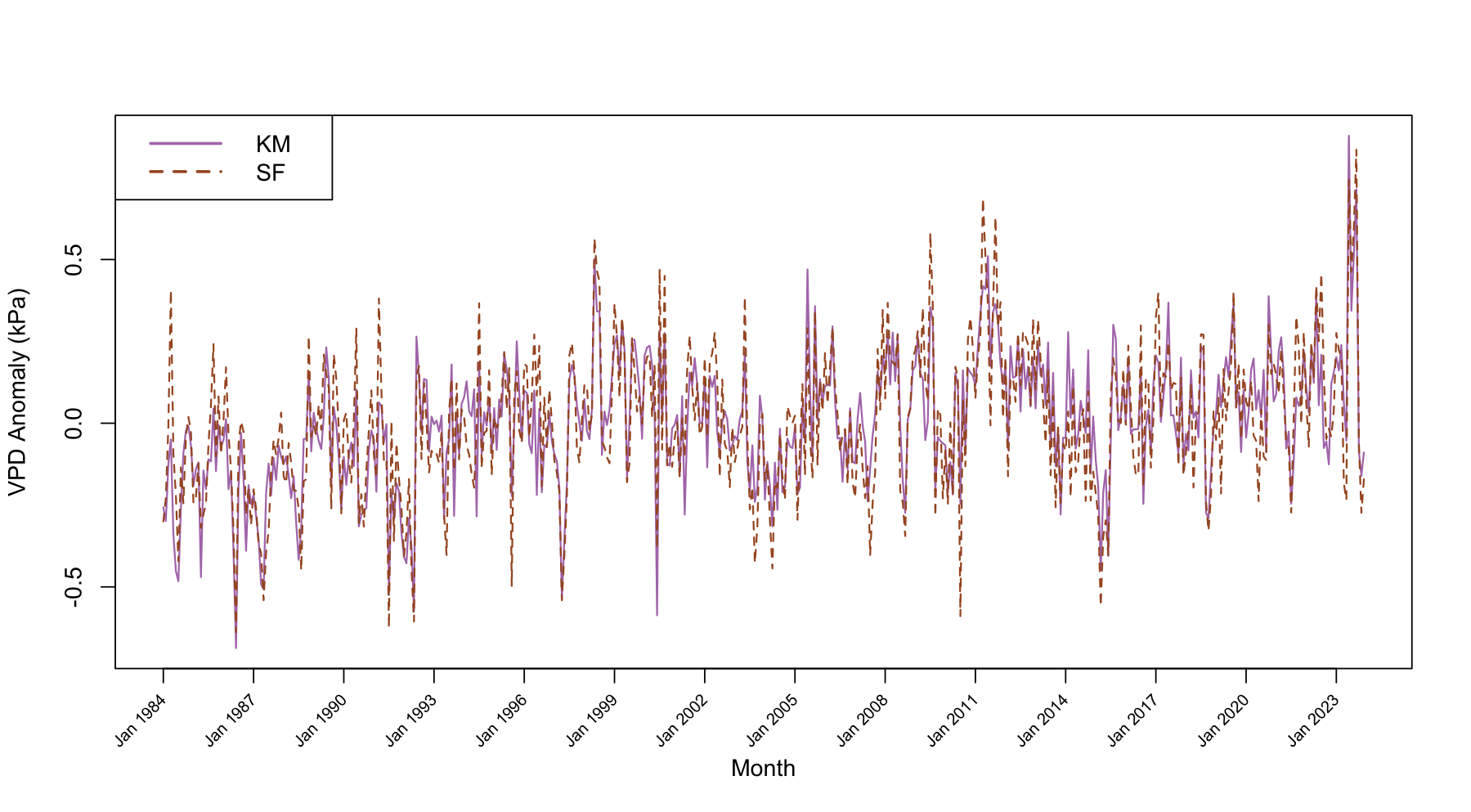}
    \caption{Monthly VPDa in KM (solid purple) and SF (dashed brown).}
    \label{fig:avg_vpd}
\end{figure}

Numerous ERA5-derived climate variables are considered, including measures of temperature, precipitation, wind speed, humidity, vapor pressure deficit, and drought conditions, together with associated climatologies, anomalies, and lagged values. Many of these variables capture related physical processes and exhibit strong collinearity. 
Our exploratory analysis of various candidate climate covariates, paired with literature precedent, identified average monthly vapor pressure deficit anomaly (VPDa) as a useful feature for the model \citep{abatzoglou2016impact, williams2019observed}.
Figure \ref{fig:avg_vpd} shows monthly VPDa for two representative pyromes: KM and SF. Both series exhibit substantial interannual variability and a gradual upward trend, indicating increasingly anomalous atmospheric dryness over the study period relative to the historical monthly climatology. 

VPD captures the effect of current atmospheric dryness on wildfire likelihood and is computed from daily ERA5 temperature $T$ ($^\circ C$) and relative humidity $RH$ (\%) data using 
\[
    VPDa = e_s(T)\bigg(  1 - \frac{RH}{100}\bigg)
\]
where $e_s(T)$ is the saturation vapor pressure computed using the Magnus--Tetens approximation:
\[
e_s(T) = 0.6108 \exp\left(\frac{17.27T}{T + 237.3}\right),
\]
with $e_s(T)$ in kPa. VPD values are then averaged to the monthly scale. VPDa are defined as deviations from the long-term monthly climatological mean, representing departures from the typical seasonal cycle.

In the following sections, we highlight the Klamath Mountains (KM) and Northern Sierran Foothills and Tuscan Flows (SF) pyromes as examples of regions exhibiting contrasting wildfire activity patterns. Results for all other pyromes are in the supplement, Section S5.

\subsection{Point Process Model Illustration} 

Estimation proceeds in two stages. 
First, parameters controlling the baseline space--time intensity function $\lambda(s,t)$ are estimated under an inhomogeneous Poisson process model described in Section \ref{sec:point_process_estimation}. 
Table \ref{tab:results_table} reports maximum likelihood estimates, standard errors, and Wald statistics for the KM pyrome. 

All covariates are standardized prior to fitting allowing effect sizes to be compared on a common scale. 
Spatial coefficient estimates represent the multiplicative change in baseline intensity from a one standard deviation increase in a covariate, given by $\exp(\hat{\beta})$, while temporal effects similarly scale intensity across all locations at a given time. 

The fitted baseline model for the KM pyrome identifies several significant spatial and climatic drivers of wildfire occurrence intensity. 
Monthly VPDa exhibits a positive and statistically significant association with wildfire occurrence, conditional on the other covariates in the model, indicating that anomalously dry atmospheric conditions correspond to increased wildfire activity. 
Elevation is negatively associated with wildfire occurrence intensity, suggesting reduced fire activity at higher elevations. 
Significant longitude and quadratic longitude effects also indicate persistent large-scale spatial gradients across the region. 

The harmonic seasonal covariates are highly significant, indicating strong annual wildfire seasonality throughout all pyromes in Northern California. 
The sine and cosine terms jointly capture smooth within-year temporal variation, representing the primary wildfire season from approximately June through October or November. 

\begin{figure}[htbp]
    \centering
    \includegraphics[width=0.7\textwidth]{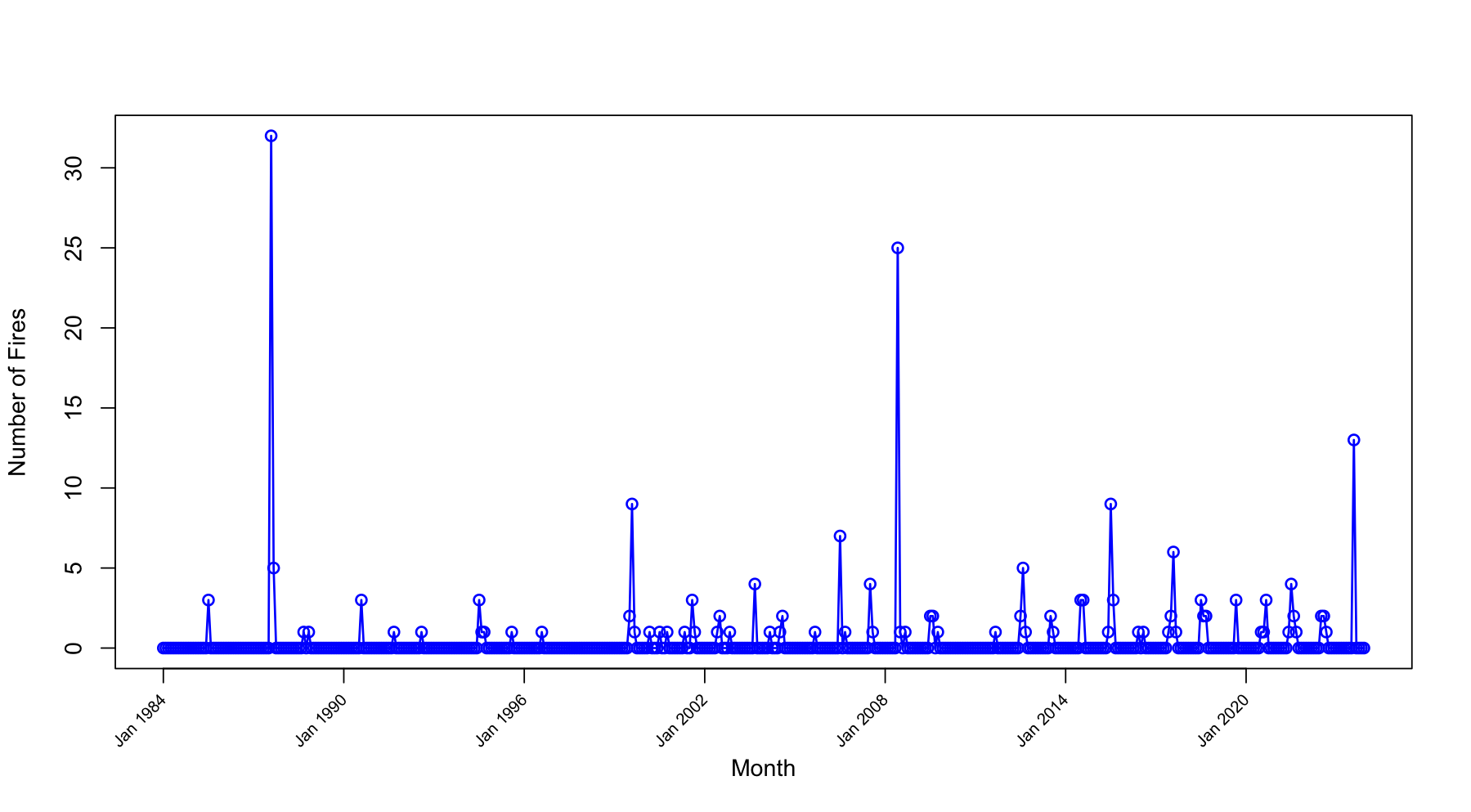}
    
    \caption{Monthly wildfire counts in the KM pyrome, 1984--2023.}
    
    \label{fig:monthlyCounts}
\end{figure}

Despite capturing broad spatial and seasonal structure, the baseline model does not fully reproduce the observed distribution of monthly wildfire counts displayed in Figure \ref{fig:monthlyCounts}.
In particular, months of unusually high wildfire counts and periods of unusually low activity occur more frequently than expected under a deterministic Poisson intensity surface where the count variance equals the mean. 
The stochastic shock term addresses this limitation by allowing the intensity to fluctuate randomly through time while retaining the estimated spatial and seasonal structure encoded in $\lambda(\bs,t)$. 

Across pyromes, VPDa estimates are generally positive and physically interpretable, reflecting increased wildfire likelihood under anomalously dry atmospheric conditions. 
However, several pyromes (NSN, SF, NCCFACM, and SNIRHB) produce negative but generally non-significant VPDa coefficients under the baseline specification, likely reflecting residual collinearity, sparse wildfire occurrence, and limitations of the deterministic Poisson intensity surface in capturing localized temporal variability. 
These inconsistencies further motivate the inclusion of the stochastic shock process in the second-stage model. 

In the second stage of estimation, a multiplicative shock term $U_t$ is introduced to capture overdispersion beyond the Poisson baseline. 
Months with unusually high counts relative to baseline intensity are common, and the shock term allows additional variance while retaining $\lambda(s,t)$ as the baseline structure. 

Parameters $(\alpha, \gamma)$ are estimated using simulation-based inference described in Section \ref{sec:point_process_estimation}.
Model performance is evaluated using a composite goodness-of-fit score based on distributional summaries of observed monthly counts.
The GOF components address Poisson limitations: $D_Q$ emphasizes upper quantiles to capture high-fire months, $D_C$ preserves total fire magnitude, $D_0$ and $D_1$ capture low-activity periods, and $D_{\max}$ ensures accurate reproduction of extreme counts. 
Together, these metrics evaluate both typical and extreme fire patterns in simulated point patterns.

Conditional on the optimal weights per pyrome (Supplement Section S3), we perform a grid search over shock parameters $(\alpha, \gamma)$ using $\alpha \in [0.1,1]$ and $\gamma \in [0,1]$ with increments of 0.1. For every parameter combination, 200 simulations of the spatio-temporal point process are generated by applying independent gamma-distributed multiplicative shocks to the baseline intensity field \citep{cox1955some, diggle2013lgcp, wikle2003hierarchical}. 
We fit a thin plate spline to the grid evaluations and identify the optimum using the resulting smooth surface, as described in Section \ref{sec:point_process_estimation}.
Table \ref{tab:shock_params} reports the estimates of $(\alpha,\gamma)$ per pyrome. 

While $\alpha > 0$ is required for a valid model, all estimated values are also less than 1. 
Since $\mathrm{Var}(G_t) = \frac{1}{\alpha}$, smaller values of $\alpha$ correspond to increased dispersion in the latent gamma shock process for each pyrome. 
In this sense, $\alpha$ controls the variability of the latent multiplicative environment that drives departures from a standard Poisson process. 

Likewise, all estimated $\gamma$ values satisfy $0 < \gamma < 1$, indicating a dampening transformation of the latent shock that reduces the influence of extreme $G_t$ values while preserving stochastic temporal variation.

Consequently, $\alpha$ determines the variability of the latent gamma process, while $\gamma$ controls how strongly fluctuations in the latent process are transmitted to the wildfire intensity surface. 
Together, these parameters determine the magnitude of temporal variability and the degree to which realized wildfire activity departs from the deterministic baseline intensity. 

Estimated shock parameters vary across pyromes as shown in Table \ref{tab:shock_params}, reflecting differing wildfire regimes and environmental conditions. ONCRR exhibits the smallest $\alpha$ estimate, indicating strong temporal variability in the latent intensity process, consistent with the intermittently extreme  fire behavior observed in coastal environments. In contrast, HLP has a larger $\alpha$ estimate, suggesting a comparatively stable baseline intensity over time. KM and EC exhibit larger $\gamma$ values, implying weaker dampening of extreme shocks. These regions are characterized by highly seasonal fire regimes and occasional extreme fire years associated with hot and dry atmospheric conditions. 

\begin{figure}[htbp]
    \centering
    \begin{subfigure}[b]{0.45\textwidth}
        \centering
        \includegraphics[width=\textwidth]{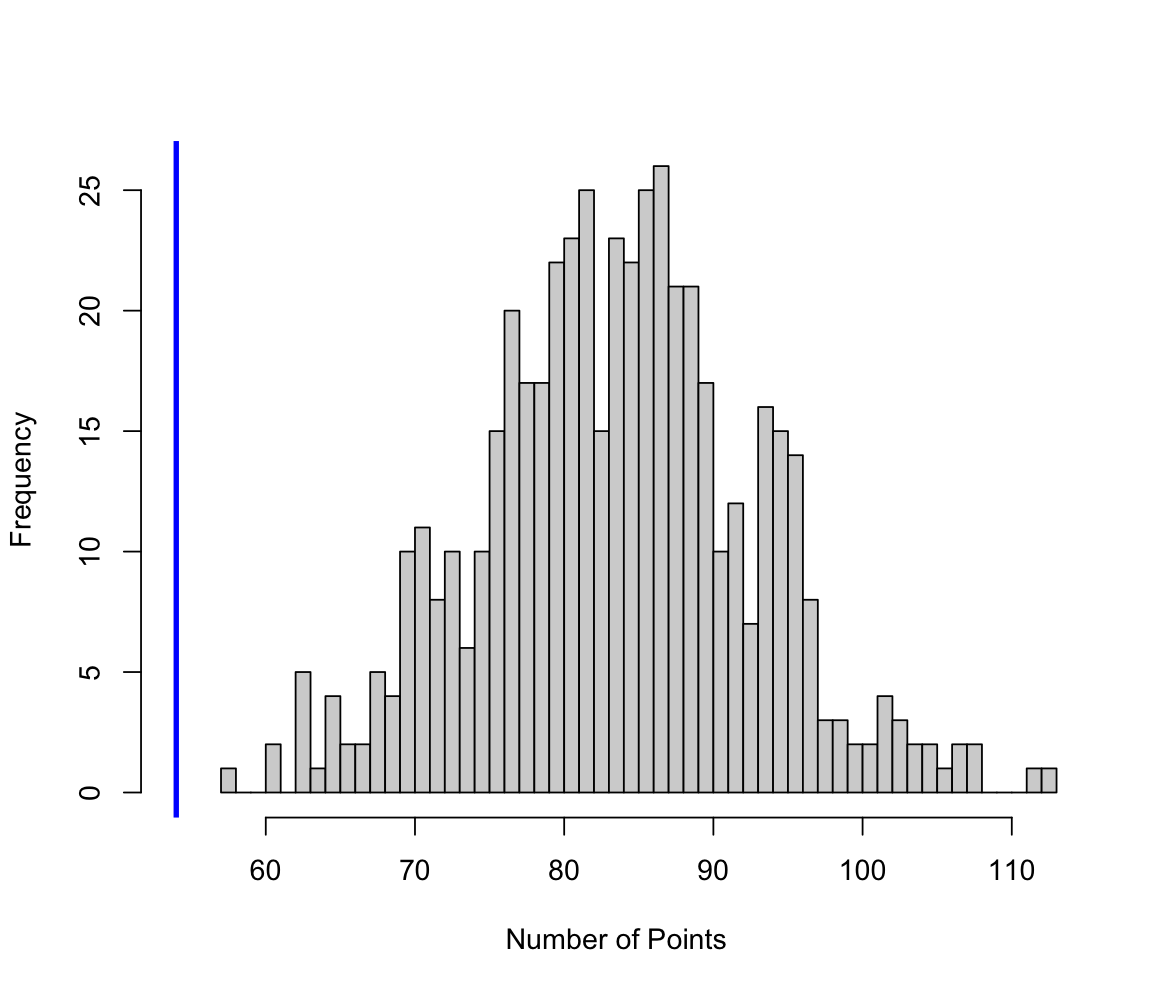}
        \caption{}
        \label{fig:no_shock}
    \end{subfigure}
    \hfill
    \begin{subfigure}[b]{0.45\textwidth}
        \centering
        \includegraphics[width=\textwidth]{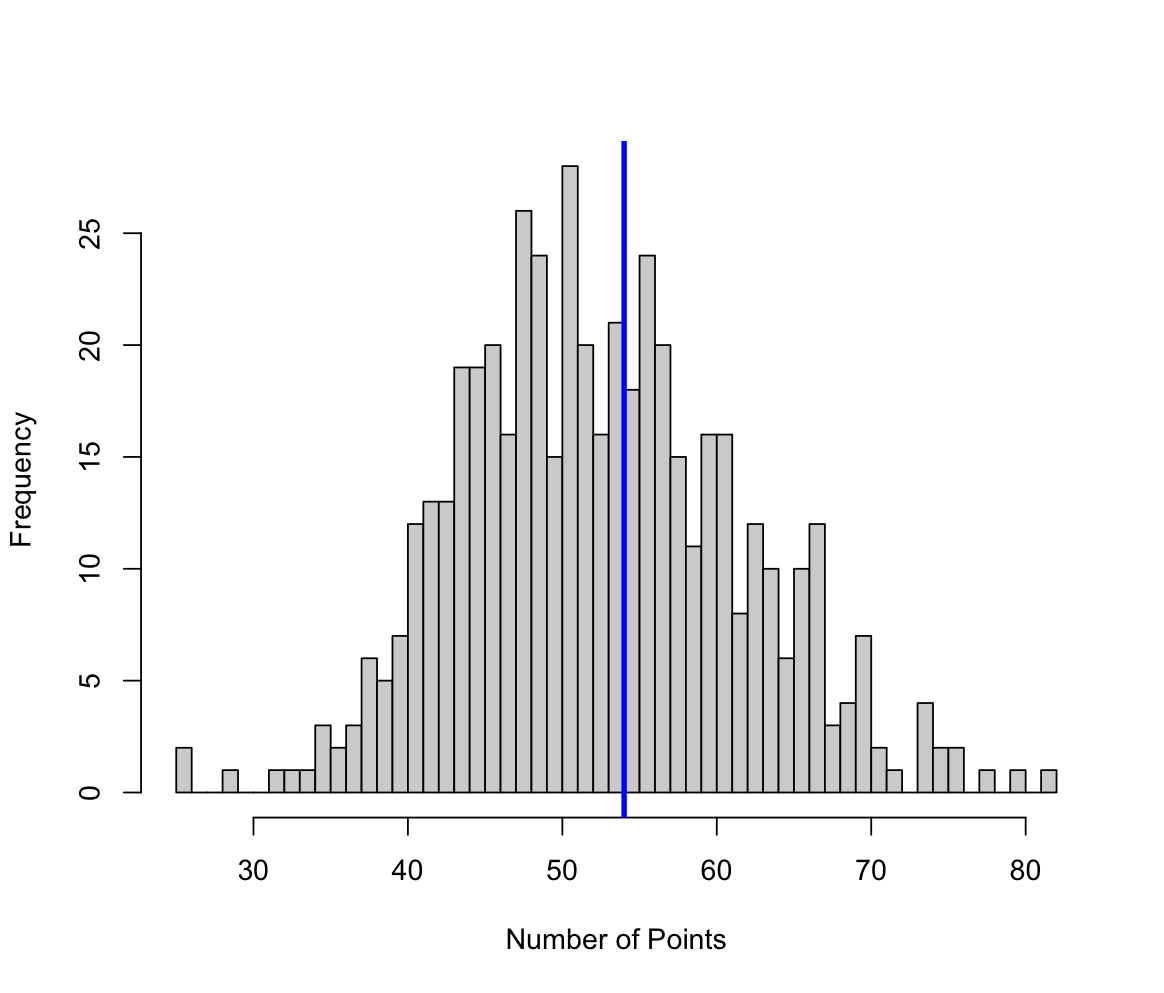}
        \caption{}
        \label{fig:with_shock}
    \end{subfigure}

    \caption{SF pyrome observed (blue line) and simulated monthly fire counts (histograms from 500 realizations) over 1984--2023. (a) Baseline model $\lambda$; (b) Full model $U_t \, \lambda$.}

    \label{fig:monthly_counts}
\end{figure}

Importantly, the shock term substantially improves the model's ability to reproduce the observed distribution of monthly fire counts, particularly in capturing total event counts that are systematically overestimated by the baseline Poisson formulation. 
This improvement is illustrated in Figure \ref{fig:monthly_counts}, where simulations with the shock term better match the observed number of points. 

Additionally, the spread of the simulated distribution remains comparable between panels (a) and (b), even though the full model produces substantially fewer fires on average. 
This illustrates a key advantage of the shock term: it allows the model to reproduce observed variability without requiring the elevated mean counts implied by a standard Poisson formulation. 
In other words, the random shock relaxes the rigid mean--variance relationship of the Poisson model by introducing an additional source of temporal variability. 

\begin{figure}[htbp]
    \centering
    \includegraphics[width=0.45\textwidth]{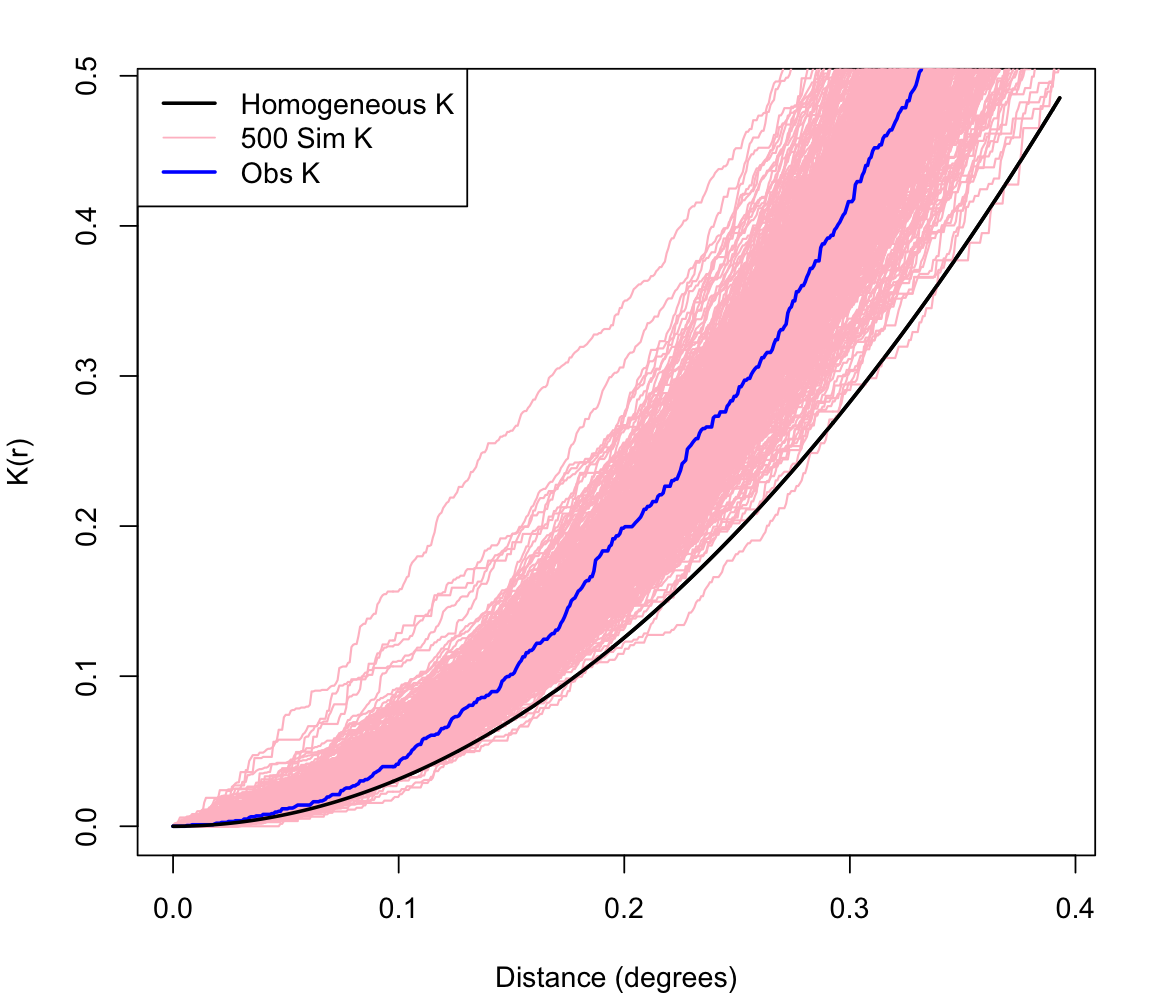}
    \caption{Ripley’s K function validation for the SF pyrome.}
    \label{fig:ripleys_k}
\end{figure}

To assess whether the final fitted model $U_t\, \lambda$ captures the observed spatial clustering of wildfire centroids, we computed Ripley's K function for each pyrome \citep{baddeley2015spatial}. 
Observed K functions are compared against simulation envelopes generated from 500 realizations under the full fitted model $U_t \, \lambda$ (Figure \ref{fig:ripleys_k}). 
Across all pyromes, the observed K functions lie within the simulation envelopes, indicating that the model adequately reproduces the observed spatial clustering structure of the wildfire centroids. 

\begin{figure}[htbp]
    \centering
    \includegraphics[width=0.75\textwidth]{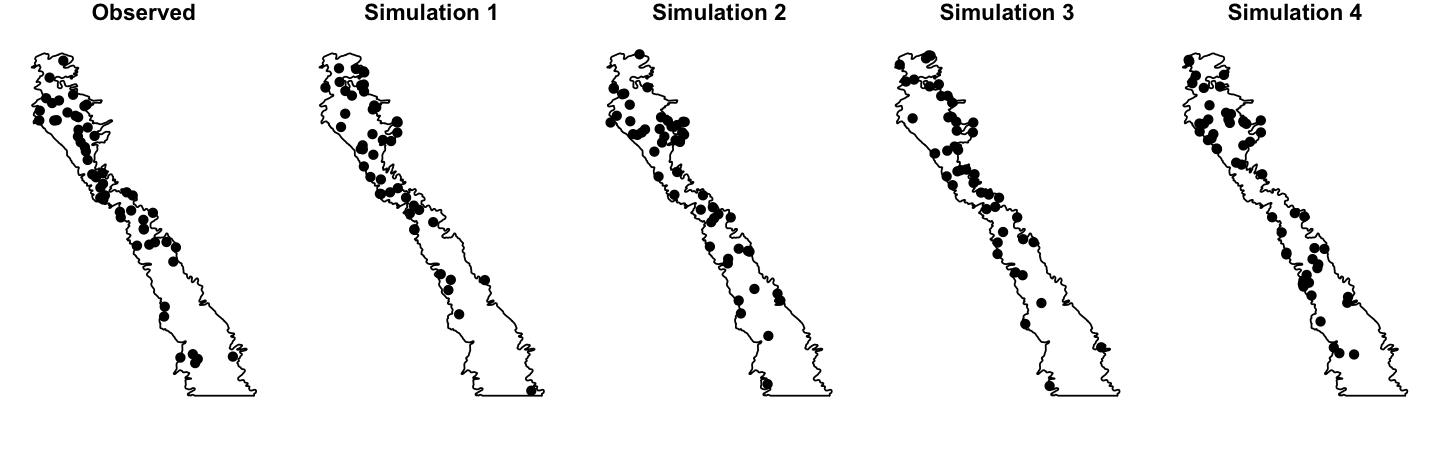}
    \caption{Observed and simulated wildfire point patterns for the SF pyrome under the fitted model with intensity $U_t \, \lambda$.}
    \label{fig:sf_point_process_sims}
\end{figure}

Figure \ref{fig:sf_point_process_sims} shows the 40-year observed point pattern of the data from the SF pyrome followed by 4 simulations from the full model. 
Across realizations, the simulations reproduce both the observed clustering structure and the overall magnitude of point counts, suggesting that the fitted intensity surface captures key first- and second-order properties of the data. 

\subsection{Burned Extent Model Illustration}

\begin{figure}[htbp]
    \centering
    \includegraphics[width=0.65\textwidth]{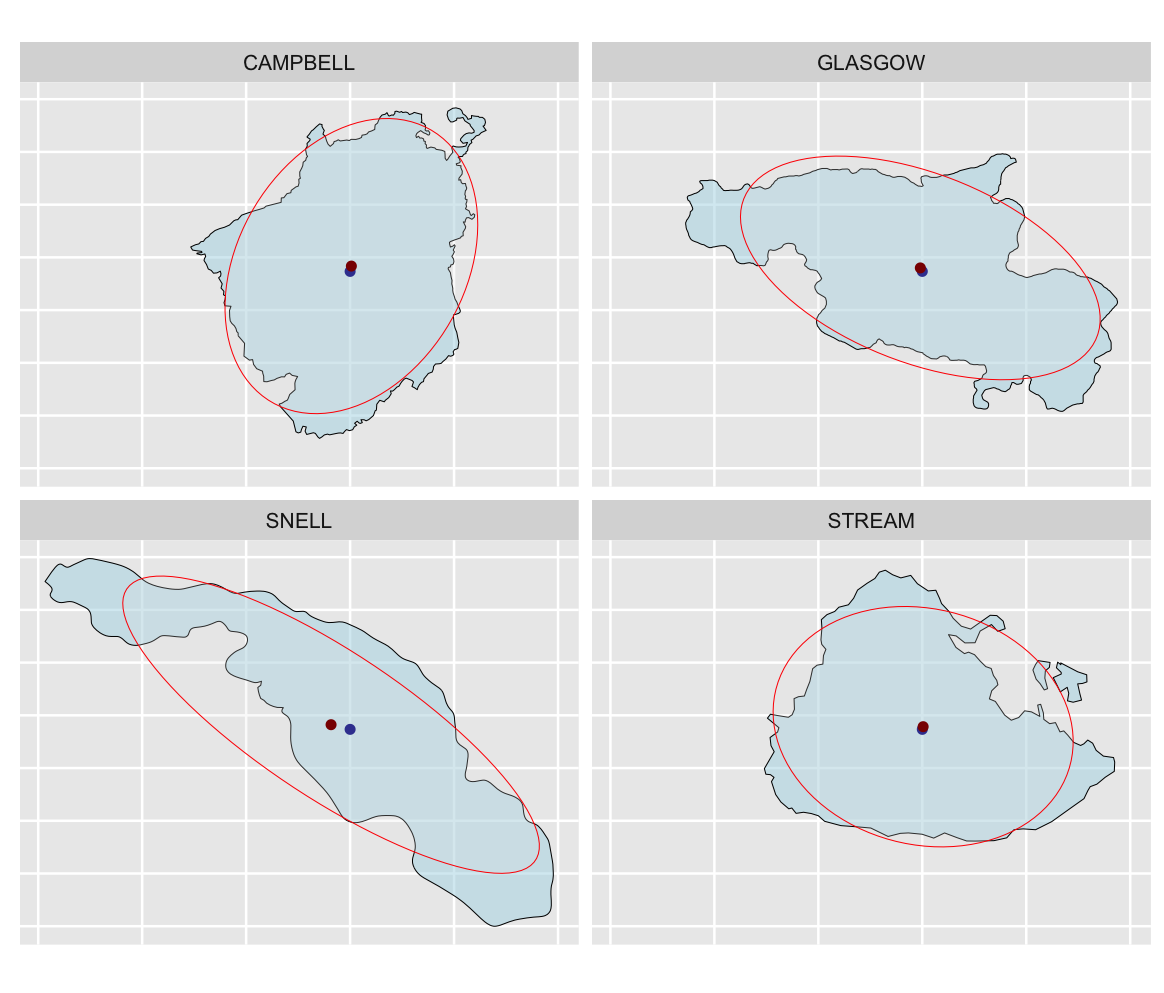}
    \caption{Examples of standardized wildfire polygons with fitted ellipses.}
    \label{fig:unit_area_polygons_and_ellipses}
\end{figure}

The second level of hierarchy in the model focuses on the area and shape of the burn pattern. 
The shapes are assumed to be elliptical, where area is abstracted out and modeled separately. 
The burn shape model relies on elliptical fits of the empirical fire extents. 
Figure \ref{fig:unit_area_polygons_and_ellipses} illustrates several representative standardized polygons (some multipolygons) with their fitted ellipses. 
These fits capture large-scale geometric characteristics, including elongation, orientation, and overall spatial footprint, while smoothing highly irregular boundary structure. 
Although standardized polygons are centered at the origin prior to fitting, optimized ellipse centroids do not always align exactly with polygon centroids because centroid location is a free optimization parameter. 
In practice, these shifts are typically small and do not materially affect spatial fire placement, so no additional jittering or relocation of fire centers is introduced in the point process simulations.

Each fitted ellipse is parameterized by its semi-major axis and rotational angle. 
Exploratory analysis suggests that we can pool observations across all 10 pyromes and all 480 months to estimate the empirical distributions of these parameters, treating them as spatially and temporally invariant for the purposes of model specification.
See spatial variograms in supplement Section S4.

\begin{figure}[!htbp]
    \centering
    \begin{subfigure}[b]{0.45\textwidth}
        \centering
        \includegraphics[width=\textwidth]{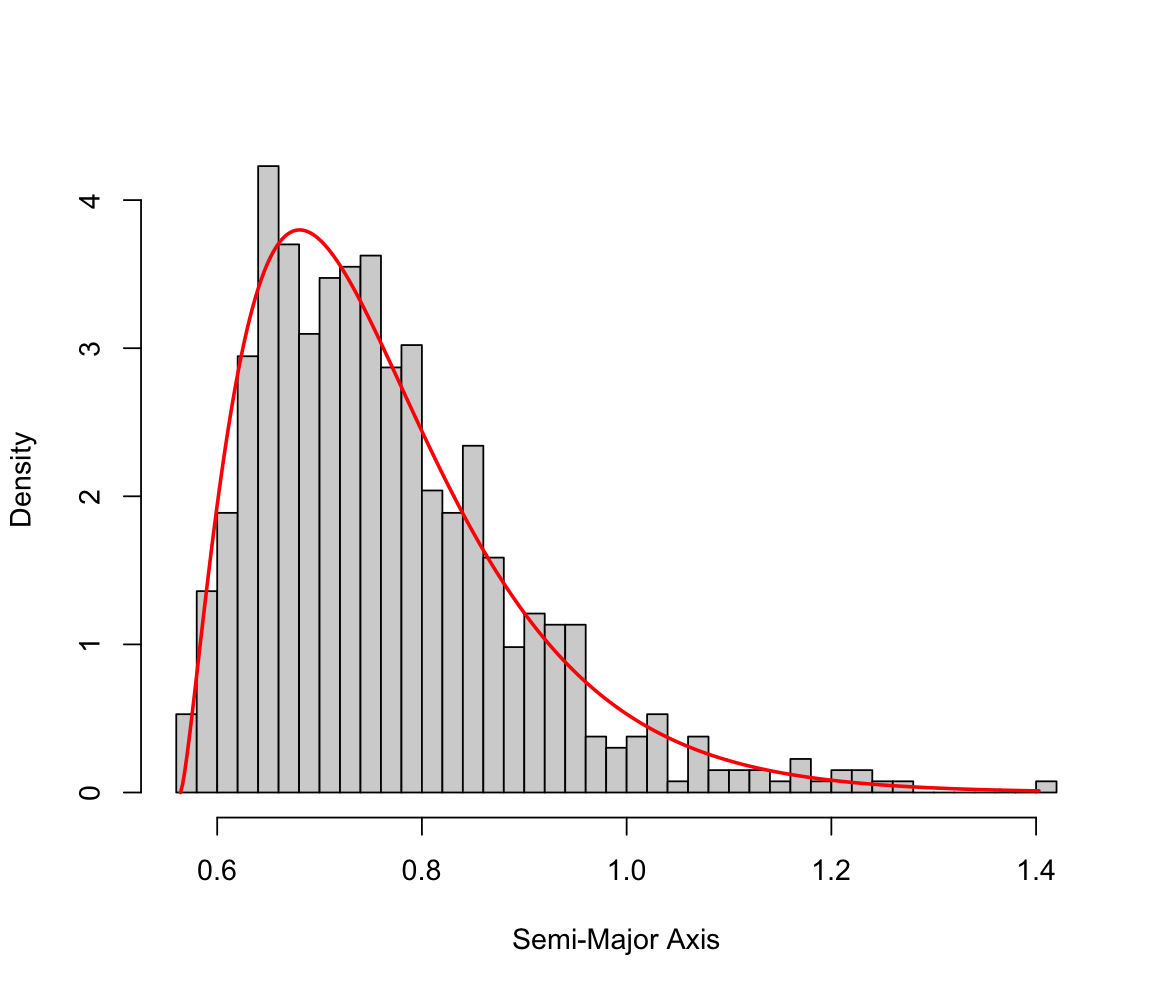}
        \label{fig:semi_major}
    \end{subfigure}
    \hfill
    \begin{subfigure}[b]{0.45\textwidth}
        \centering
        \includegraphics[width=\textwidth]{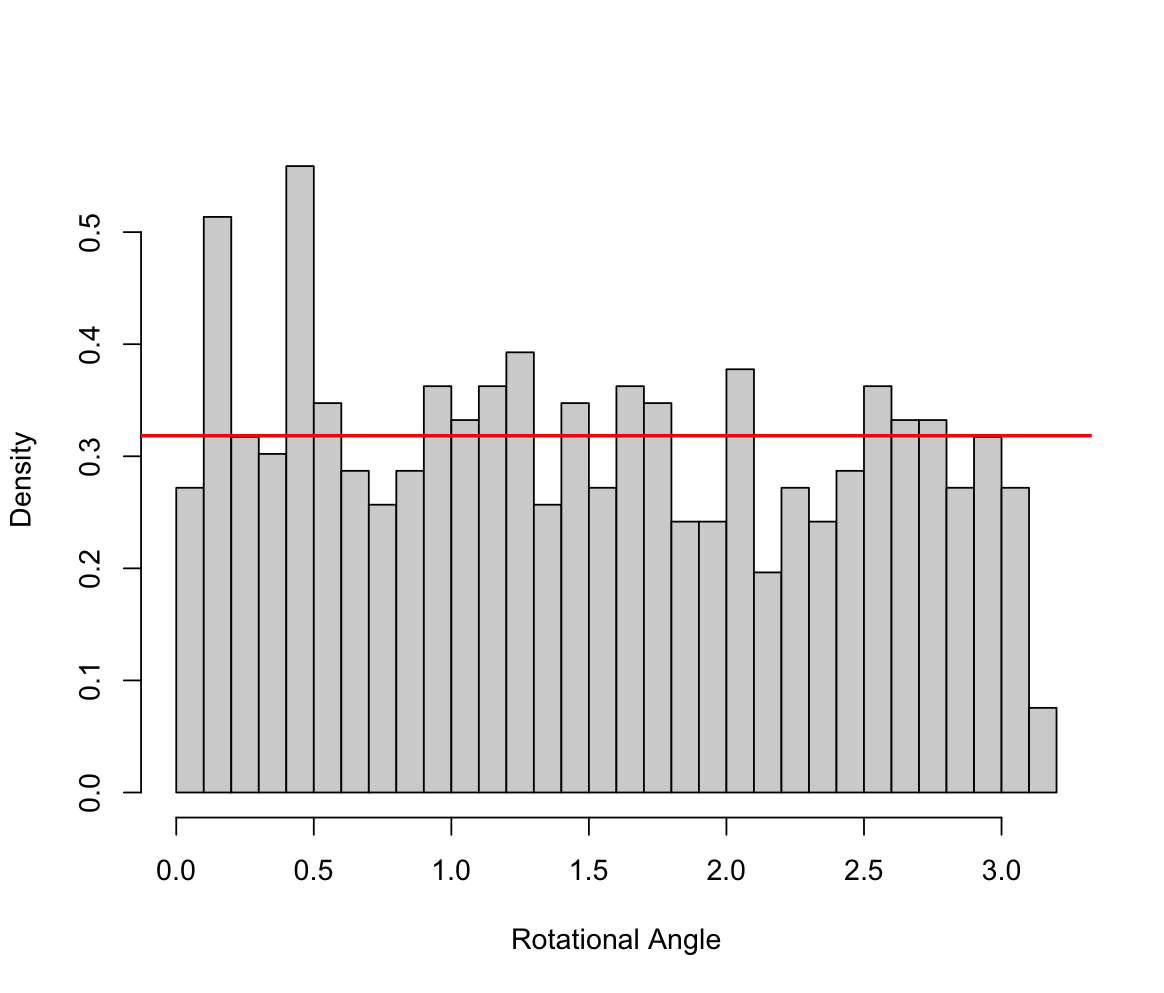}
        \label{fig:rot_angle}
    \end{subfigure}
    \caption{Distributions of fitted ellipse parameters: (a) semi-major axis length (shifted gamma) and (b) rotation angle (uniform).}
    \label{fig:ellipse_params}
\end{figure}

The estimated standardized semi-major axes are right-skewed, indicating that most fires are moderately elongated while fewer have highly elongated geometries. 
A shifted gamma distribution provides a reasonable fit to the empirical distribution of standardized semi-major axes, with estimated shape parameter $2.4$ and rate parameter $11.9$ under the lower-bound shift $1/\sqrt{\pi}$, as shown in Figure \ref{fig:ellipse_params}(a). 
The rotation angles $\phi_i$ are approximately uniformly distributed on $[0,\pi)$ in Figure \ref{fig:ellipse_params}(b), suggesting no strong preferred orientation in fire elongation in Northern California.

\subsection{Burned Area Model Illustration}

\begin{figure}[htbp]
    \centering
    \includegraphics[width=0.9\textwidth]{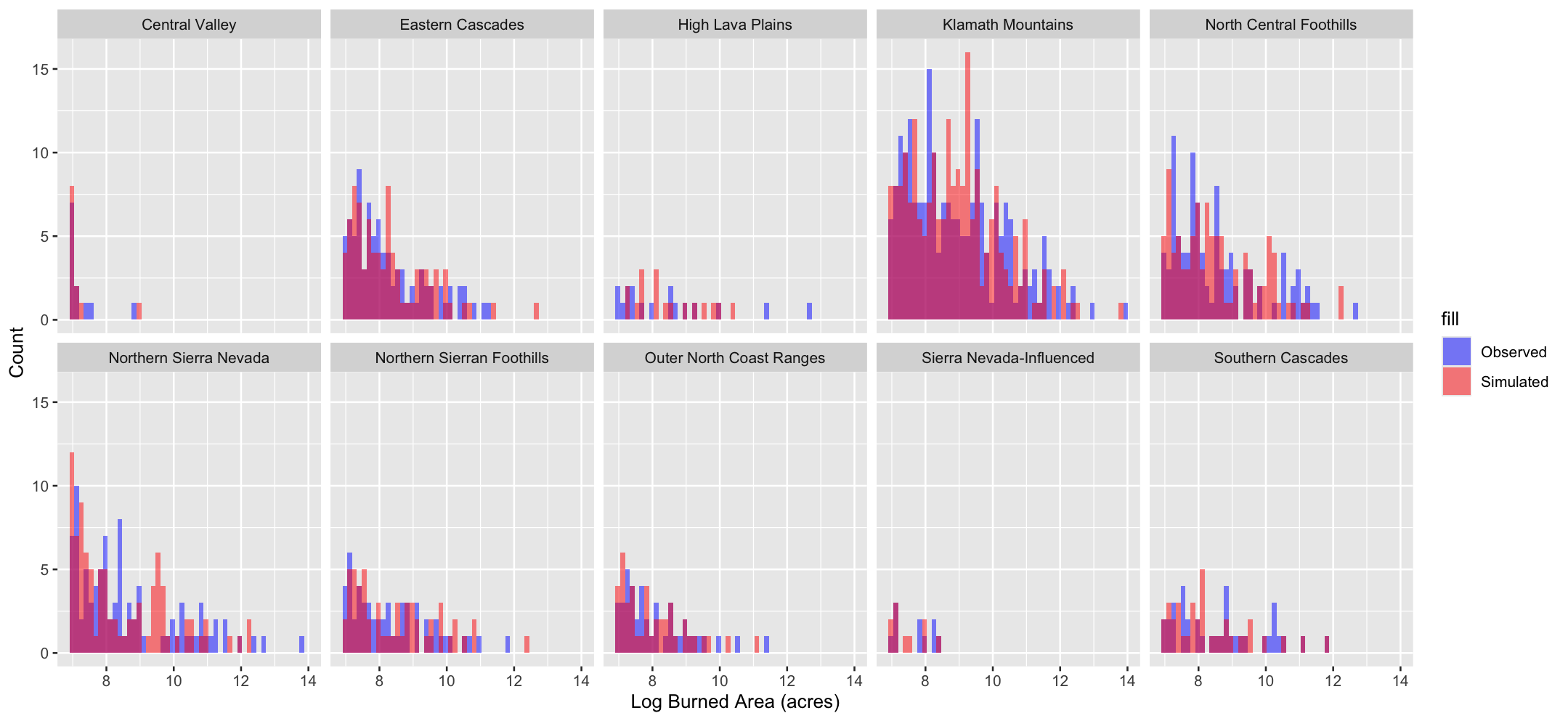}
    \caption{Observed and simulated log burned areas by pyrome.}
    \label{fig:log_burned_area}
\end{figure}

Wildfire burned area is modeled using a heteroskedastic lognormal generalized least squares (GLS) model with Gaussian errors on the log burned area scale and pyrome-specific residual variances. 
The mean structure includes VPDa, seasonal harmonic terms, and pyrome indicators to capture both temporal and regional variation in wildfire size.
Pyrome-specific effects reveal regional differences in baseline fire size distributions, with CCV tending toward smaller fires and pyromes such as EC exhibiting larger fire extents (Figure \ref{fig:log_burned_area}).

The fitted model identifies significant temporal and regional structure in wildfire burned area (Table \ref{tab:burned_area_gls}).
Positive VPDa effects indicate that anomalously dry atmospheric conditions increase both wildfire occurrence and extent. 
Seasonal harmonic terms capture within-year variation in fire size, while significant pyrome effects reveal persistent regional differences after accounting for climate and seasonality. Notably, CCV exhibits substantially lower burned areas, whereas KM and SNIRHB show significant positive and negative deviations, respectively.

The heteroskedastic variance structure allows residual variability to differ across pyromes, capturing regional differences in wildfire size dispersion. 
Some pyromes exhibit relatively stable fire behavior, while others show substantial variability around their regional mean. 
Figure \ref{fig:log_burned_area} highlights this contrast, with KM exhibiting the greatest variability in observed fire sizes and CCV showing comparatively low variability. 
Simulated burned areas reproduce the strong right-skew and variability observed in empirical fire sizes, as shown in Figure \ref{fig:log_burned_area}.

\subsection{Full Model Validation}

\begin{figure}[!htbp]
    \centering

    \begin{subfigure}[t]{0.85\textwidth}
        \centering
        \includegraphics[width=\textwidth]{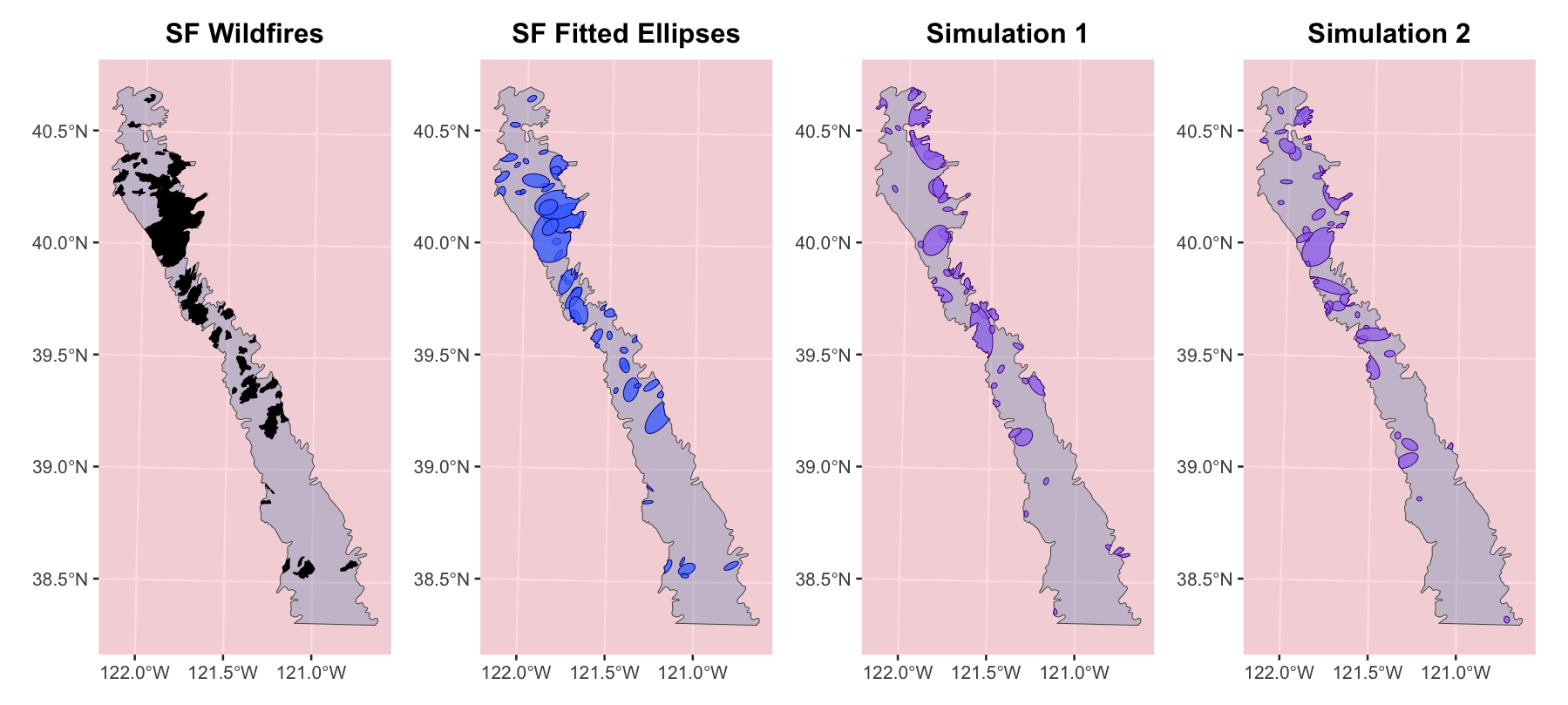}
        \label{fig:sf_final_sims}
    \end{subfigure}

    \vspace{-2cm}

    \begin{subfigure}[t]{0.85\textwidth}
        \centering
        \includegraphics[width=\textwidth]{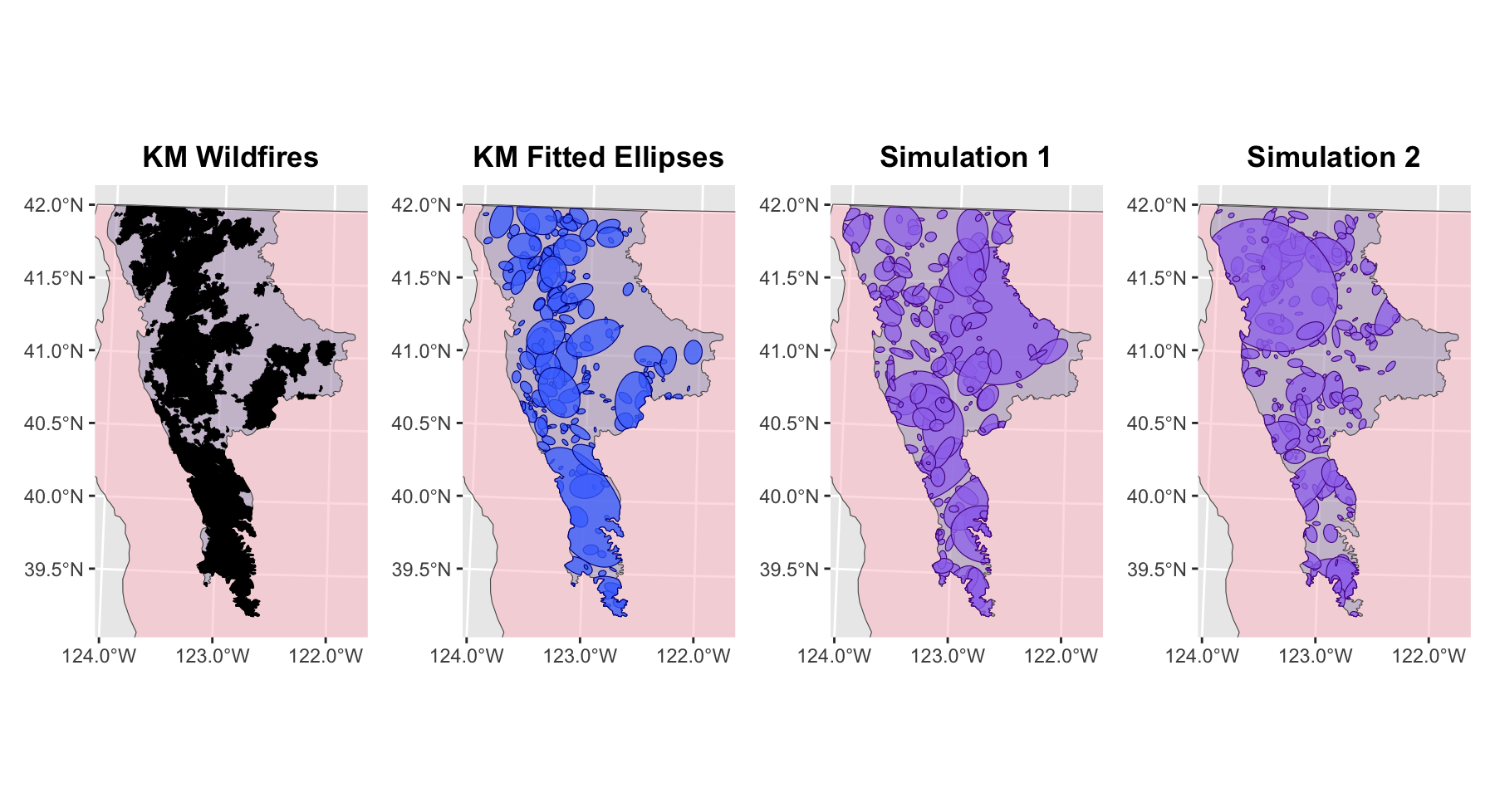}
        \label{fig:km_final_sims}
    \end{subfigure}

    \vspace{-1.5cm}

    \caption{Observed and simulated wildfire perimeters in the SF and KM pyromes.}
    \label{fig:sf_km_final_sims}
\end{figure}

For each month, simulated wildfires are generated iteratively. 
Fire counts and locations are first sampled from the point process model, followed by independent sampling of ellipse shape parameters and burned areas from their fitted models.
Unit area ellipses are then generated using simulated semi-major axes and rotational angles, rescaled according to the simulated burned area, and centered at the simulated points from the point process model. Figure \ref{fig:sf_km_final_sims} shows SF and KM wildfires, fitted ellipses, and simulations from the full hierarchical model. 

In the preceding sections, we show that the component pieces of the full model adequately fit the empirical wildfire distribution. 
However, it remains to be seen whether, when simulated together, the resulting distribution of wildfire characteristics reasonably captures empirical behavior. 
For example, a key validation statistic concerns the long-term cumulative area burned (that is, over many years). 
To assess this, we generate 500 independent simulations spanning the full 40-year study period (1984--2023) for the SF pyrome. 
Within each simulation and the observed data, all simulated wildfires are intersected with the corresponding pyrome boundary to ensure spatial consistency with the study region. 

\begin{figure}[htbp]
    \centering

    \begin{subfigure}[t]{0.48\textwidth}
        \centering
        \includegraphics[width=\textwidth]{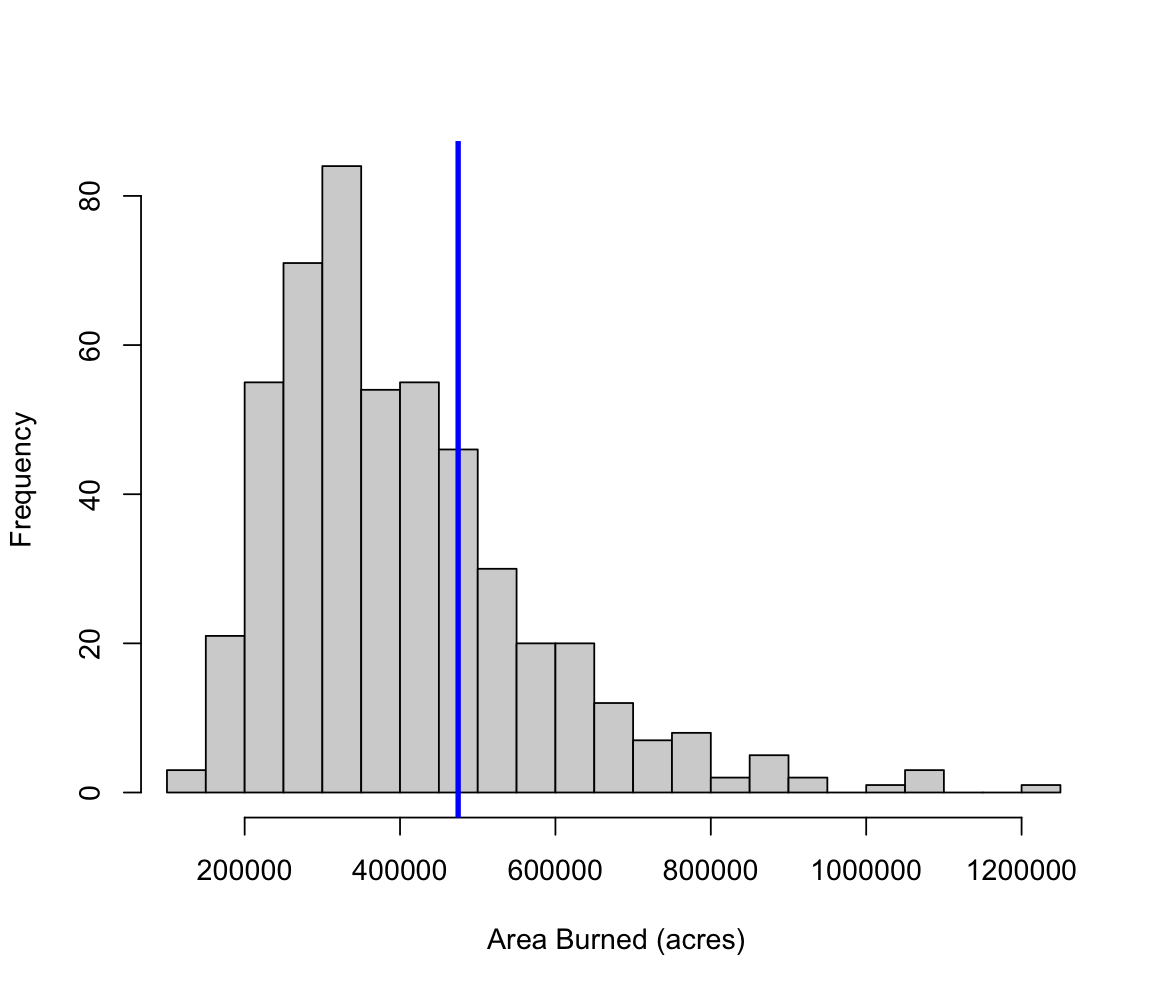}
        \caption{}
        \label{fig:sf_total_area}
    \end{subfigure}
    \hfill
    \begin{subfigure}[t]{0.48\textwidth}
        \centering
        \includegraphics[width=\textwidth]{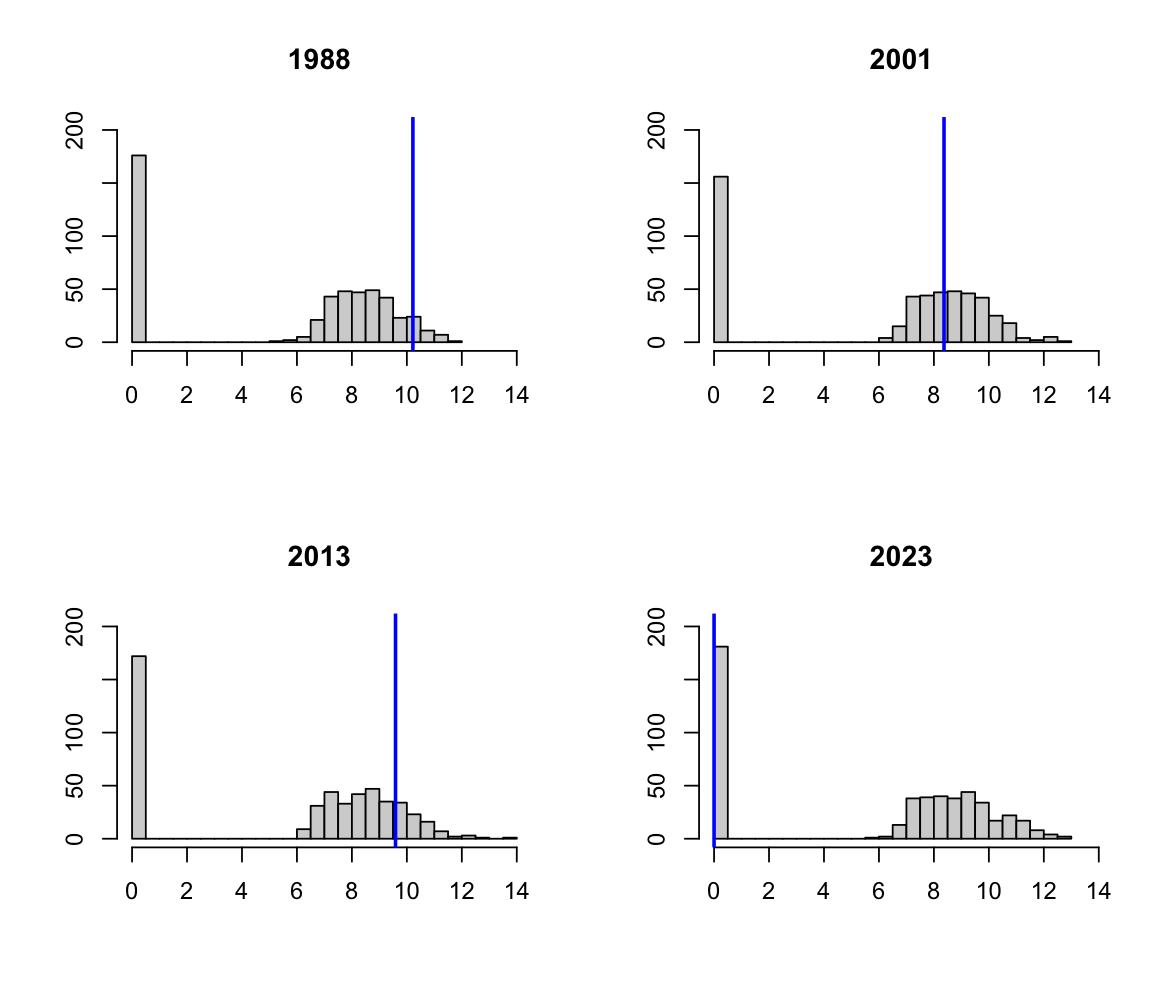}
        \caption{}
        \label{fig:sf_yearly_area}
    \end{subfigure}

    \caption{Observed and simulated burned area in the SF pyrome: (a) 1984--2023 cumulative burned area; (b) annual log burned area.}

    \label{fig:burned_area_comparison}
\end{figure}

Figure \ref{fig:burned_area_comparison} compares observed and simulated burned area in the SF pyrome at cumulative (1984--2023) and annual time scales using 500 realizations from the fitted model. 
In Figure \ref{fig:burned_area_comparison}(a), the blue line denotes the cumulative burned area of the fitted ellipses rescaled to their original sizes, while in Figure \ref{fig:burned_area_comparison}(b) it shows the observed annual $\log(\text{burned area}+1)$ values. 
Across both temporal scales, the observed values fall within the range of simulated outcomes, indicating that the model captures long-term burned area accumulation, year-to-year variability, and extreme fire years.

A key advantage of the statistical framework is that it enables explicit uncertainty quantification via simulation. Because the fitted model generates full stochastic realizations of wildfire activity, probabilities of extreme events can be estimated using Monte Carlo methods. Here, we focus on the probability that total annual burned area across Northern California exceeds 685,000 acres, approximately the 90th percentile of the historical distribution.

\begin{figure}[htbp]
\centering
\includegraphics[width=0.85\textwidth]{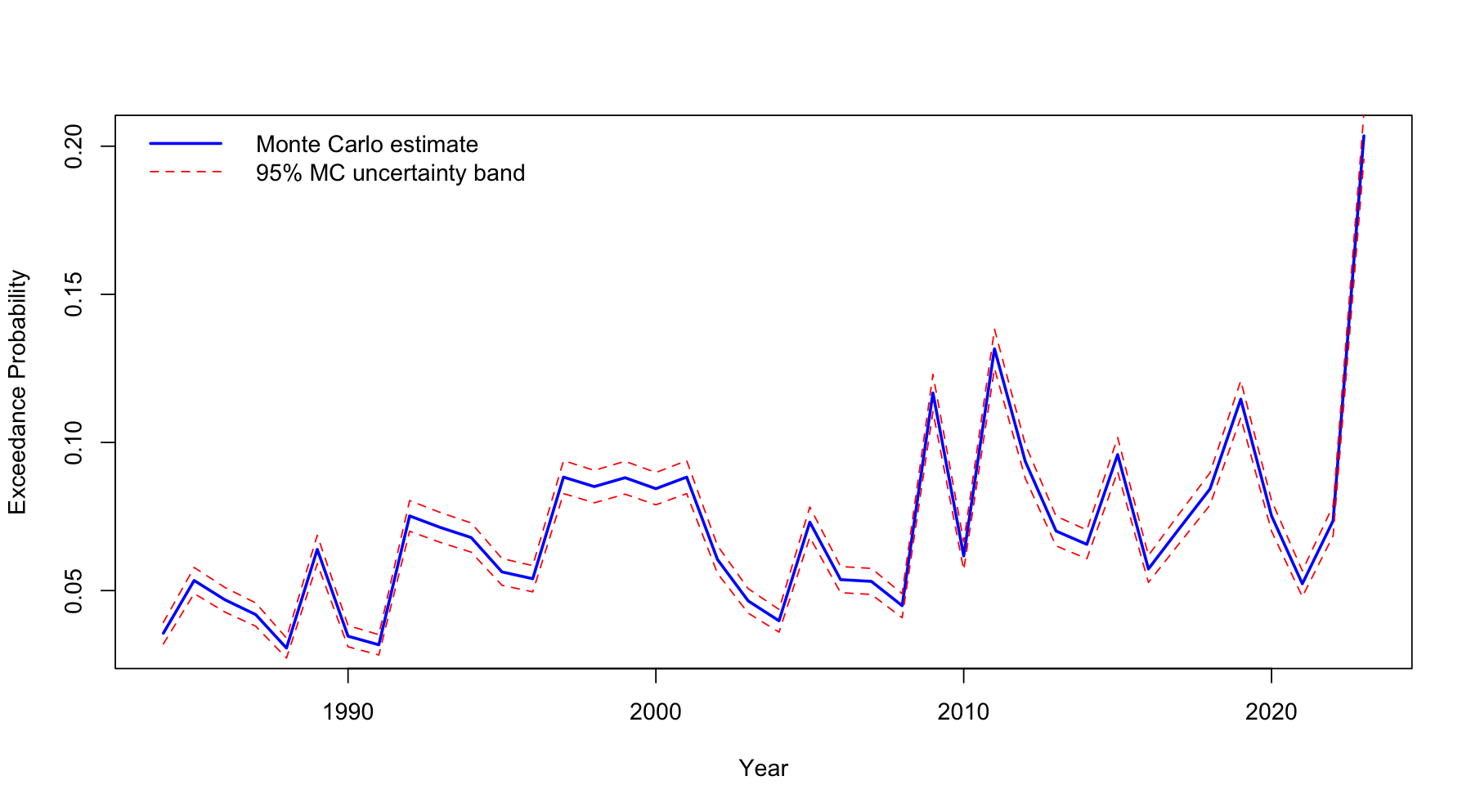}
\caption{Probability of exceeding 685,000 acres of annual burned area (90th percentile) estimated from 10,000 Monte Carlo simulations per year. Dashed lines indicate 95\% Monte Carlo intervals.}
\label{fig:monte_carlo}
\end{figure}

For each year, exceedance probabilities are computed from 10,000 simulated realizations as the fraction exceeding this threshold (Figure \ref{fig:monte_carlo}). Probabilities vary substantially over time, ranging from roughly 3--5\% in low-risk years to over 20\% in 2023, with most years between 5--10\%. This indicates that extreme burned-area outcomes are uncommon but not rare under the fitted model.

The high probability in 2023 is consistent with that year exhibiting the largest VPDa across all pyromes and months in the record, indicating exceptionally fire-conducive conditions. In the lognormal burned-area model, positive VPDa effects shift the distribution toward larger fires and increase tail probability, while residual log-scale variability further amplifies uncertainty. Notably, observed exceedances of the threshold happened in 2007, 2017, 2019, and 2020, aligning with the upper tail of the simulated distribution. Overall, these results illustrate how the framework can be used to quantify the likelihood of rare but consequential wildfire outcomes.

\section{Discussion}   \label{sec:discussion}

This study develops a generative hierarchical statistical modeling framework for wildfire occurrence, burned area, and fire geometry in a unified probabilistic setting. 
Wildfire counts and locations are modeled as an inhomogeneous Poisson process with a stochastic shock in the intensity to better capture observed mean and variability of fire counts. 
Given the location of a fire occurrence, the burn pattern is modeled as a randomly-oriented and randomly-sized ellipse. 
Each component of the model includes physiographic and climate covariates, allowing the framework to generalize across regions with differing climate drivers.

A key contribution of the model is its ability to simulate wildfire regimes under observed and future climate scenarios, enabling Monte Carlo approximation of wildfire risk and probabilistic characterization of impacts such as total burned area \citep{williams2026wulffss}. By simulating the full data-generating process, including event locations, temporal intensity, ellipse-based geometries, and burned areas, the model produces empirical distributions of quantities such as total burned area and fire counts. The inclusion of VPDa in both the occurrence and burned area components allows changes in atmospheric dryness to propagate through both fire frequency and size, supporting simulations under changing climate conditions derived from climate models or reanalysis products. These simulations can also serve as inputs to economic, infrastructure, and insurance risk models.

Importantly, this simulation-based approach provides a direct representation of uncertainty in wildfire outcomes across spatial and temporal scales. 
Repeated simulations reveal substantial variability in rare, high-impact events, including extreme fire seasons and large cumulative burned areas. 
This variability reflects the heavy-tailed nature of wildfire activity and highlights that risk cannot be characterized by mean outcomes alone. 
By generating ensembles of plausible wildfire realizations, the framework enables direct estimation of extreme-event probabilities and provides a more informative risk assessment than deterministic or point-estimate approaches.  


The results also highlight meaningful heterogeneity across pyromes. 
Differences in shock parameters suggest that pyromes vary in how variable wildfire occurrence is over time, in addition to differences in mean intensity and burned-area variance. 
Pyromes such as KM and EC show relatively higher $\gamma$ values in the shock term, consistent with more variable fire activity, while NCCFACM and SNIRHB exhibit lower values, indicating more stable regimes. 
The burned-area model further reflects this heterogeneity, with KM associated with higher mean burned area, CCV and SNIRHB with substantially lower levels, and additional variation in variance structure across regions. 
These differences are naturally captured in the Monte Carlo simulations, which preserves region-specific uncertainty rather than imposing a homogeneous risk structure.  

Several limitations of the current model suggest directions for future research. 
The elliptical approximation of burned area polygons simplifies complex fire geometries and may underrepresent boundary irregularities, suggesting that more flexible representations of fire shape could improve realism, particularly in heterogeneous terrain or under strong wind-driven spread. 
The use of temporally independent shock variables captures overdispersion in monthly fire counts but does not reflect persistence or memory in extreme fire conditions, indicating that incorporating higher-frequency meteorological information could help better resolve short-term drivers of wildfire activity and reduce reliance on latent stochasticity. 
In addition, the set of climate covariates considered in the current specification may not fully capture the range of environmental controls on fire activity, motivating further exploration of alternative or higher-resolution climate variables and temporal aggregation scales.

Future work could extend the framework in several directions. 
More flexible representations of burned area geometry could better capture complex perimeter structure beyond the elliptical approximation, improving realism and allowing fire shapes to respond to environmental features such as land cover. 
Incorporating dynamic feedback between wildfire occurrence and landscape would allow dependence between successive fire events and evolving fire regimes over time. 
In addition, integrating higher-frequency meteorological information into the shock process may improve representation of short-term drivers of extreme wildfire activity and reduce reliance on latent stochasticity. 
Finally, further exploration of alternative climate covariates and temporal aggregation scales could help identify more robust predictors of wildfire activity that are consistently significant across pyromes. 

Overall, the proposed model demonstrates that a hierarchical stochastic representation of wildfire processes can simultaneously capture spatial occurrence, temporal variability, geometric structure, and burned area dynamics. 
Most importantly, it provides a coherent framework for Monte Carlo approximation of wildfire risk, enabling full uncertainty quantification of both typical and extreme fire outcomes. 
This represents a substantive step toward probabilistic wildfire modeling that is directly aligned with risk assessment and decision-making under uncertainty.





\bibliographystyle{jasa}
\bibliography{references}

\section{Tables}

\begin{table}[htbp]
\centering
\small
\setlength{\tabcolsep}{3pt}
\renewcommand{\arraystretch}{0.9}
\caption{Wildfire occurrence summary by pyrome, 1984--2023.}
\vspace{-0.25cm}
\label{tab:pyrome_summary}

\begin{tabular}{llccc}
\hline
\textbf{Pyrome} & \textbf{Abbr.} & \textbf{Total Fires} & \textbf{Mean Annual} & \textbf{Max Monthly} \\
\hline
Central California Valley                           & CCV     & 12  & 0.30 & 2 \\
Eastern Cascades                                    & EC      & 84  & 2.10 & 4 \\
High Lava Plains                                    & HLP     & 17  & 0.43 & 2 \\
Klamath Mountains                                   & KM      & 216 & 5.40 & 32 \\
North Central California Foothills and Coastal Mountains & NCCFACM & 99  & 2.48 & 5 \\
Northern Sierra Nevada                              & NSN     & 91  & 2.28 & 8 \\
Northern Sierran Foothills and Tuscan Flows         & SF      & 54  & 1.35 & 8 \\
Outer North Coast Ranges and Redwoods               & ONCRR   & 42  & 1.05 & 12 \\
Sierra Nevada-Influenced Ranges, Hills, and Basins  & SNIRHB  & 10  & 0.25 & 1 \\
Southern Cascades                                   & SC      & 36  & 0.90 & 6 \\
\hline
\end{tabular}
\end{table}

\begin{table}[htbp]
\centering
\small
\caption{KM point process model estimates and standard errors.}
\vspace{-0.25cm}
\label{tab:results_table}
\begin{tabular}{lrrrr}
\hline
\textbf{Covariate} & \textbf{Estimate} & \textbf{Std. Error} & \textbf{z-value} & \textbf{p-value} \\
\hline
Intercept      & -4.30 & 0.30 & -14.18 & $<0.001$ \\
Latitude       & -0.16 & 0.13 & -1.19  & 0.23 \\
Longitude      & -0.58 & 0.21 & -2.76  & 0.006 \\
Lat--Lon Interaction & -0.53 & 0.48 & -1.09 & 0.27 \\
Latitude$^2$   & -0.15 & 0.17 & -0.87  & 0.38 \\
Longitude$^2$  & -1.56 & 0.46 & -3.43  & $<0.001$ \\
Elevation      & -0.20 & 0.08 & -2.60  & 0.009 \\
Monthly VPDa Anomaly & 0.15 & 0.06 & 2.44 & 0.015 \\
$\sin(2\pi (\text{month}) / 12)$ & -2.07 & 0.19 & -10.81 & $<0.001$ \\
$\cos(2\pi (\text{month}) / 12)$ & -1.81 & 0.18 & -10.35 & $<0.001$ \\
\hline
\end{tabular}
\end{table}

\begin{table}[htbp]
\centering
\small
\caption{Shock parameter estimates ($\alpha$ and $\gamma$) for each pyrome.}
\begin{tabular}{lcc}
\hline
\textbf{Pyrome} & \boldmath$\alpha$ & \boldmath$\gamma$ \\
\hline
CCV       & 0.4980 & 0.5025 \\
EC        & 0.3080 & 0.7940 \\
HLP       & 0.6925 & 0.5025 \\
KM        & 0.4075 & 0.8744 \\
NCCFACM   & 0.1724 & 0.1357 \\
NSN       & 0.3894 & 0.6884 \\
ONCRR     & 0.0874 & 0.7789 \\
SC        & 0.2176 & 0.4221 \\
SF        & 0.2040 & 0.4673 \\
SNIRHB    & 0.3533 & 0.1759 \\
\hline
\end{tabular}
\label{tab:shock_params}
\end{table}

\begin{table}[htbp]
\centering
\small
\setlength{\tabcolsep}{5pt}
\renewcommand{\arraystretch}{0.9}
\caption{Parameter estimates for the heteroskedastic lognormal GLS burned-area model.}
\vspace{-0.25cm}
\label{tab:burned_area_gls}

\begin{tabular}{lrrrr}
\hline
\textbf{Covariate} & \textbf{Estimate} & \textbf{Std. Error} & \textbf{t-value} & \textbf{p-value} \\
\hline
Intercept & 7.07 & 0.22 & 31.76 & $<0.001$ \\
VPDa Anomaly & 0.85 & 0.32 & 2.65 & 0.008 \\
$\sin(2\pi(\text{month})/12)$ & -0.85 & 0.18 & -4.61 & $<0.001$ \\
$\cos(2\pi(\text{month})/12)$ & -0.30 & 0.17 & -1.75 & 0.081 \\
KM & 0.72 & 0.22 & 3.27 & 0.001 \\
SC & 0.28 & 0.37 & 0.77 & 0.444 \\
HLP & -0.17 & 0.57 & -0.29 & 0.771 \\
NSN & 0.22 & 0.30 & 0.73 & 0.466 \\
SF & -0.11 & 0.31 & -0.35 & 0.725 \\
CCV & -3.11 & 0.57 & -5.44 & $<0.001$ \\
NCCFACM & 0.36 & 0.25 & 1.44 & 0.152 \\
ONCRR & -0.41 & 0.31 & -1.34 & 0.181 \\
SNIRHB & -1.06 & 0.54 & -1.98 & 0.048 \\
\hline
\end{tabular}
\end{table}

\end{document}


\sloppy
\baselineskip=20pt

\begin{center}{
  \Large \bf
  Supplementary Material for\\[0.2cm]
  FireGen: Quantifying Wildfire Risk by Simulation
}

\bigskip
\bigskip

{\bf Allyson Hineman,
William Kleiber,
Stephan Sain,
and Alexis Hoffman}

\bigskip
\bigskip

{\bf \today}

\end{center}

\bigskip

This supplement provides additional methodological details, theoretical results, model-fitting procedures, and pyrome-specific diagnostics that support the analyses presented in the main manuscript. 
Included are proofs and derivations for components of the FireGen framework, optimal weighting parameters used in model calibration, additional spatial diagnostics, and supplementary validation results for each pyrome.


\section{Moment Properties of the Multiplicative Shock Model}
\label{sec:supp_shock_proof}

This section derives the first and second moments of the wildfire count process under the multiplicative shock model and shows how the shock term modifies the mean--variance relationship of a standard Poisson process.

Let
\[
    \mathcal{W}_t = \mathcal{S} \times \{t\}
\]
denote the spatial domain at time \(t\), and define the baseline integrated intensity
\[
    \Lambda(t)
    =
    \int_{\mathcal{S}}
    \lambda(\bs,t)\, d\bs.
\]

Under the baseline model,
\[
    N_t = N(X \cap \mathcal{W}_t)
\]
follows a Poisson distribution with parameter \(\Lambda(t)\). Consequently,
\[
    \mathbb{E}[N_t]
    =
    \Lambda(t),
    \qquad
    \mathrm{Var}(N_t)
    =
    \Lambda(t),
\]
so that the variance is constrained to equal the mean.

Under the full model, the intensity is modified by a multiplicative shock yielding $U_t\,\lambda(\bs,t)$, where
\[
    U_t = G_t^\gamma,
    \qquad
    G_t \sim \mathrm{Gamma}(\alpha,\alpha).
\]
Conditional on \(U_t\), the process remains Poisson with integrated intensity
\[
    U_t\,\Lambda(t).
\]
Therefore,
\[
    N_t \mid U_t
    \sim
    \mathrm{Poisson}\!\left(U_t\,\Lambda(t)\right),
\]
implying
\[
    \mathbb{E}[N_t \mid U_t]
    =
    U_t\,\Lambda(t),
    \qquad
    \mathrm{Var}(N_t \mid U_t)
    =
    U_t\,\Lambda(t).
\]

Applying the laws of total expectation and total variance yields
\[
    \mathbb{E}[N_t]
    =
    \mathbb{E}[U_t]\,\Lambda(t),
\]
and
\[
    \mathrm{Var}(N_t)
    =
    \mathbb{E}\!\left[
    \mathrm{Var}(N_t \mid U_t)
    \right]
    +
    \mathrm{Var}\!\left[
    \mathbb{E}(N_t \mid U_t)
    \right].
\]

Substituting the conditional moments gives
\[
    \mathrm{Var}(N_t)
    =
    \mathbb{E}[U_t]\,\Lambda(t)
    +
    \mathrm{Var}(U_t)\,\Lambda(t)^2.
\]

Using the expression for the marginal mean,
\[
    \mathrm{Var}(N_t)
    =
    \mathbb{E}[N_t]
    +
    \mathrm{Var}(U_t)\,\Lambda(t)^2.
\]

The first term is the usual Poisson variance, while the second term arises from temporal fluctuations in the latent shock process. 
It follows that
\[
    \mathrm{Var}(N_t)
    >
    \mathbb{E}[N_t];
\]
thus the multiplicative shock model exhibits overdispersion relative to a standard Poisson process.

The degree of overdispersion can be expressed through the dispersion index
\[
    D_t = \frac{\mathrm{Var}(N_t)}{\mathbb{E}[N_t]} = 1 + \frac{\mathrm{Var}(U_t)}{\mathbb{E}[U_t]}\,\Lambda(t),
\]
which exceeds one whenever \(\mathrm{Var}(U_t)>0\). Consequently, the shock process relaxes the rigid Poisson mean--variance relationship by introducing an additional source of variability through the random intensity surface.

The multiplicative shock also modifies the marginal mean. Since
\[
    U_t = G_t^\gamma,
    \qquad
    G_t \sim \mathrm{Gamma}(\alpha,\alpha),
\]
we have
\[
    \mathbb{E}[U_t] = \frac{\Gamma(\alpha+\gamma)}{\Gamma(\alpha)\alpha^\gamma}.
\]

For the parameter estimates considered in this study,
\[
    0 < \gamma < 1.
\]

Since the function \(g(x)=x^\gamma\) is concave on \((0,\infty)\), Jensen's inequality implies
\[
    \mathbb{E}[U_t]
    =
    \mathbb{E}[G_t^\gamma]
    <
    \left(\mathbb{E}[G_t]\right)^\gamma.
\]

Because \(G_t \sim \mathrm{Gamma}(\alpha,\alpha)\) satisfies
\[
    \mathbb{E}[G_t]=1,
\]
it follows that
\[
    \mathbb{E}[U_t] < 1.
\]

Therefore,
\[
    \mathbb{E}[N_t]
    =
    \mathbb{E}[U_t]\Lambda(t)
    <
    \Lambda(t).
\]

In summary, the multiplicative shock model affects both the first and second moments of the count process. Relative to the baseline Poisson model, the shock term can reduce the average number of events while simultaneously introducing additional variability through a random intensity surface. This additional flexibility allows the model to reproduce observed wildfire count distributions without requiring the deterministic baseline intensity to absorb all temporal heterogeneity.
\qed

\section{Gradients of Ellipse Objective Function} 
\label{sec:supp_ellipse_gradients}

This section includes closed-form gradients of the function $f(\btheta)$ from Section 3.2.2. 
Gradients of the objective with respect to all parameters are computed analytically. 
\[
    \frac{\partial f}{\partial \btheta_k} = \sum_j 2 r_j \frac{\partial r_j}{\partial \btheta_k}, \quad \theta_k \in \{x_0,y_0,a,b,\phi\}
\]
with
\[
\begin{aligned}
\frac{\partial r_j}{\partial x_0} &= \frac{2 x_{\mathrm{rot}, j}}{a^2} (-\cos\phi) + \frac{2 y_{\mathrm{rot}, j}}{b^2} (\sin\phi), \\
\frac{\partial r_j}{\partial y_0} &= \frac{2 x_{\mathrm{rot}, j}}{a^2} (-\sin\phi) + \frac{2 y_{\mathrm{rot}, j}}{b^2} (-\cos\phi), \\
\frac{\partial r_j}{\partial a} &= -2 \frac{x_{\mathrm{rot}, j}^2}{a^3}, \quad
\frac{\partial r_j}{\partial b} = -2 \frac{y_{\mathrm{rot}, j}^2}{b^3}, \\
\frac{\partial r_j}{\partial \phi} &= \frac{2 x_{\mathrm{rot}, j}}{a^2} \big(-\sin\phi (x_j - x_0) + \cos\phi (y_j - y_0)\big) + \frac{2 y_{\mathrm{rot}, j}}{b^2} \big(-\cos\phi (x_j - x_0) - \sin\phi (y_j - y_0)\big).
\end{aligned}
\]

The unit area constraint is imposed as a nonlinear equality: 
\[
  g(\btheta) = a b - \frac{1}{\pi} = 0, 
\]
with Jacobian 
\[
  \nabla g = (0,0,b,a,0). 
\]


\section{Optimal weights per pyrome}
\label{sec:supp_optimal_weights}
Table \ref{tab:optimal_weights} displays the optimal GOF weights per pyrome. 

\begin{table}[htbp]
\centering
\caption{Optimal weighting design parameters for each pyrome based on minimum MeanFocused rank.}
\begin{tabular}{lccccc}
\hline
\textbf{Pyrome} & \boldmath$w_Q$ & \boldmath$w_{Count}$ & \boldmath$w_{Prop0}$ & \boldmath$w_{Prop1}$ & \boldmath$w_{Max}$ \\
\hline
CCV      & 11.6009 & 5.3132  & 12.8925 & 4.3190  & 4.2090 \\
EC       & 11.4710 & 10.2450 & 3.8085  & 12.5986 & 5.5028 \\
HLP      & 11.6009 & 5.3132  & 12.8925 & 4.3190  & 4.2090 \\
KM       & 3.3239  & 4.9410  & 4.3806  & 11.7963 & 3.9327 \\
NCCFACM  & 11.1104 & 12.4587 & 10.4859 & 12.0905 & 3.9230 \\
NSN      & 12.5931 & 4.7572  & 5.4258  & 2.8426  & 5.0909 \\
ONCRR    & 10.8371 & 12.7099 & 4.0583  & 5.1191  & 4.2522 \\
SC       & 3.3239  & 4.9410  & 4.3806  & 11.7963 & 3.9327 \\
SF       & 12.5931 & 4.7572  & 5.4258  & 2.8426  & 5.0909 \\
SNIRHB   & 2.8164  & 4.3284  & 2.1902  & 5.6936  & 11.5500 \\
\hline
\end{tabular}
\label{tab:optimal_weights}
\end{table}

\section{Spatial Variograms of Ellipse Parameters}
\label{sec:supp_variograms}

\begin{figure}[H]
    \centering

    \begin{subfigure}[b]{0.48\textwidth}
        \centering
        \includegraphics[width=\textwidth]{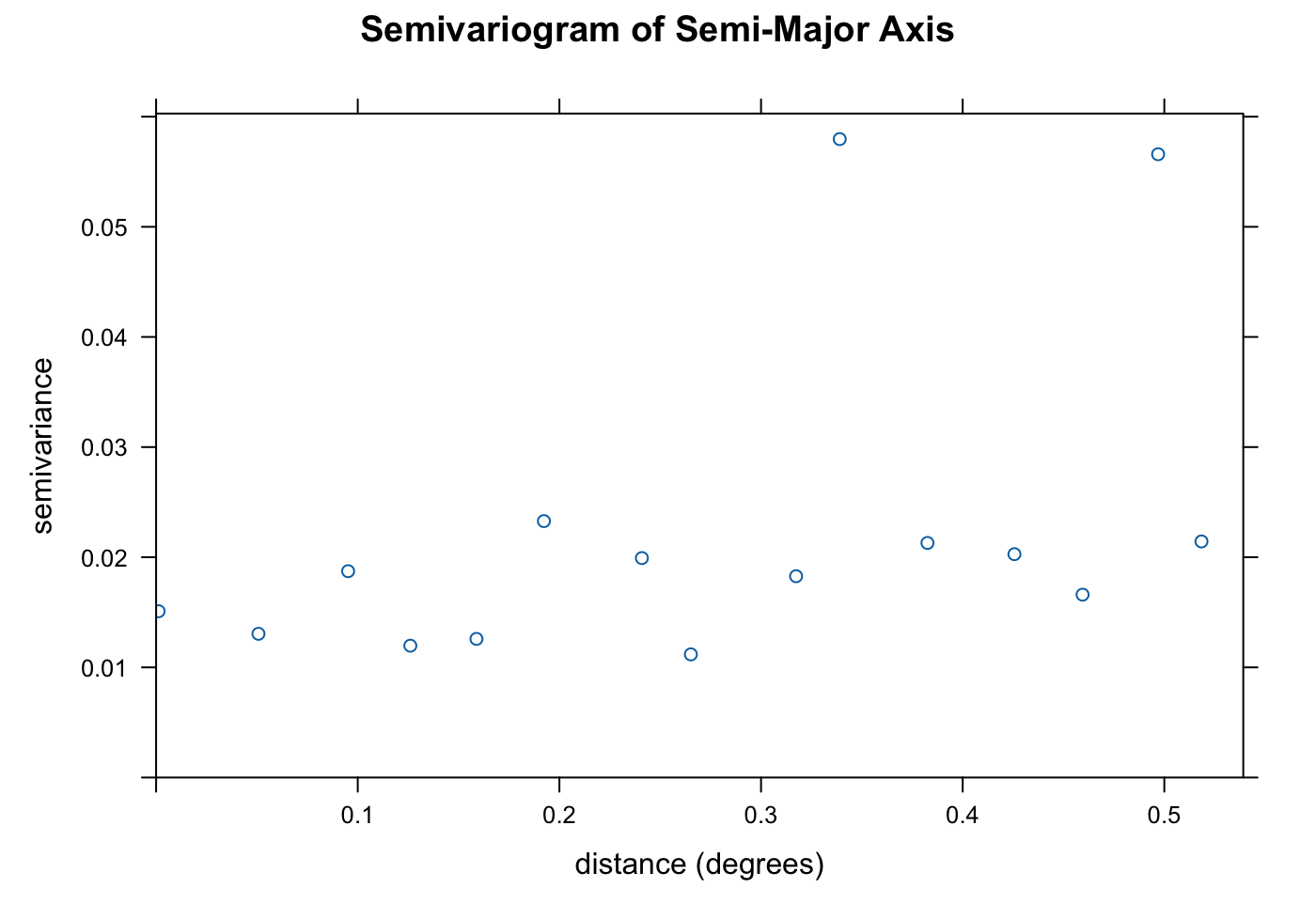}
        \caption{Semi-major axis}
    \end{subfigure}
    \hfill
    \begin{subfigure}[b]{0.48\textwidth}
        \centering
        \includegraphics[width=\textwidth]{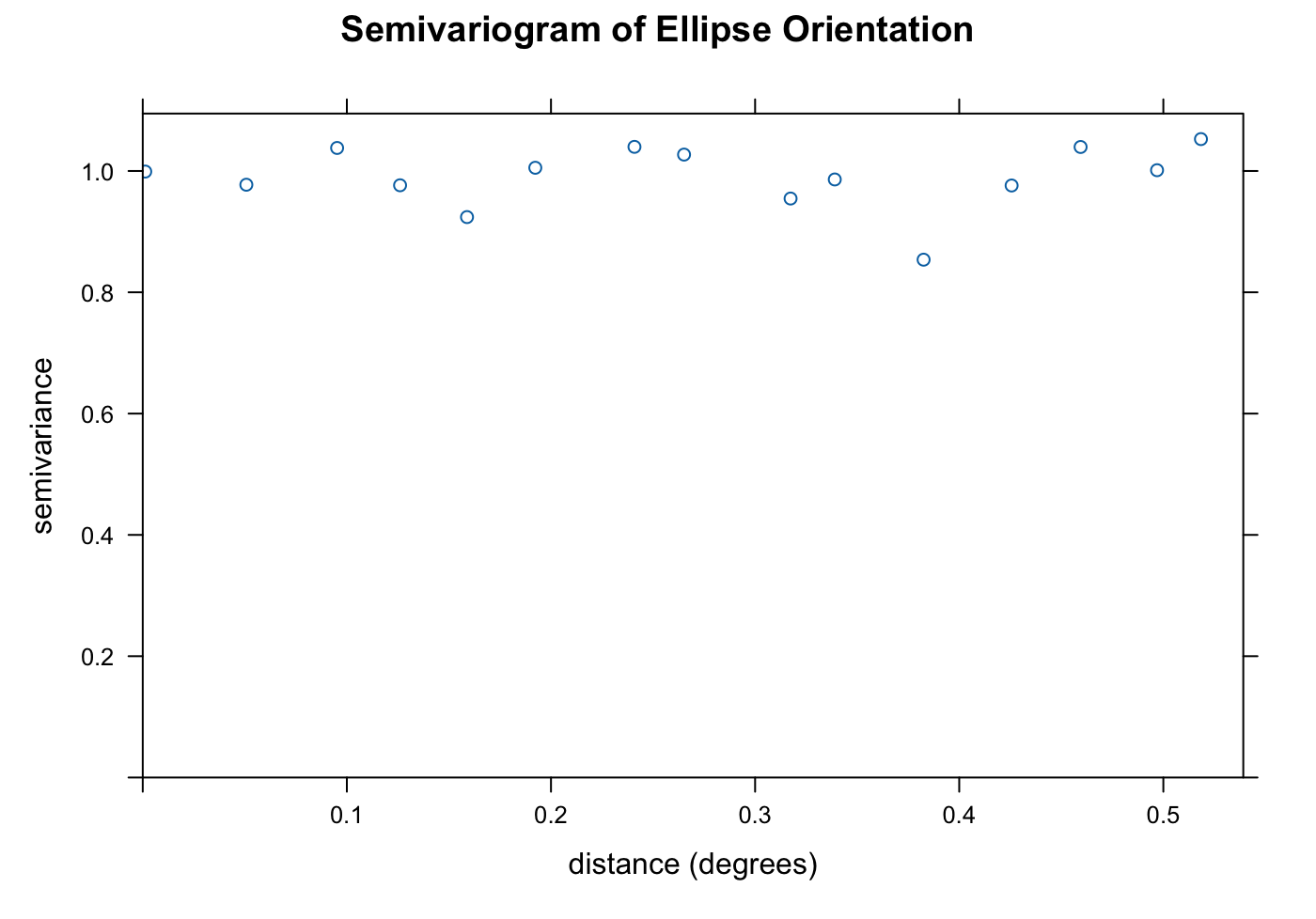}
        \caption{Ellipse orientation}
    \end{subfigure}

    \caption{
    Empirical semivariograms of fitted ellipse parameters computed across all
    pyromes and time periods.
    }
    \label{fig:ellipse_variograms_supp}
\end{figure}

Figure~\ref{fig:ellipse_variograms_supp} shows empirical semivariograms of fitted ellipse parameters computed across all pyromes and time periods.

\section{Additional Pyrome Results}
\label{sec:supp_additional_pyromes}

Figures in this section provide model fits and diagnostics for pyromes not displayed in the main manuscript.

\subsection{Temporal VPDa Anomalies}

\begin{figure}[H]
    \centering    
    \includegraphics[width=0.8\textwidth]{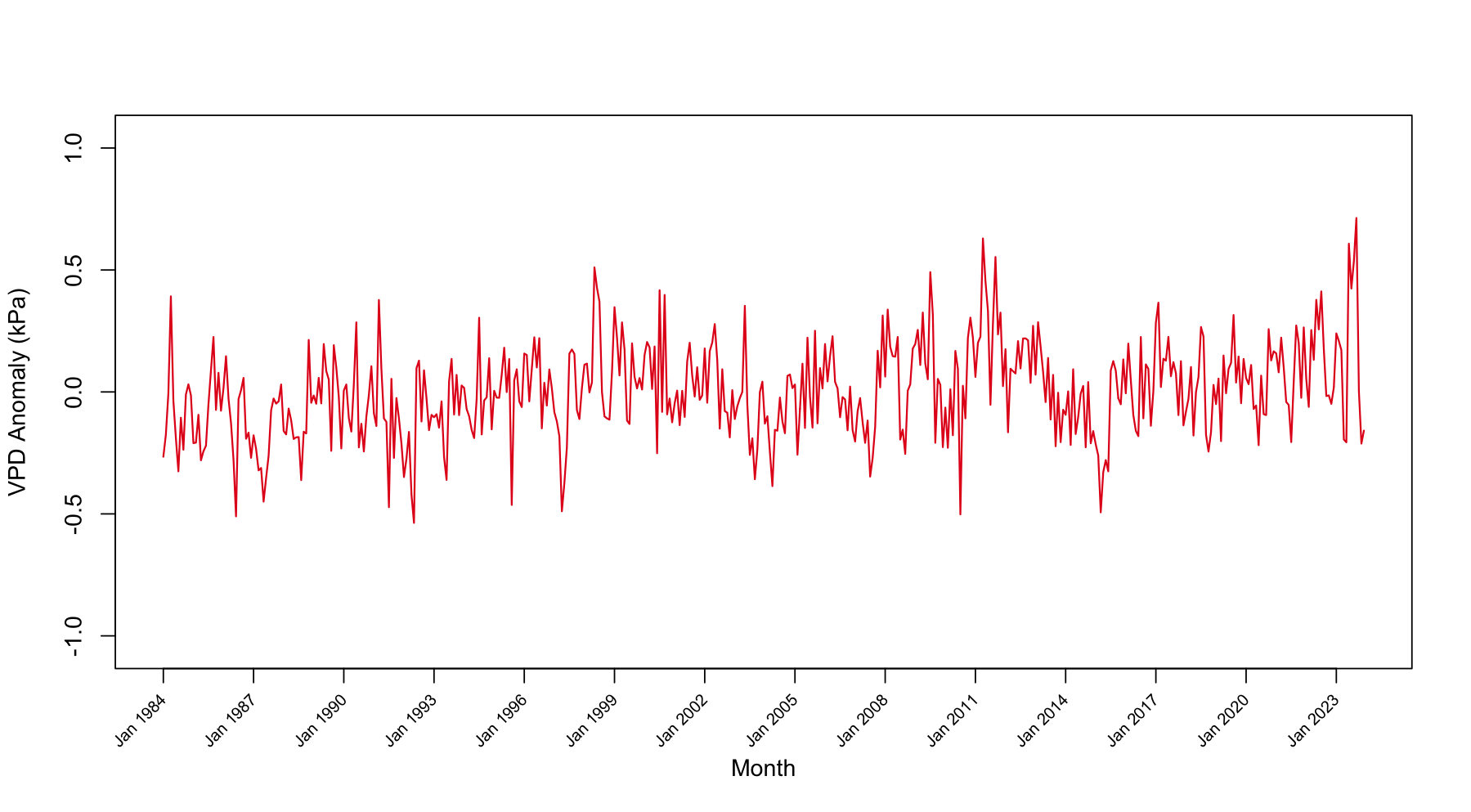}
    \caption{Monthly VPDa in CCV.}
\end{figure}

\begin{figure}[H]
    \centering    
    \includegraphics[width=0.8\textwidth]{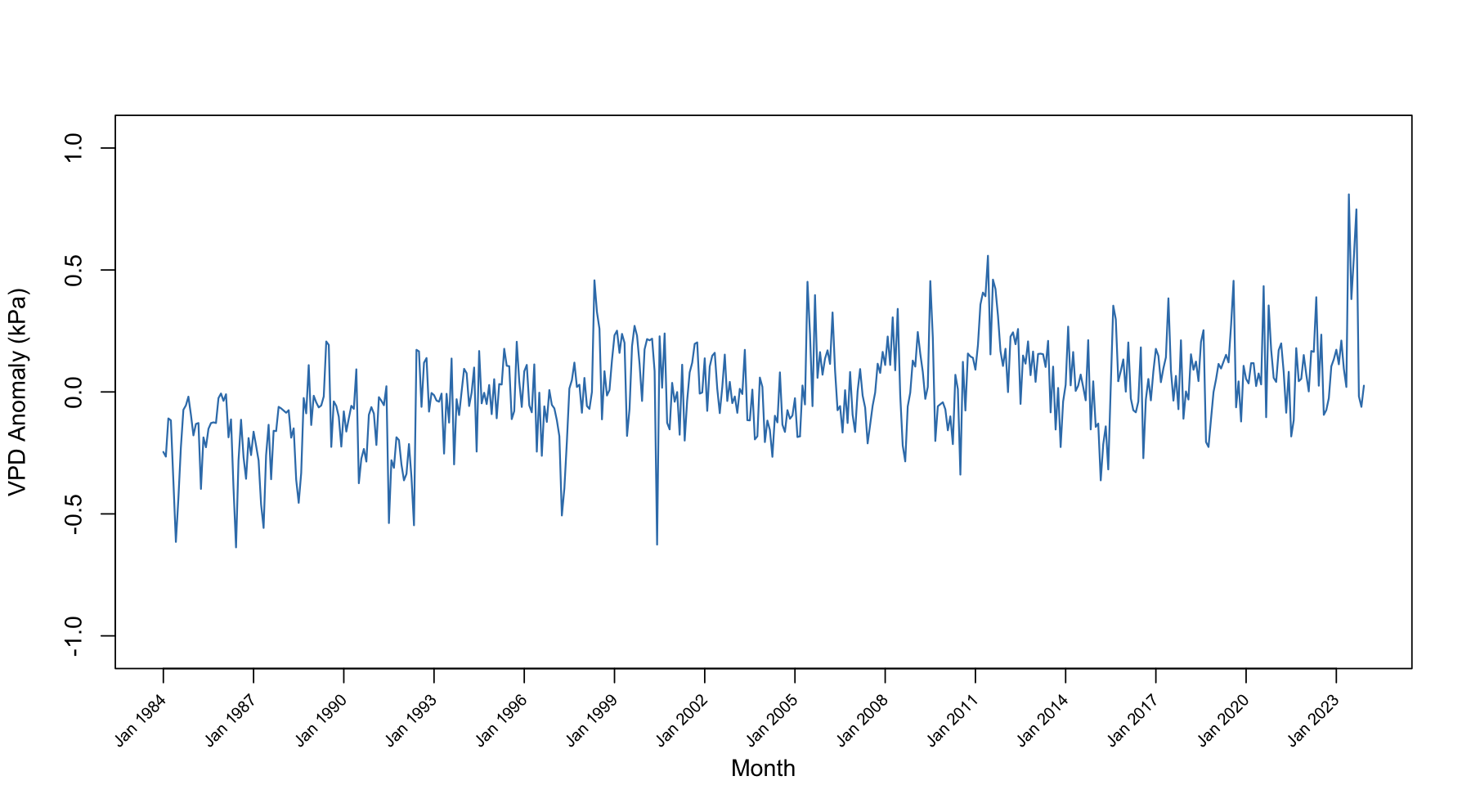}
    \caption{Monthly VPDa in EC.}
\end{figure}

\begin{figure}[H]
    \centering    
    \includegraphics[width=0.8\textwidth]{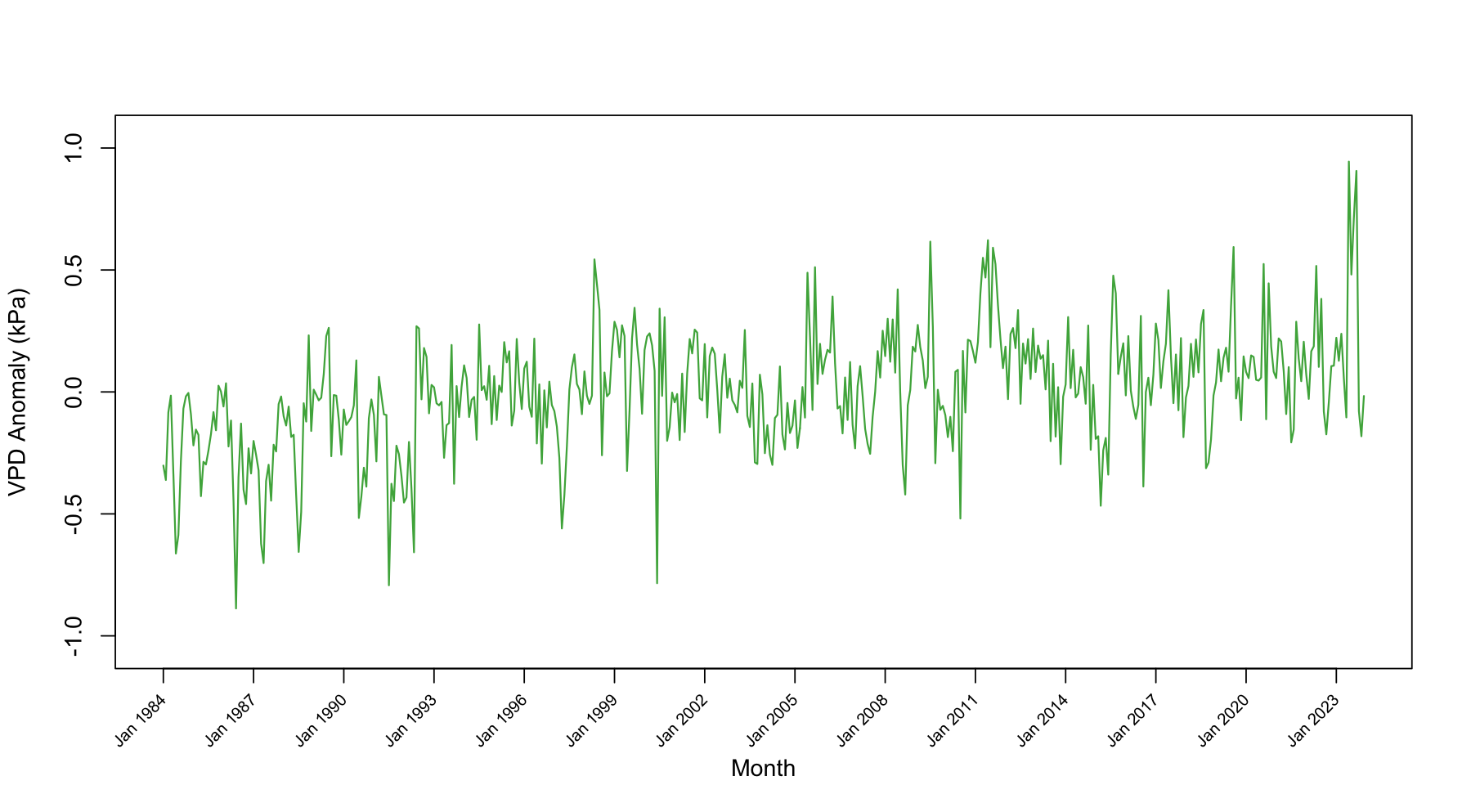}
    \caption{Monthly VPDa in HLP.}
\end{figure}

\begin{figure}[H]
    \centering    
    \includegraphics[width=0.8\textwidth]{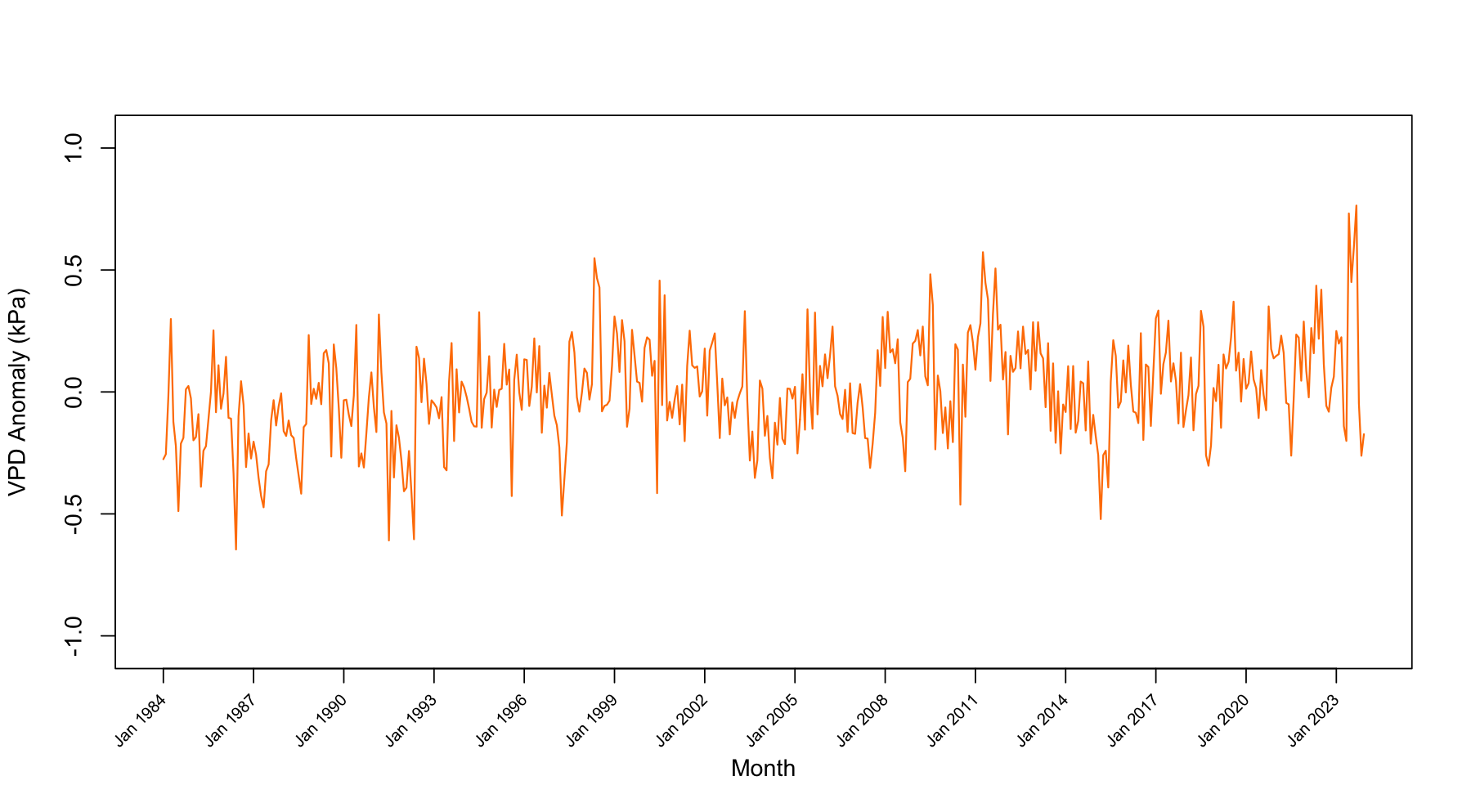}
    \caption{Monthly VPDa in NCCFACM.}
\end{figure}

\begin{figure}[H]
    \centering    
    \includegraphics[width=0.8\textwidth]{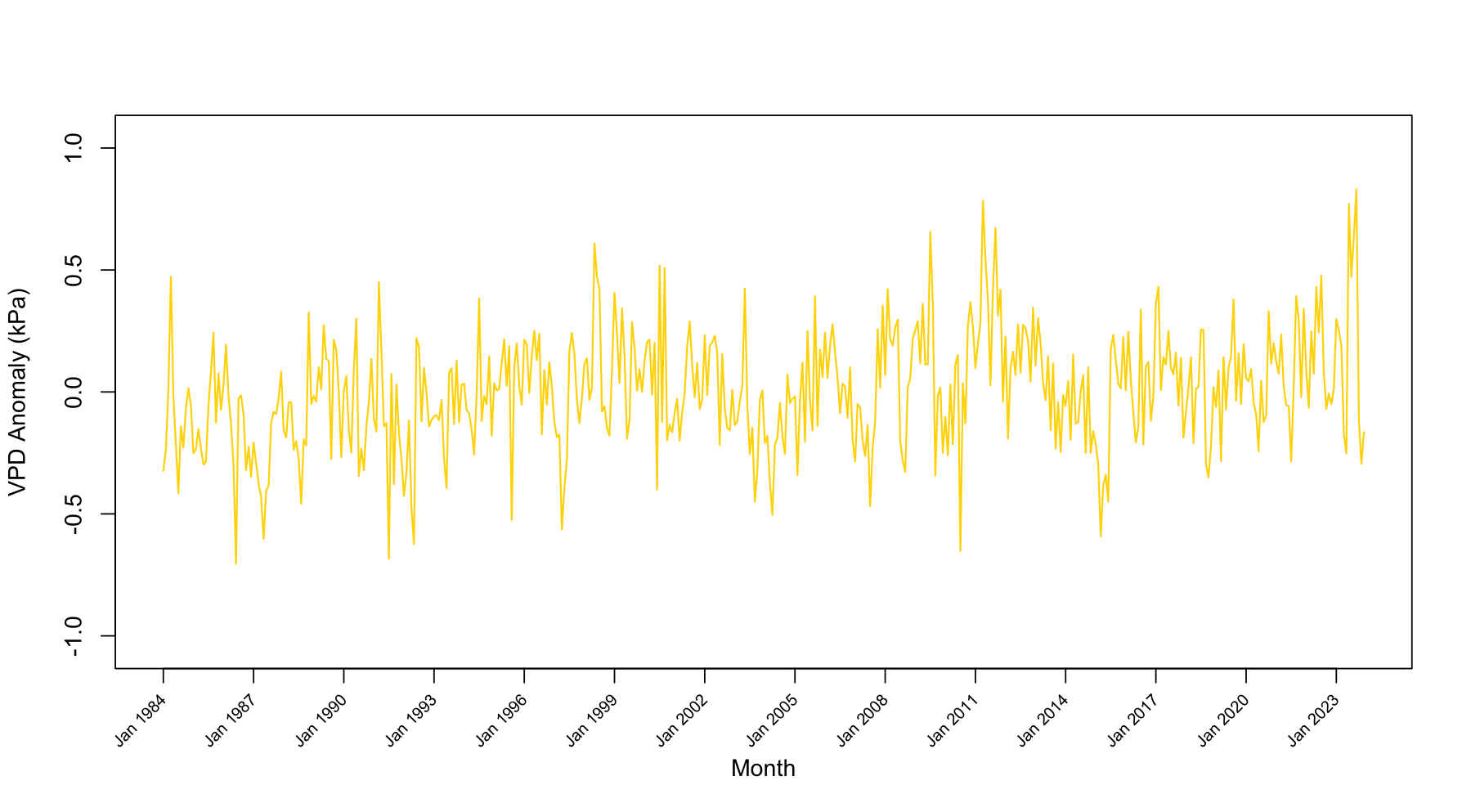}
    \caption{Monthly VPDa in NSN.}
\end{figure}

\begin{figure}[H]
    \centering    
    \includegraphics[width=0.8\textwidth]{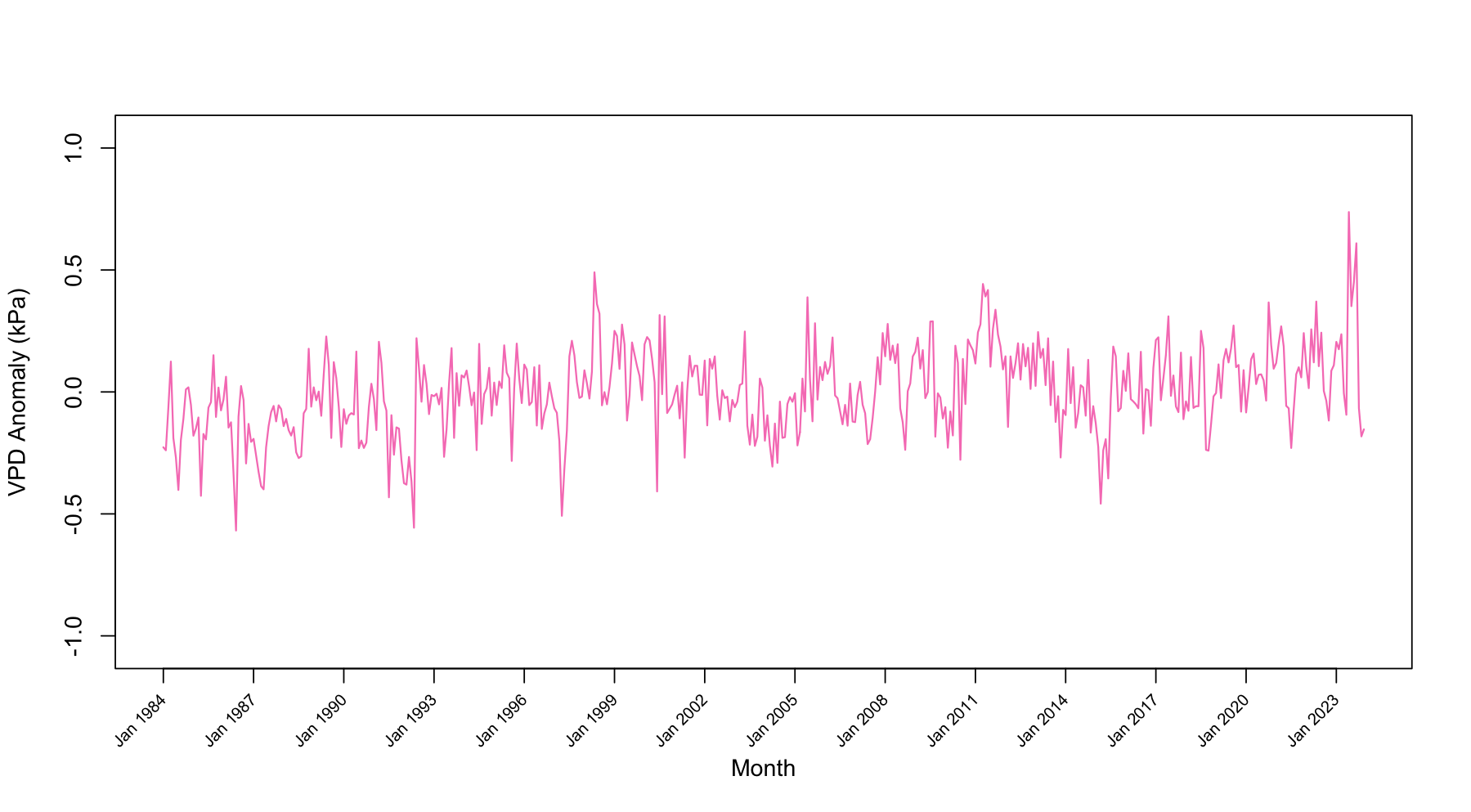}
    \caption{Monthly VPDa in ONCRR.}
\end{figure}

\begin{figure}[H]
    \centering    
    \includegraphics[width=0.8\textwidth]{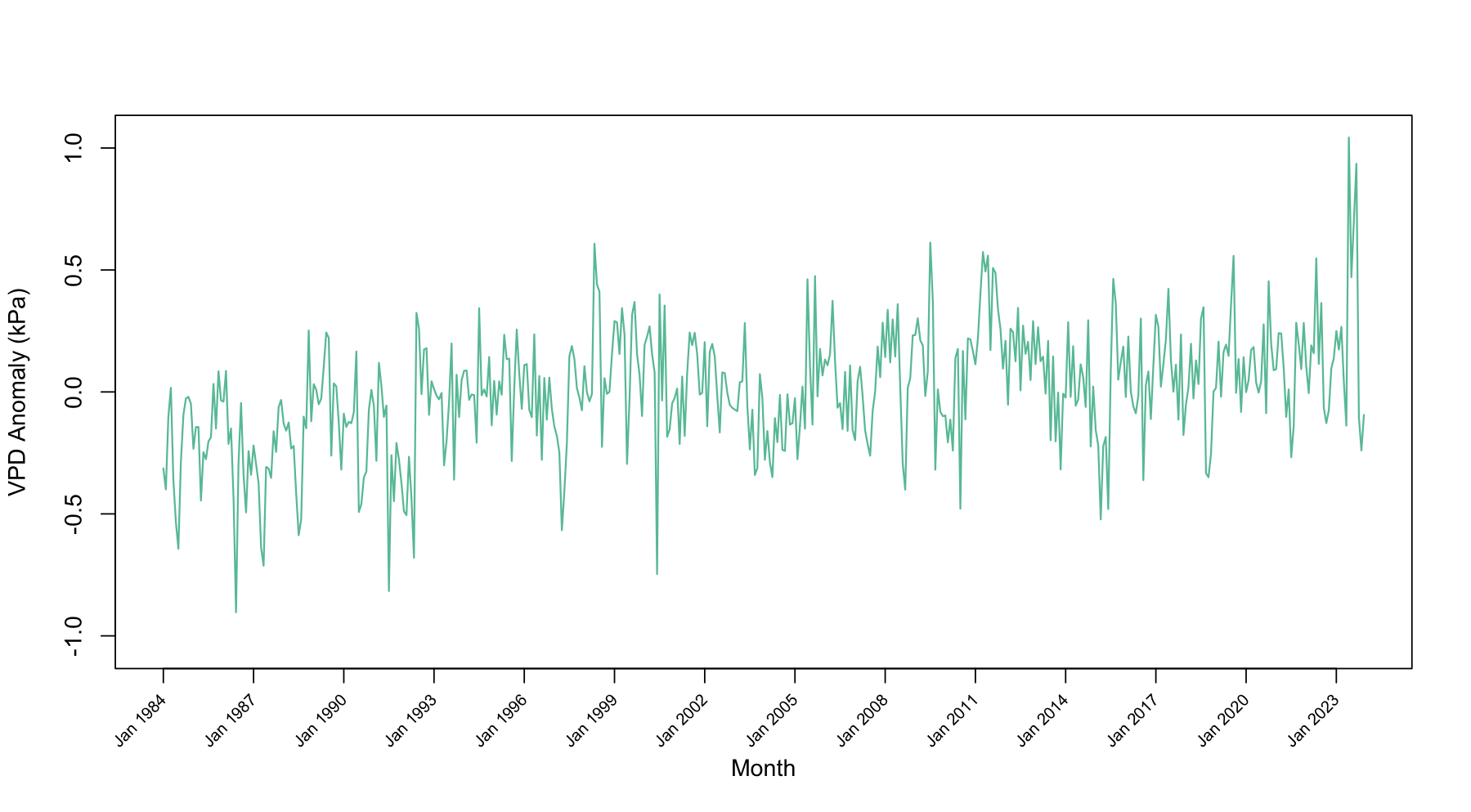}
    \caption{Monthly VPDa in SC.}
\end{figure}

\begin{figure}[H]
    \centering    
    \includegraphics[width=0.8\textwidth]{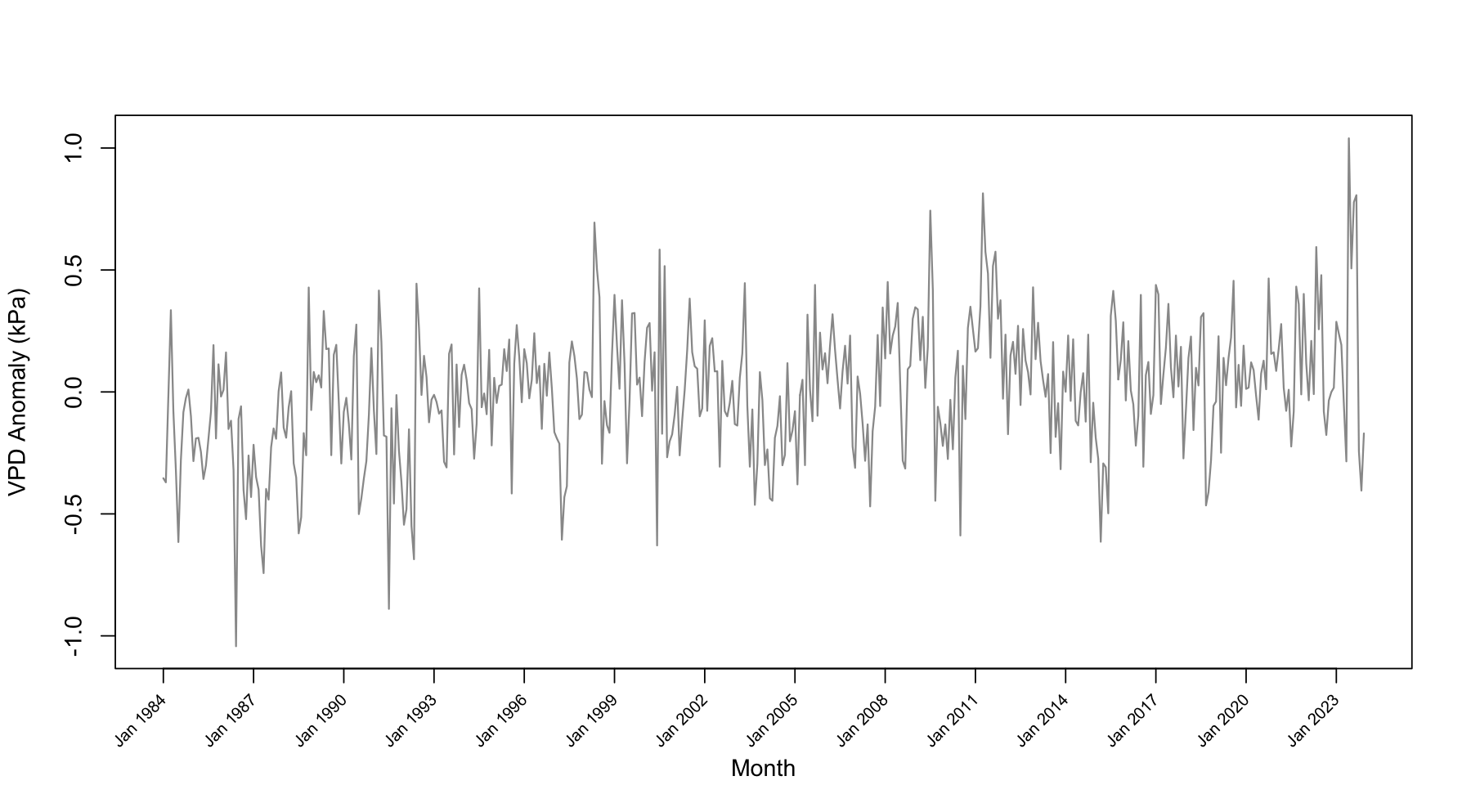}
    \caption{Monthly VPDa in SNIRHB.}
\end{figure}

\subsection{Occurrence Model Parameter Estimates}

\begin{table}[H]
\centering
\caption{CCV point process model estimates and standard errors.}
\begin{tabular}{lrrrr}
\hline
\textbf{Covariate} & \textbf{Estimate} & \textbf{Std. Error} & \textbf{z-value} & \textbf{p-value} \\
\hline
Intercept      & -7.03 & 2.39 & -2.93 & 0.003 \\
Latitude       & 5.25 & 3.61 & 1.45 & 0.146 \\
Longitude      & 7.16 & 4.60 & 1.56 & 0.119 \\
Lat--Lon Interaction & -3.98 & 12.90 & -0.31 & 0.757 \\
Latitude$^2$   & -4.72 & 5.36 & -0.88 & 0.378 \\
Longitude$^2$  & 1.49 & 10.22 & 0.15 & 0.884 \\
Elevation      & 0.38 & 0.14 & 2.67 & 0.008 \\
Monthly VPDa Anomaly & 0.25 & 0.25 & 0.98 & 0.327 \\
$\sin(2\pi (\text{month}) / 12)$ & -1.31 & 0.50 & -2.63 & 0.009 \\
$\cos(2\pi (\text{month}) / 12)$ & -0.77 & 0.41 & -1.87 & 0.061 \\
\hline
\end{tabular}
\end{table}

\begin{table}[H]
\centering
\caption{EC point process model estimates and standard errors.}
\label{tab:results_table}
\begin{tabular}{lrrrr}
\hline
\textbf{Covariate} & \textbf{Estimate} & \textbf{Std. Error} & \textbf{z-value} & \textbf{p-value} \\
\hline
Intercept      & -3.46 & 0.25 & -14.13 & $<0.001$ \\
Latitude       & 0.84 & 0.39 & 2.14 & 0.032 \\
Longitude      & -0.00 & 0.39 & -0.00 & 0.998 \\
Lat--Lon Interaction & -2.54 & 0.96 & -2.65 & 0.008 \\
Latitude$^2$   & 0.54 & 0.81 & 0.66 & 0.509 \\
Longitude$^2$  & -0.84 & 0.36 & -2.32 & 0.020 \\
Elevation      & 0.15 & 0.19 & 0.76 & 0.450 \\
Monthly VPDa Anomaly & 0.03 & 0.09 & 0.28 & 0.778 \\
$\sin(2\pi (\text{month}) / 12)$ & -0.91 & 0.15 & -5.90 & $<0.001$ \\
$\cos(2\pi (\text{month}) / 12)$ & -0.91 & 0.15 & -5.90 & $<0.001$ \\
\hline
\end{tabular}
\end{table}

\begin{table}[H]
\centering
\caption{HLP point process model estimates and standard errors.}
\label{tab:results_table}
\begin{tabular}{lrrrr}
\hline
\textbf{Covariate} & \textbf{Estimate} & \textbf{Std. Error} & \textbf{z-value} & \textbf{p-value} \\
\hline
Intercept      & -6.17 & 1.43 & -4.30 & $<0.001$ \\
Latitude       & -2.06 & 1.26 & -1.64 & 0.102 \\
Longitude      & 6.95 & 5.23 & 1.33 & 0.184 \\
Lat--Lon Interaction & 8.16 & 10.31 & 0.79 & 0.429 \\
Latitude$^2$   & 0.63 & 1.30 & 0.49 & 0.627 \\
Longitude$^2$  & -29.03 & 25.54 & -1.14 & 0.256 \\
Elevation      & 0.25 & 0.23 & 1.09 & 0.277 \\
Monthly VPDa Anomaly & 0.03 & 0.19 & 0.15 & 0.882 \\
$\sin(2\pi (\text{month}) / 12)$ & -2.17 & 0.72 & -3.01 & 0.003 \\
$\cos(2\pi (\text{month}) / 12)$ & -1.97 & 0.67 & -2.93 & 0.003 \\
\hline
\end{tabular}
\end{table}

\begin{table}[H]
\centering
\caption{NCCFACM point process model estimates and standard errors.}
\label{tab:results_table}
\begin{tabular}{lrrrr}
\hline
\textbf{Covariate} & \textbf{Estimate} & \textbf{Std. Error} & \textbf{z-value} & \textbf{p-value} \\
\hline
Intercept      & -2.96 & 0.19 & -15.70 & $<0.001$ \\
Latitude       & -0.04 & 0.20 & -0.20 & 0.842 \\
Longitude      & 1.38 & 0.46 & 2.99 & 0.003 \\
Lat--Lon Interaction & -0.88 & 0.62 & -1.43 & 0.154 \\
Latitude$^2$   & 0.12 & 0.27 & 0.42 & 0.671 \\
Longitude$^2$  & 0.83 & 1.02 & 0.81 & 0.419 \\
Elevation      & 0.49 & 0.11 & 4.62 & $<0.001$ \\
Monthly VPDa Anomaly & -0.13 & 0.09 & -1.34 & 0.181 \\
$\sin(2\pi (\text{month}) / 12)$ & -1.16 & 0.15 & -7.53 & $<0.001$ \\
$\cos(2\pi (\text{month}) / 12)$ & -0.47 & 0.12 & -3.80 & $<0.001$ \\
\hline
\end{tabular}
\end{table}

\begin{table}[H]
\centering
\caption{NSN point process model estimates and standard errors.}
\label{tab:results_table}
\begin{tabular}{lrrrr}
\hline
\textbf{Covariate} & \textbf{Estimate} & \textbf{Std. Error} & \textbf{z-value} & \textbf{p-value} \\
\hline
Intercept      & -4.14 & 0.35 & -11.68 & $<0.001$ \\
Latitude       & 0.99 & 0.33 & 3.04 & 0.002 \\
Longitude      & 1.10 & 0.64 & 1.72 & 0.085 \\
Lat--Lon Interaction & 1.06 & 1.09 & 0.97 & 0.331 \\
Latitude$^2$   & 0.41 & 0.48 & 0.84 & 0.399 \\
Longitude$^2$  & 2.07 & 0.70 & 2.95 & 0.003 \\
Elevation      & -0.61 & 0.22 & -2.74 & 0.006 \\
Monthly VPDa Anomaly & -0.05 & 0.09 & -0.52 & 0.602 \\
$\sin(2\pi (\text{month}) / 12)$ & -1.71 & 0.24 & -7.21 & $<0.001$ \\
$\cos(2\pi (\text{month}) / 12)$ & -1.33 & 0.21 & -6.48 & $<0.001$ \\
\hline
\end{tabular}
\end{table}

\begin{table}[H]
\centering
\caption{ONCRR point process model estimates and standard errors.}
\label{tab:results_table}
\begin{tabular}{lrrrr}
\hline
\textbf{Covariate} & \textbf{Estimate} & \textbf{Std. Error} & \textbf{z-value} & \textbf{p-value} \\
\hline
Intercept      & -4.00 & 0.63 & -6.38 & $<0.001$ \\
Latitude       & -0.63 & 0.37 & -1.71 & 0.087 \\
Longitude      & -0.76 & 0.78 & -0.97 & 0.332 \\
Lat--Lon Interaction & 5.07 & 2.15 & 2.35 & 0.019 \\
Latitude$^2$   & 0.93 & 0.51 & 1.81 & 0.071 \\
Longitude$^2$  & 4.39 & 2.04 & 2.15 & 0.031 \\
Elevation      & 0.59 & 0.20 & 2.97 & 0.003 \\
Monthly VPDa Anomaly & 0.09 & 0.14 & 0.68 & 0.496 \\
$\sin(2\pi (\text{month}) / 12)$ & -1.50 & 0.32 & -4.63 & $<0.001$ \\
$\cos(2\pi (\text{month}) / 12)$ & -1.57 & 0.33 & -4.73 & $<0.001$ \\
\hline
\end{tabular}
\end{table}

\begin{table}[H]
\centering
\caption{SC point process model estimates and standard errors.}
\label{tab:results_table}
\begin{tabular}{lrrrr}
\hline
\textbf{Covariate} & \textbf{Estimate} & \textbf{Std. Error} & \textbf{z-value} & \textbf{p-value} \\
\hline
Intercept      & -5.74 & 0.83 & -6.87 & $<0.001$ \\
Latitude       & -0.21 & 0.50 & -0.42 & 0.676 \\
Longitude      & 0.29 & 0.69 & 0.42 & 0.675 \\
Lat--Lon Interaction & -0.37 & 2.26 & -0.16 & 0.870 \\
Latitude$^2$   & -1.15 & 0.95 & -1.21 & 0.225 \\
Longitude$^2$  & -0.50 & 1.67 & -0.30 & 0.764 \\
Elevation      & -0.37 & 0.22 & -1.68 & 0.093 \\
Monthly VPDa Anomaly & 0.22 & 0.13 & 1.66 & 0.097 \\
$\sin(2\pi (\text{month}) / 12)$ & -2.03 & 0.47 & -4.27 & $<0.001$ \\
$\cos(2\pi (\text{month}) / 12)$ & -2.12 & 0.49 & -4.33 & $<0.001$ \\
\hline
\end{tabular}
\end{table}

\begin{table}[H]
\centering
\caption{SF point process model estimates and standard errors.}
\label{tab:results_table}
\begin{tabular}{lrrrr}
\hline
\textbf{Covariate} & \textbf{Estimate} & \textbf{Std. Error} & \textbf{z-value} & \textbf{p-value} \\
\hline
Intercept      & -1.86 & 3.68 & -0.51 & 0.612 \\
Latitude       & -0.46 & 1.22 & -0.38 & 0.704 \\
Longitude      & -3.09 & 1.94 & -1.59 & 0.112 \\
Lat--Lon Interaction & 6.28 & 12.44 & 0.50 & 0.614 \\
Latitude$^2$   & 1.28 & 3.57 & 0.36 & 0.721 \\
Longitude$^2$  & 2.25 & 10.71 & 0.21 & 0.834 \\
Elevation      & 0.13 & 0.18 & 0.73 & 0.466 \\
Monthly VPDa Anomaly & -0.03 & 0.12 & -0.27 & 0.786 \\
$\sin(2\pi (\text{month}) / 12)$ & -1.58 & 0.28 & -5.56 & $<0.001$ \\
$\cos(2\pi (\text{month}) / 12)$ & -1.22 & 0.25 & -4.91 & $<0.001$ \\
\hline
\end{tabular}
\end{table}

\begin{table}[H]
\centering
\caption{SNIRHB point process model estimates and standard errors.}
\label{tab:results_table}
\begin{tabular}{lrrrr}
\hline
\textbf{Covariate} & \textbf{Estimate} & \textbf{Std. Error} & \textbf{z-value} & \textbf{p-value} \\
\hline
Intercept      & -5.62 & 1.47 & -3.83 & $<0.001$ \\
Latitude       & 2.00 & 3.08 & 0.65 & 0.515 \\
Longitude      & 0.21 & 4.13 & 0.05 & 0.960 \\
Lat--Lon Interaction & -53.53 & 32.72 & -1.64 & 0.102 \\
Latitude$^2$   & -19.58 & 16.72 & -1.17 & 0.242 \\
Longitude$^2$  & -36.54 & 27.18 & -1.34 & 0.179 \\
Elevation      & 0.89 & 0.35 & 2.57 & 0.010 \\
Monthly VPDa Anomaly & -0.47 & 0.26 & -1.77 & 0.077 \\
$\sin(2\pi (\text{month}) / 12)$ & -1.45 & 0.63 & -2.31 & 0.021 \\
$\cos(2\pi (\text{month}) / 12)$ & -1.30 & 0.61 & -2.15 & 0.032 \\
\hline
\end{tabular}
\end{table}

\subsection{Monthly Wildfire Count Distributions}

\begin{figure}[H]
    \centering
    \includegraphics[width=0.7\textwidth]{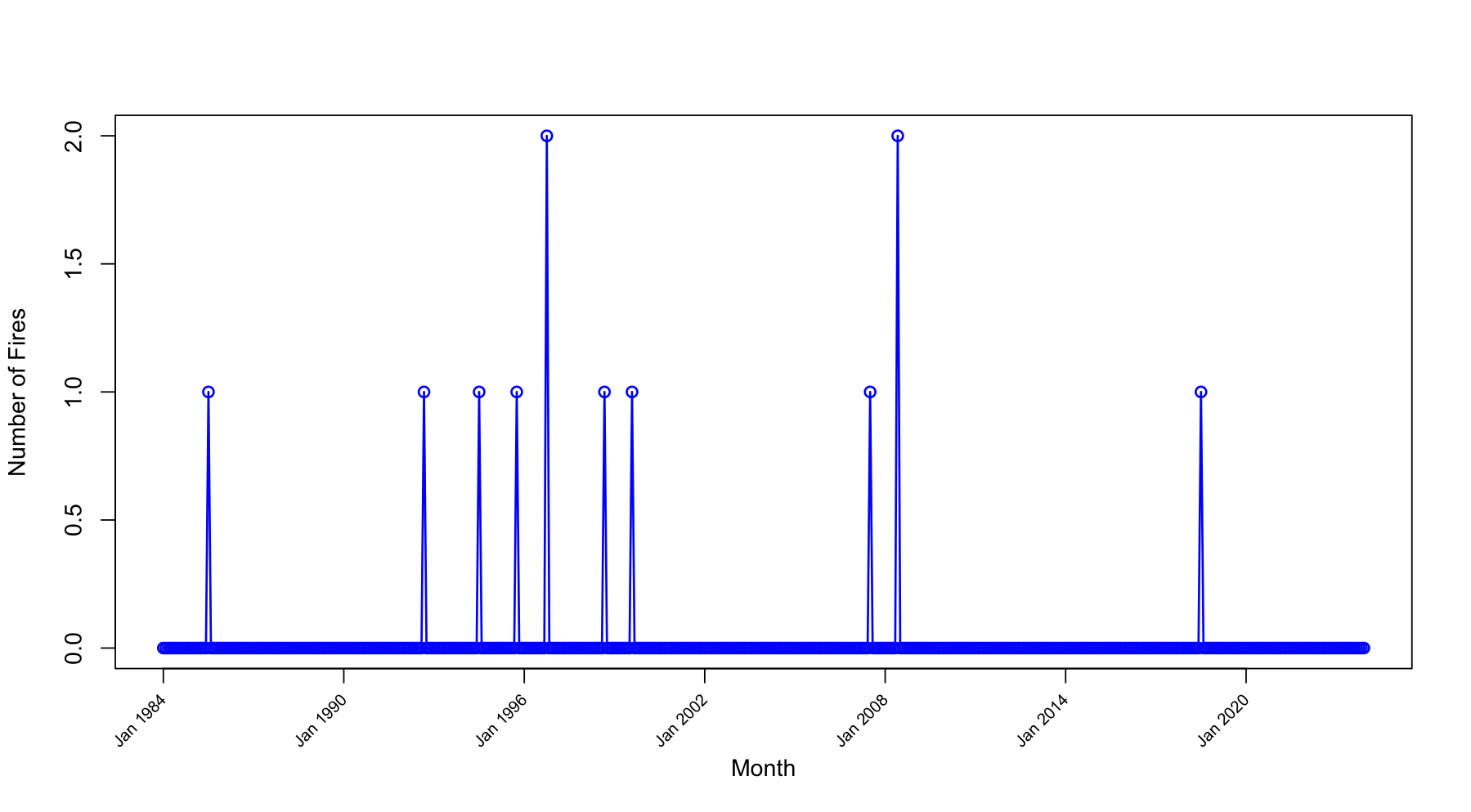}
    \caption{Monthly wildfire counts in the CCV pyrome, 1984--2023.}
\end{figure}

\begin{figure}[H]
    \centering
    \includegraphics[width=0.7\textwidth]{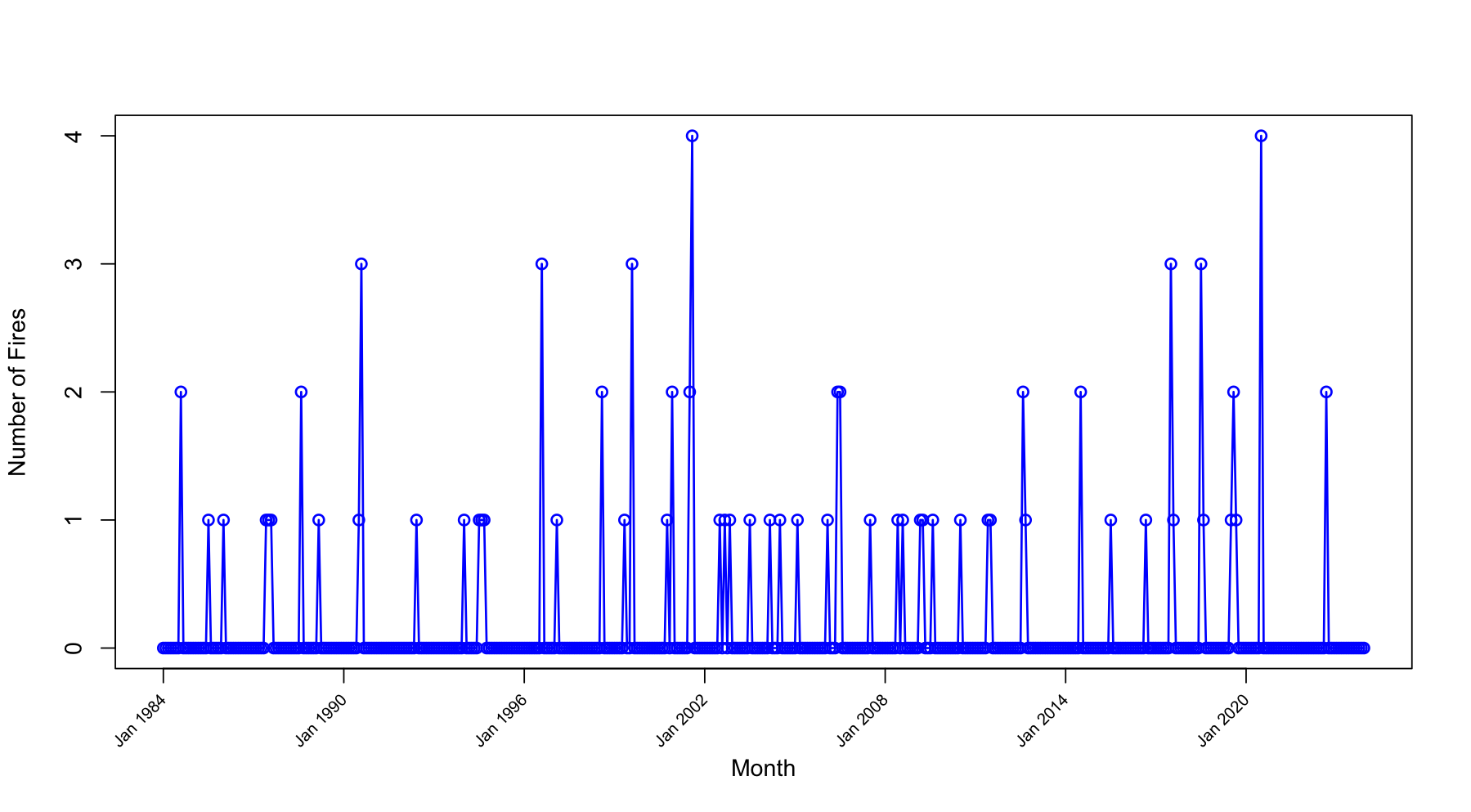}
    \caption{Monthly wildfire counts in the EC pyrome, 1984--2023.}
\end{figure}

\begin{figure}[H]
    \centering
    \includegraphics[width=0.7\textwidth]{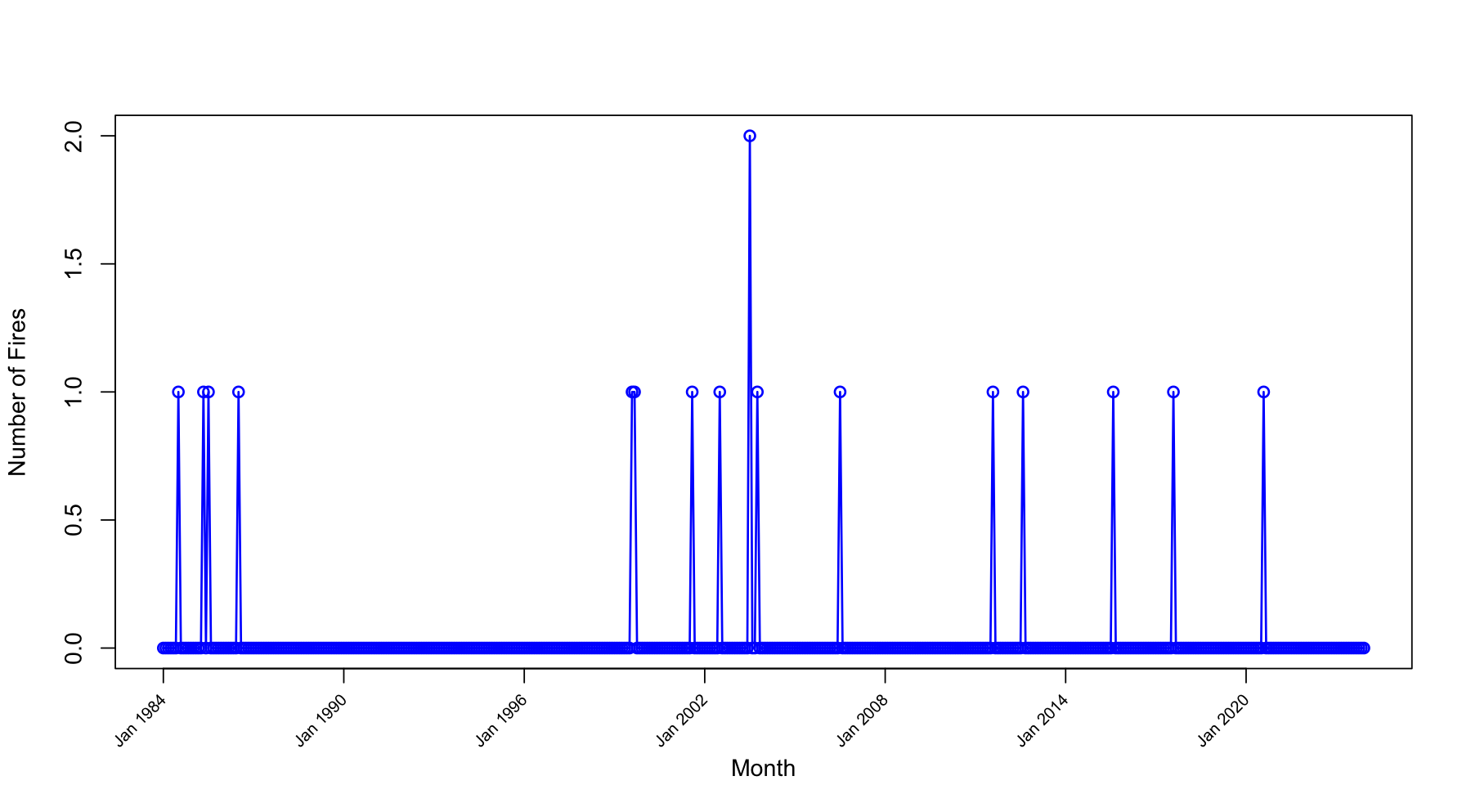}
    \caption{Monthly wildfire counts in the HLP pyrome, 1984--2023.}
\end{figure}

\begin{figure}[H]
    \centering
    \includegraphics[width=0.7\textwidth]{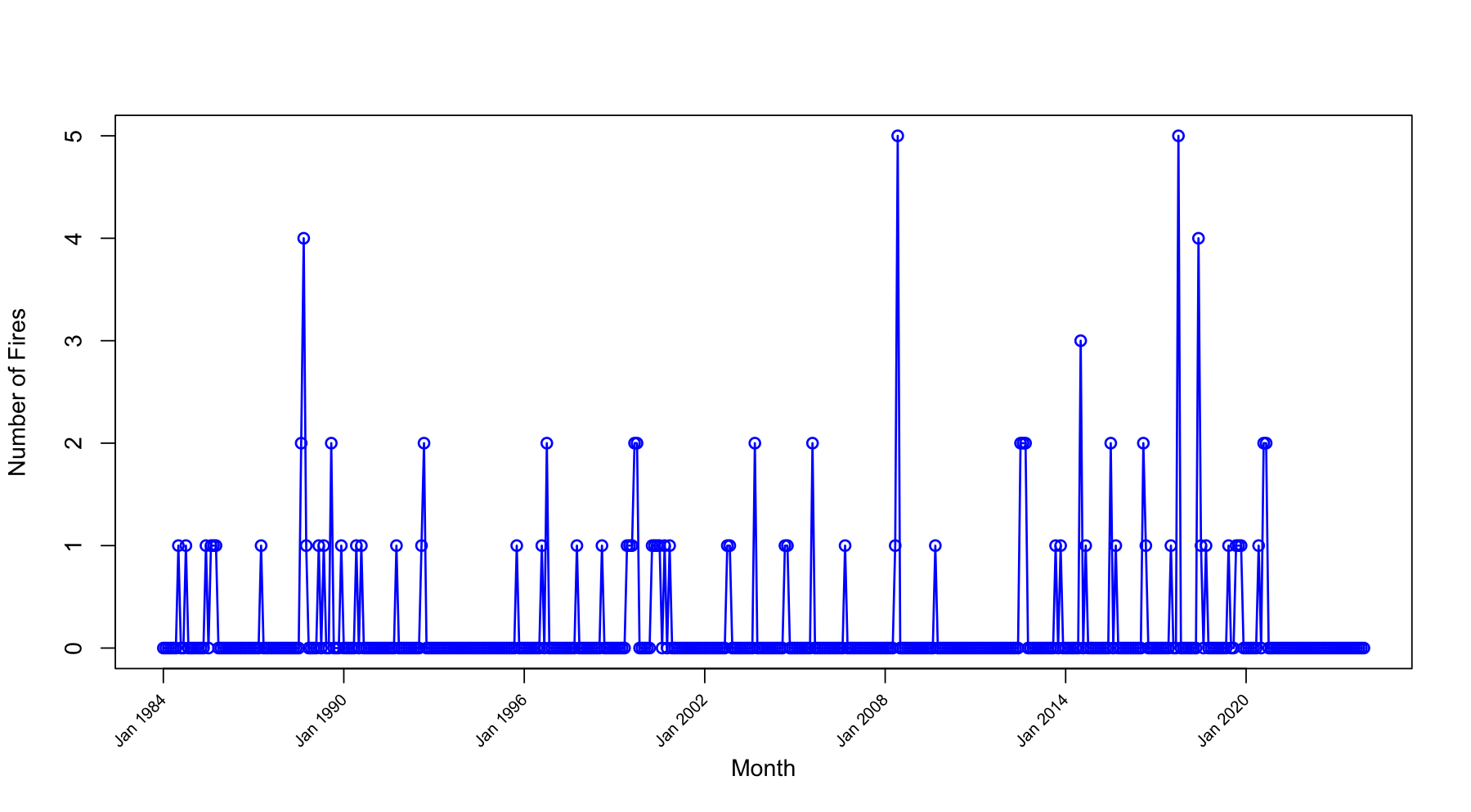}
    \caption{Monthly wildfire counts in the NCCFACM pyrome, 1984--2023.}
\end{figure}

\begin{figure}[H]
    \centering
    \includegraphics[width=0.7\textwidth]{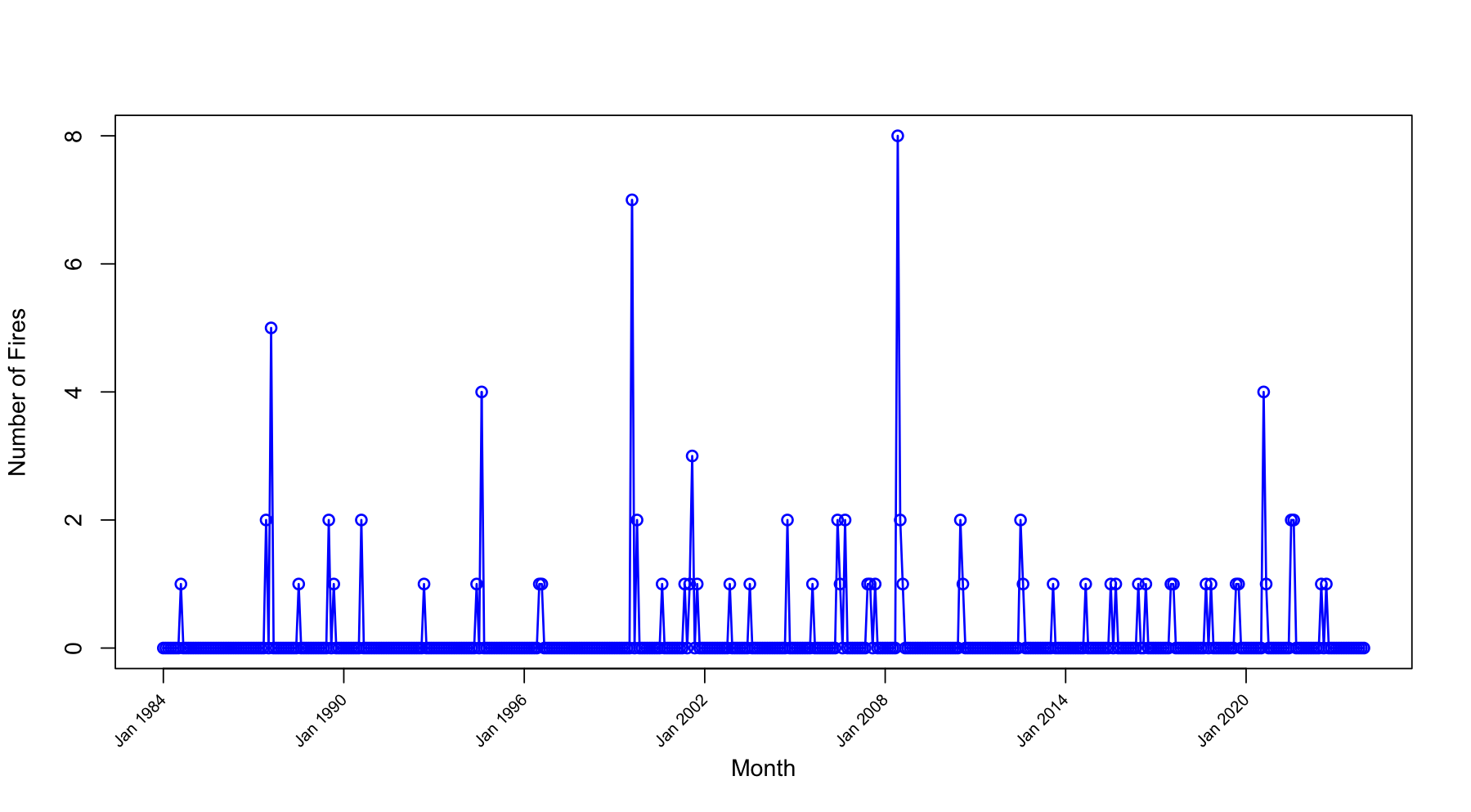}
    \caption{Monthly wildfire counts in the NSN pyrome, 1984--2023.}
\end{figure}

\begin{figure}[H]
    \centering
    \includegraphics[width=0.7\textwidth]{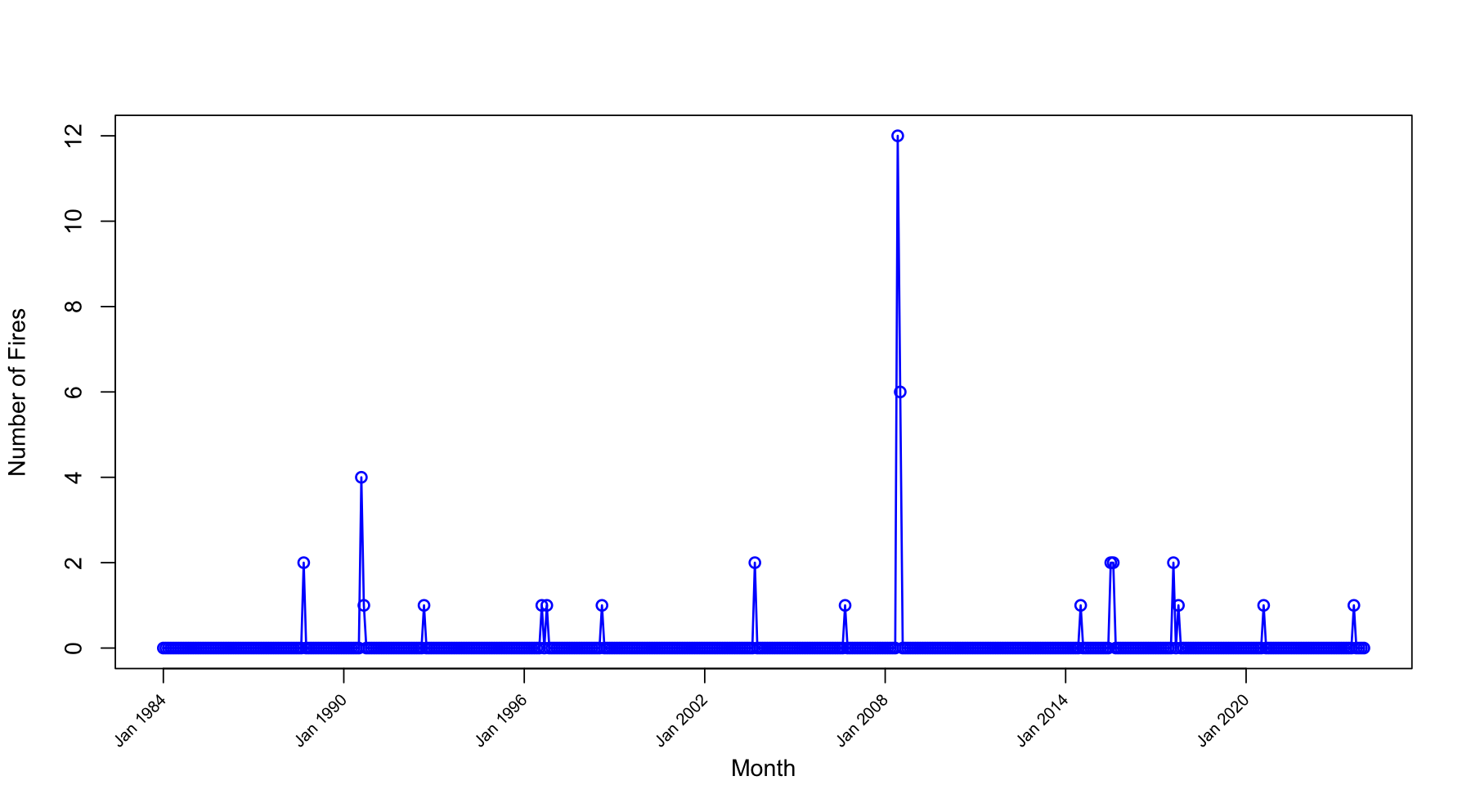}
    \caption{Monthly wildfire counts in the ONCRR pyrome, 1984--2023.}
\end{figure}

\begin{figure}[H]
    \centering
    \includegraphics[width=0.7\textwidth]{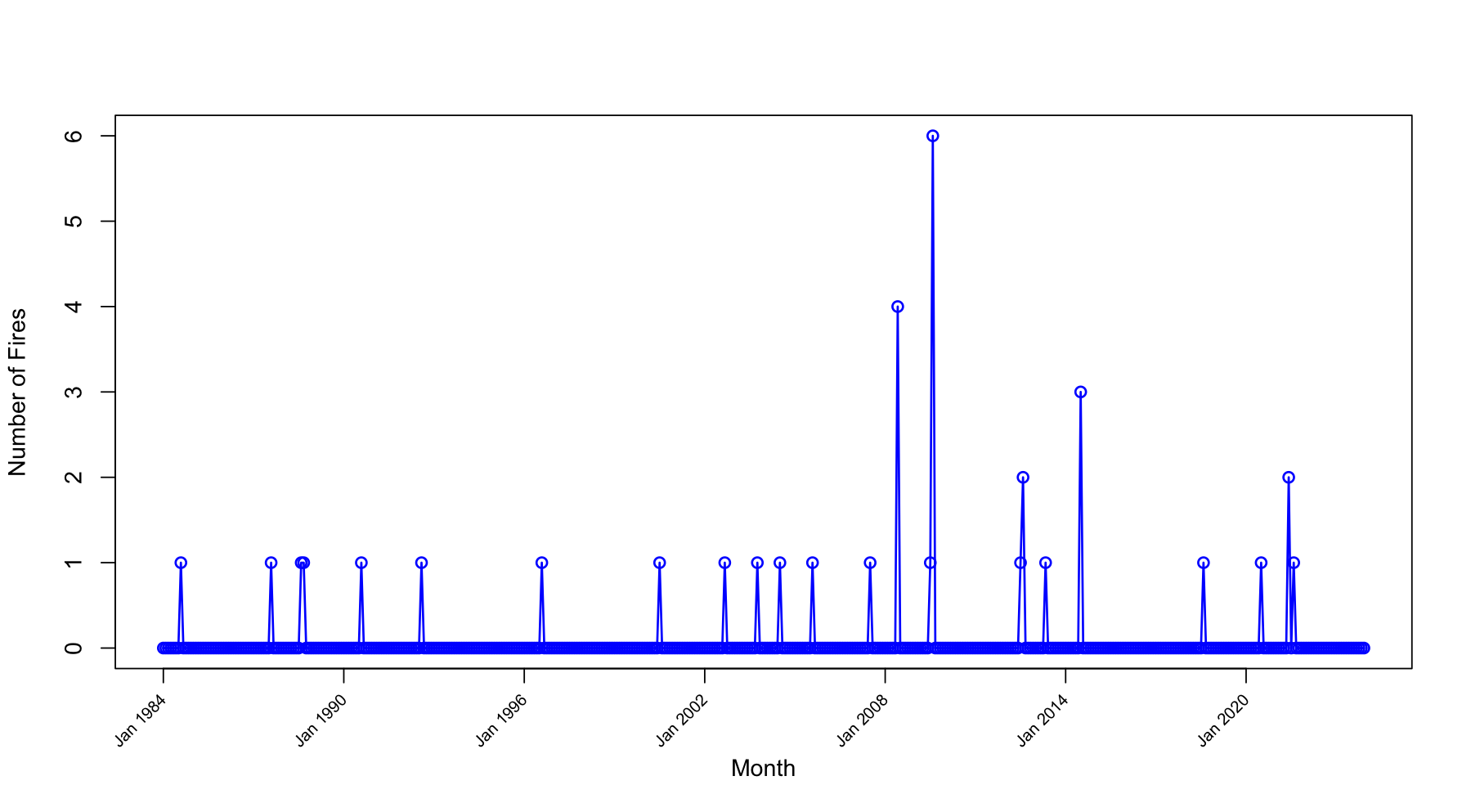}
    \caption{Monthly wildfire counts in the SC pyrome, 1984--2023.}
\end{figure}

\begin{figure}[H]
    \centering
    \includegraphics[width=0.7\textwidth]{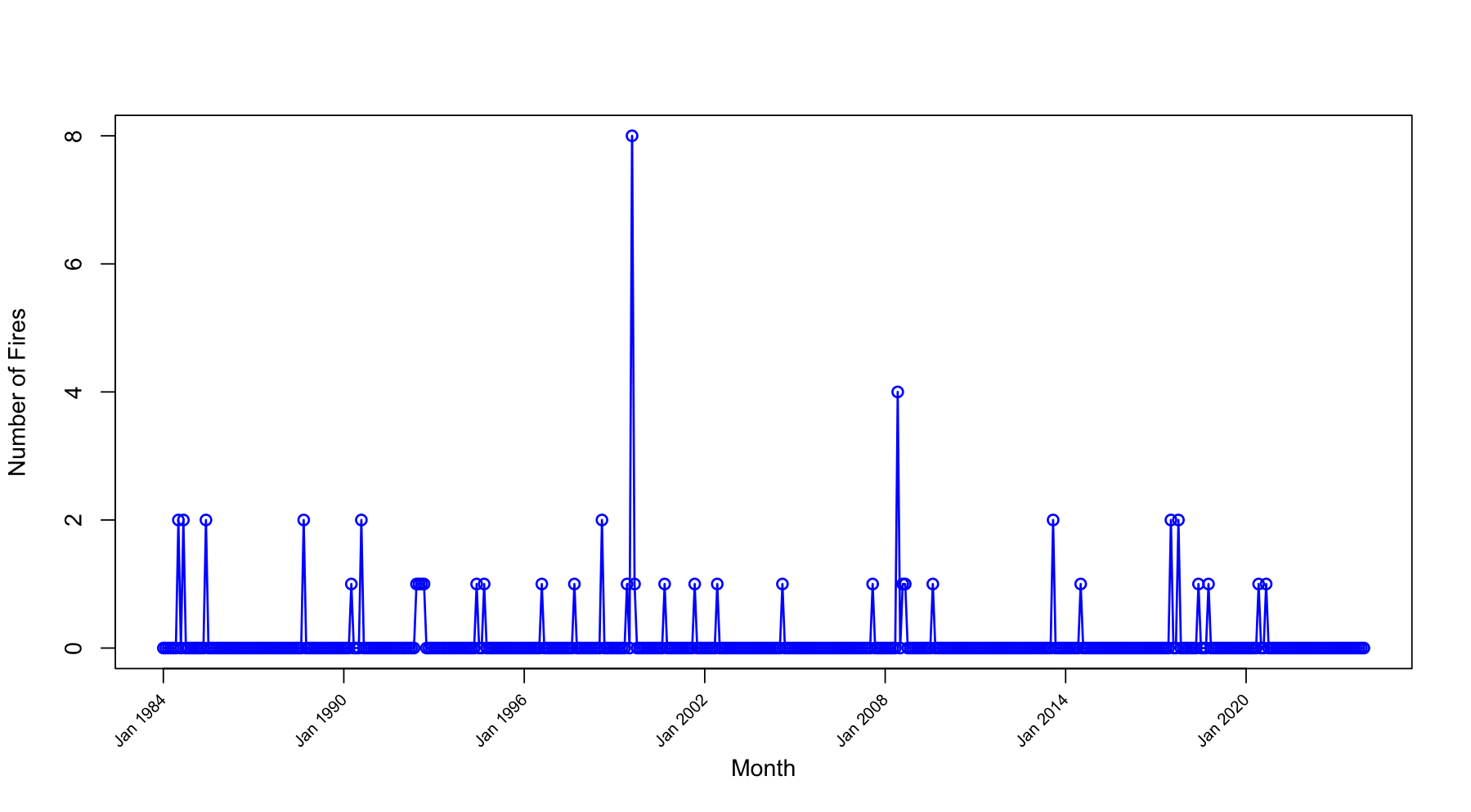}
    \caption{Monthly wildfire counts in the SF pyrome, 1984--2023.}
\end{figure}

\begin{figure}[H]
    \centering
    \includegraphics[width=0.7\textwidth]{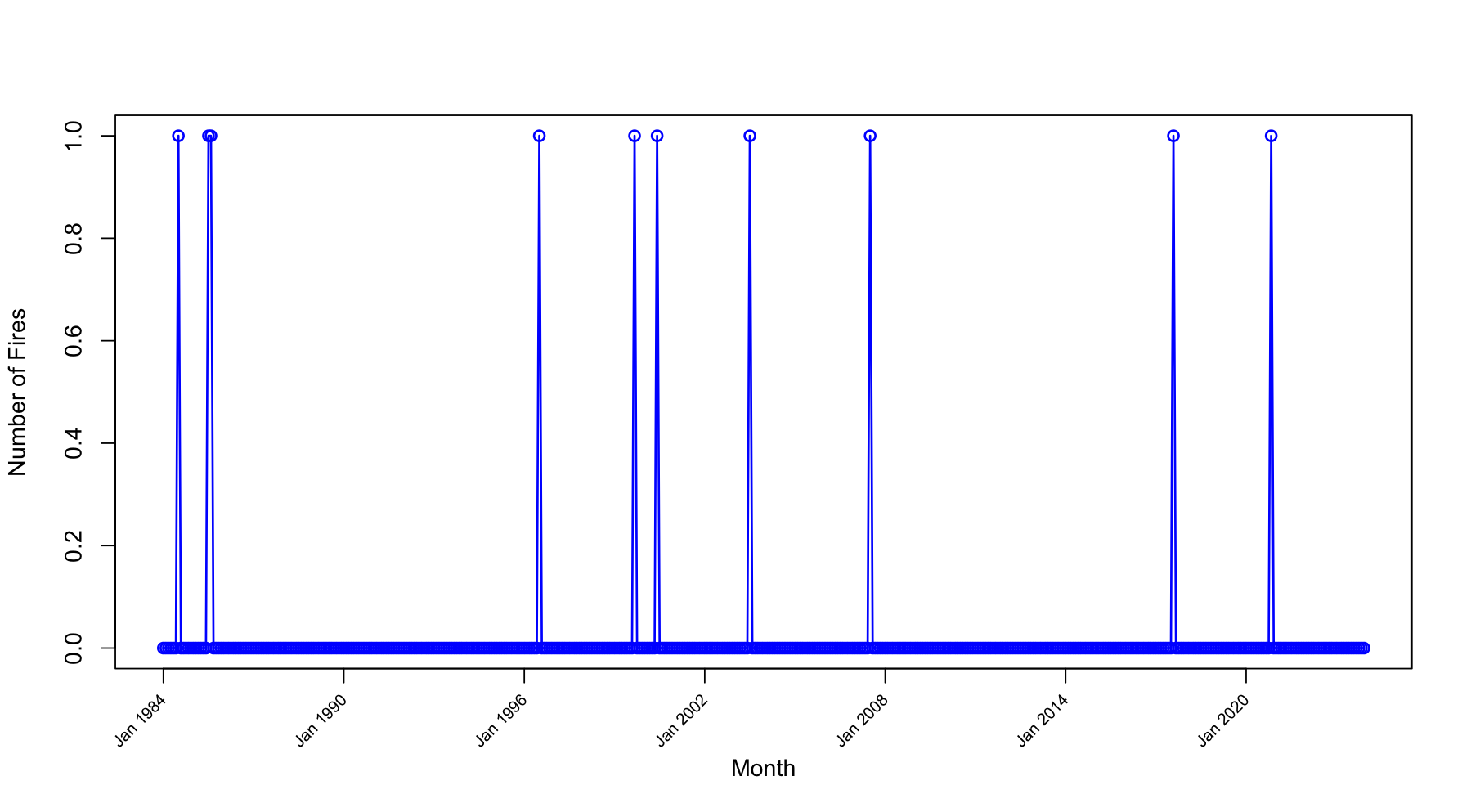}
    \caption{Monthly wildfire counts in the SNIRHB pyrome, 1984--2023.}
\end{figure}

\subsection{Shock Model Validation}

\begin{figure}[H]
    \centering
    \begin{subfigure}[b]{0.48\textwidth}
        \centering
        \includegraphics[width=\textwidth]{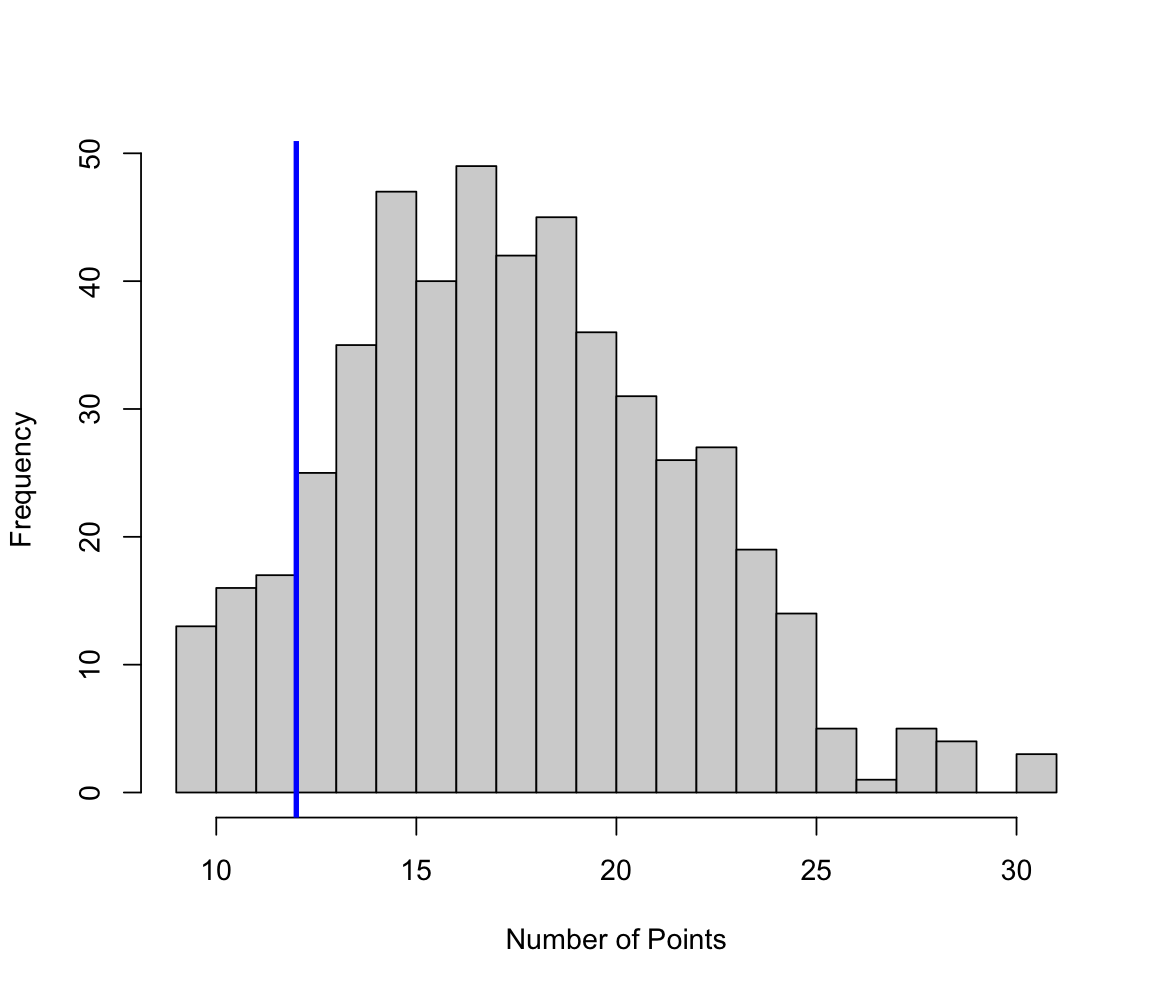}
        \caption{}
    \end{subfigure}
    \hfill
    \begin{subfigure}[b]{0.48\textwidth}
        \centering
        \includegraphics[width=\textwidth]{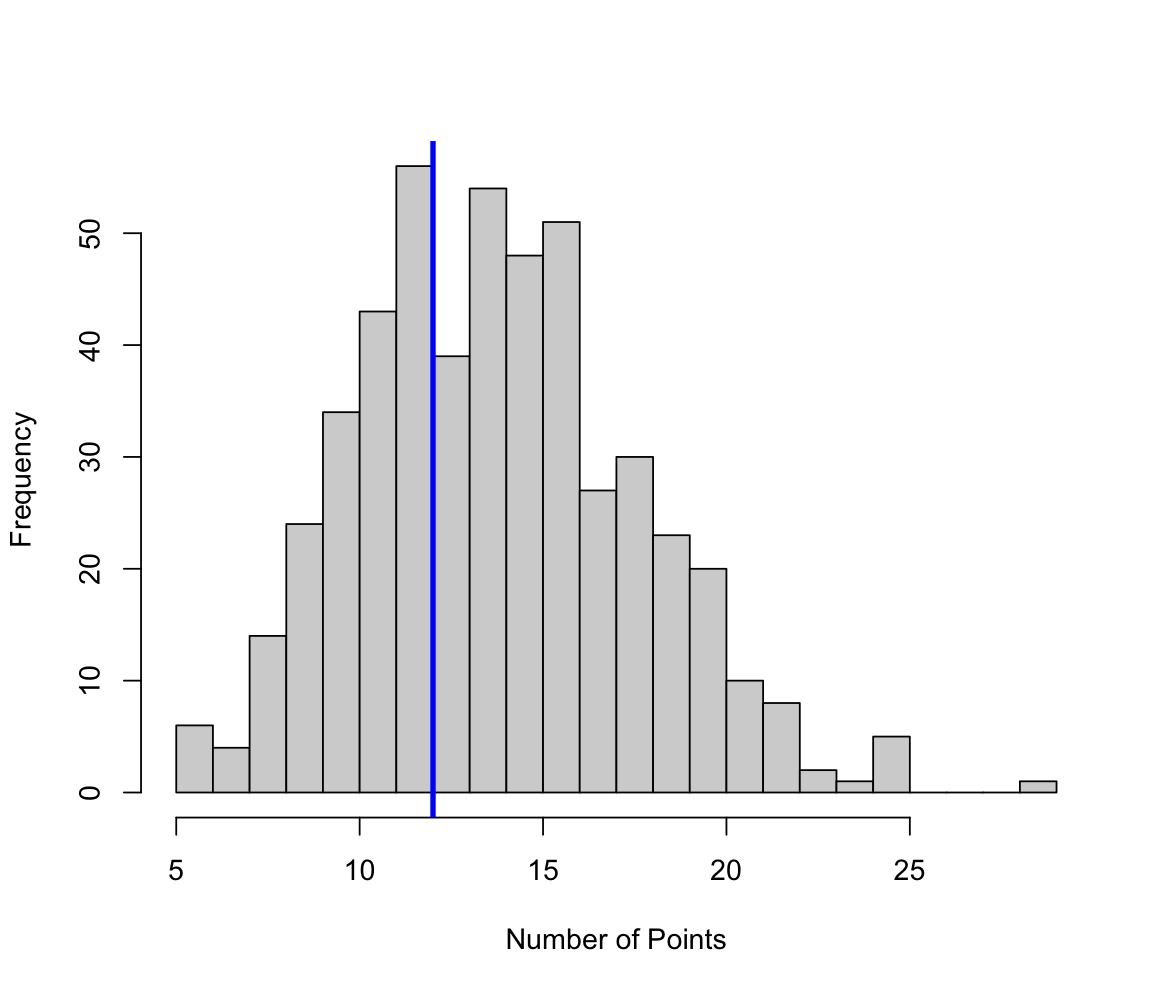}
        \caption{}
    \end{subfigure}

    \caption{CCV pyrome observed (blue line) and simulated monthly fire counts (histograms from 500 realizations) over 1984--2023. (a) Baseline model $\lambda$; (b) Full model $U_t \, \lambda$.}
\end{figure} 

\begin{figure}[H]
    \centering
    \begin{subfigure}[b]{0.48\textwidth}
        \centering
        \includegraphics[width=\textwidth]{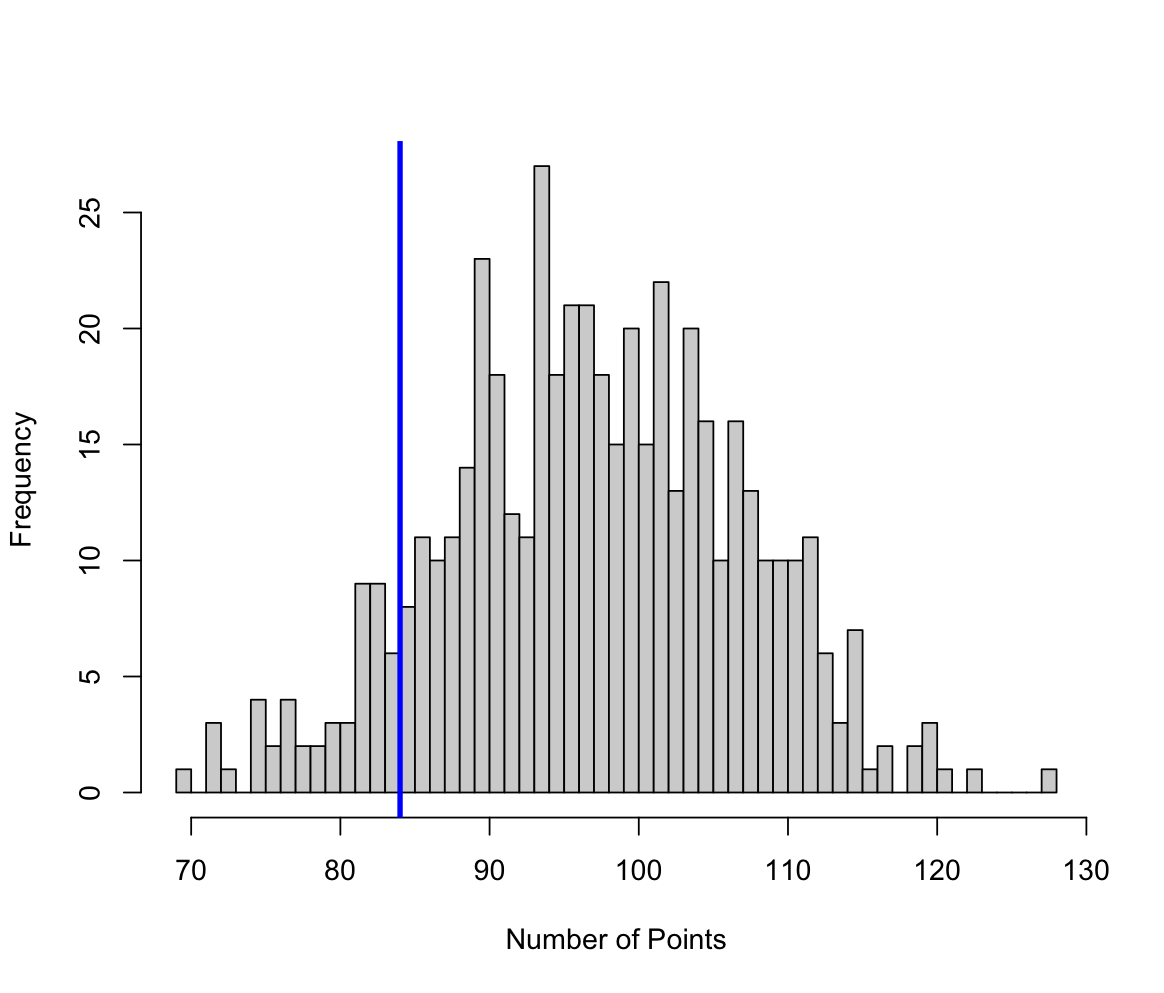}
        \caption{}
    \end{subfigure}
    \hfill
    \begin{subfigure}[b]{0.48\textwidth}
        \centering
        \includegraphics[width=\textwidth]{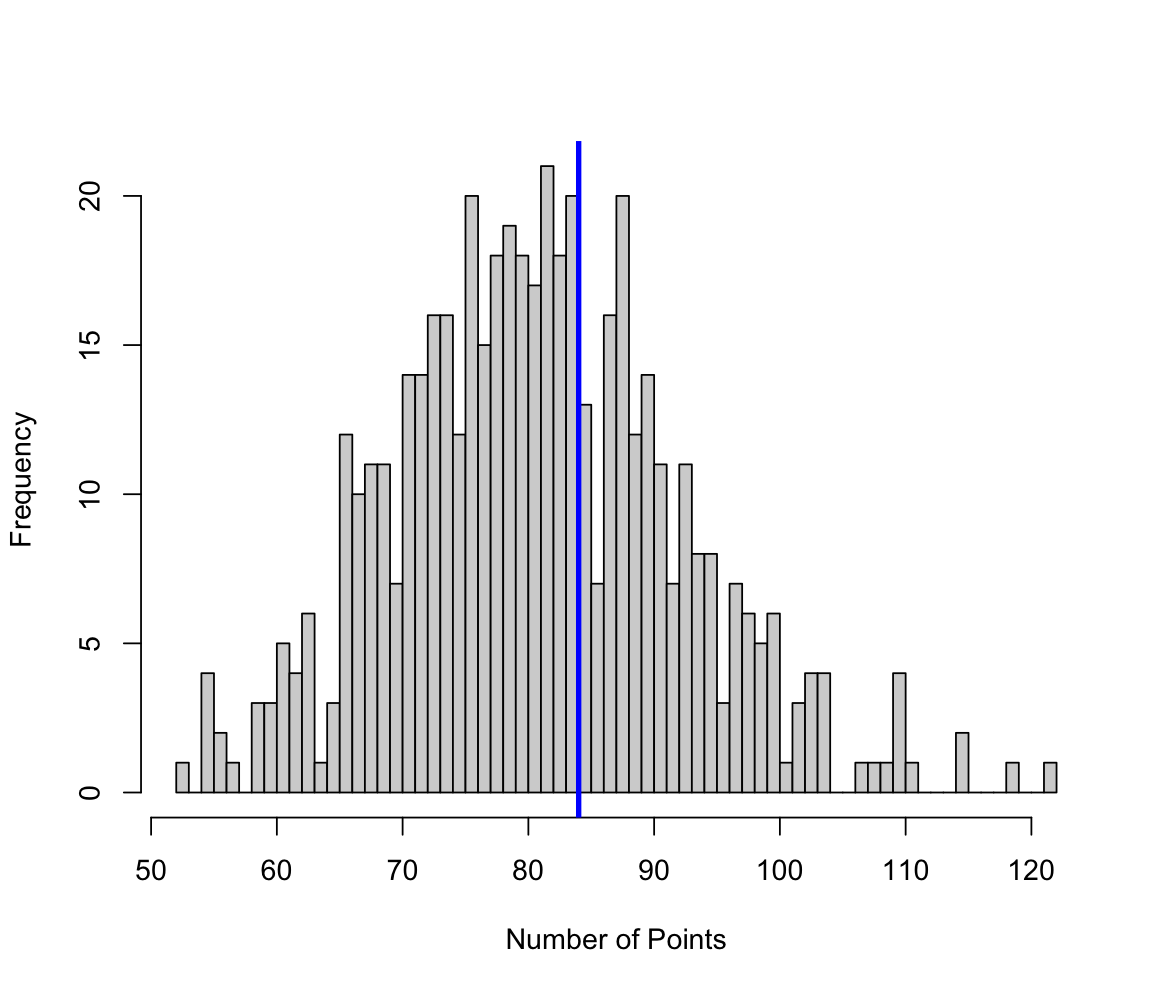}
        \caption{}
    \end{subfigure}

    \caption{EC pyrome observed (blue line) and simulated monthly fire counts (histograms from 500 realizations) over 1984--2023. (a) Baseline model $\lambda$; (b) Full model $U_t \, \lambda$.}
\end{figure} 

\begin{figure}[H]
    \centering
    \begin{subfigure}[b]{0.48\textwidth}
        \centering
        \includegraphics[width=\textwidth]{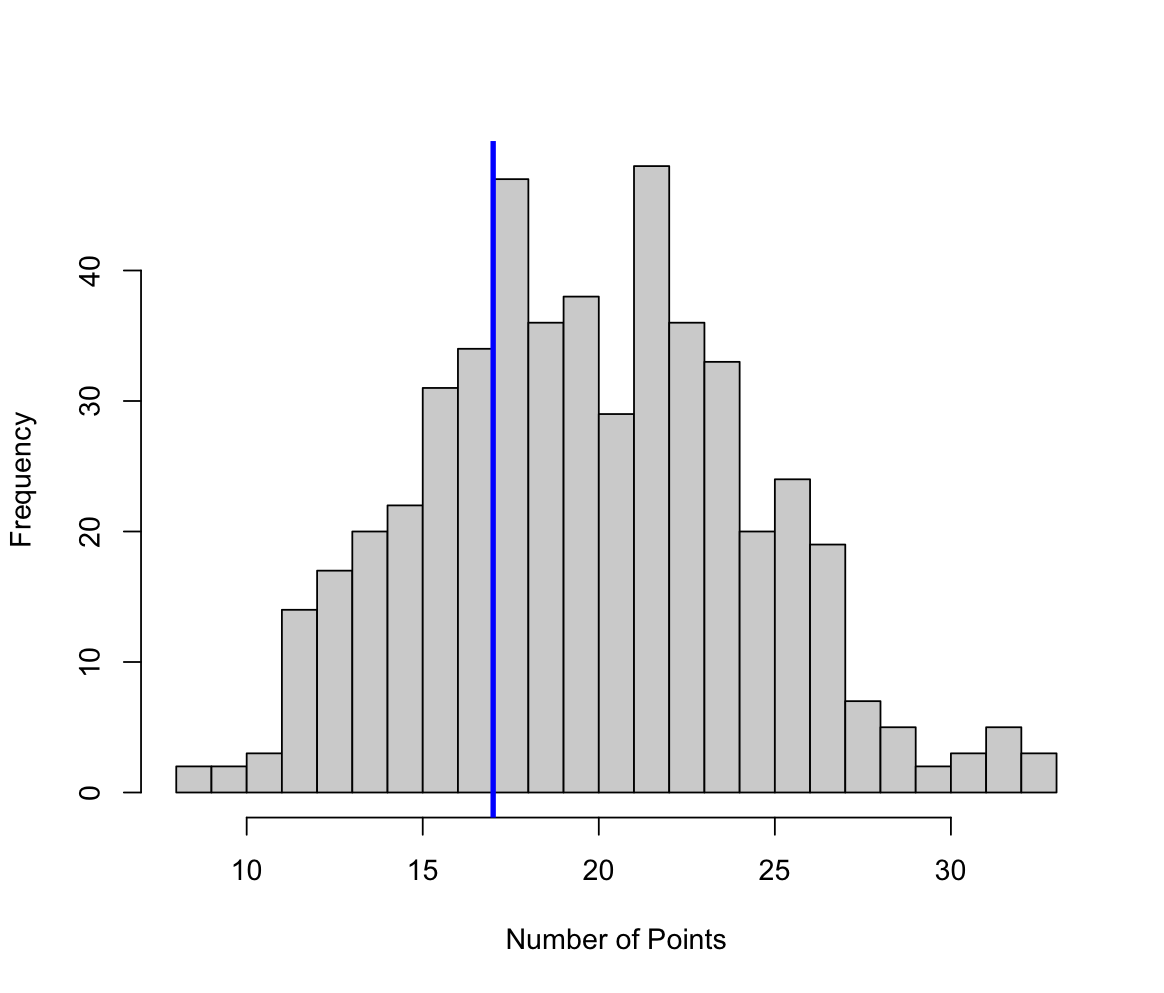}
        \caption{}
    \end{subfigure}
    \hfill
    \begin{subfigure}[b]{0.48\textwidth}
        \centering
        \includegraphics[width=\textwidth]{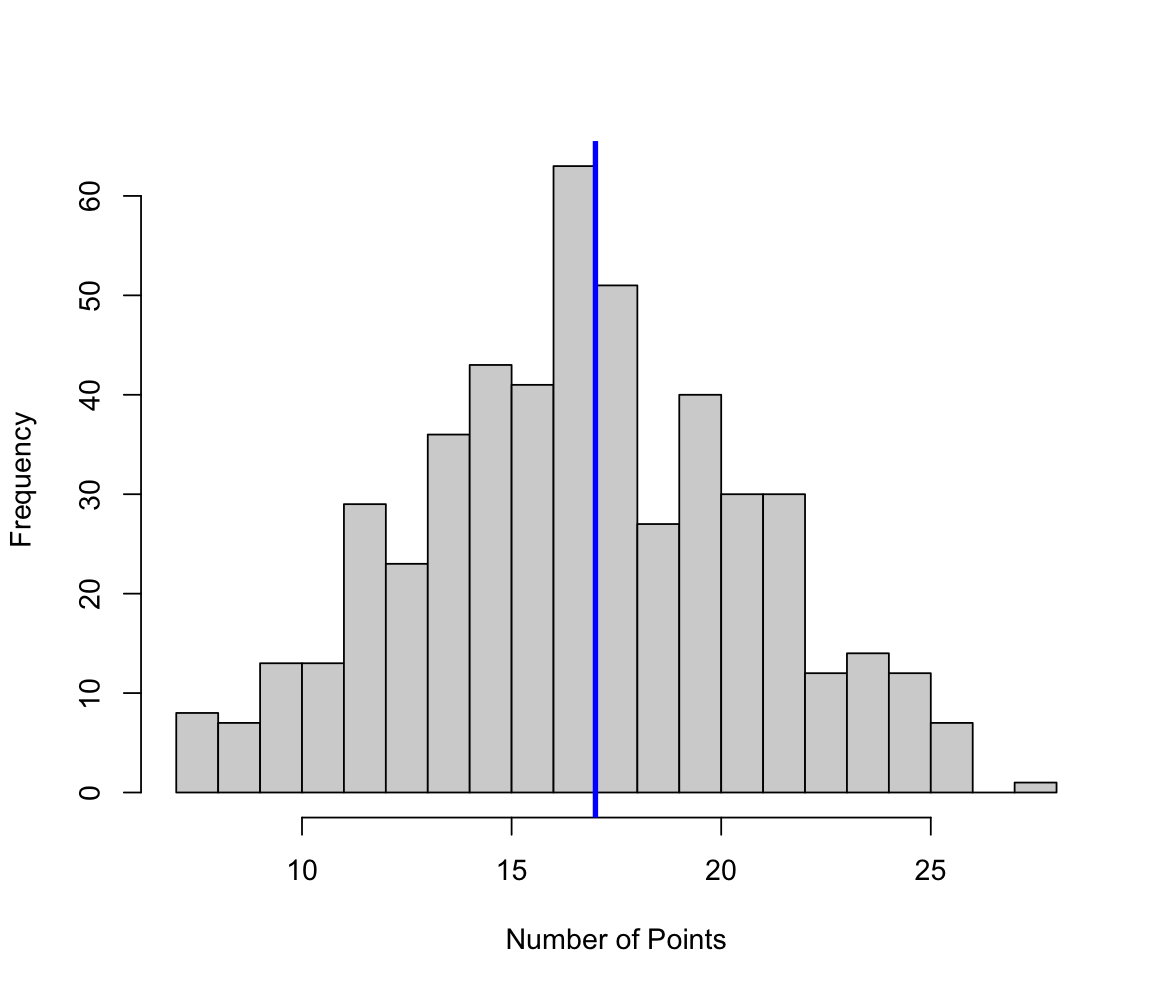}
        \caption{}
    \end{subfigure}

    \caption{HLP pyrome observed (blue line) and simulated monthly fire counts (histograms from 500 realizations) over 1984--2023. (a) (a) Baseline model $\lambda$; (b) Full model $U_t \, \lambda$.}
\end{figure} 

\begin{figure}[H]
    \centering
    \begin{subfigure}[b]{0.48\textwidth}
        \centering
        \includegraphics[width=\textwidth]{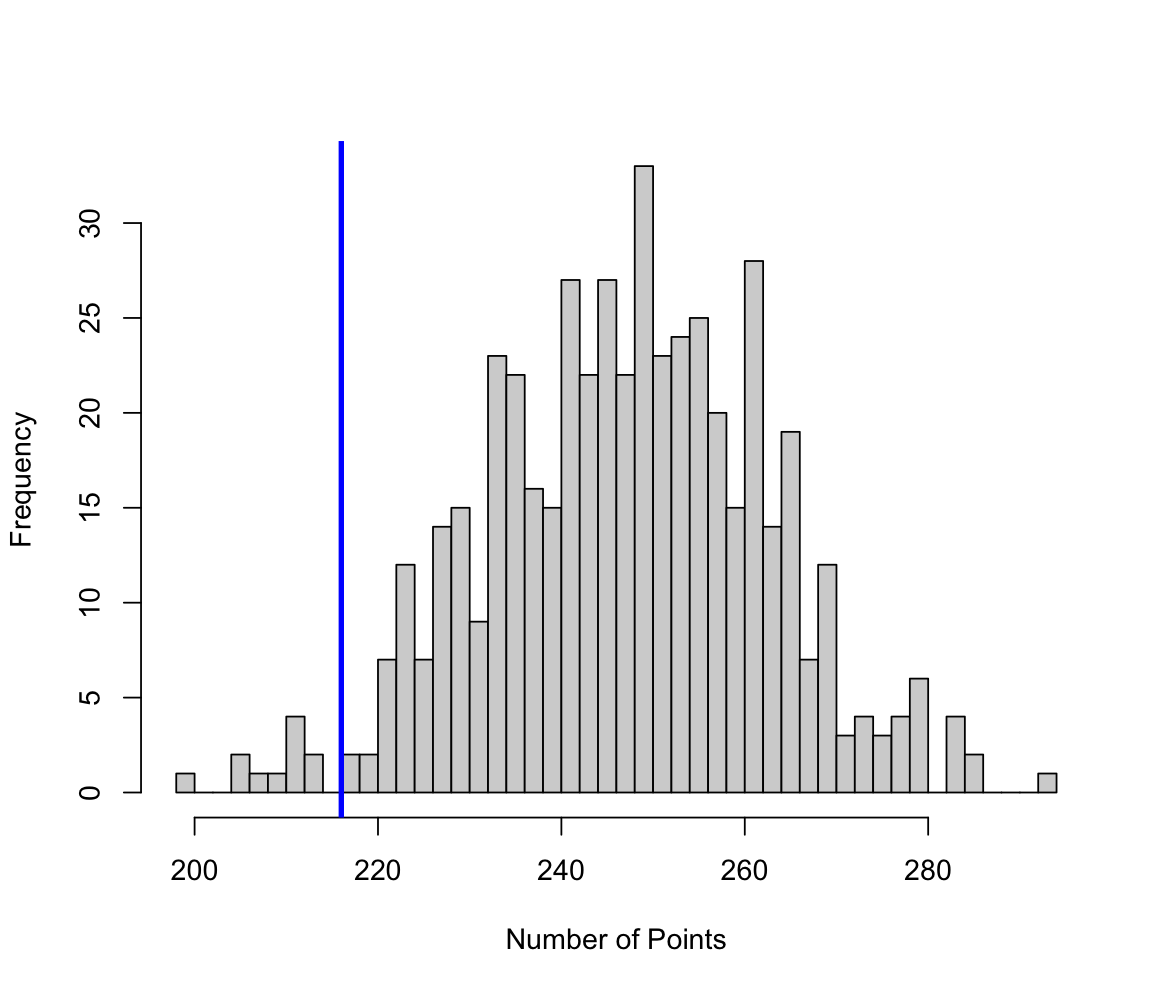}
        \caption{}
    \end{subfigure}
    \hfill
    \begin{subfigure}[b]{0.48\textwidth}
        \centering
        \includegraphics[width=\textwidth]{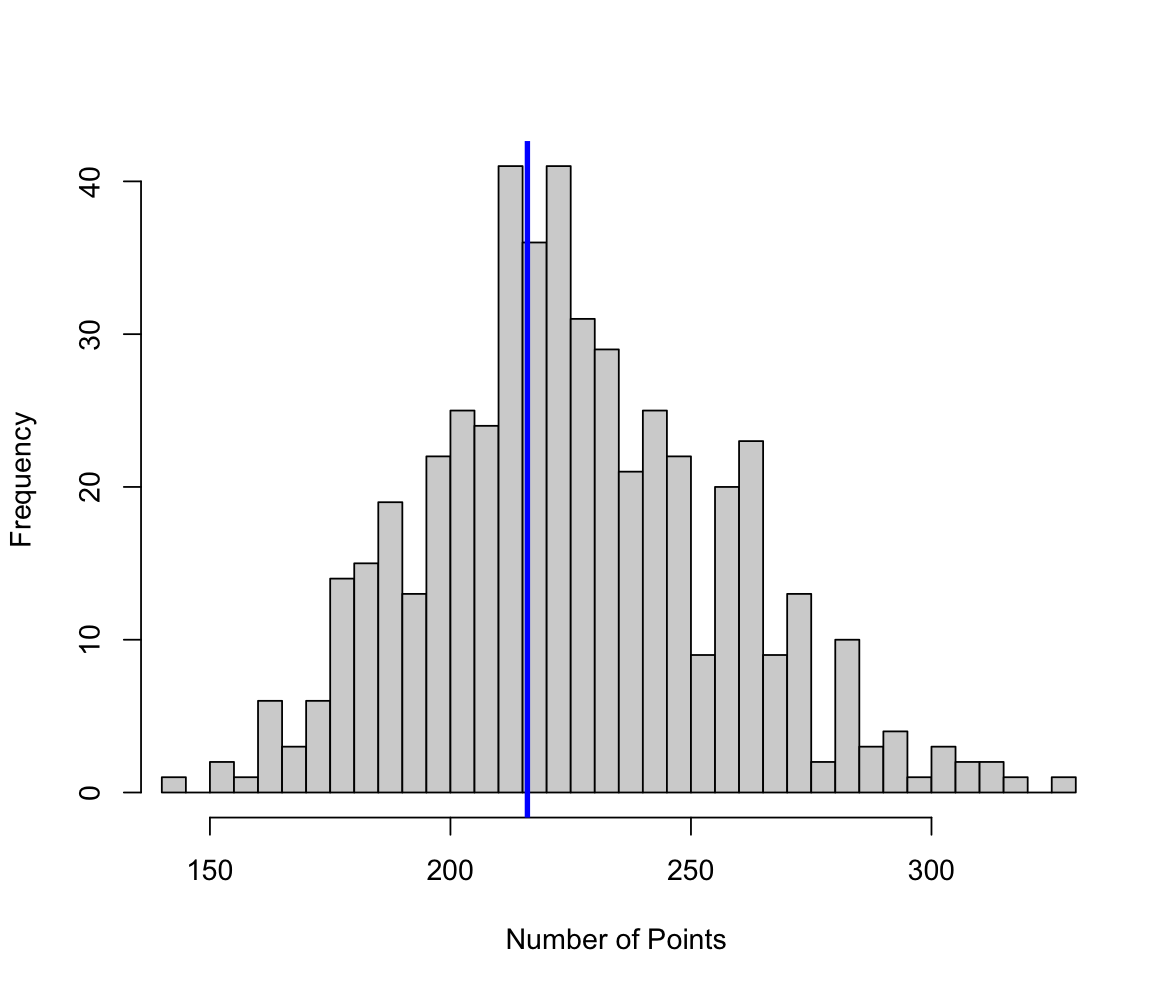}
        \caption{}
    \end{subfigure}

    \caption{KM pyrome observed (blue line) and simulated monthly fire counts (histograms from 500 realizations) over 1984--2023. (a) (a) Baseline model $\lambda$; (b) Full model $U_t \, \lambda$.}
\end{figure} 

\begin{figure}[H]
    \centering
    \begin{subfigure}[b]{0.48\textwidth}
        \centering
        \includegraphics[width=\textwidth]{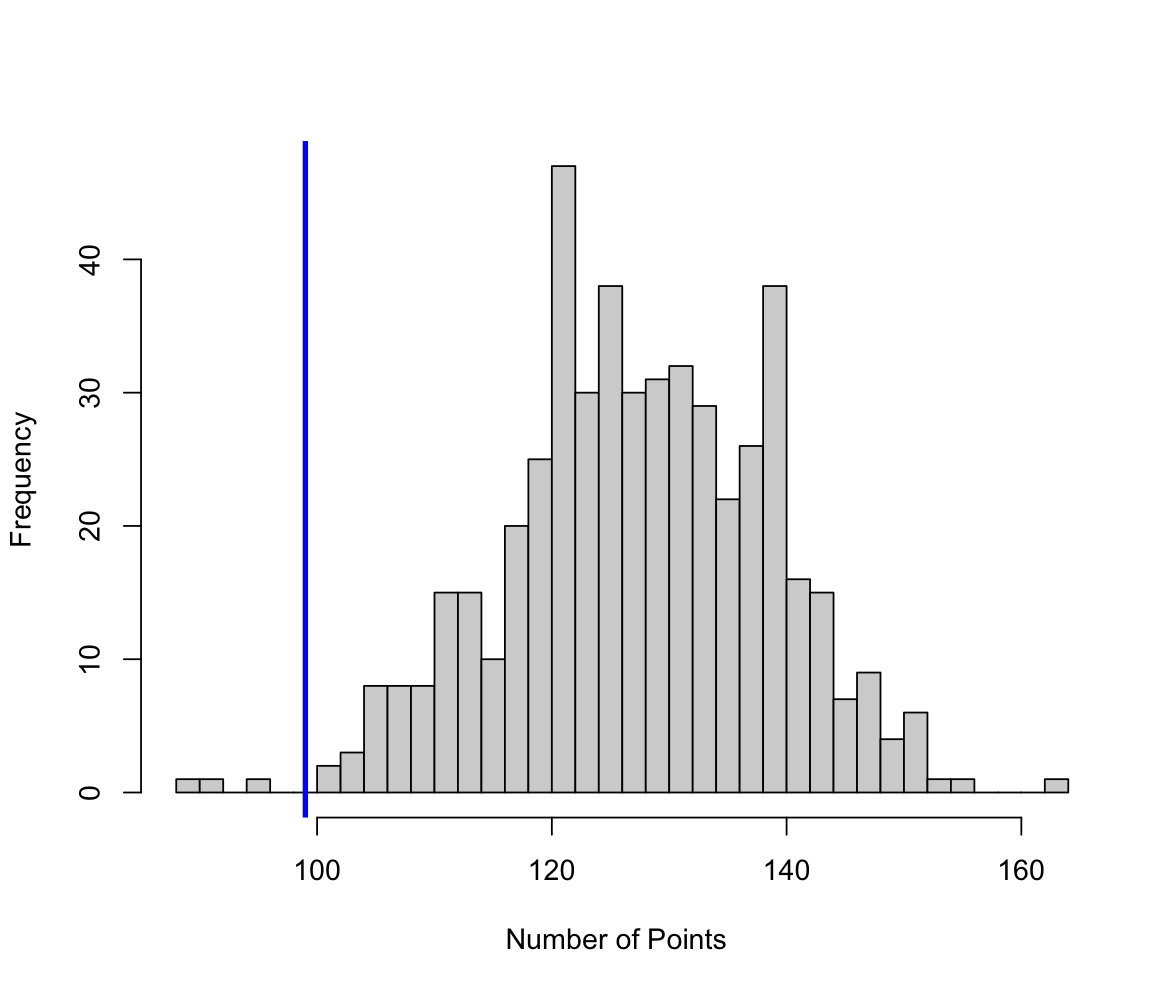}
        \caption{}
    \end{subfigure}
    \hfill
    \begin{subfigure}[b]{0.48\textwidth}
        \centering
        \includegraphics[width=\textwidth]{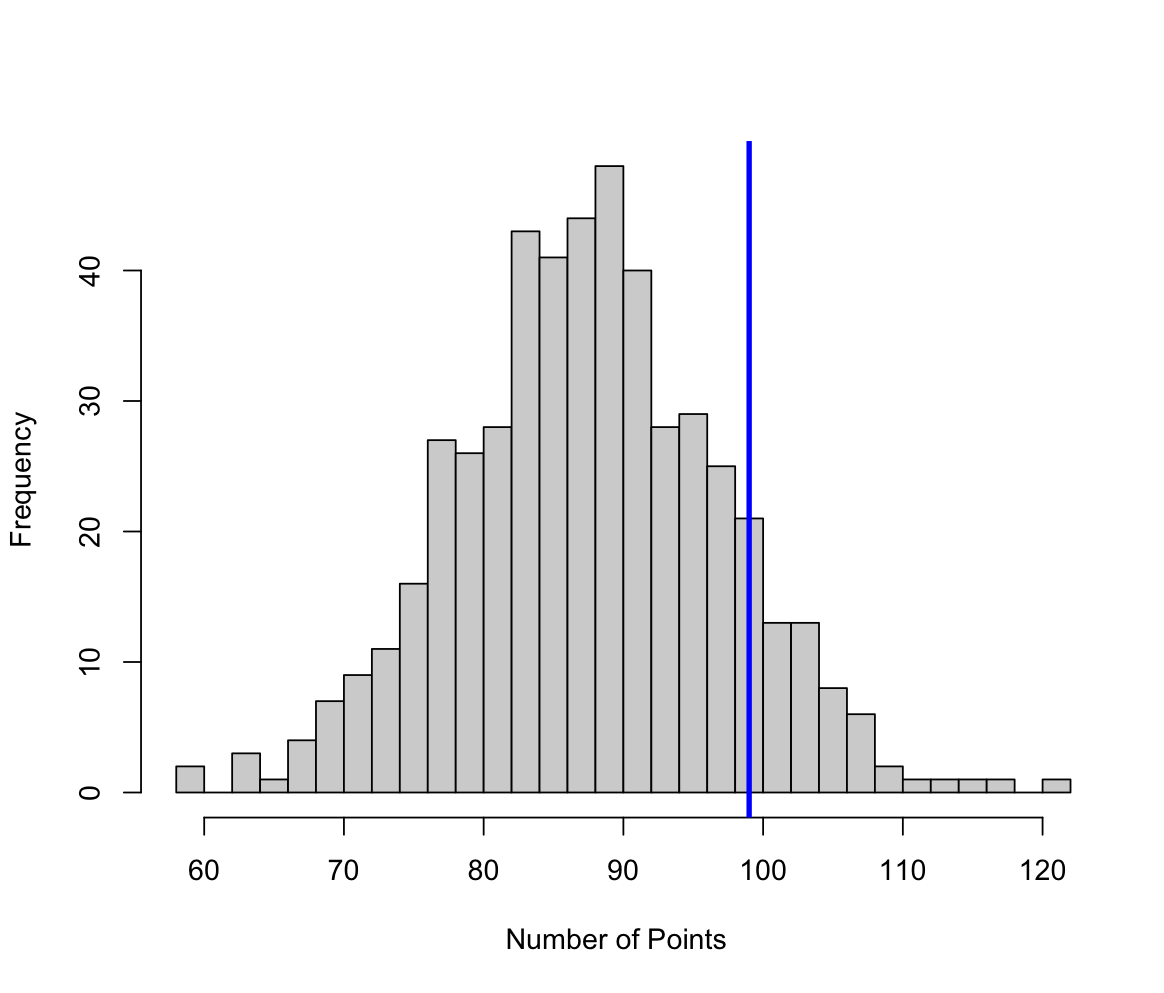}
        \caption{}
    \end{subfigure}

    \caption{NCCFACM pyrome observed (blue line) and simulated monthly fire counts (histograms from 500 realizations) over 1984--2023. (a) (a) Baseline model $\lambda$; (b) Full model $U_t \, \lambda$.}
\end{figure} 

\begin{figure}[H]
    \centering
    \begin{subfigure}[b]{0.48\textwidth}
        \centering
        \includegraphics[width=\textwidth]{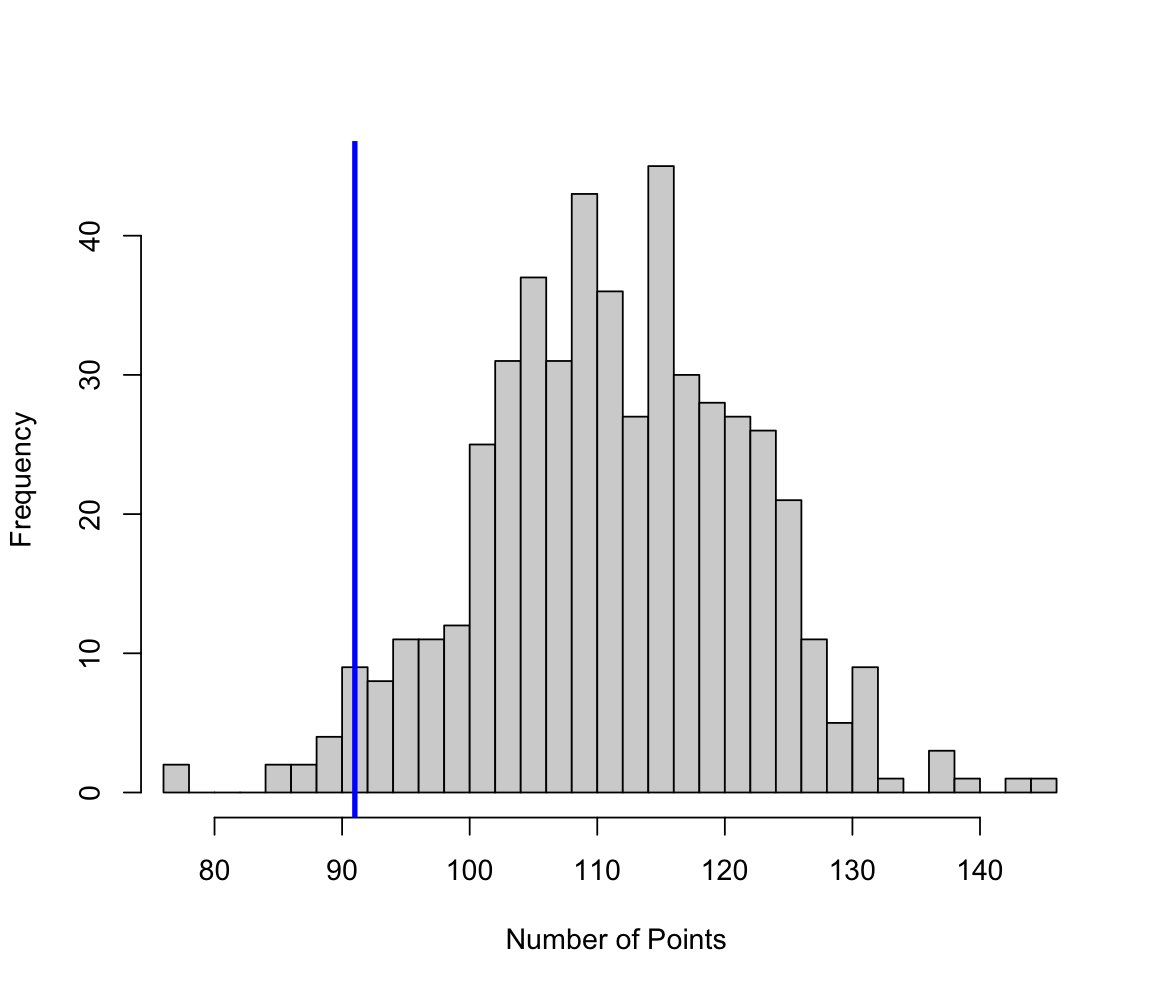}
        \caption{}
    \end{subfigure}
    \hfill
    \begin{subfigure}[b]{0.48\textwidth}
        \centering
        \includegraphics[width=\textwidth]{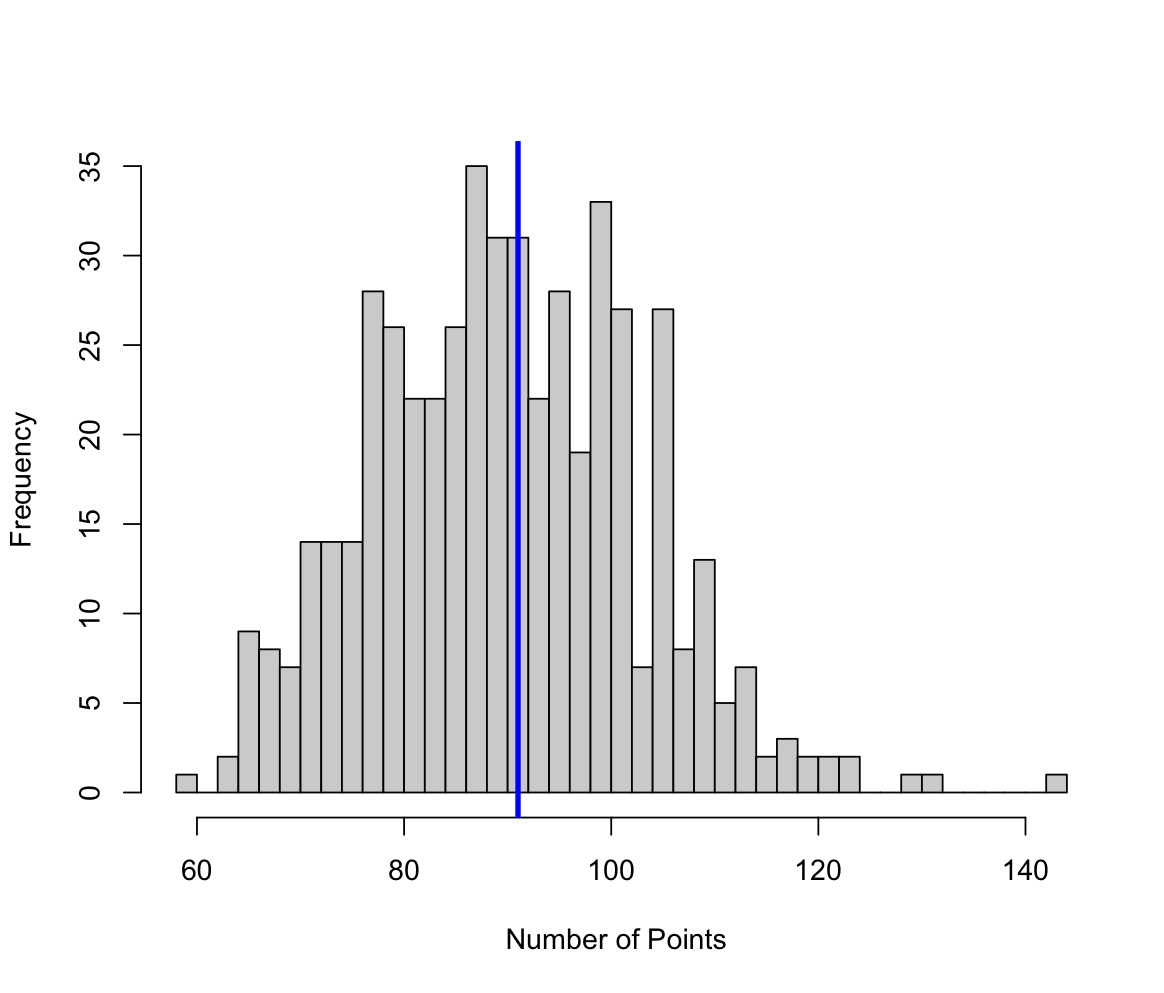}
        \caption{}
    \end{subfigure}

    \caption{NSN pyrome observed (blue line) and simulated monthly fire counts (histograms from 500 realizations) over 1984--2023. (a) (a) Baseline model $\lambda$; (b) Full model $U_t \, \lambda$.}
\end{figure} 

\begin{figure}[H]
    \centering
    \begin{subfigure}[b]{0.48\textwidth}
        \centering
        \includegraphics[width=\textwidth]{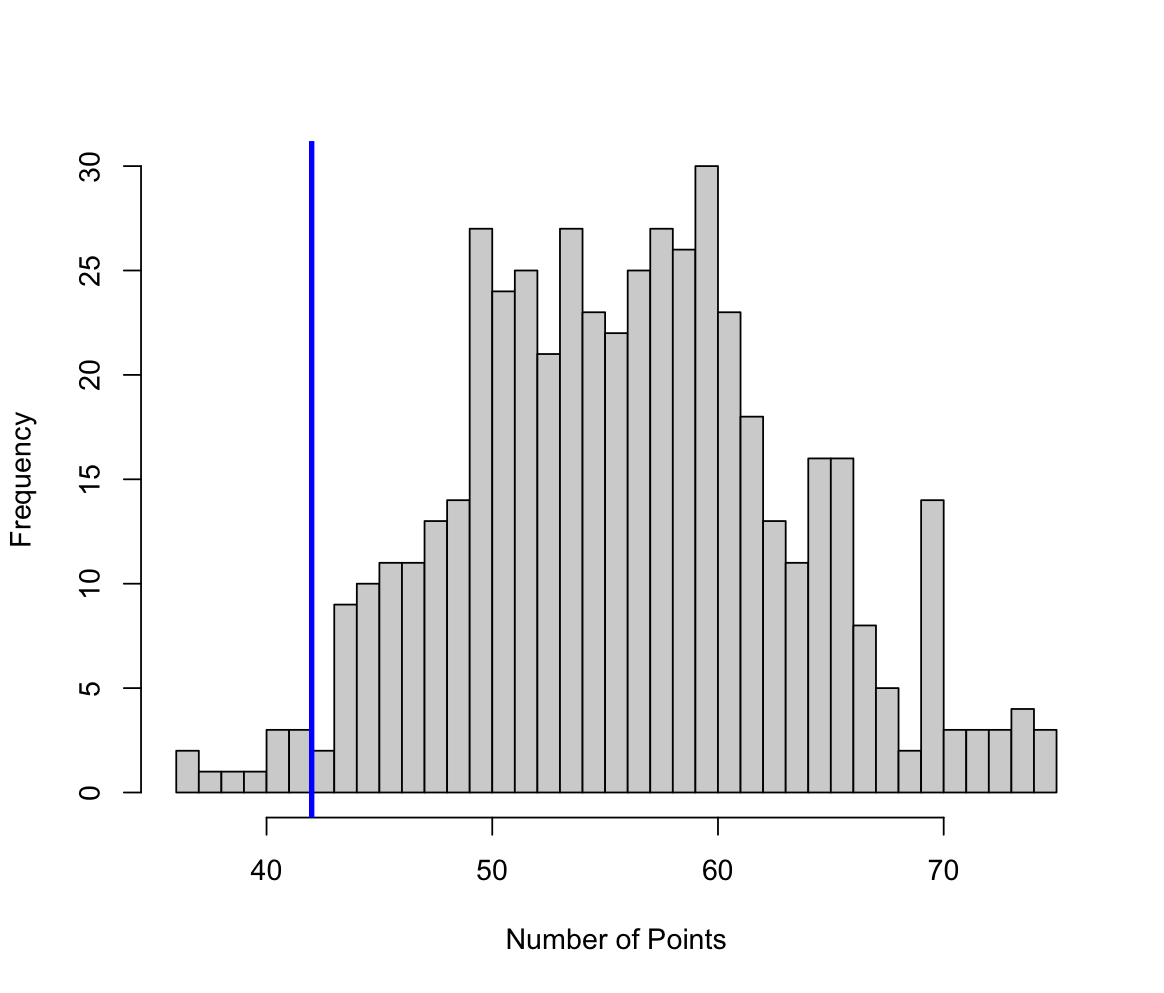}
        \caption{}
    \end{subfigure}
    \hfill
    \begin{subfigure}[b]{0.48\textwidth}
        \centering
        \includegraphics[width=\textwidth]{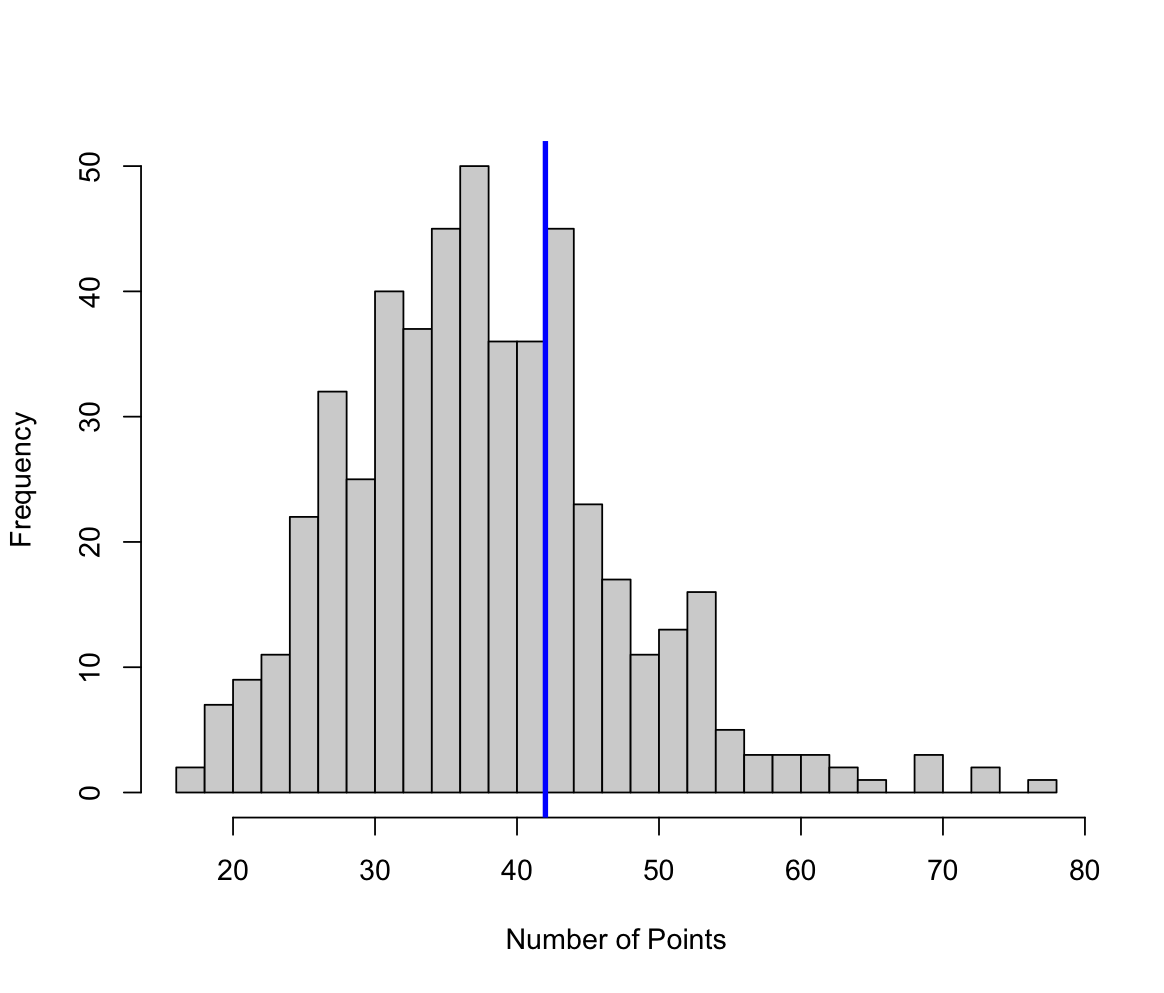}
        \caption{}
    \end{subfigure}

    \caption{ONCRR pyrome observed (blue line) and simulated monthly fire counts (histograms from 500 realizations) over 1984--2023. (a) (a) Baseline model $\lambda$; (b) Full model $U_t \, \lambda$.}
\end{figure} 

\begin{figure}[H]
    \centering
    \begin{subfigure}[b]{0.48\textwidth}
        \centering
        \includegraphics[width=\textwidth]{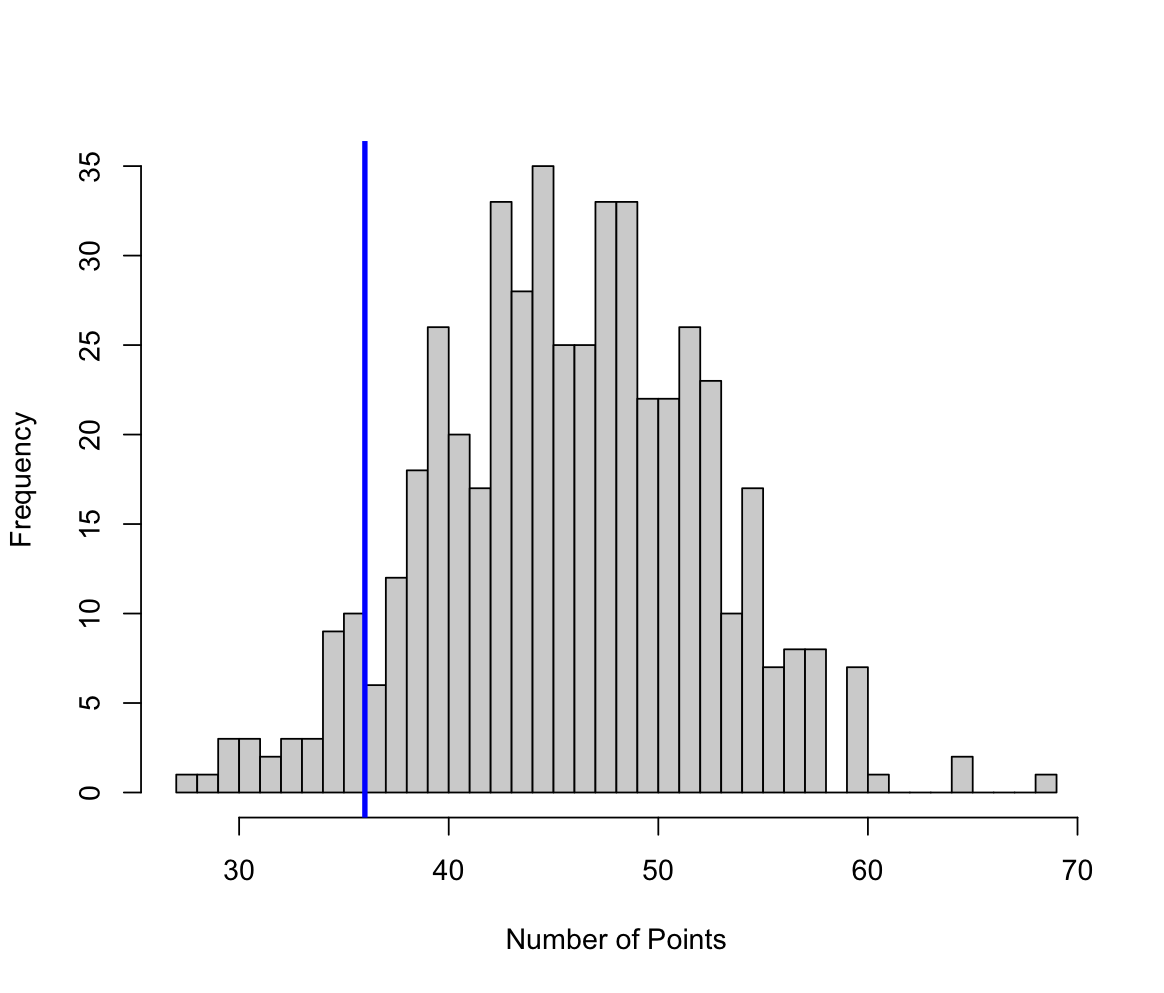}
        \caption{}
    \end{subfigure}
    \hfill
    \begin{subfigure}[b]{0.48\textwidth}
        \centering
        \includegraphics[width=\textwidth]{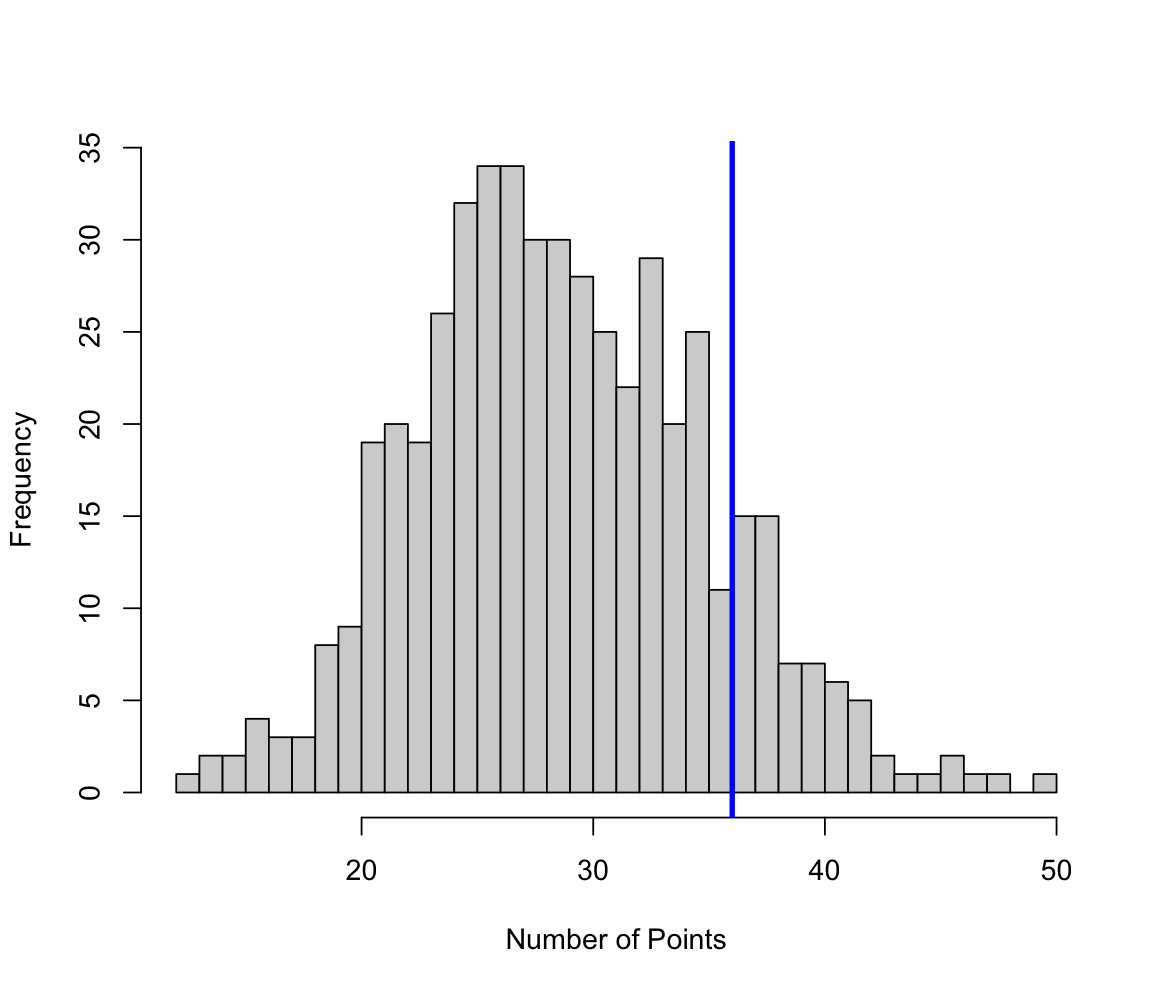}
        \caption{}
    \end{subfigure}

    \caption{SC pyrome observed (blue line) and simulated monthly fire counts (histograms from 500 realizations) over 1984--2023. (a) (a) Baseline model $\lambda$; (b) Full model $U_t \, \lambda$.}
\end{figure} 

\begin{figure}[H]
    \centering
    \begin{subfigure}[b]{0.48\textwidth}
        \centering
        \includegraphics[width=\textwidth]{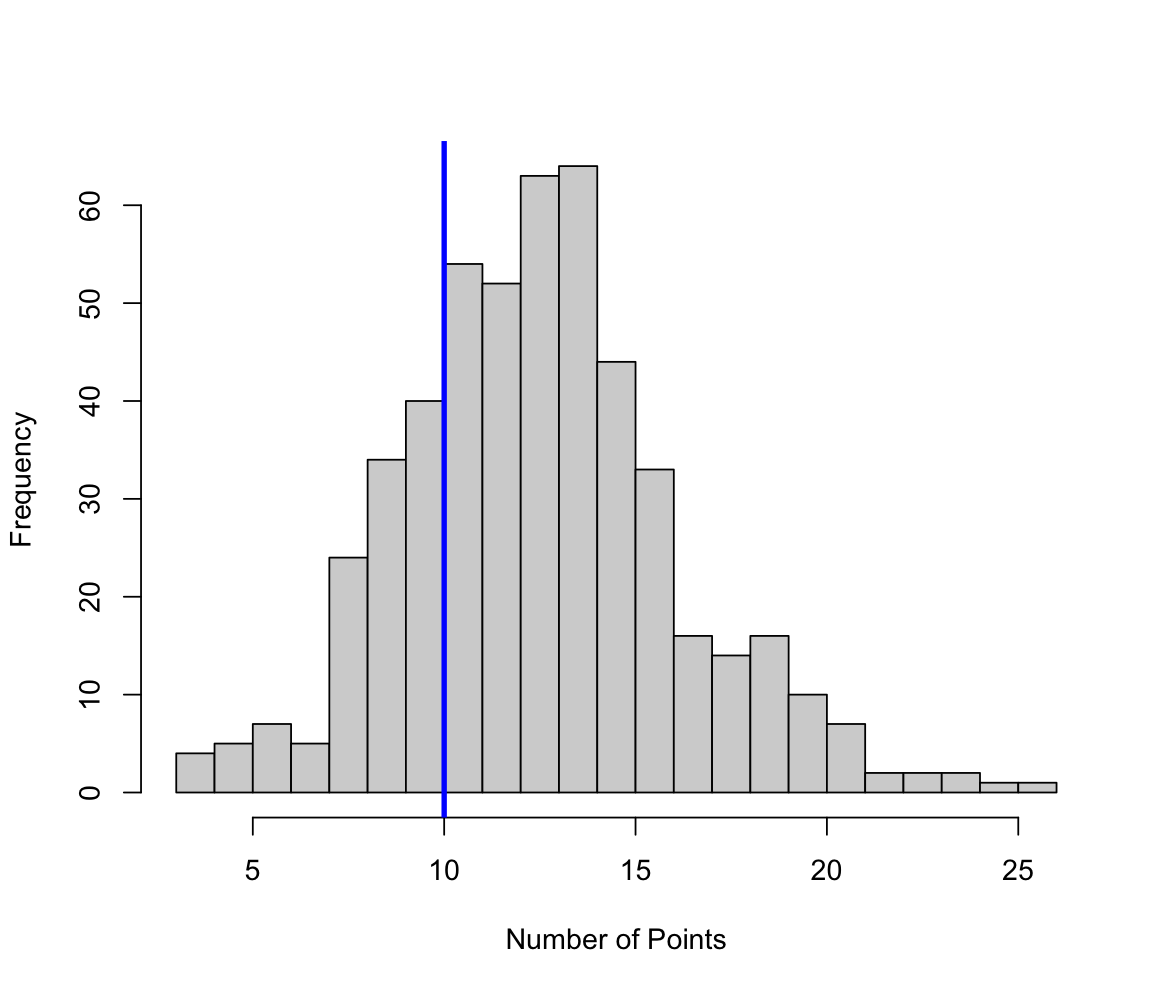}
        \caption{}
    \end{subfigure}
    \hfill
    \begin{subfigure}[b]{0.48\textwidth}
        \centering
        \includegraphics[width=\textwidth]{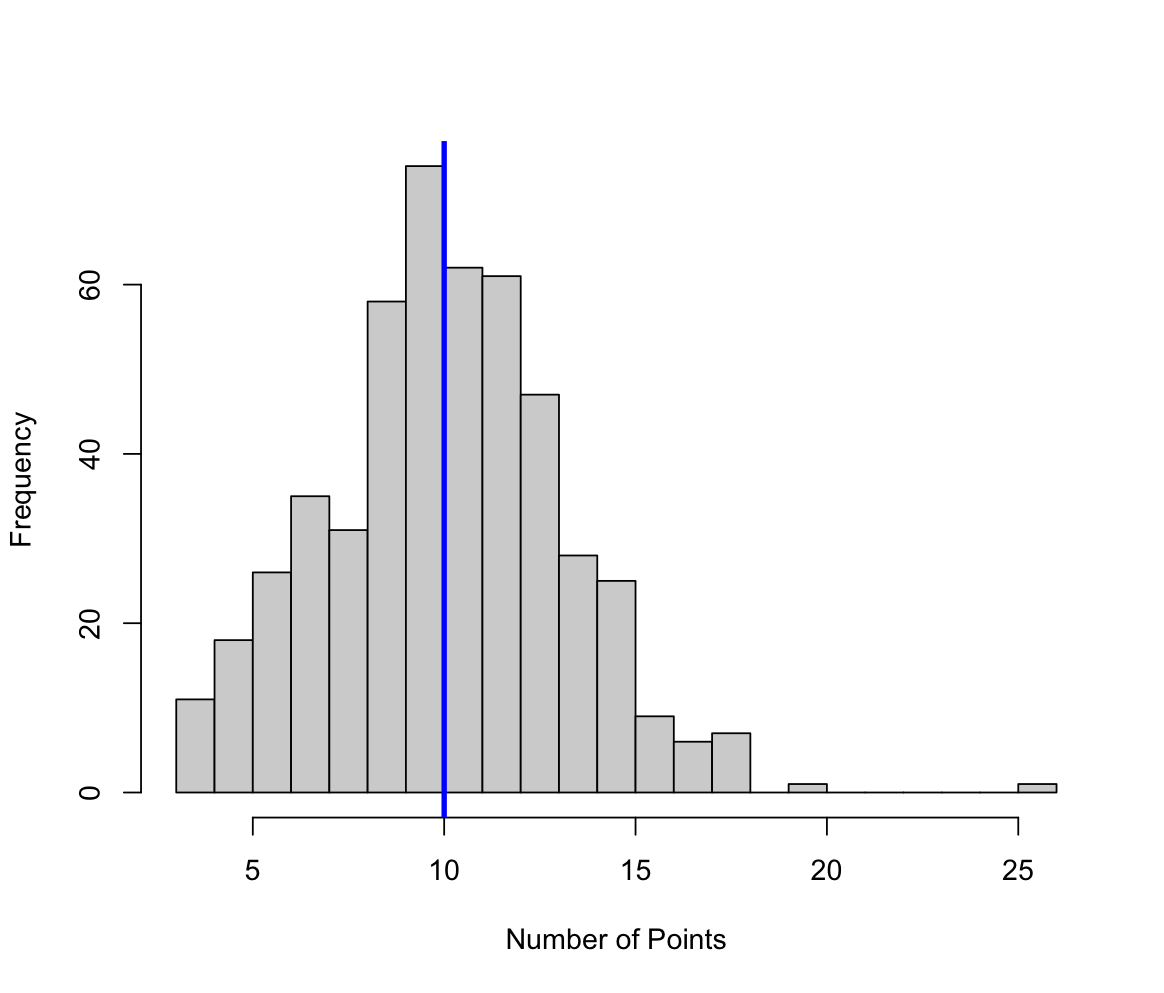}
        \caption{}
    \end{subfigure}

    \caption{SNIRHB pyrome observed (blue line) and simulated monthly fire counts (histograms from 500 realizations) over 1984--2023. (a) (a) Baseline model $\lambda$; (b) Full model $U_t \, \lambda$.}
\end{figure} 

\subsection{Ripley's K Diagnostics}

\begin{figure}[H]
    \centering
    \includegraphics[width=0.5\textwidth]{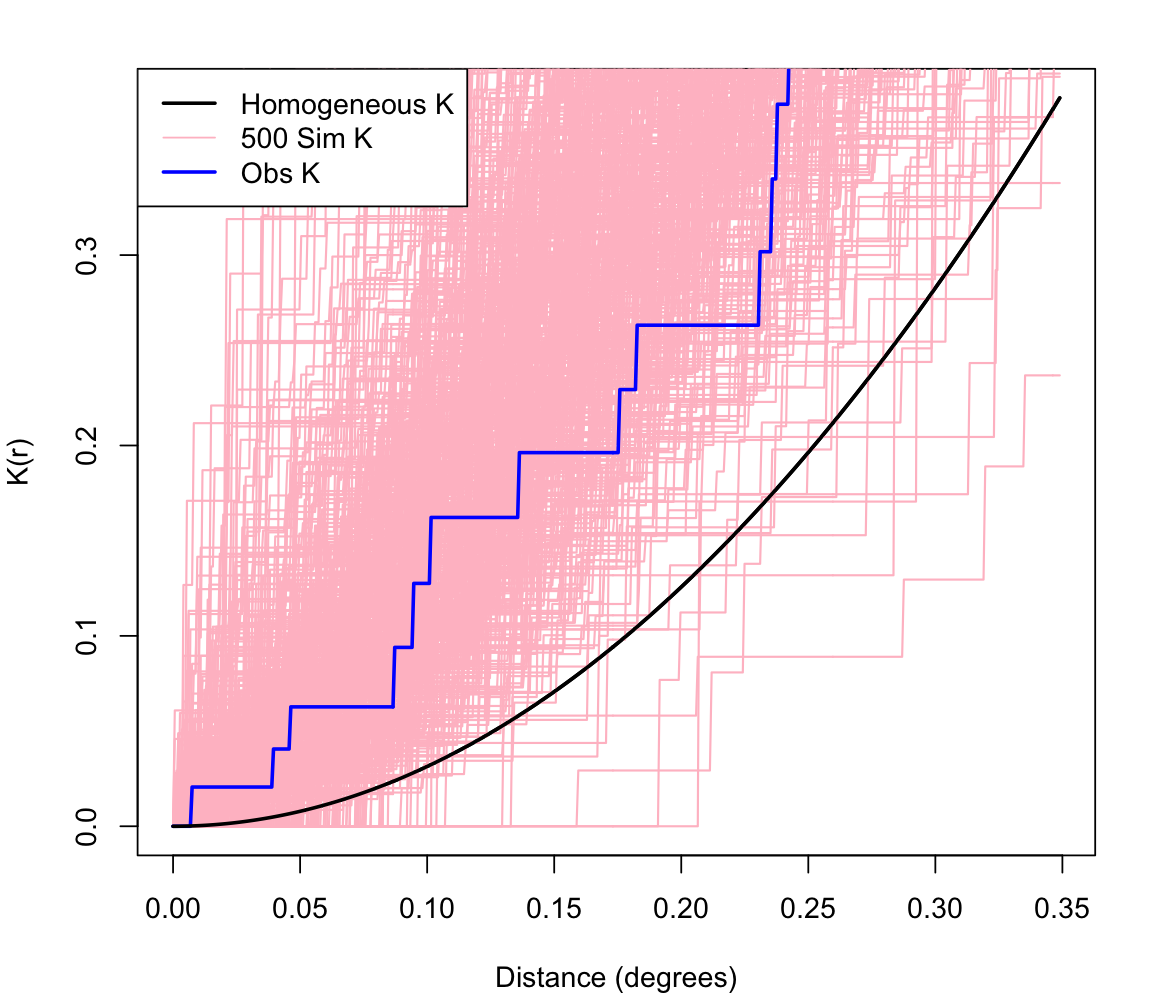}
    \caption{Ripley’s K function validation for the CCV pyrome.}
\end{figure}

\begin{figure}[H]
    \centering
    \includegraphics[width=0.5\textwidth]{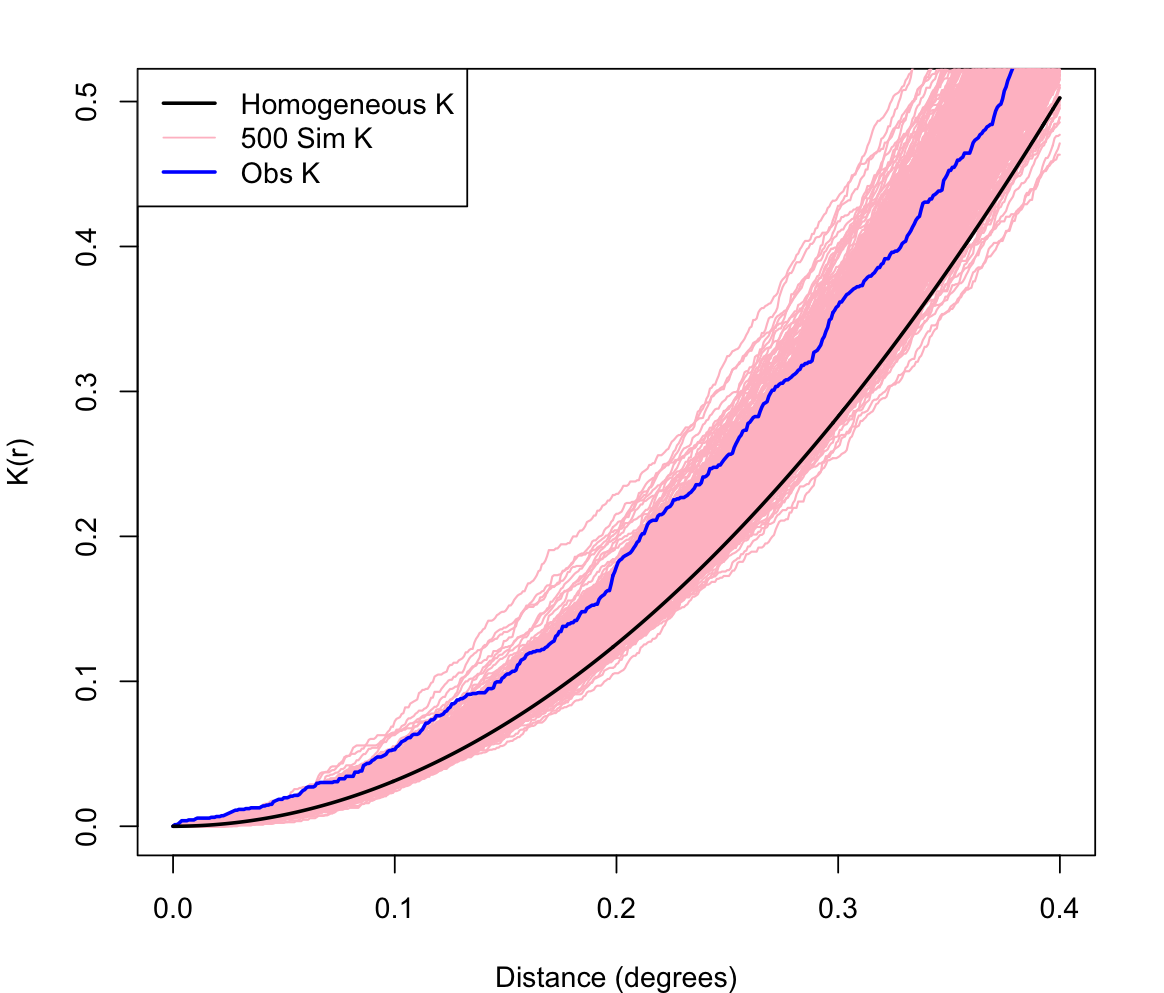}
    \caption{Ripley’s K function validation for the EC pyrome.}
\end{figure}

\begin{figure}[H]
    \centering
    \includegraphics[width=0.5\textwidth]{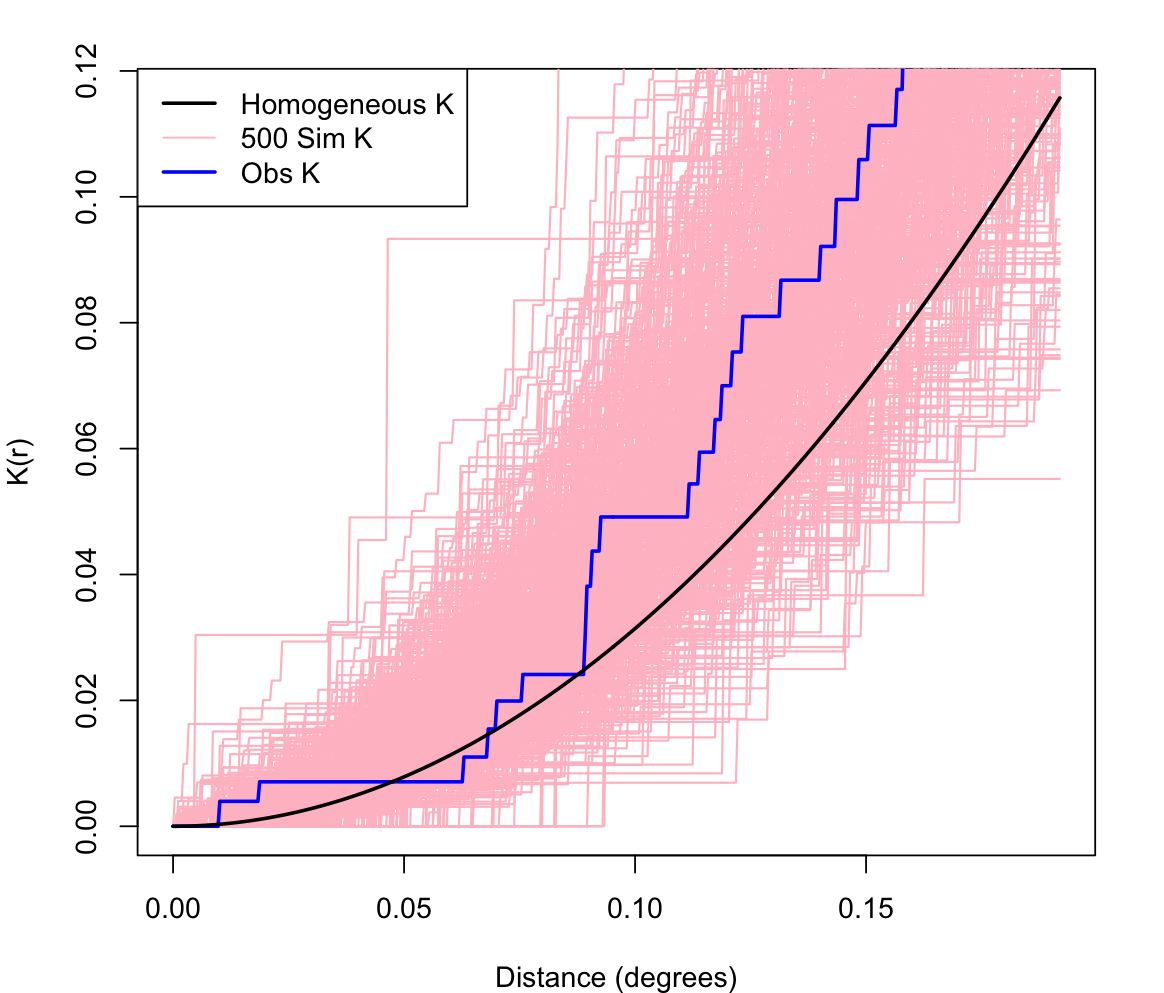}
    \caption{Ripley’s K function validation for the HLP pyrome.}
\end{figure}

\begin{figure}[H]
    \centering
    \includegraphics[width=0.5\textwidth]{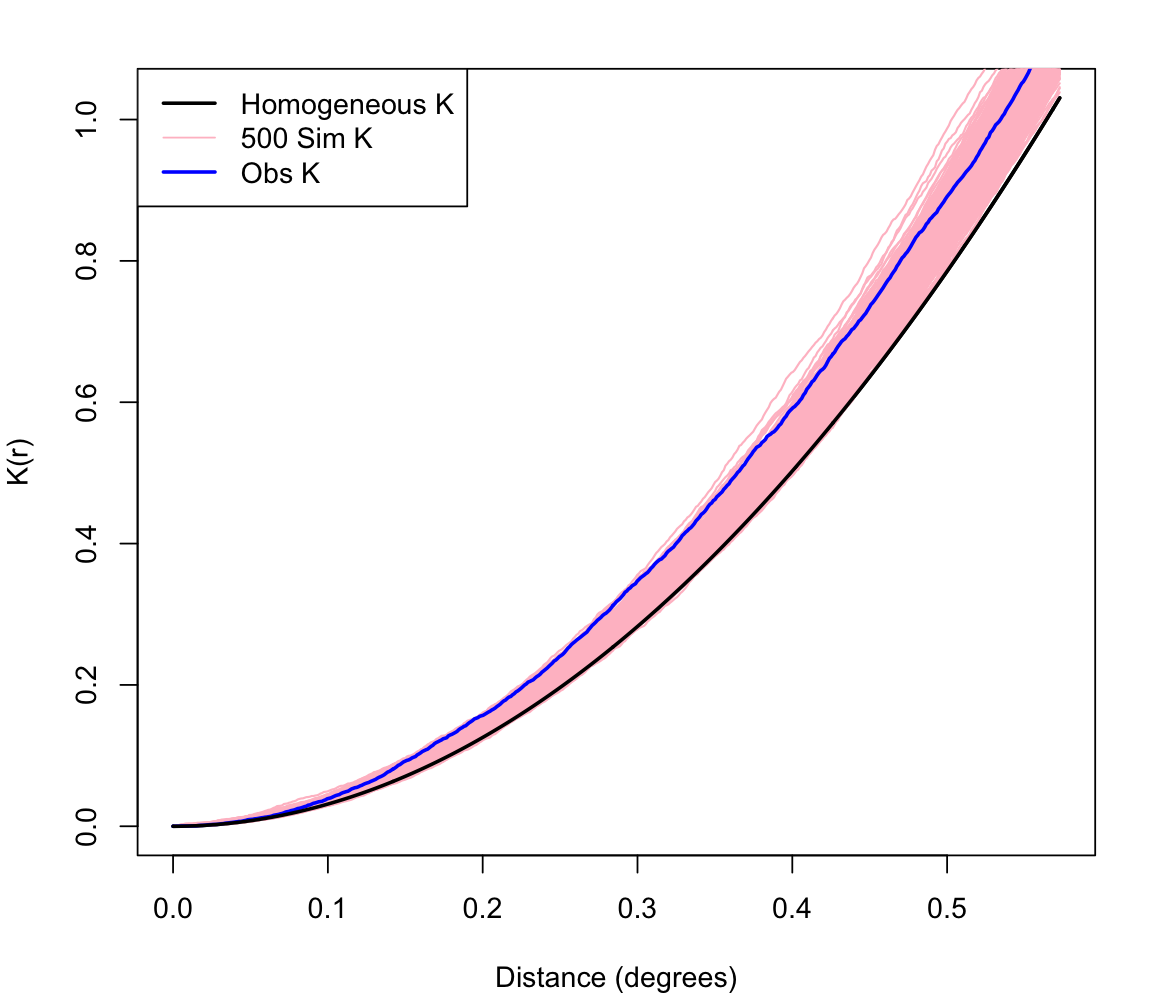}
    \caption{Ripley’s K function validation for the KM pyrome.}
\end{figure}

\begin{figure}[H]
    \centering
    \includegraphics[width=0.5\textwidth]{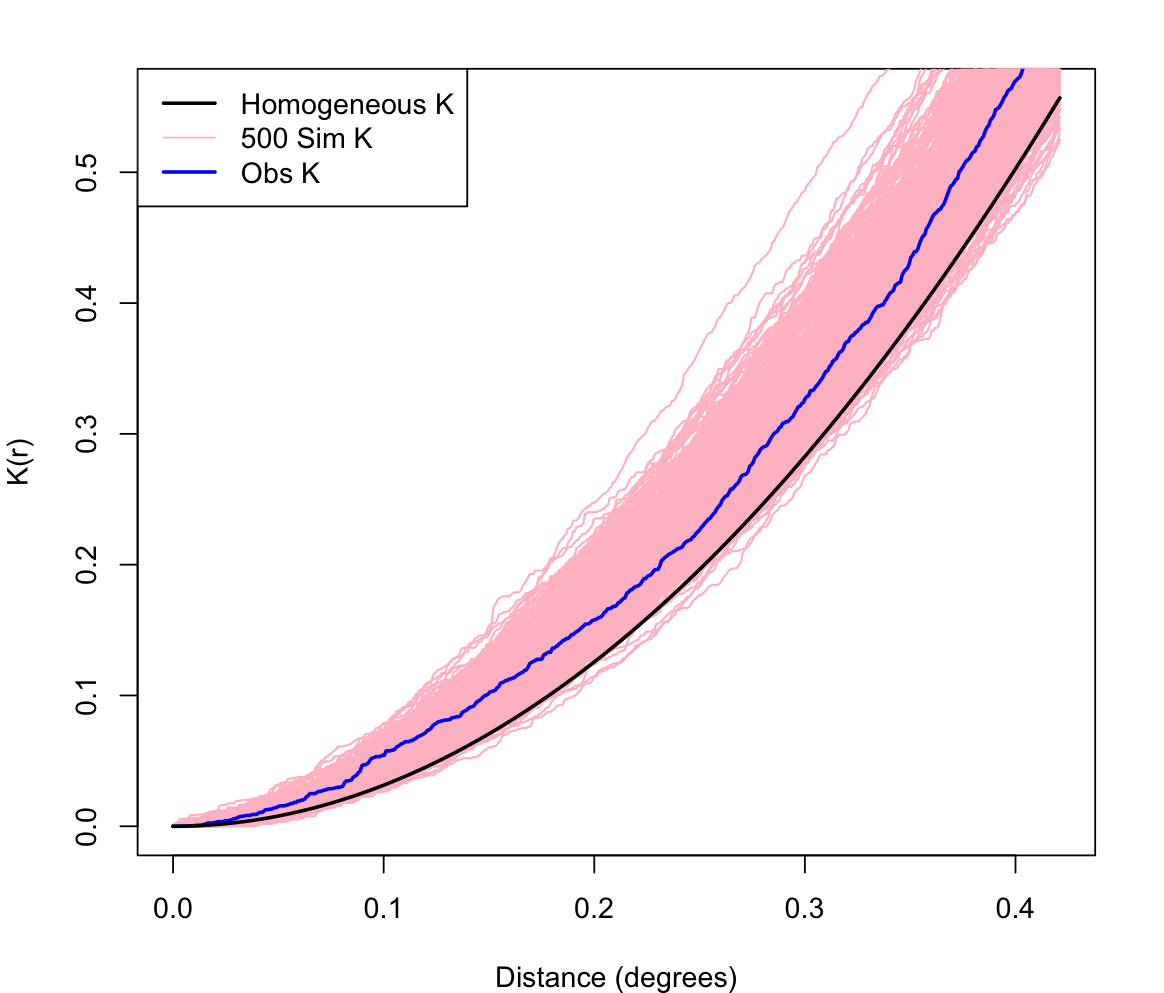}
    \caption{Ripley’s K function validation for the NCCFACM pyrome.}
\end{figure}

\begin{figure}[H]
    \centering
    \includegraphics[width=0.5\textwidth]{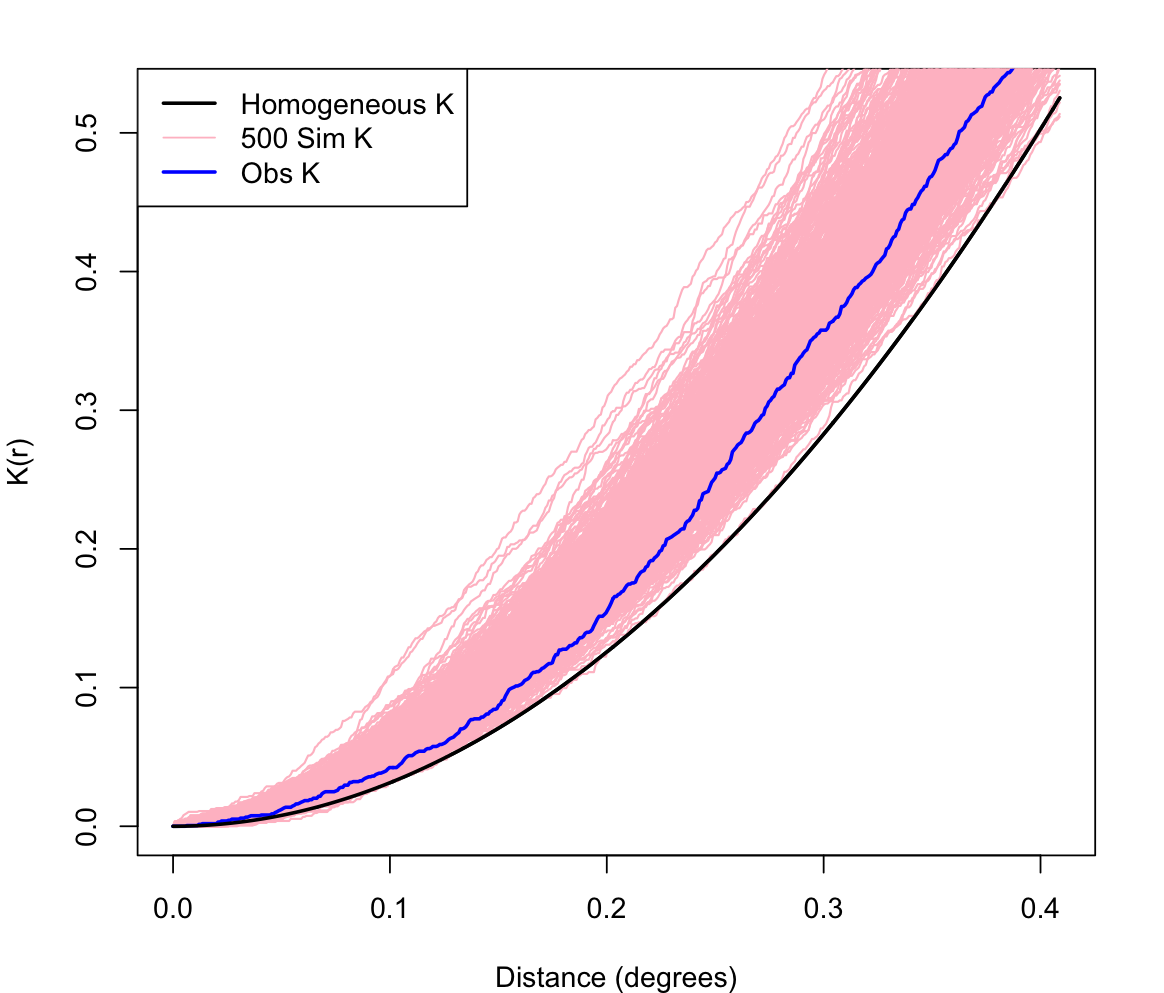}
    \caption{Ripley’s K function validation for the NSN pyrome.}
\end{figure}

\begin{figure}[H]
    \centering
    \includegraphics[width=0.5\textwidth]{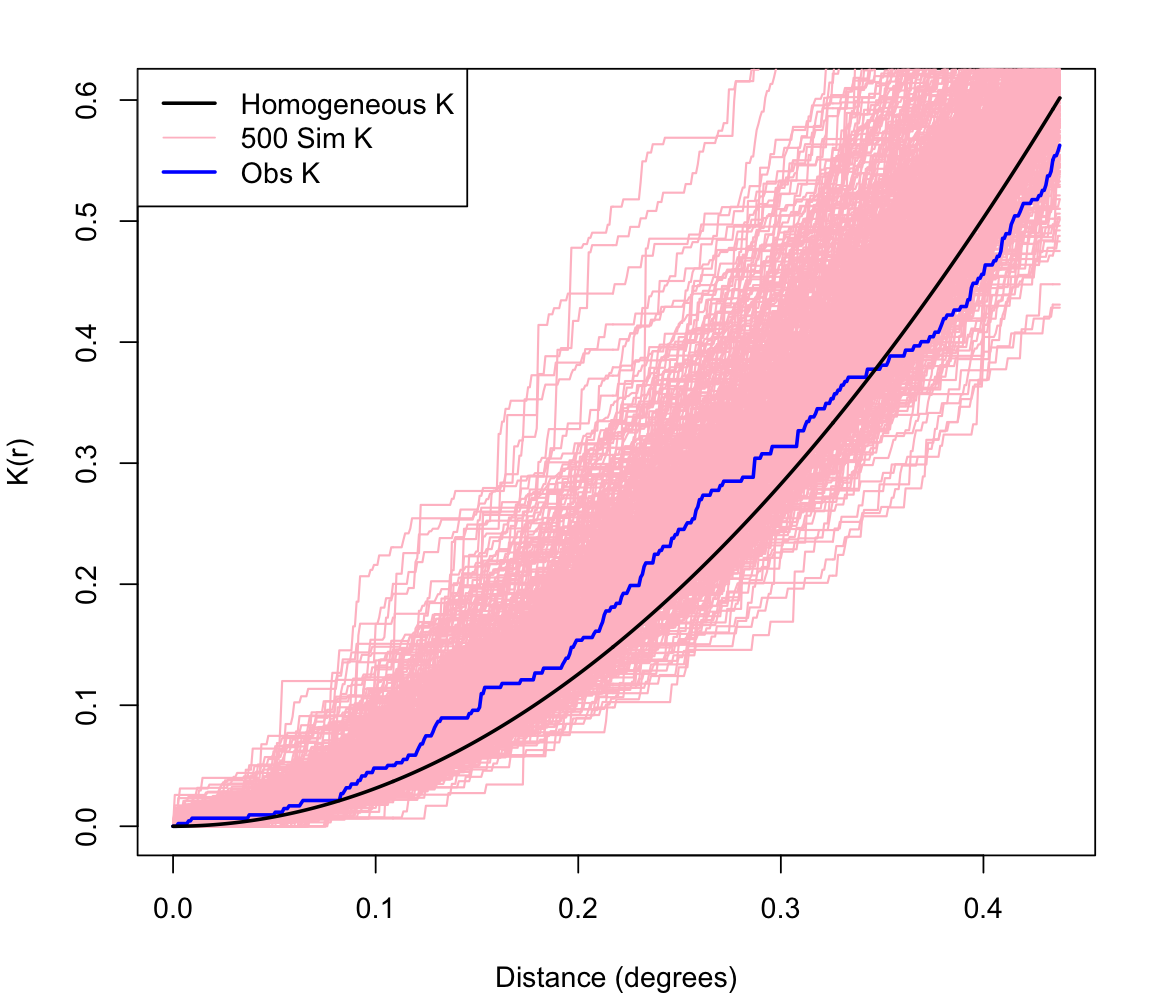}
    \caption{Ripley’s K function validation for the ONCRR pyrome.}
\end{figure}

\begin{figure}[H]
    \centering
    \includegraphics[width=0.5\textwidth]{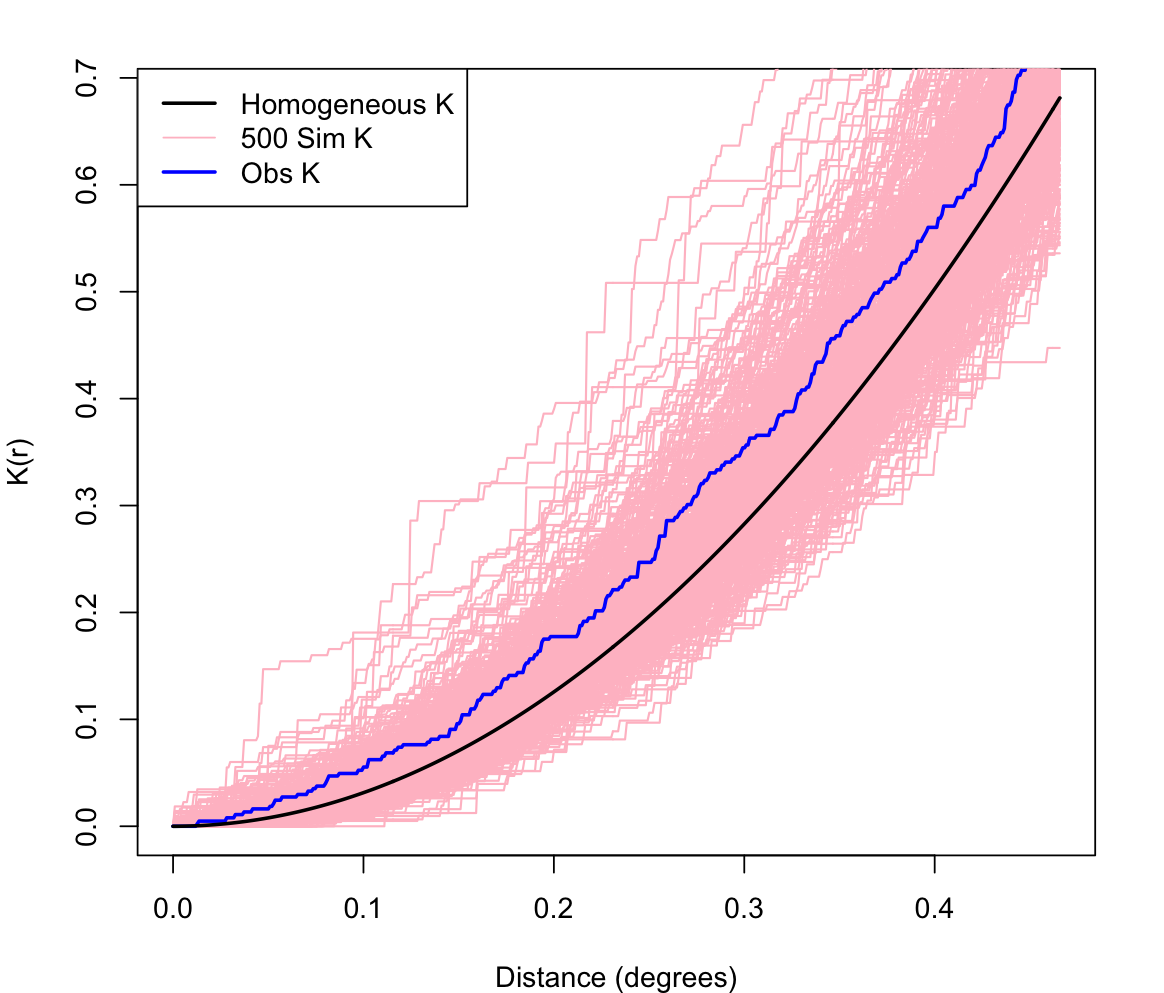}
    \caption{Ripley’s K function validation for the SC pyrome.}
\end{figure}

\begin{figure}[H]
    \centering
    \includegraphics[width=0.5\textwidth]{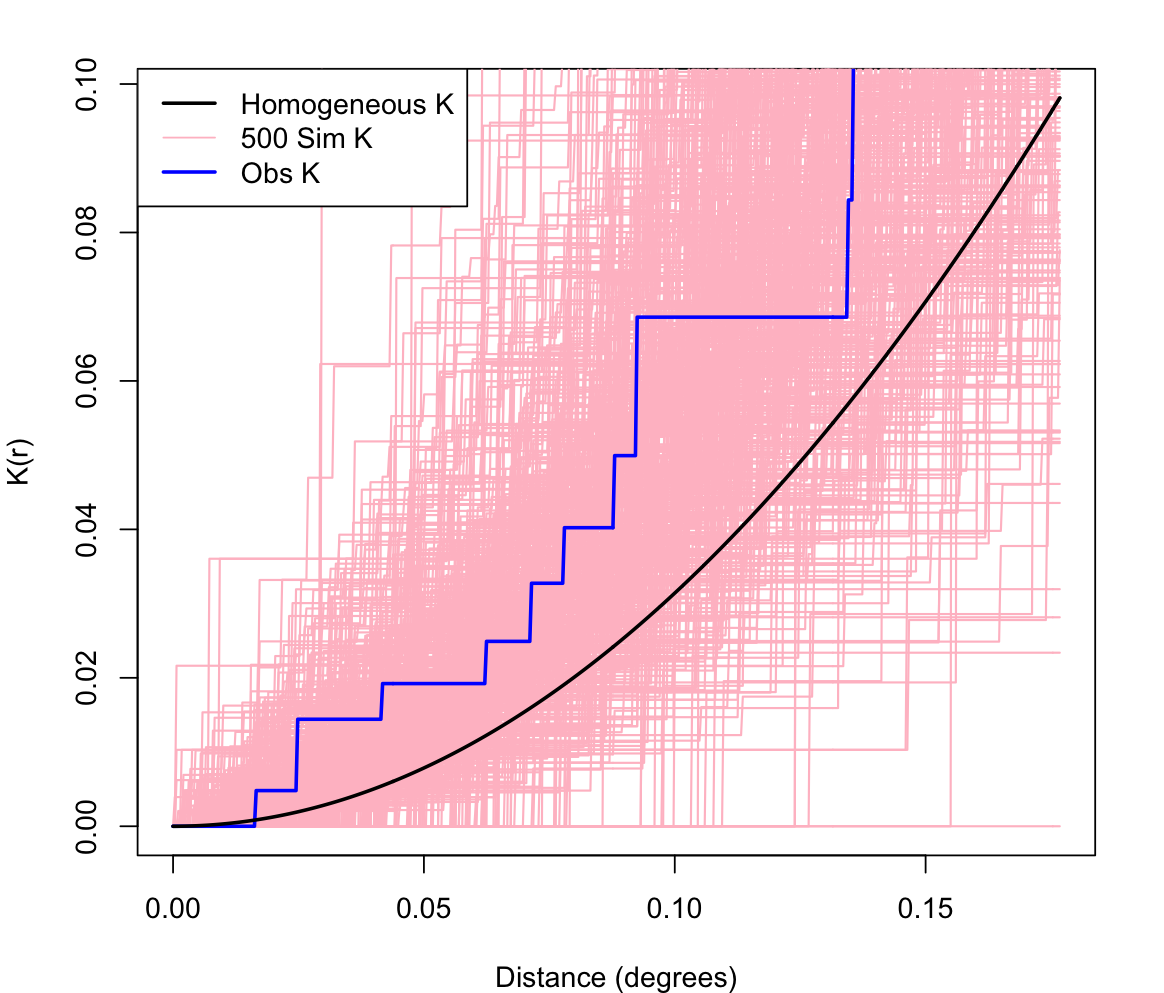}
    \caption{Ripley’s K function validation for the SNIRHB pyrome.}
\end{figure}

\subsection{Point Process Simulation Diagnostics}

\begin{figure}[H]
    \centering
    \includegraphics[width=0.8\textwidth]{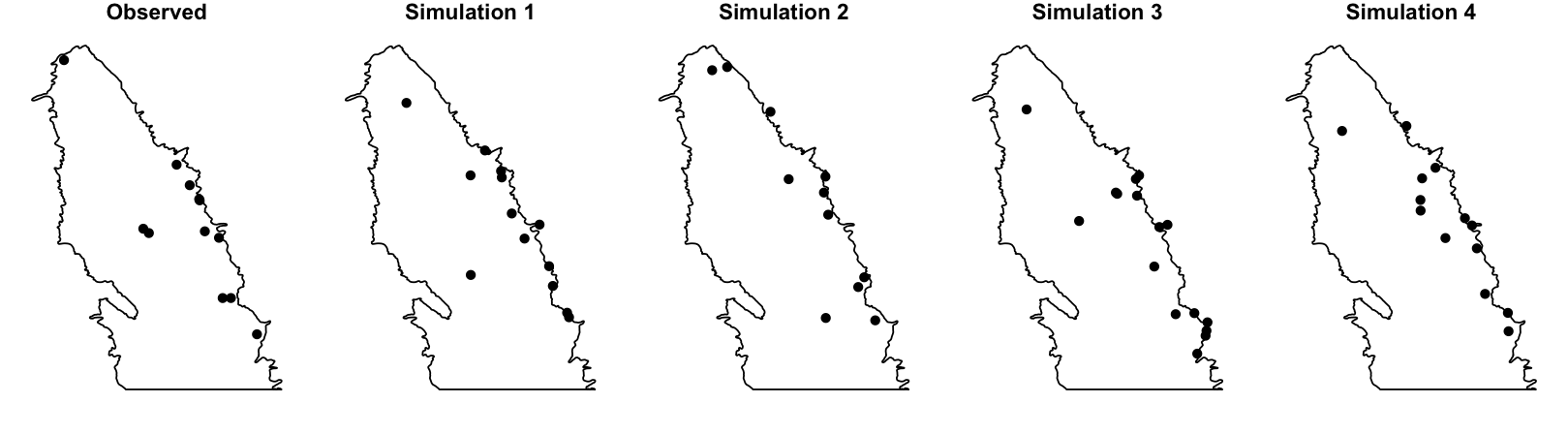}
    \caption{Observed and simulated wildfire point patterns for the CCV pyrome under the fitted model with intensity $U_t \, \lambda$.}
\end{figure}

\begin{figure}[H]
    \centering
    \includegraphics[width=0.8\textwidth]{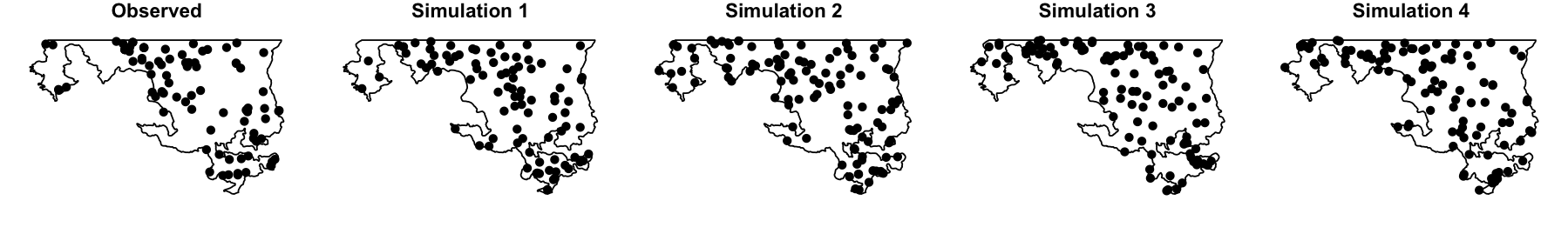}
    \caption{Observed and simulated wildfire point patterns for the EC pyrome under the fitted model with intensity $U_t \, \lambda$.}
\end{figure}

\begin{figure}[H]
    \centering
    \includegraphics[width=0.8\textwidth]{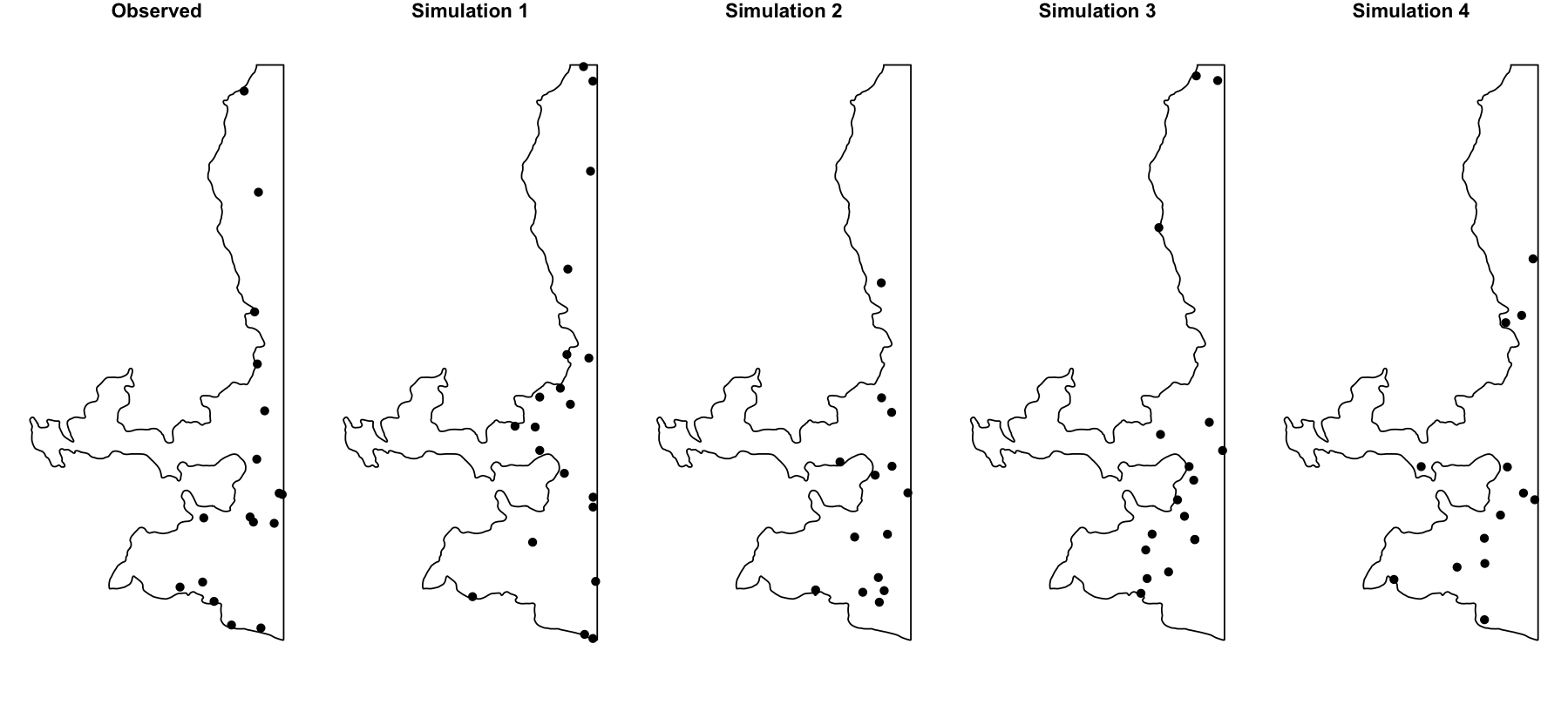}
    \caption{Observed and simulated wildfire point patterns for the HLP pyrome under the fitted model with intensity $U_t \, \lambda$.}
\end{figure}

\begin{figure}[H]
    \centering
    \includegraphics[width=0.8\textwidth]{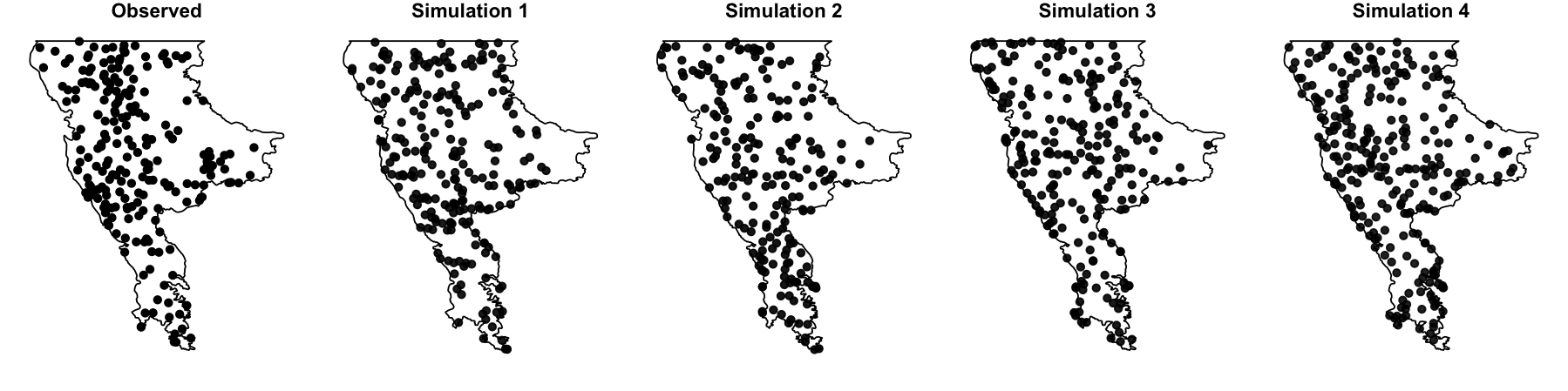}
    \caption{Observed and simulated wildfire point patterns for the KM pyrome under the fitted model with intensity $U_t \, \lambda$.}
\end{figure}

\begin{figure}[H]
    \centering
    \includegraphics[width=0.8\textwidth]{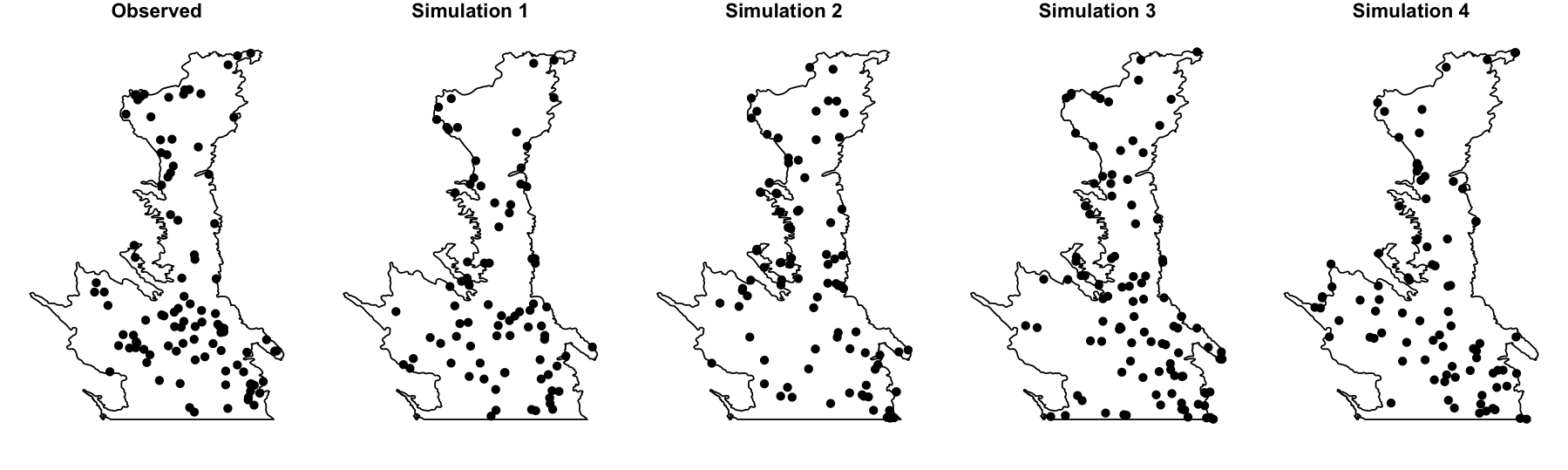}
    \caption{Observed and simulated wildfire point patterns for the NCCFACM pyrome under the fitted model with intensity $U_t \, \lambda$.}
\end{figure}

\begin{figure}[H]
    \centering
    \includegraphics[width=0.8\textwidth]{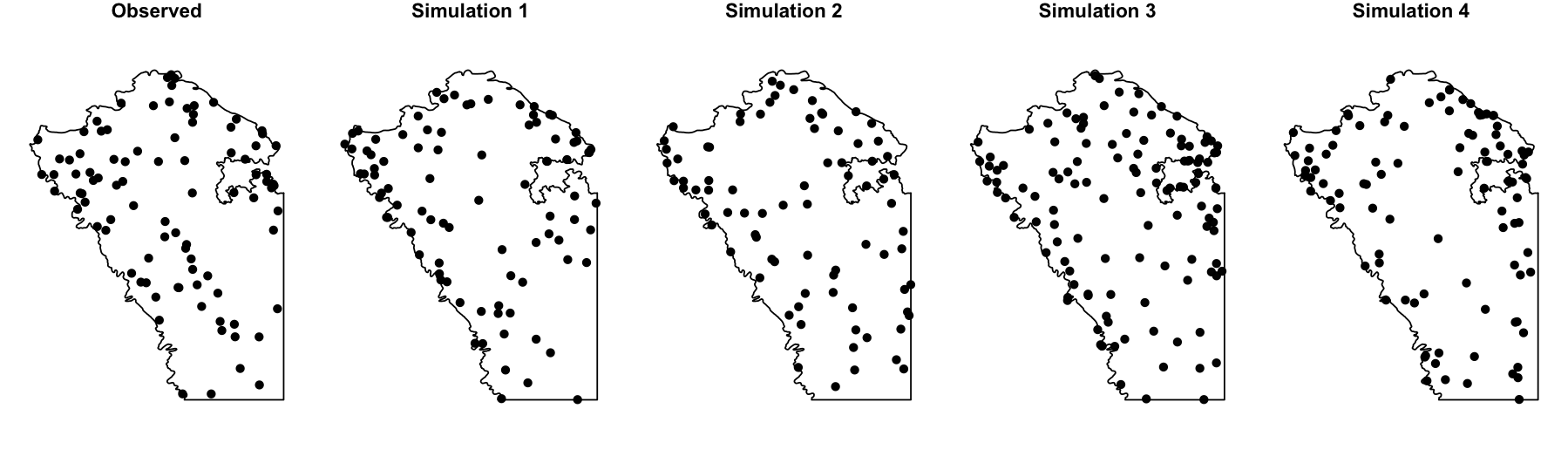}
    \caption{Observed and simulated wildfire point patterns for the NSN pyrome under the fitted model with intensity $U_t \, \lambda$.}
\end{figure}

\begin{figure}[H]
    \centering
    \includegraphics[width=0.8\textwidth]{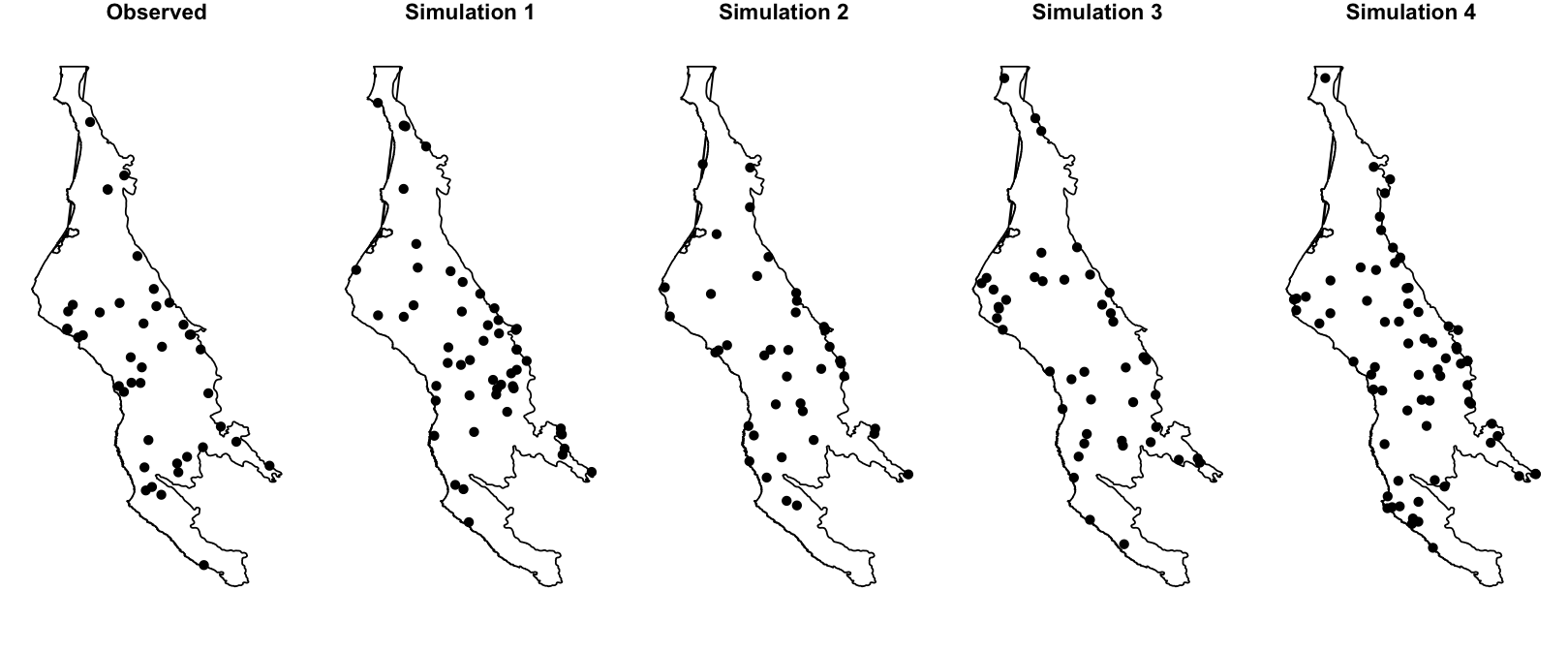}
    \caption{Observed and simulated wildfire point patterns for the ONCRR pyrome under the fitted model with intensity $U_t \, \lambda$.}
\end{figure}

\begin{figure}[H]
    \centering
    \includegraphics[width=0.8\textwidth]{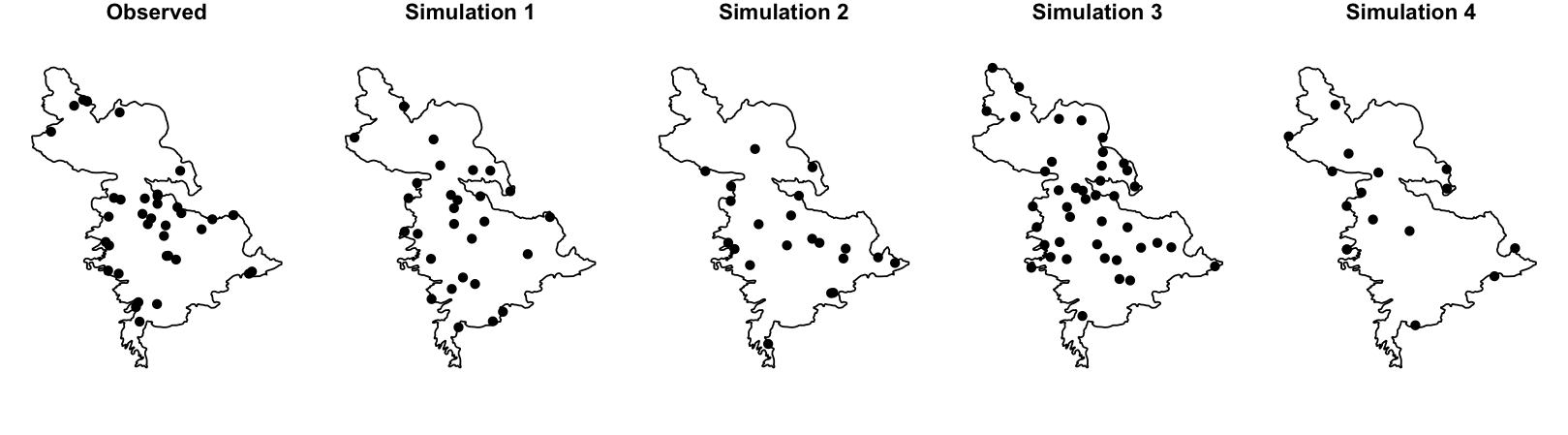}
    \caption{Observed and simulated wildfire point patterns for the SC pyrome under the fitted model with intensity $U_t \, \lambda$.}
\end{figure}

\begin{figure}[H]
    \centering
    \includegraphics[width=0.8\textwidth]{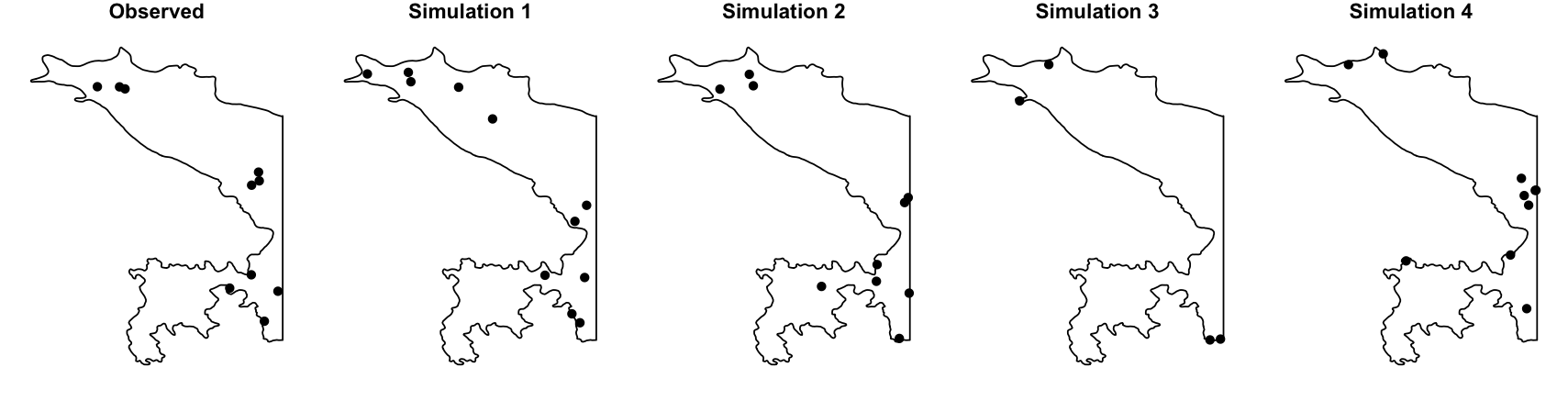}
    \caption{Observed and simulated wildfire point patterns for the SNIRHB pyrome under the fitted model with intensity $U_t \, \lambda$.}
\end{figure}

\subsection{Full Area Simulations}

\begin{figure}[H]
    \centering
    \includegraphics[width=\textwidth]{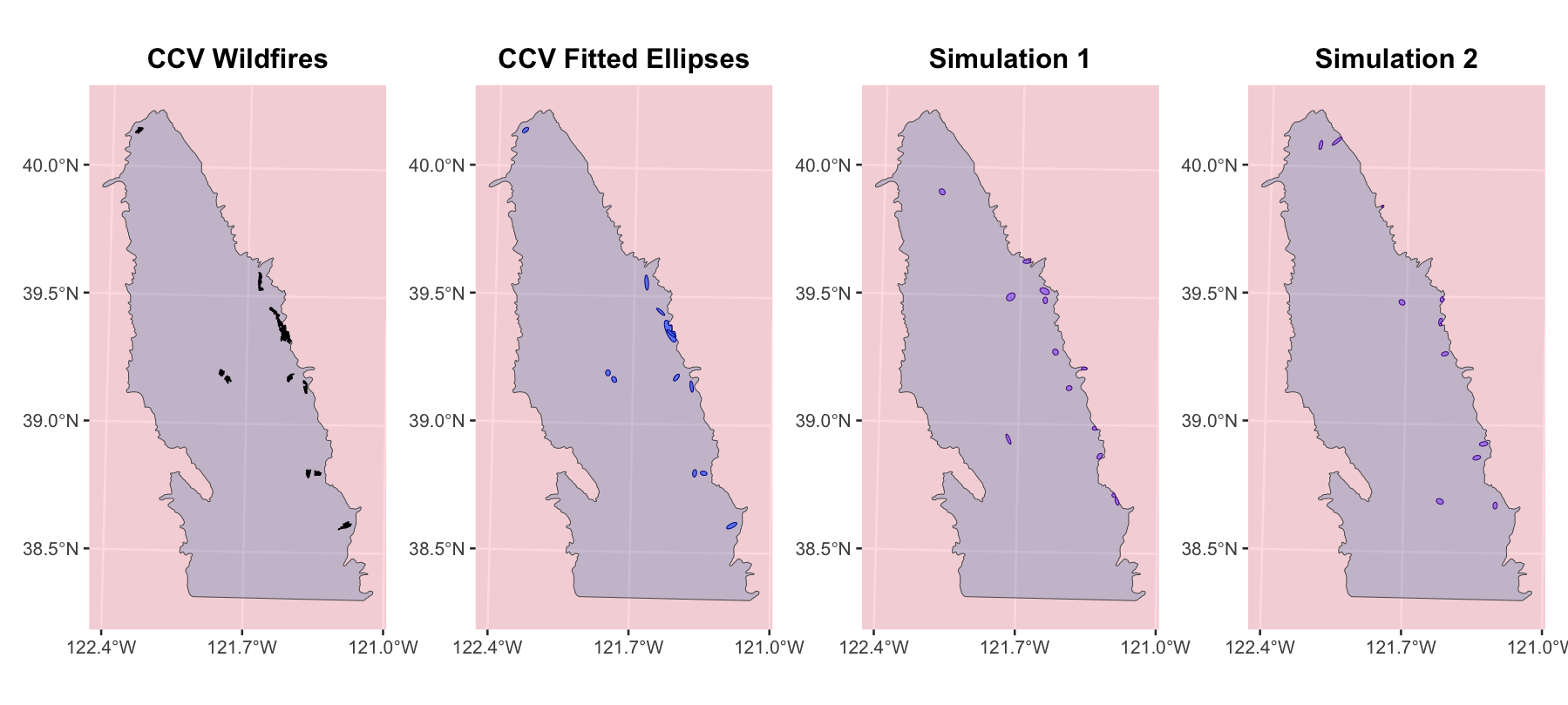}
    \caption{Observed and simulated wildfire perimeters in the CCV pyrome.}
\end{figure}

\begin{figure}[H]
    \centering
    \includegraphics[width=\textwidth]{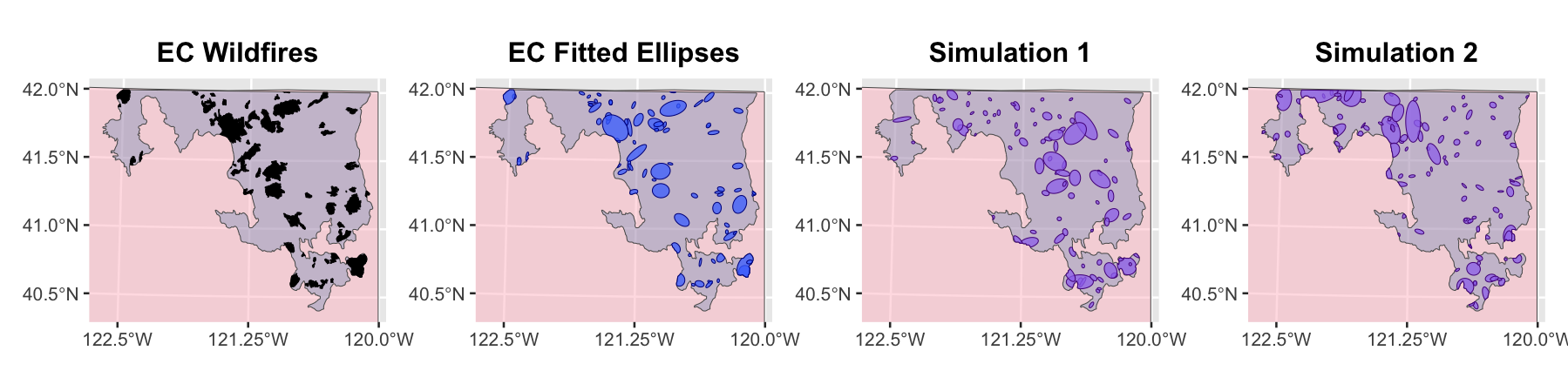}
    \caption{Observed and simulated wildfire perimeters in the EC pyrome.}
\end{figure}

\begin{figure}[H]
    \centering
    \includegraphics[width=\textwidth]{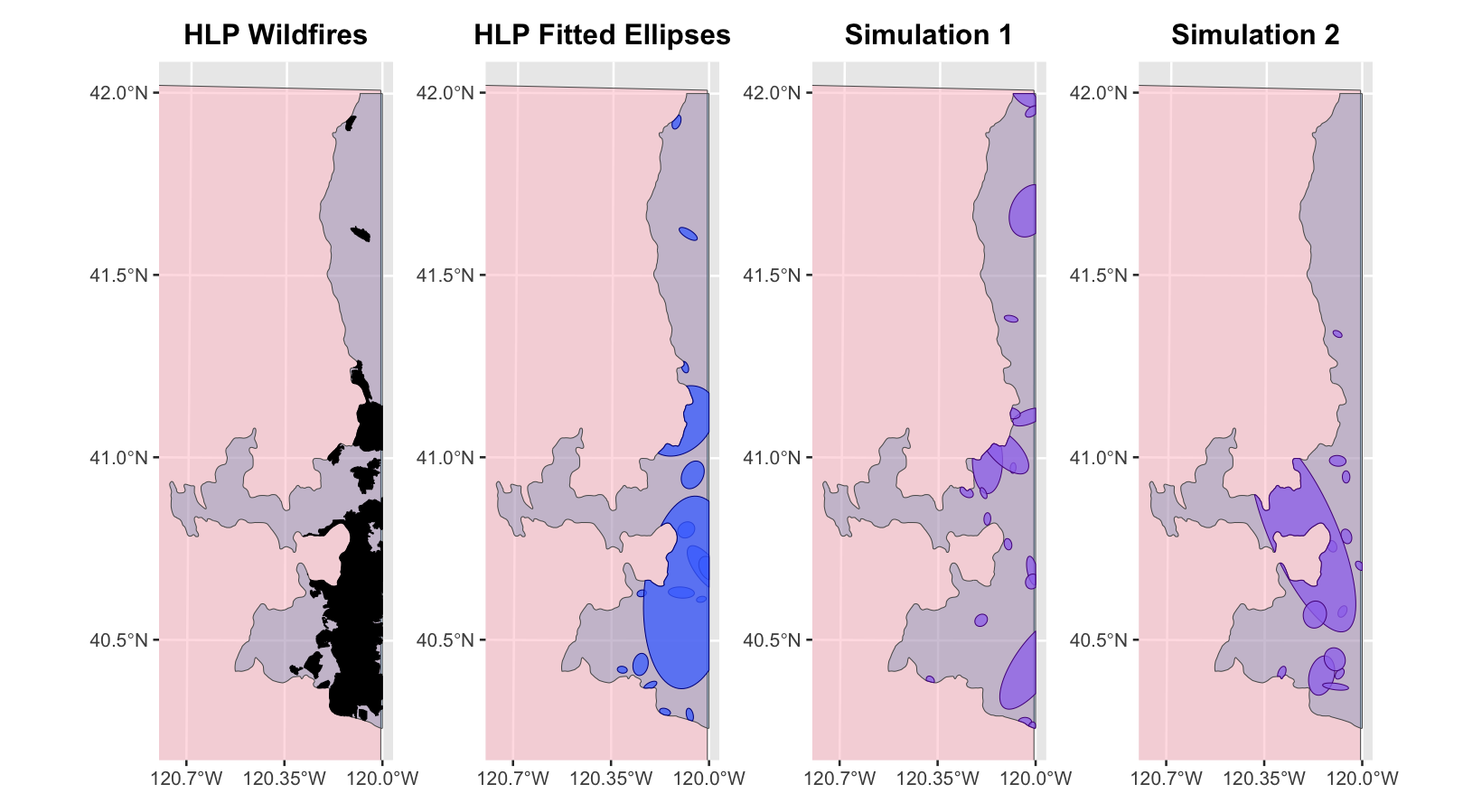}
    \caption{Observed and simulated wildfire perimeters in the HLP pyrome.}
\end{figure}

\begin{figure}[H]
    \centering
    \includegraphics[width=\textwidth]{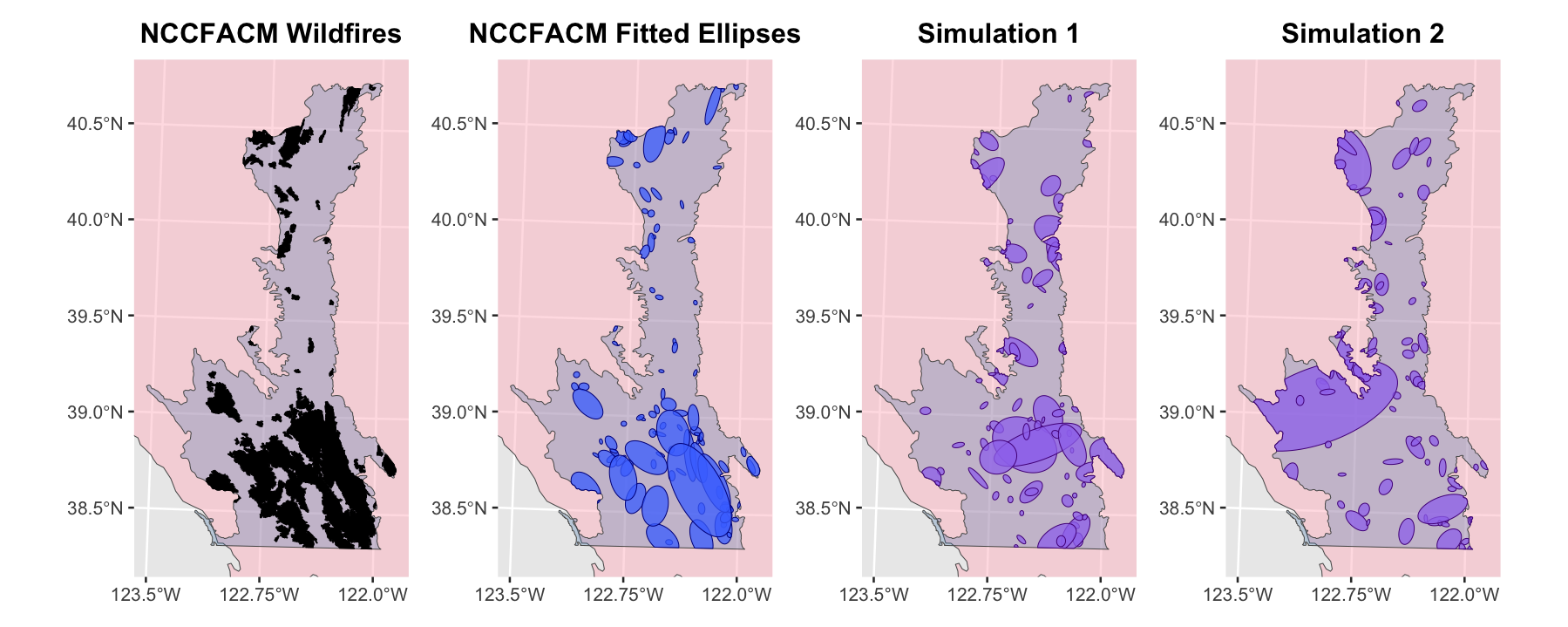}
    \caption{Observed and simulated wildfire perimeters in the NCCFACM pyrome.}
\end{figure}

\begin{figure}[H]
    \centering
    \includegraphics[width=\textwidth]{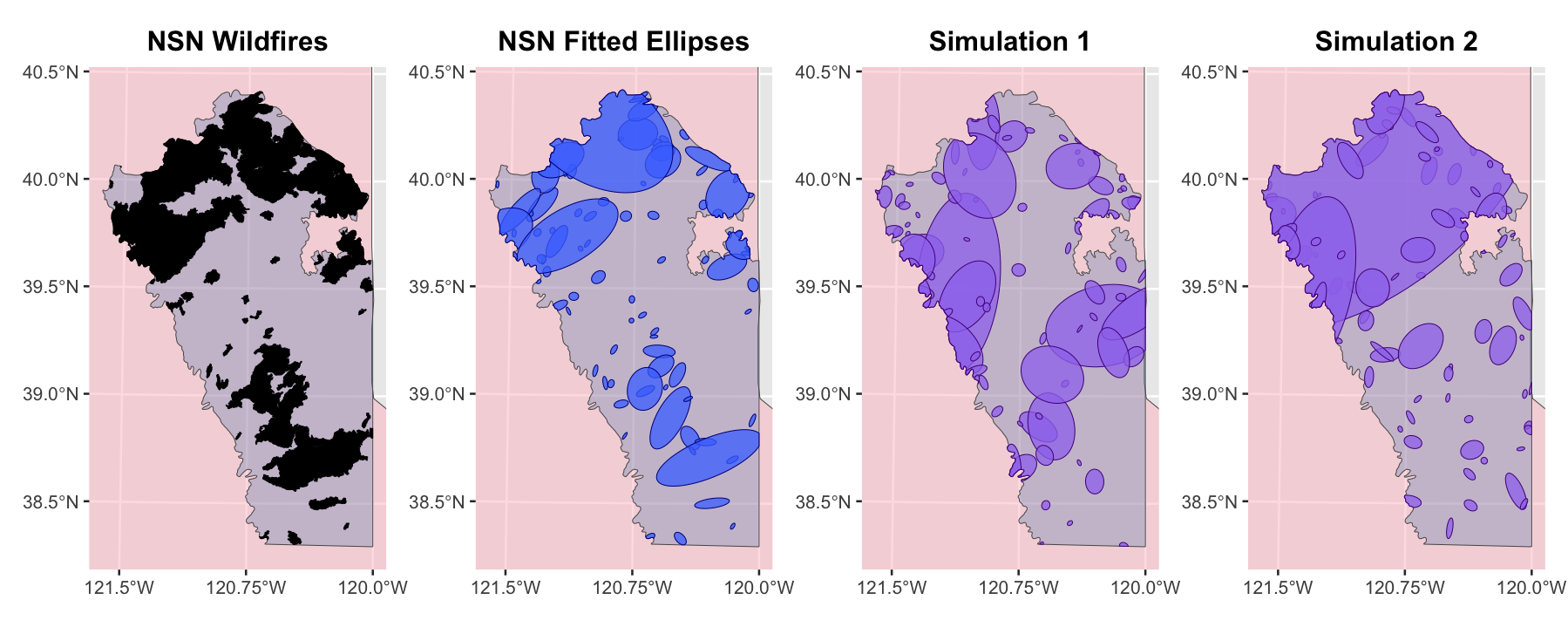}
    \caption{Observed and simulated wildfire perimeters in the NSN pyrome.}
\end{figure}

\begin{figure}[H]
    \centering
    \includegraphics[width=\textwidth]{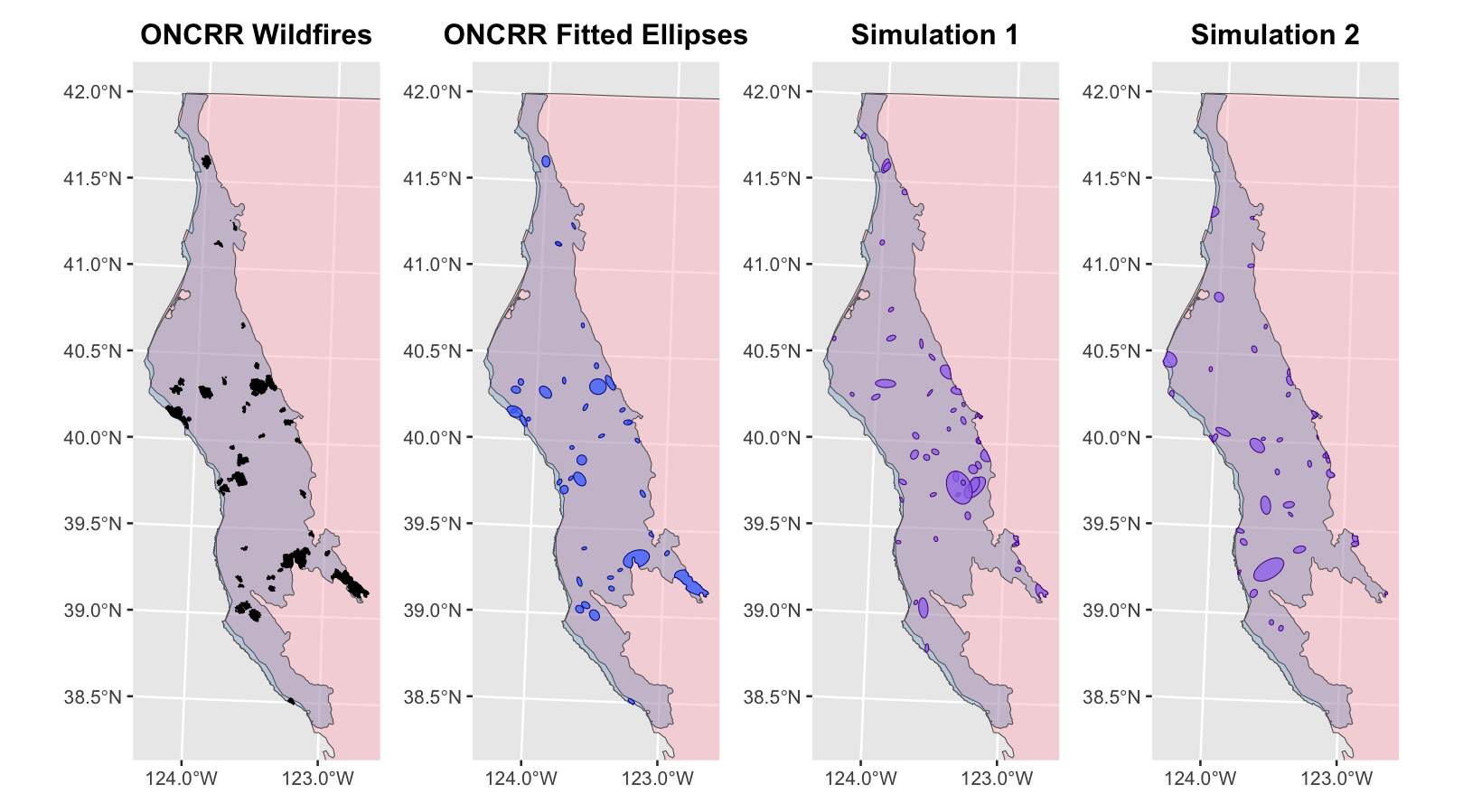}
    \caption{Observed and simulated wildfire perimeters in the ONCRR pyrome.}
\end{figure}

\begin{figure}[H]
    \centering
    \includegraphics[width=\textwidth]{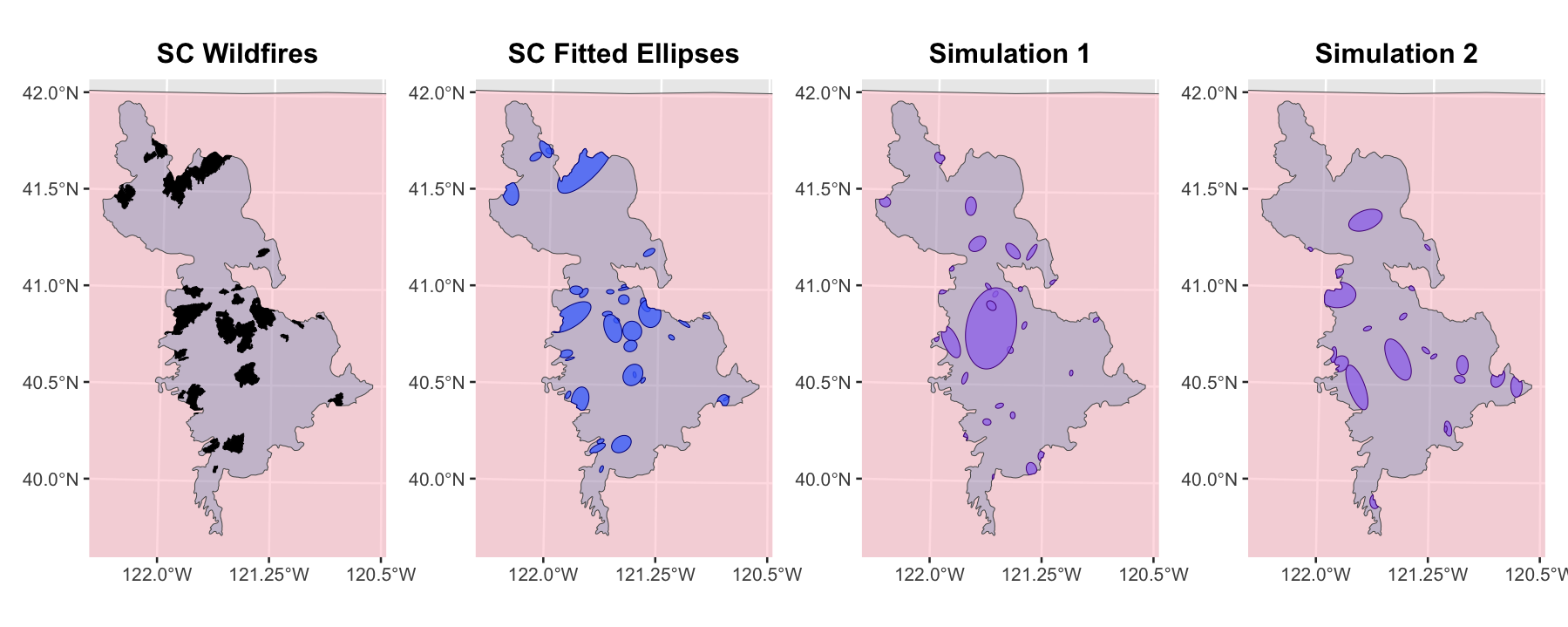}
    \caption{Observed and simulated wildfire perimeters in the SC pyrome.}
\end{figure}

\begin{figure}[H]
    \centering
    \includegraphics[width=\textwidth]{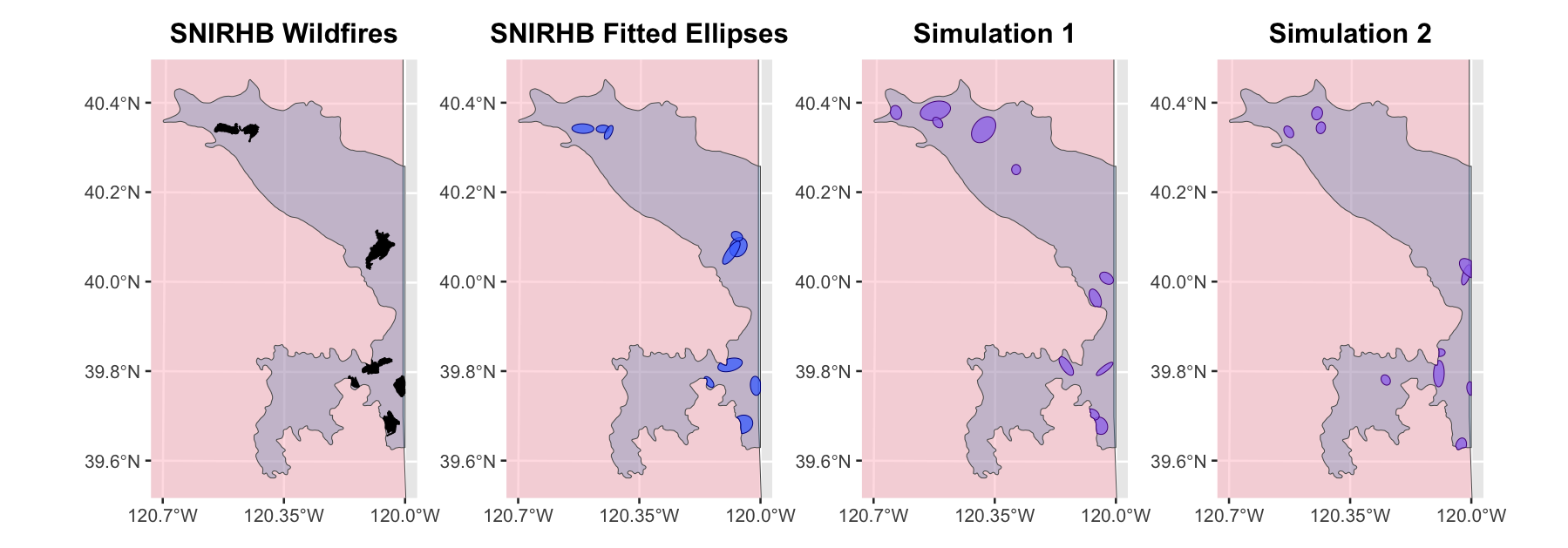}
    \caption{Observed and simulated wildfire perimeters in the SNIRHB pyrome.}
\end{figure}

\subsection{Burned Area Validation}

\begin{figure}[H]
    \centering
    \begin{subfigure}[t]{0.49\textwidth}
        \centering
        \includegraphics[width=\textwidth]{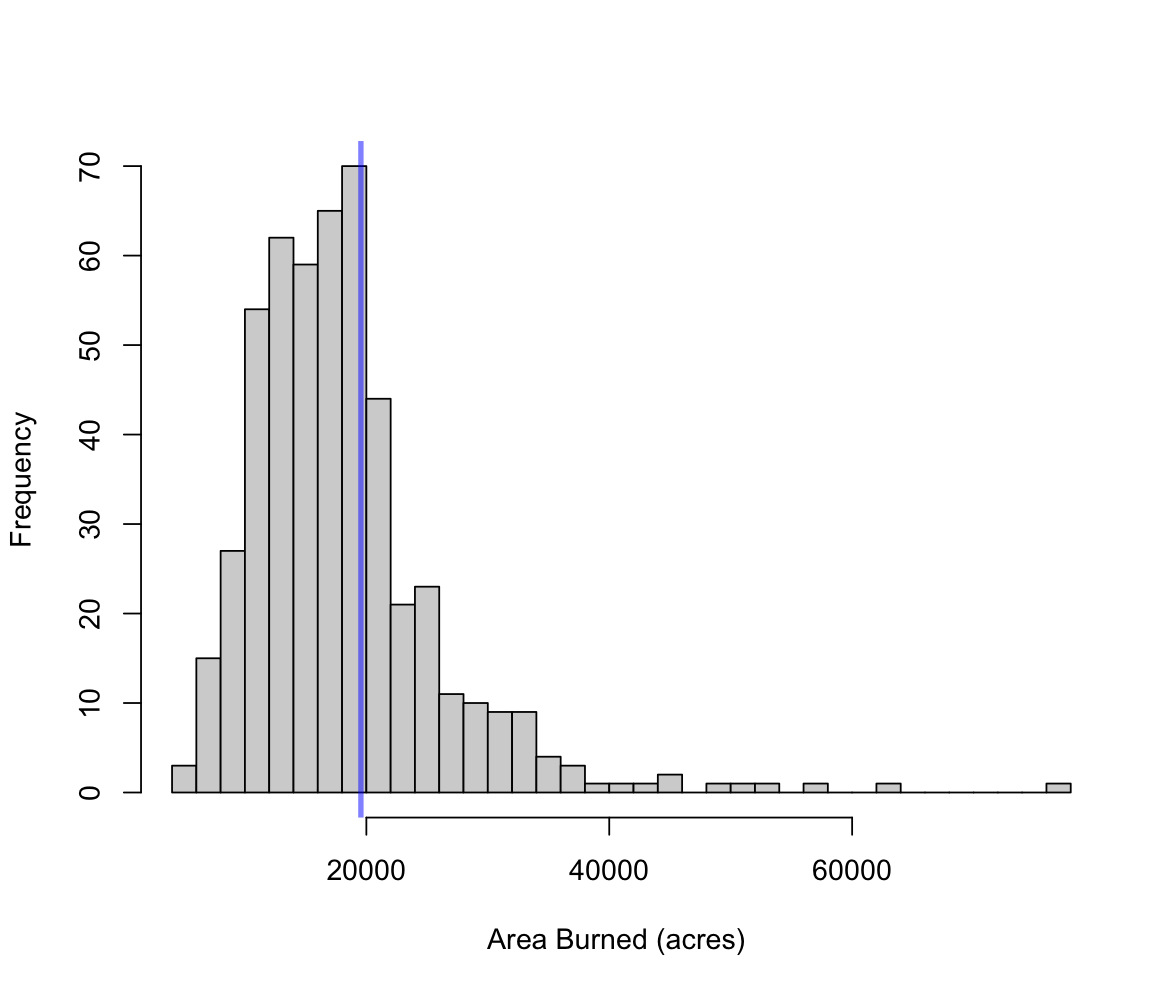}
        \caption{}
    \end{subfigure}
    \hfill
    \begin{subfigure}[t]{0.49\textwidth}
        \centering
        \includegraphics[width=\textwidth]{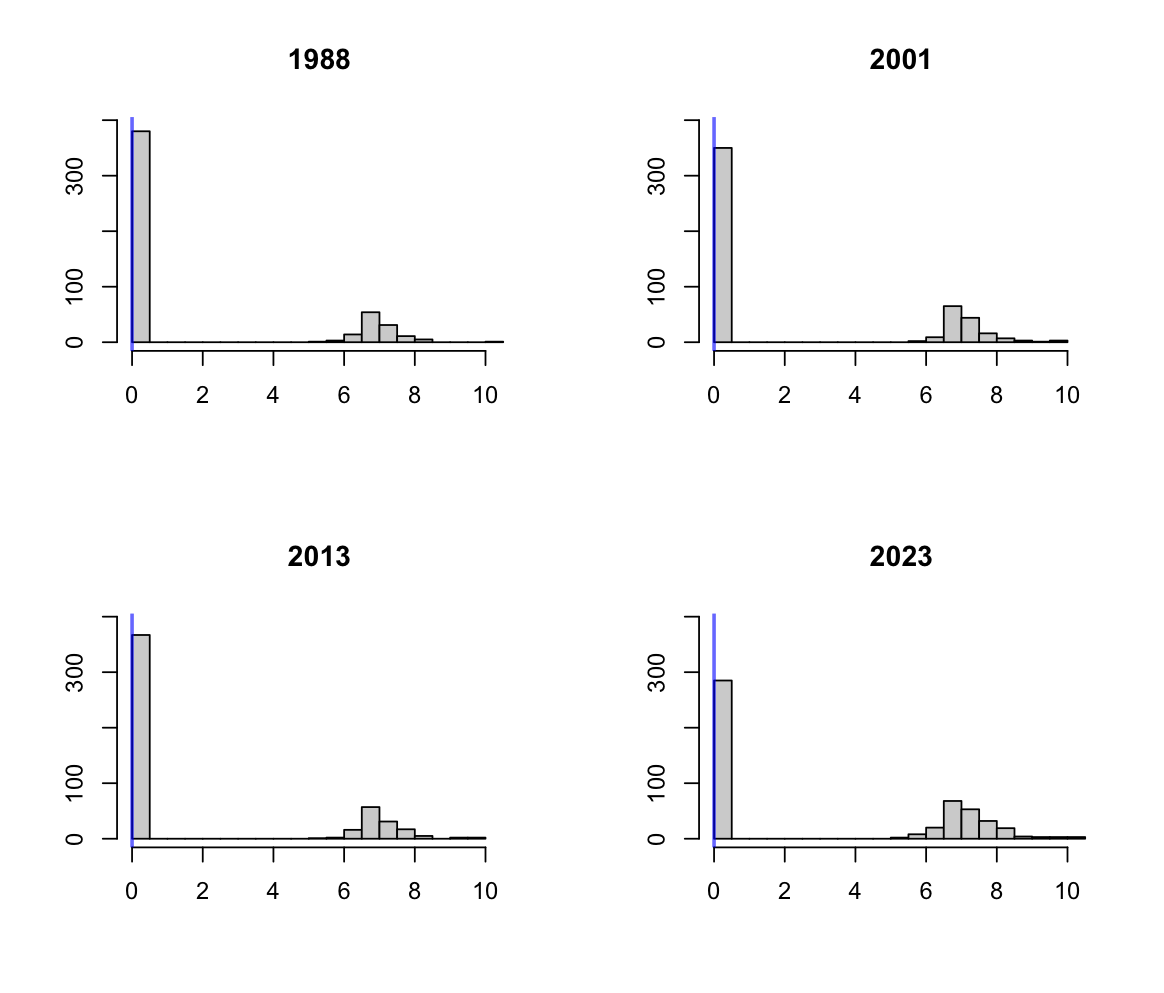}
        \caption{}
    \end{subfigure}
    \caption{Observed and simulated burned area in the CCV pyrome: (a) 1984--2023 cumulative burned area; (b) annual log burned area.}
\end{figure}

\begin{figure}[H]
    \centering
    \begin{subfigure}[t]{0.49\textwidth}
        \centering
        \includegraphics[width=\textwidth]{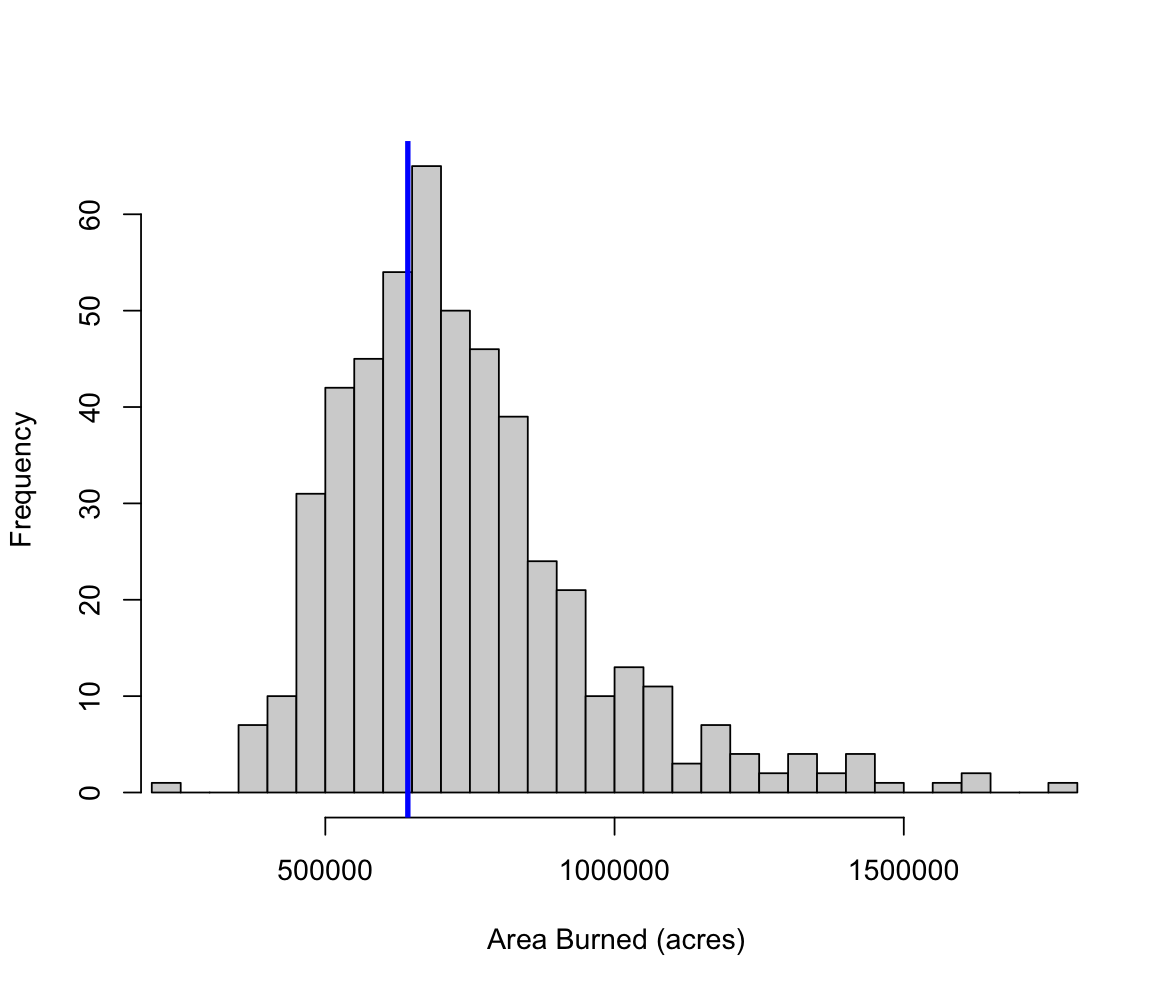}
        \caption{}
    \end{subfigure}
    \hfill
    \begin{subfigure}[t]{0.49\textwidth}
        \centering
        \includegraphics[width=\textwidth]{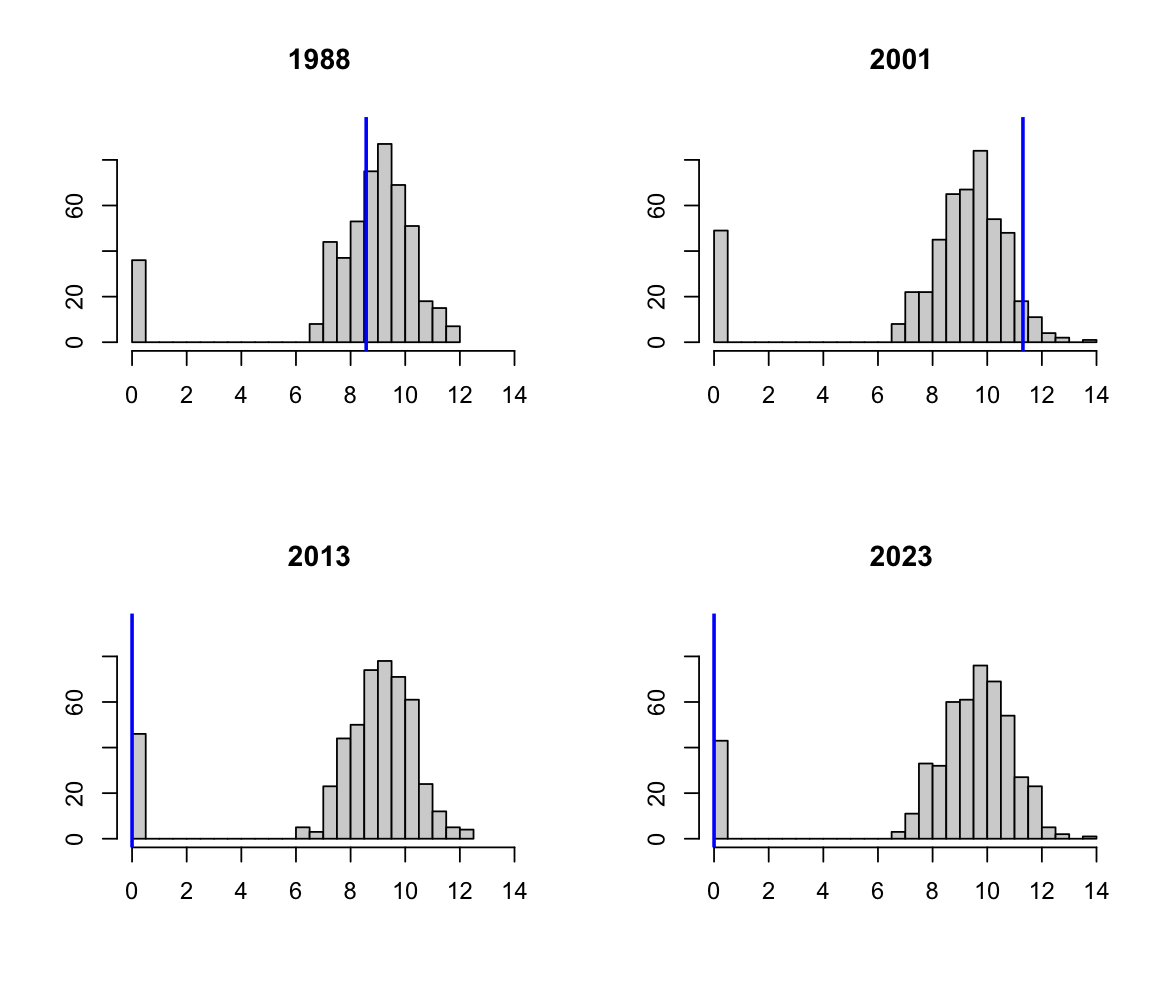}
        \caption{}
    \end{subfigure}
    \caption{Observed and simulated burned area in the EC pyrome: (a) 1984--2023 cumulative burned area; (b) annual log burned area.}
\end{figure}

\begin{figure}[H]
    \centering
    \begin{subfigure}[t]{0.49\textwidth}
        \centering
        \includegraphics[width=\textwidth]{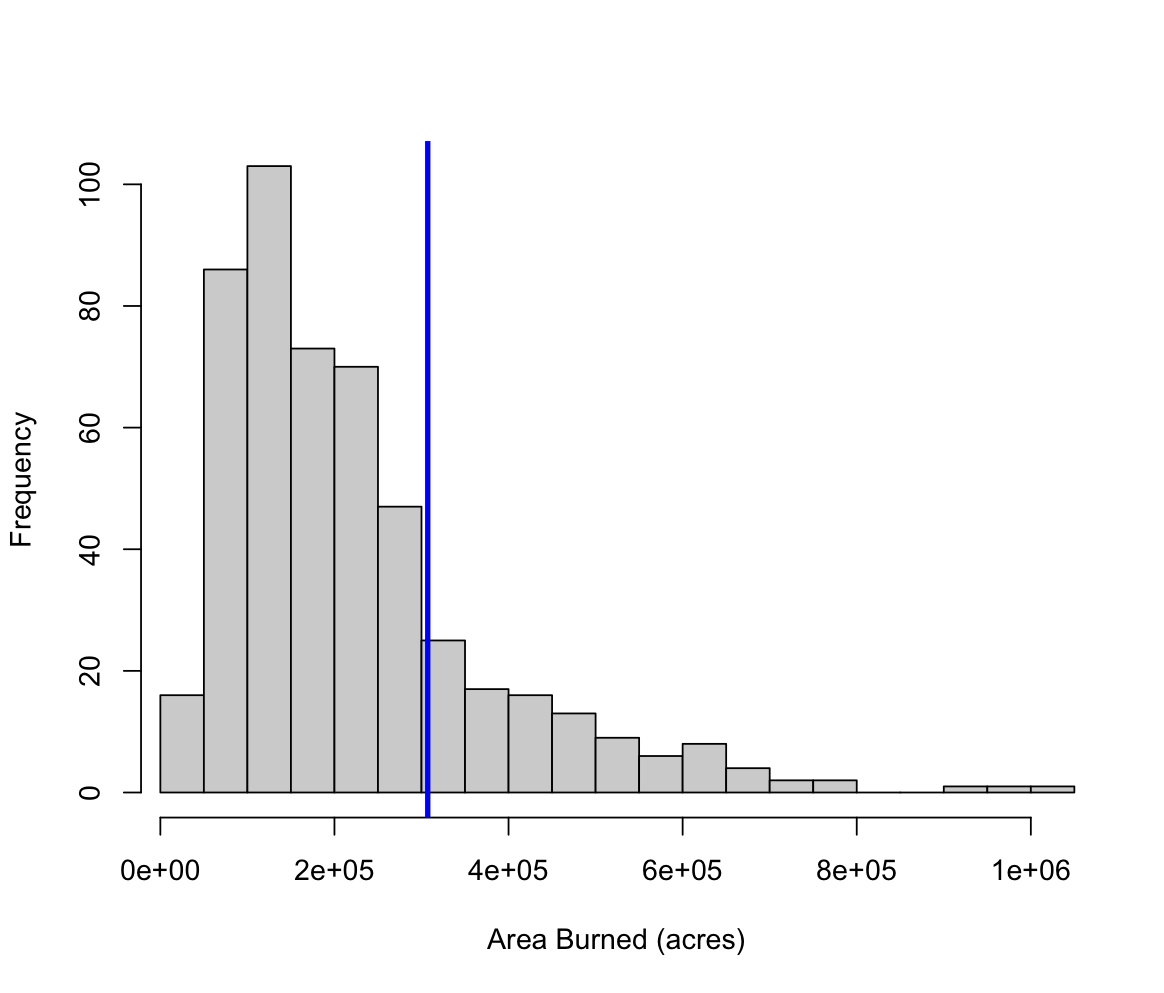}
        \caption{}
    \end{subfigure}
    \hfill
    \begin{subfigure}[t]{0.49\textwidth}
        \centering
        \includegraphics[width=\textwidth]{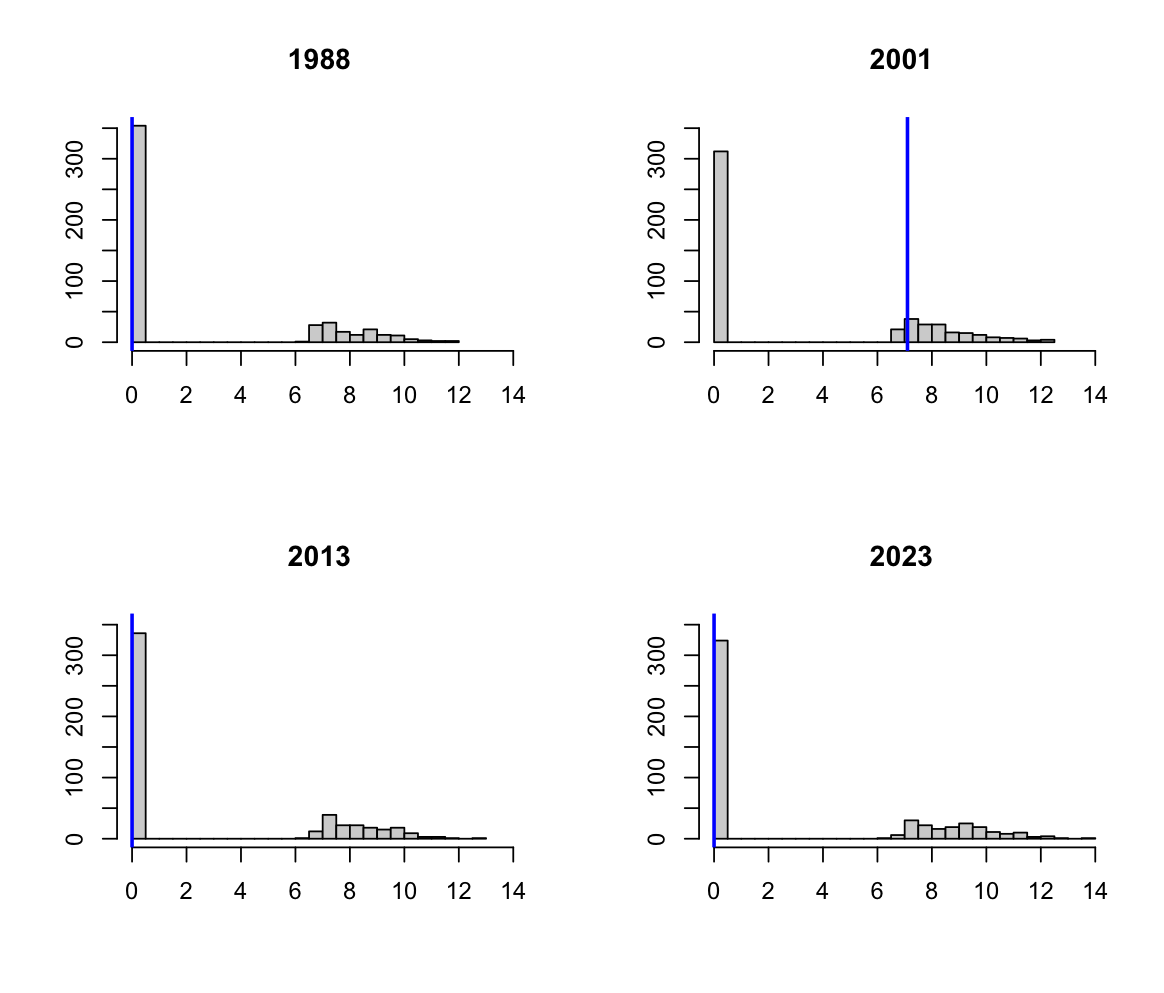}
        \caption{}
    \end{subfigure}
    \caption{Observed and simulated burned area in the HLP pyrome: (a) 1984--2023 cumulative burned area; (b) annual log burned area.}
\end{figure}

\begin{figure}[H]
    \centering
    \begin{subfigure}[t]{0.49\textwidth}
        \centering
        \includegraphics[width=\textwidth]{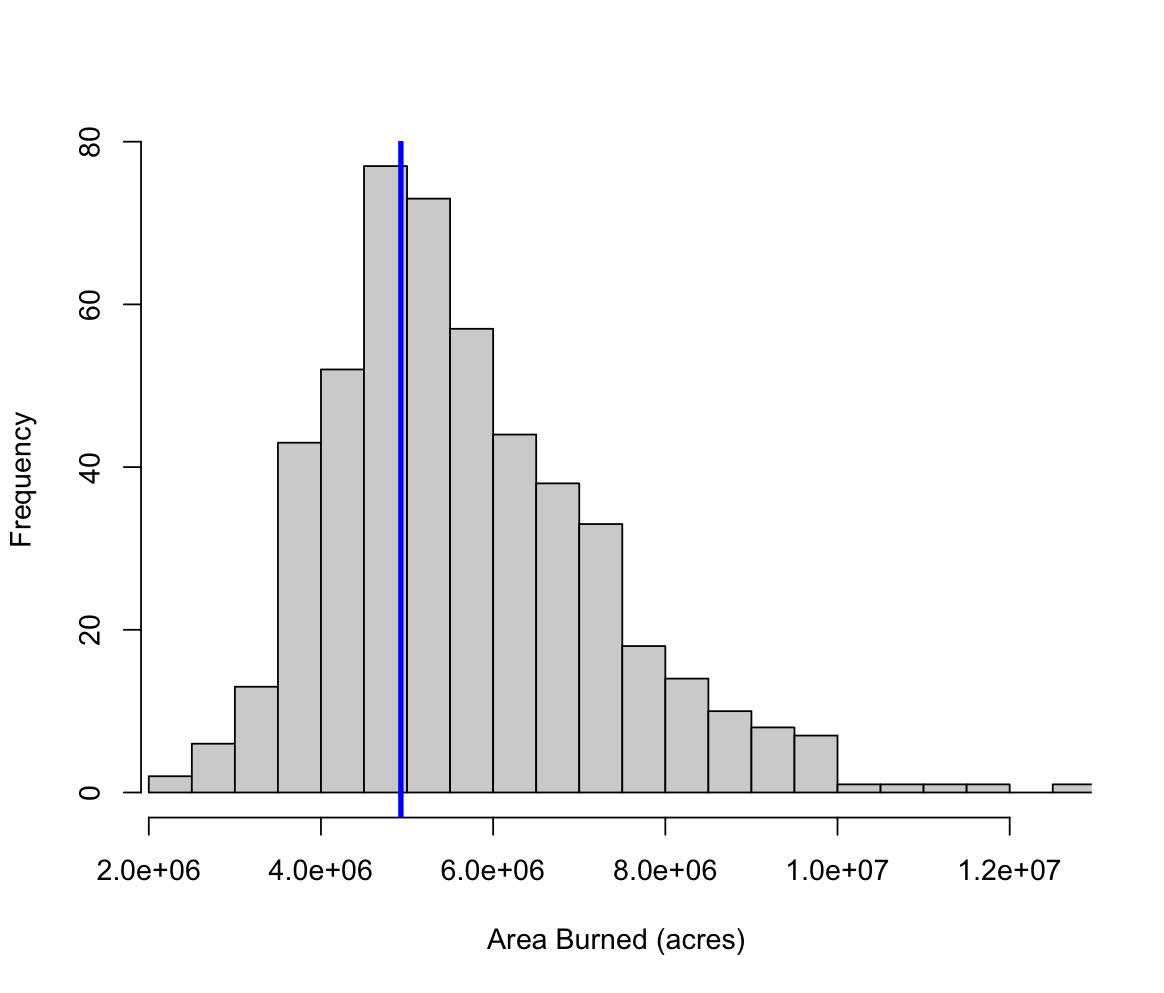}
        \caption{}
    \end{subfigure}
    \hfill
    \begin{subfigure}[t]{0.49\textwidth}
        \centering
        \includegraphics[width=\textwidth]{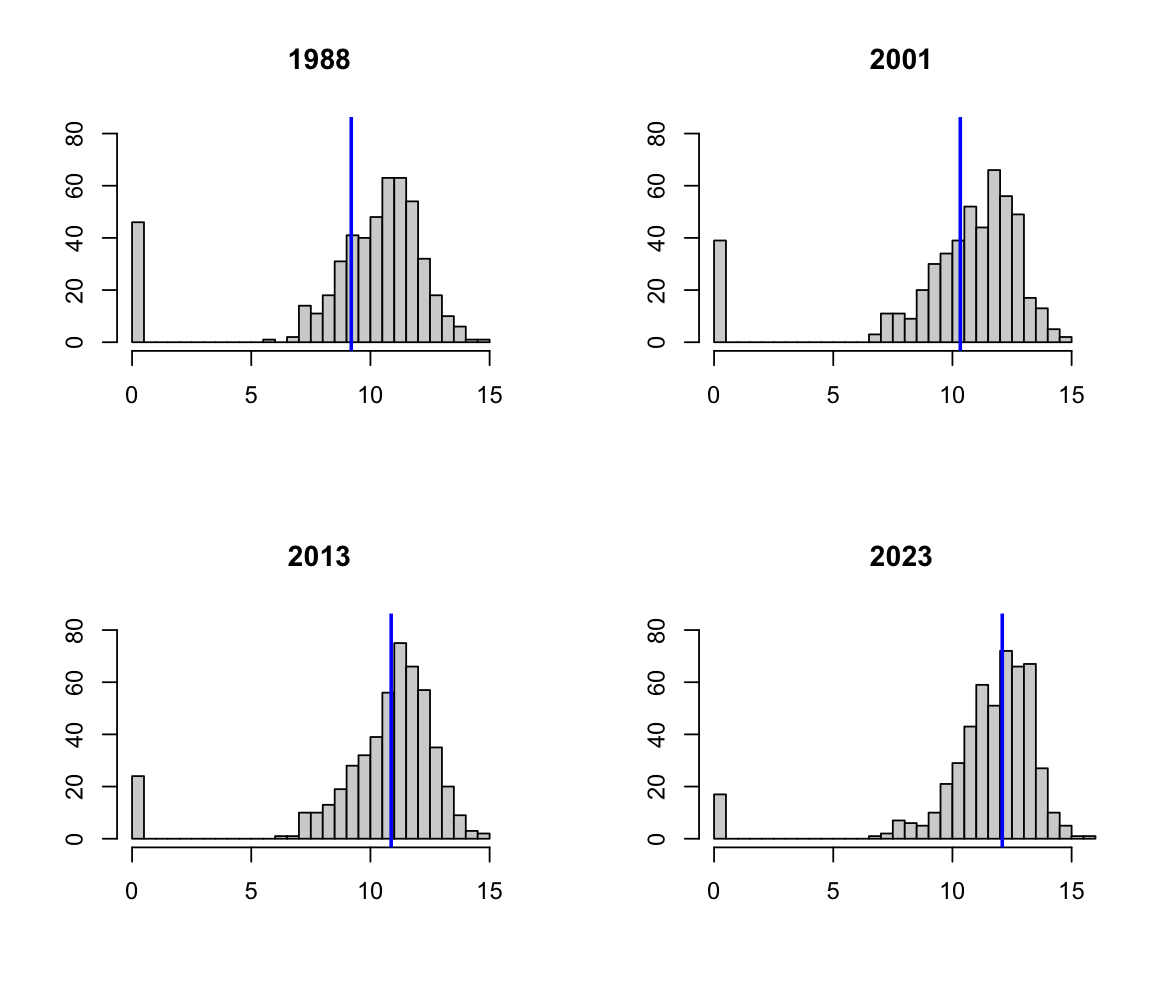}
        \caption{}
    \end{subfigure}
    \caption{Observed and simulated burned area in the KM pyrome: (a) 1984--2023 cumulative burned area; (b) annual log burned area.}
\end{figure}

\begin{figure}[H]
    \centering
    \begin{subfigure}[t]{0.49\textwidth}
        \centering
        \includegraphics[width=\textwidth]{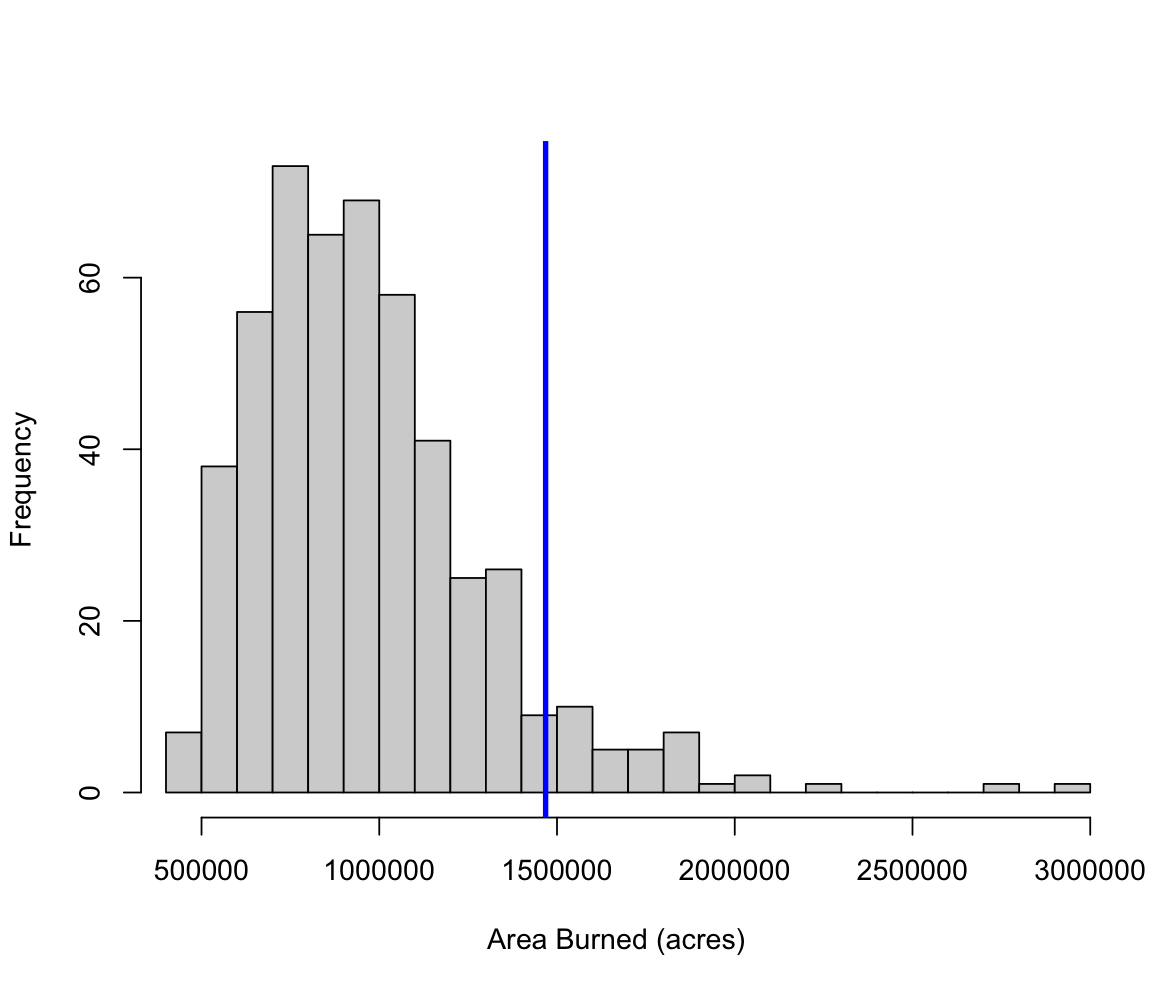}
        \caption{}
    \end{subfigure}
    \hfill
    \begin{subfigure}[t]{0.49\textwidth}
        \centering
        \includegraphics[width=\textwidth]{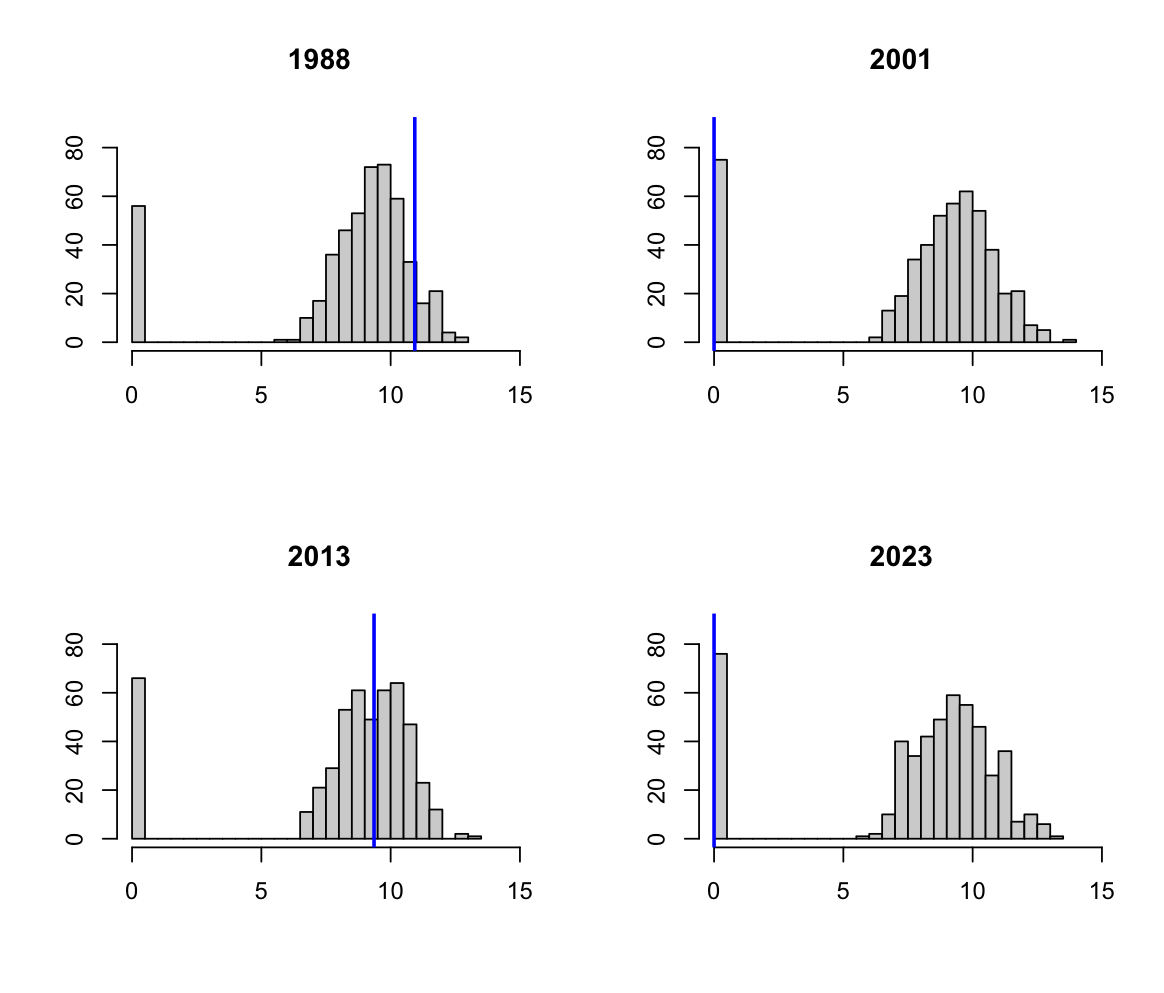}
        \caption{}
    \end{subfigure}
    \caption{Observed and simulated burned area in the NCCFACM pyrome: (a) 1984--2023 cumulative burned area; (b) annual log burned area.}
\end{figure}

\begin{figure}[H]
    \centering
    \begin{subfigure}[t]{0.49\textwidth}
        \centering
        \includegraphics[width=\textwidth]{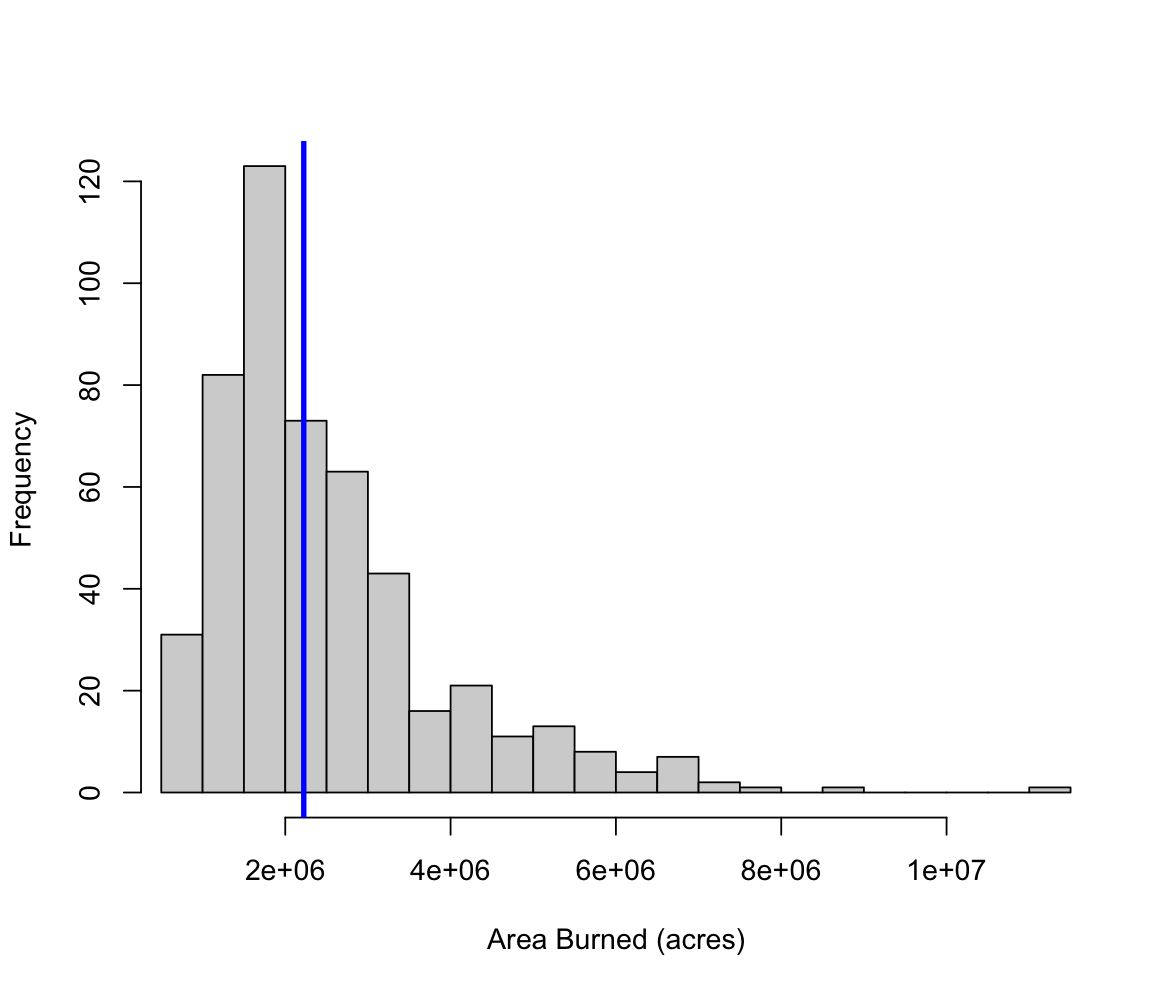}
        \caption{}
    \end{subfigure}
    \hfill
    \begin{subfigure}[t]{0.49\textwidth}
        \centering
        \includegraphics[width=\textwidth]{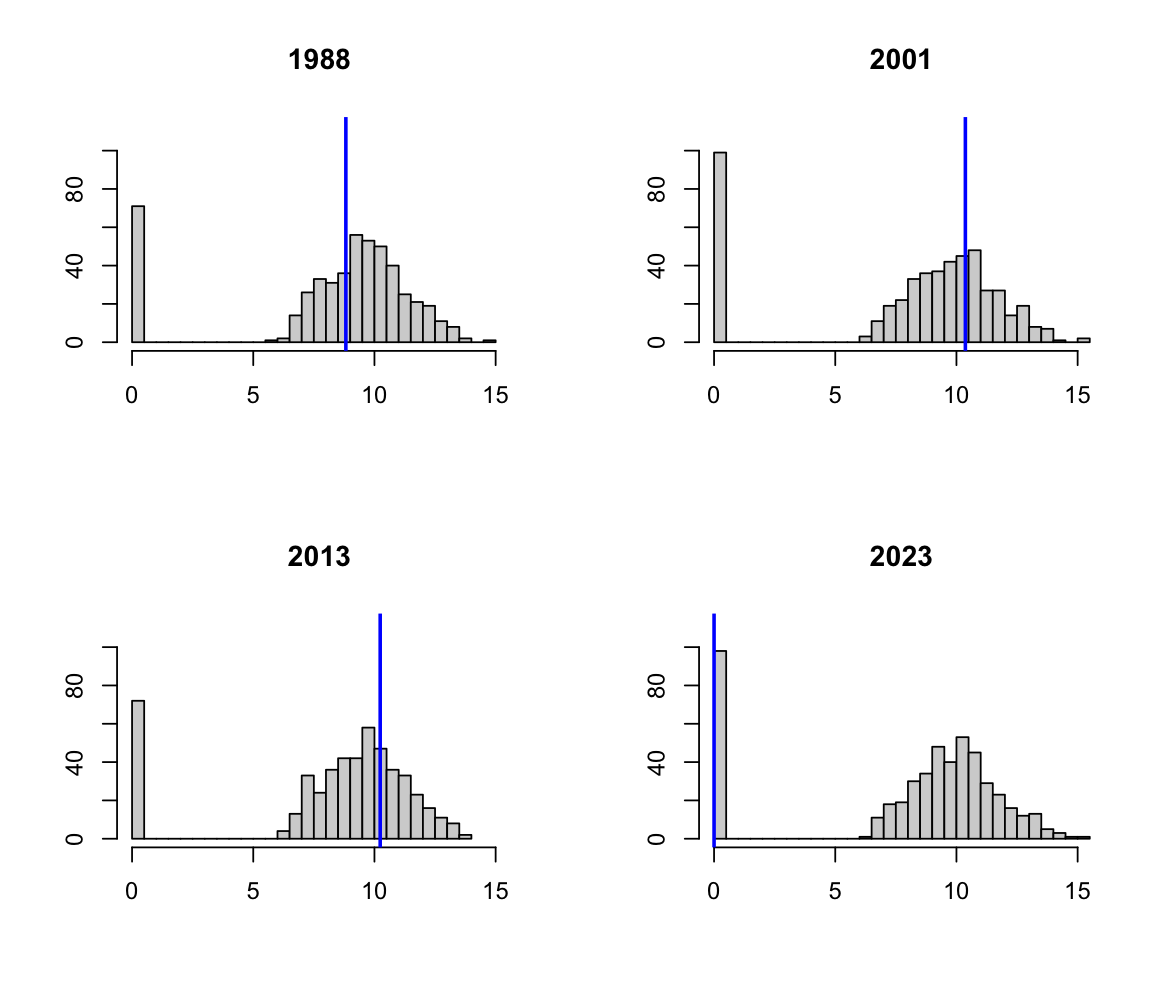}
        \caption{}
    \end{subfigure}
    \caption{Observed and simulated burned area in the NSN pyrome: (a) 1984--2023 cumulative burned area; (b) annual log burned area.}
\end{figure}

\begin{figure}[H]
    \centering
    \begin{subfigure}[t]{0.49\textwidth}
        \centering
        \includegraphics[width=\textwidth]{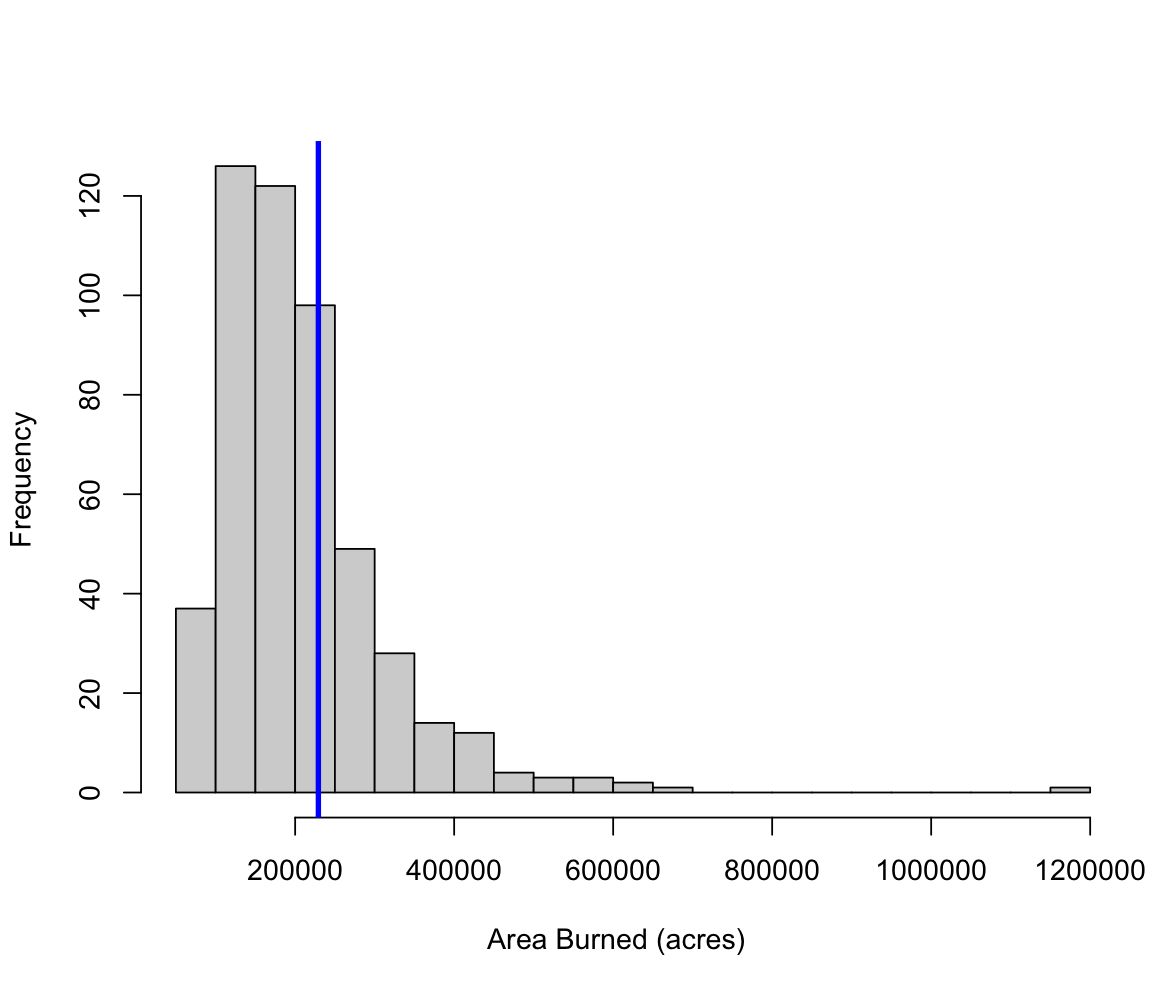}
        \caption{}
    \end{subfigure}
    \hfill
    \begin{subfigure}[t]{0.49\textwidth}
        \centering
        \includegraphics[width=\textwidth]{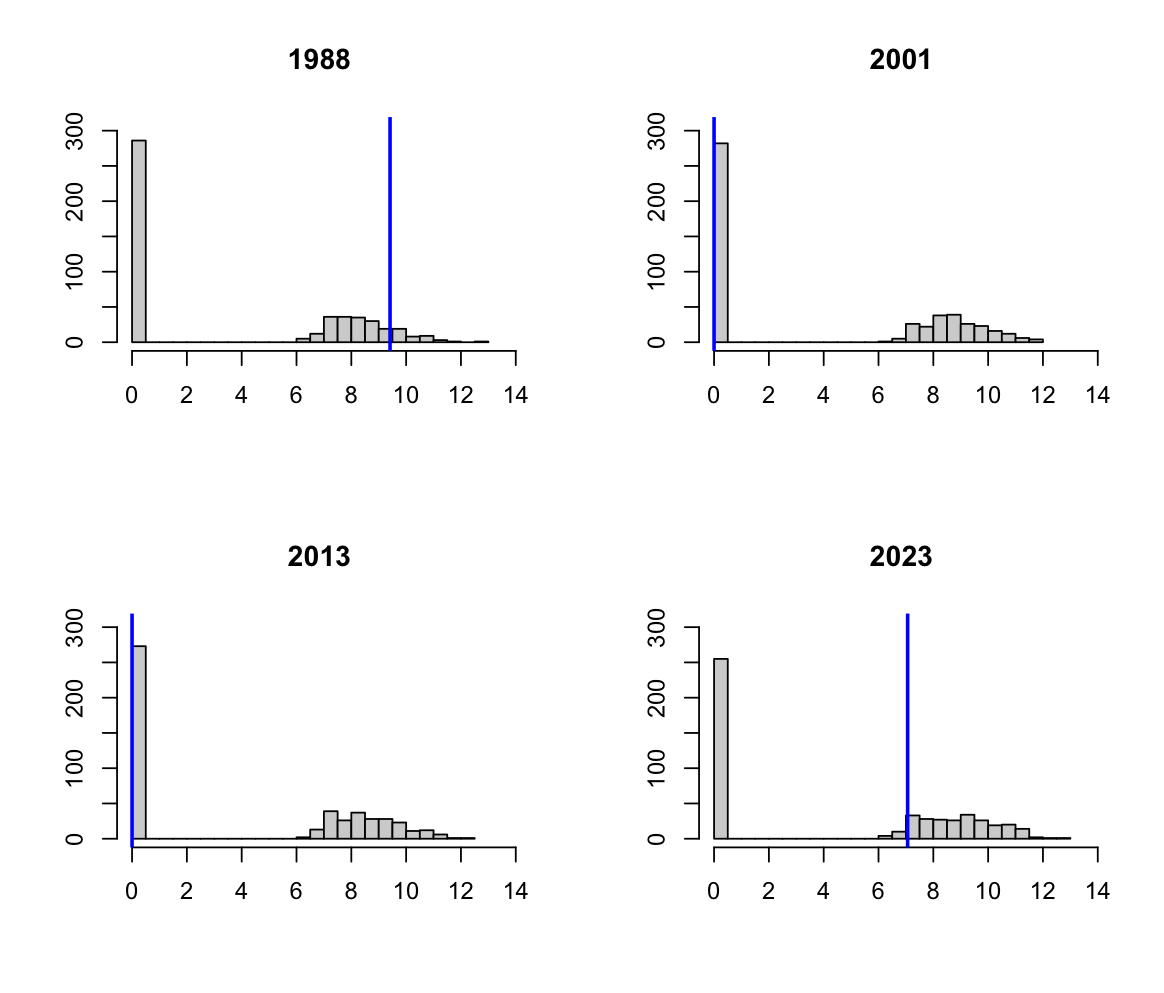}
        \caption{}
    \end{subfigure}
    \caption{Observed and simulated burned area in the ONCRR pyrome: (a) 1984--2023 cumulative burned area; (b) annual log burned area.}
\end{figure}

\begin{figure}[H]
    \centering
    \begin{subfigure}[t]{0.49\textwidth}
        \centering
        \includegraphics[width=\textwidth]{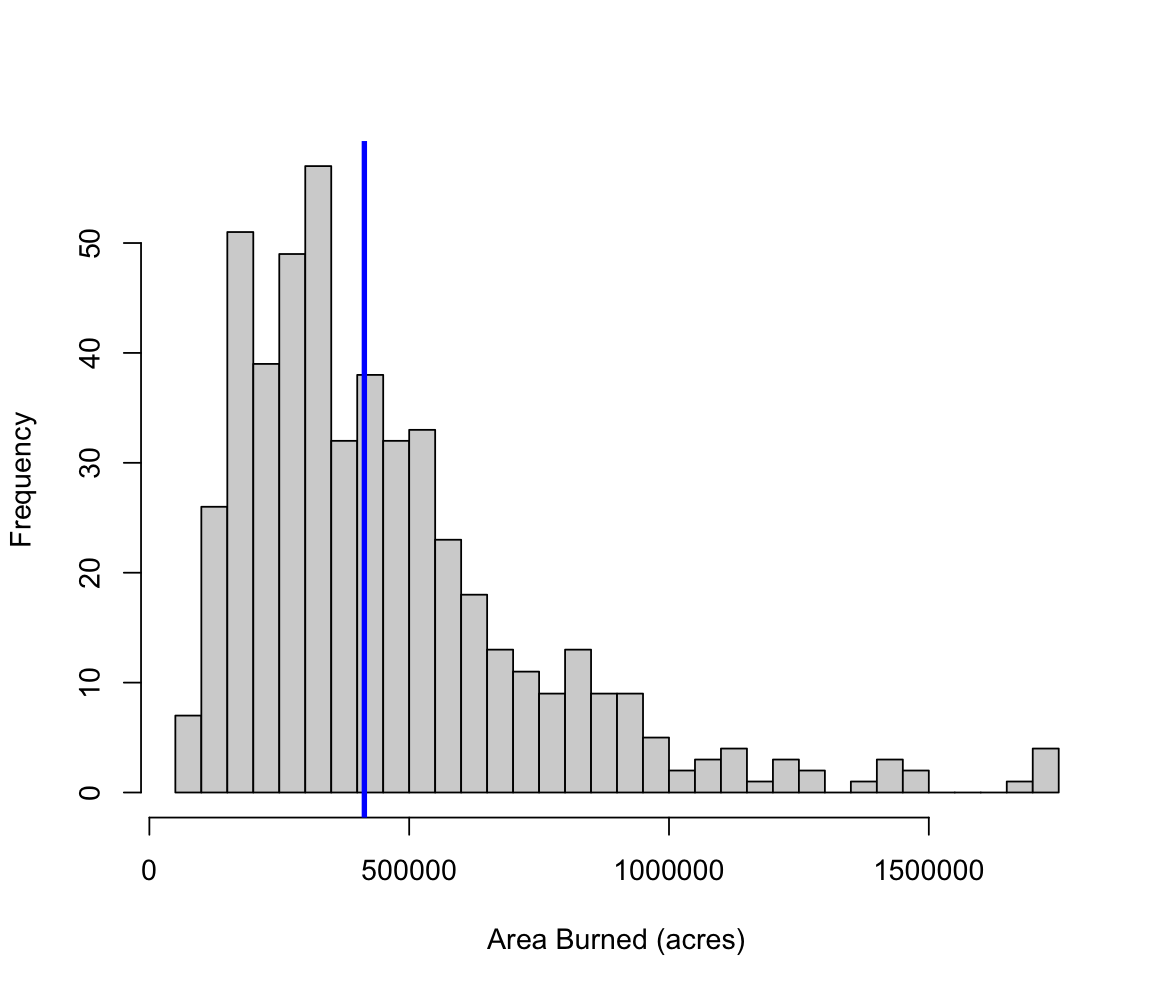}
        \caption{}
    \end{subfigure}
    \hfill
    \begin{subfigure}[t]{0.49\textwidth}
        \centering
        \includegraphics[width=\textwidth]{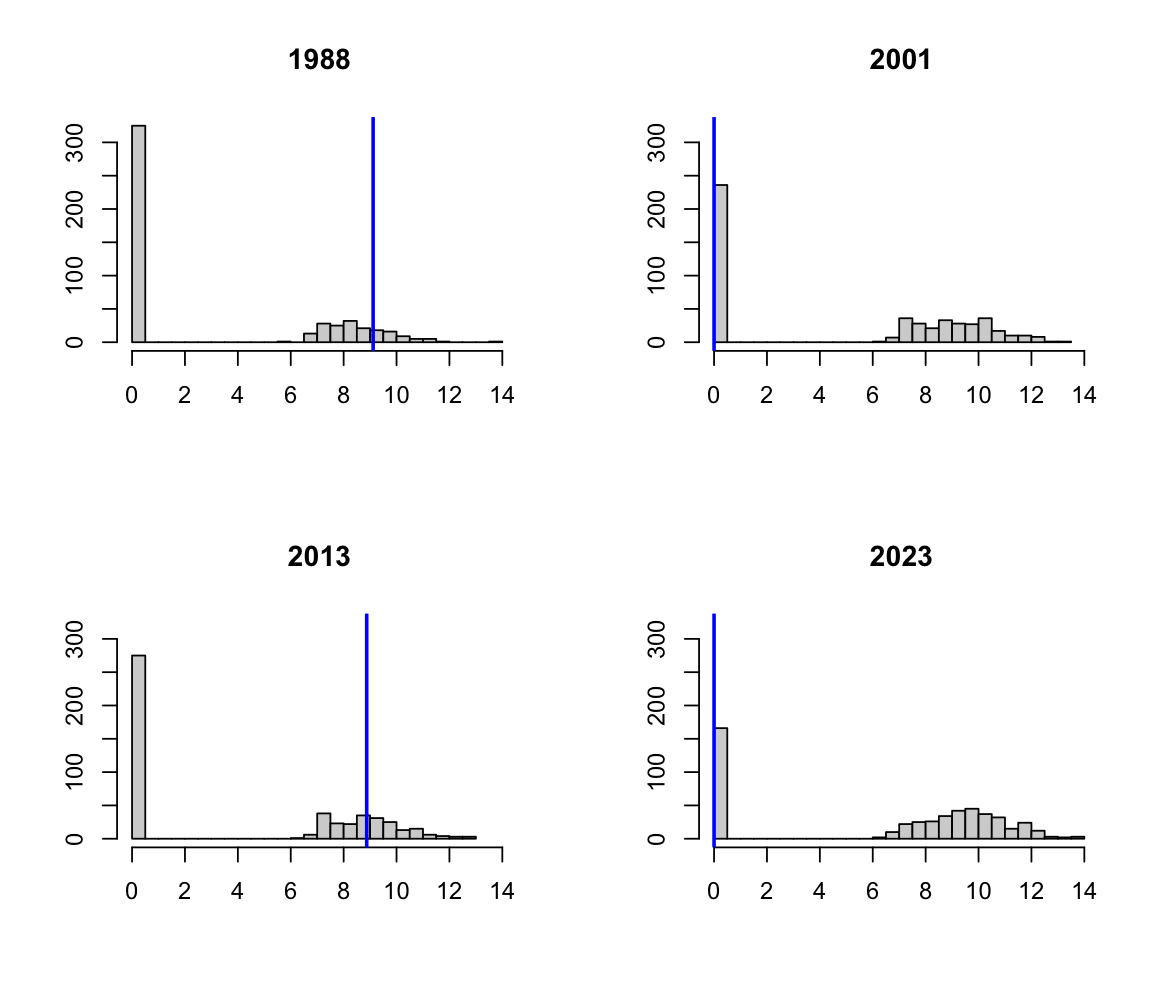}
        \caption{}
    \end{subfigure}
    \caption{Observed and simulated burned area in the SC pyrome: (a) 1984--2023 cumulative burned area; (b) annual log burned area.}
\end{figure}

\begin{figure}[H]
    \centering
    \begin{subfigure}[t]{0.49\textwidth}
        \centering
        \includegraphics[width=\textwidth]{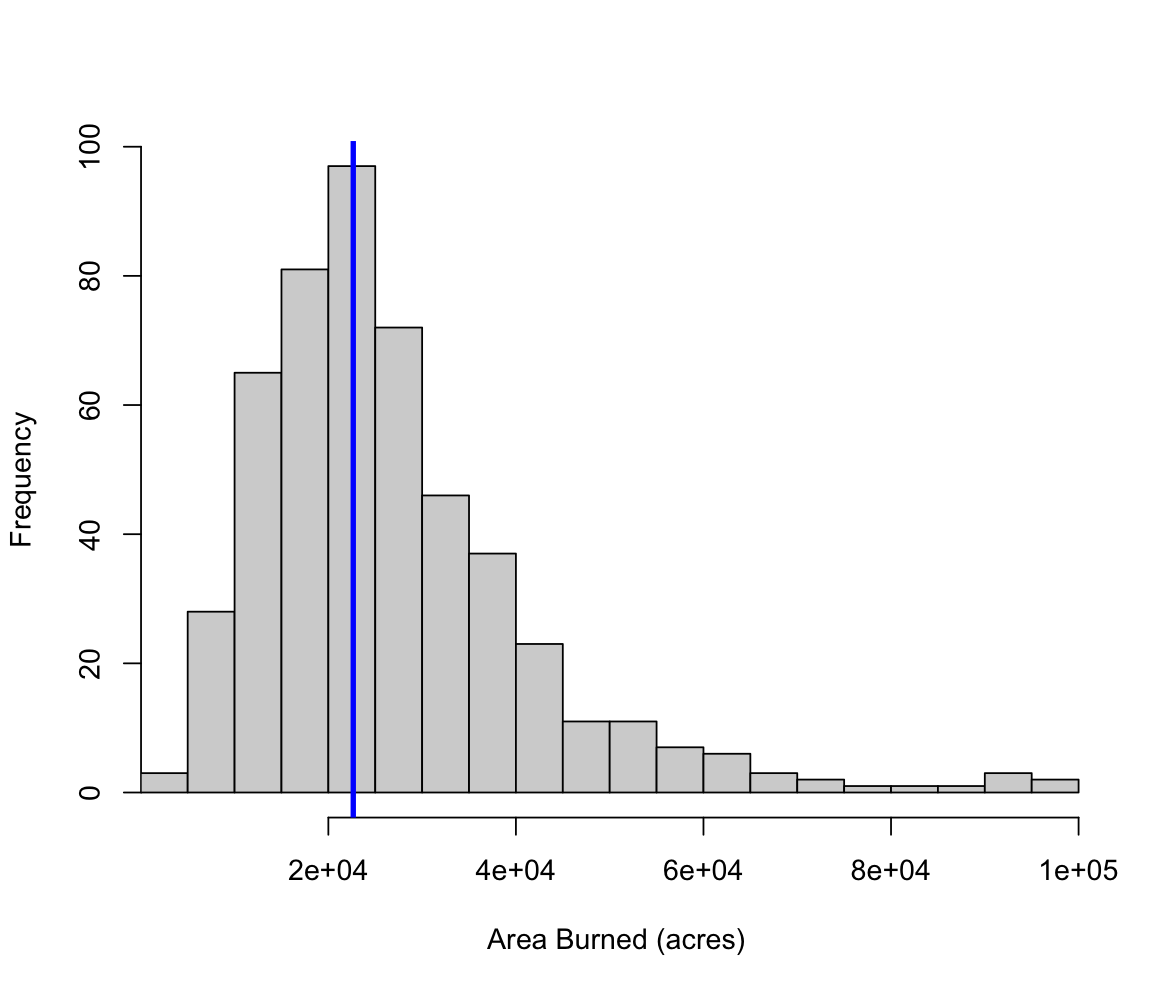}
        \caption{}
    \end{subfigure}
    \hfill
    \begin{subfigure}[t]{0.49\textwidth}
        \centering
        \includegraphics[width=\textwidth]{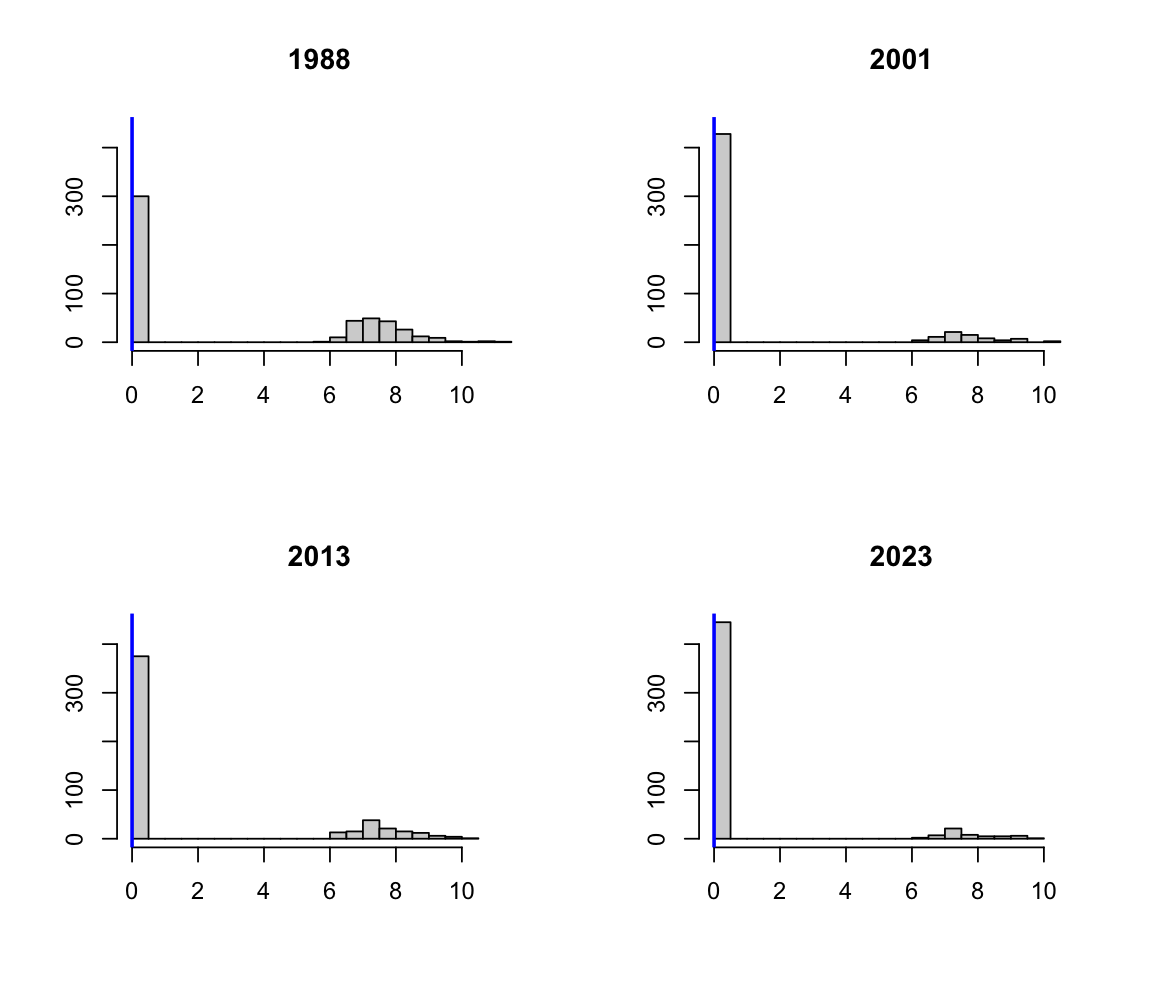}
        \caption{}
    \end{subfigure}
    \caption{Observed and simulated burned area in the SNIRHB pyrome: (a) 1984--2023 cumulative burned area; (b) annual log burned area.}
\end{figure}

